\documentclass[10pt, a4paper]{article}
\usepackage{hyperref}
\usepackage{epstopdf}
\usepackage{stackrel}
\usepackage{graphicx}
\usepackage[utf8]{inputenc}
\usepackage{amssymb}
\usepackage{amsthm}
\usepackage{amsmath, calc}
\usepackage[authoryear]{natbib}

\usepackage{xcolor}
\usepackage{subcaption}
\usepackage[boxruled,linesnumbered]{algorithm2e}

\makeatletter
\newcommand*{\centerfloat}{%
  \parindent \z@
  \leftskip \z@ \@plus 1fil \@minus \textwidth
  \rightskip\leftskip
  \parfillskip \z@skip}
\makeatother

\DeclareMathOperator*{\sign}{sign}
\DeclareMathOperator*{\RSS}{RSS}

\begin{document}


\title{Cross validating extensions of kernel, sparse or regular partial least squares regression models to censored data.}

\author{Fr\'ed\'eric Bertrand, \\ IRMA, CNRS UMR~7501, Labex IRMIA, \\ Universit\'e de Strasbourg, Strasbourg, France \\ fbertran@math.unistra.fr\\
Philippe Bastien, \\ L'Or\'eal Recherche, Aulnay, France \\ pbastien@rd.loreal.com\\
Myriam Maumy-Bertrand, \\ IRMA, CNRS UMR~7501, Labex IRMIA, \\ Universit\'e de Strasbourg, Strasbourg, France \\  mmaumy@math.unistra.fr}

\maketitle

\begin{abstract}
When cross-validating standard or extended Cox models, the commonly used criterion is the cross-validated partial loglikelihood using a naive or a van Houwelingen scheme -to make efficient use of the death times of the left out data in relation to the death times of all the data-. Quite astonishingly, we will show, using a strong simulation study involving three different data simulation algorithms, that these two cross-validation methods fail with the extensions, either straightforward or more involved ones, of partial least squares regression to the Cox model. 

This is quite an interesting result for at least two reasons. Firstly, several nice features of PLS based models, including regularization, interpretability of the components, missing data support, data visualization thanks to biplots of individuals and variables -and even parsimony for SPLS based models-, account for a common use of these extensions by statisticians who usually select their hyperparameters using cross-validation. Secondly, they are almost always featured in benchmarking studies to assess the performance of a new estimation technique used in a high dimensional context and often show poor statistical properties. 

We carried out a vast simulation study to evaluate more than a dozen of potential cross-validation criteria, either AUC or prediction error based. Several of them lead to the selection of a reasonable number of components. Using these newly found cross-validation criteria to fit extensions of partial least squares regression to the Cox model, we performed a benchmark reanalysis that showed enhanced performances of these techniques.

The R-package {\tt plsRcox} used in this article is available on the CRAN, \url{http://cran.r-project.org/web/packages/plsRcox/index.html}.
\end{abstract}





\section{Introduction}
Regular PLS regression is used to find the fundamental relations between two matrices ($X$ and $Y$), \textit{i.e.} a latent variable approach to modeling the covariance structures in these two spaces. A PLS regression model will try to find iteratively the multidimensional direction in the $X$ space that explains the maximum multidimensional variance direction in the $Y$ space. A key step in PLSR, is to select the right unknown number of these latent variables (called components) to use. PLS regression is particularly suited when the matrix of predictors has more variables than observations, and when there is multicollinearity among $X$ values. By contrast, standard regression will fail in these cases (unless it is regularized).

PLS has become an established tool in various experimental -including chemometric, networks, or systems biology- modeling, primarily because it is often possible to interpret the extracted components in terms of the underlying physical system -that is, to derive ``hard'' modeling information from the soft model: chemical components for NIR spectra, gene subnetwork for GRN or biological function for systems biology-.
As a consequence, choosing the right number of components is not only a major aim to avoid under or overfitting and ensure a relevant modeling or good predicting ability but also \textit{per se}.

A vast literature from the last decade is devoted to relating gene profiles and subject survival or time to cancer recurrence. Biomarker discovery from high-dimensional data, such as transcriptomic or SNP profiles, is a major challenge in the search for more precise diagnoses. The proportional hazard regression model suggested by \citealp{cox72} to study the relationship between the time to event and a set of covariates in the presence of censoring is the most commonly used model for the analysis of survival data. However, like multivariate regression, it supposes that more observations than variables, complete data, and not strongly correlated variables are available. In practice when dealing with high-dimensional data, these constraints are crippling.

In this article we deal with several PLS regression based extensions of the Cox model. These extensions share features praised by practionners, including regularization, interpretability of the components, missing data support, biplots of individuals and variables -and even parsimony for SPLS based models-, and allow to deal with highly correlated predictors or even rectangular datasets, which is especially relevant for high dimensional datasets.

\section{Models}
\subsection{{Modeling censored data}}
\subsubsection{{The Cox proportional hazards model}}

The model assumes the following hazard function for the occurrence of an event at time $t$ in the presence of censoring:
\begin{equation}
\lambda(t)=\lambda_0(t)\exp{(\beta'X)},
\end{equation}
where $\lambda_0(t)$ is an unspecified baseline hazard function, $\beta$ the vector of the regression coefficients and $X$ the matrix of prognosis factors which will be the gene expressions in the following. The event could be death or cancer relapse. Based on the available data, the Cox's partial likelihood can be written as:
\begin{equation}
PL(\beta)=\prod_{k\in D}{\frac{\exp{(\beta'x_k)}}{\sum_{j\in R_k}\exp{(\beta'x_j)}}},
\end{equation}
where $D$ is the set of indices of the events and $R_k$ denotes the set of indices of the individuals at risk at time $t_k$.

The goal is to find the coefficients $\hat \beta$  which maximize the log partial likelihood function 
\begin{equation}
l(\beta)=\log{PL(\beta)}.
\end{equation}
The vector $\hat \beta$ is the solution of the equation:  
\begin{equation}
u(\beta)=\frac{\partial l}{\partial \beta}=0
\end{equation}
with $u(\beta)$ the vector of efficient scores.

However, there is no explicit solution and the minimization is generally accomplished using the Newton-Raphson procedure. An estimate of the vector of $\beta$-parameters at the $(k+1)^\textup{th}$ cycle of the iterative procedure is:
\begin{equation}
\hat\beta_{k+1}=\hat\beta_{k}+I^{-1}(\hat\beta_{k})u(\hat\beta_{k})
\end{equation}
where $I(\beta)=-\frac{\partial^2 l}{\partial \beta\partial \beta'}$ is the observed information matrix. The process can be started by taking $\hat\beta_0=0$ and iterated up to convergence, \textit{i.e.} when the change in the log likelihood function is small enough. When the iterative procedure has converged, the variance-covariance matrix of the parameter estimates can be approximated by the inverse of the observed information matrix $I^{-1}(\hat\beta)$.

Note that when $p>n$, there is no unique $\hat\beta$ to maximize this log partial likelihood function. Even when $p\leqslant n$, covariates could be highly correlated and regularization may still be required in order to reduce the variances of the estimates and to improve the predictive performance.

\subsubsection{{
Deviance Residuals}}\label{matingaleresidualsdef}
For the Cox model with no time-dependent explanatory variables and at most one event per patient, the martingale residuals for the $i^\textup{th}$ subject with observation time $t_i$ and event status $\delta_i$, where $\delta_i=0$ if $t_i$ is a censored time, and $\delta_i=1$ otherwise is:
\begin{equation}
\hat M_i=\delta_i-\hat E_i=\delta_i-\hat\Delta_0(t_i)\exp{(\hat \beta'x_i)}
\end{equation}
with $\hat \Delta_0(t_i)$ the estimated cumulative hazard function at time $t_i$.

Martingale residuals are highly skewed. The deviance residuals $d_i$ are a normalized transform of the martingale residuals. For the Cox model, the deviance residuals (\citealp{coll94}) amount to the form:
\begin{equation}
d_i=\sign(\hat M_i)\cdot\left[2\left\{-\hat M_i-\delta_i\log{\left(\frac{\delta_i-\hat M_i}{\delta_i}\right)}\right\}\right]^{1/2}\cdot
\end{equation}
The $\sign{}$ function is to ensure that the deviance residuals have the same sign as the martingale residuals. Martingale residuals take values between $-\infty$ and $1$. The square root shrinks large negative martingale residuals, while the logarithmic transformation expands towards $+\infty$ martingale residuals that are close to $1$. As such, the deviance residuals are more symmetrically distributed around zero than the martingale residuals. The deviance residual is a measure of excess of death and can therefore be interpreted as a measure of hazard. Moreover, Segal showed that the expression to be minimized in step 3 of the Cox-Lasso procedure of Tibshirani can be approximated, in a first order Taylor-series approximation sense, by the deviance residual sum of squares:
\begin{equation}
(z-X\beta)'A(z-X\beta) \approx \RSS(\hat D),
\end{equation}
with $\eta=\beta'X$, $\mu=\frac{\partial l}{\partial \eta}$, $A=-\frac{\partial^2 l}{\partial \eta\eta'}$, and $z=\eta+A^-\mu$.

\subsection{{
PLS regression models and extensions}}
\subsubsection{{
PLSR}}
Prediction in high-dimensional and low-sample size settings already arose in  chemistry in the eighties. PLS regression, that can be viewed as a regularization method based on dimension reduction, was developed as a chemometric tool in an attempt to find reliable predictive models with spectral data (\citealp{wold82,tenen98}). Nowadays, the difficulty encountered with the use of genomic or proteomic data for classification or prediction, using very large matrices, is of comparable nature. It was thus natural to use PLS regression principles in this new context.
The method starts by constructing latent components, using linear combinations of the original variables, which are then used as new descriptors in standard regression analysis. Different from the principal components analysis (PCA), this method makes use of the response variable in constructing the latent components. The PLS regression can be viewed as a regularized approach searching the solution in a sub-space named Krylov space giving biased regression coefficients but with lower variance. In the framework of censored genomic data, the PLS regression operates a reduction of the dimensionality of the gene's space oriented towards the explanation of the hazard function. It allows transcriptomic signatures correlated to survival to be determined.

\subsubsection{{
Sparse PLSR}}
Recently, \citealp{chun10}, provided both empirical and theoretical results showing that the performance of PLS regression was ultimately affected by the large number of predictors. In parti\-cular, a higher number of irrelevant variables leads to inconsistency of coefficient estimates in linear regression setting. There is a need to filter the descriptors as a preprocessing step before PLS fit. However, commonly used variables filtering approaches are all univariate and ignore correlation between variables. To solve these issues, Chun and Keles proposed "sparse PLS regression" which promotes variables selection within the course of PLS dimension reduction. sPLS has the ability to include variables that variable filtering would select in the construction of the first direction vector. Moreover, it can select additional variables, \textit{i.e.}, variables that become significant once the response is adjusted for other variables in the construction of the subsequent direction vectors. This is the case of "proxy genes" acting as suppressor variables which do not predict the outcome variable directly but improve the overall prediction by enhancing the effects of prime genes despite having no direct predictive power, \citealp{magi10}.

A direct extension of PLS regression to sPLS regression could be provided by imposing $L_1$ constraint on PLS direction vector $w$:
\begin{align*}
&\max_w w'Mw \quad \textup{subject to } w'w=\|w\|_2=1, \|w\|_1\leqslant \lambda,\\
&\textup{where } M=X'YY'X.
\label{eq:}
\end{align*}
When $Y$=$X$, the objective function coincides with that of sPCA (\citealp{jolli03}). However in that case Jolliffe {\it et al.} pointed out that the solution tends not to be sparse enough and the problem is not convex. To solve these issues, Chun and Keles provided an efficient implementation of sPLS based on the LARS algorithm by generalizing the regression formulation of sPCA of \citealp{zou06}:
\begin{align*}
&\min_{w,c} -\kappa w'Mw+(1-\kappa)(c-w)'M(c-w)+\lambda_1\|c\|_1+\lambda_2\|c\|_2\\
&\textup{subject to } w'w=1, \textup{where } M=X'YY'X.
\end{align*}
This formulation promotes exact zero property by imposing $L_1$ penalty onto a surrogate of the direction vector $c$ instead of the original direction $w$ while keeping $w$ and $c$ close to each other. The $L_2$ penalty takes care of the potential singularity of $M$. Moreover, they demonstrated that for univariate PLS, $y$ regressed on $X$, the first direction vector of the sparse PLS algorithm was obtained by soft thresholding of the original PLS direction vector:
\begin{equation}
(\lvert Z\rvert-\frac{\lambda}2)_+\sign{(Z)}, \textup{ where } Z=X'y/\|X'y\|_2.
\end{equation}
In order to inherit the property of the Krylov subsequences which is known to be crucial for the correctness of the algorithm (\citealp{kramer07}), the thresholding phase is followed by a PLS regression on the previously selected variables. The algorithm is then iterated with $y$ replaced by $y-X\hat \beta$, the residuals of the PLS regression based on the variables selected from the previous steps. The sPLS algorithm leads therefore to sparse solutions by keeping the Krylov subsequence structure of the direction vectors in a restricted $X$ space which is composed of the selected variables. The thresholding parameter $\lambda$ and the number of hidden components are tuned by cross validation.

sPLS has connections to other variable selection algorithms including the elastic net method (\citealp{zou05}) and the threshold gradient method (\citealp{friedm04}). The elastic net algorithm deals with the collinearity issue in variable selection problem by incorporating the ridge regression method into the LARS algorithm. In a way, sPLS handles the same issue by fusing the PLS technique into the LARS algorithm. sPLS can also be related to threshold gradient method in that both algorithms use only thresholded gradient and not the Hessian. However, sPLS achieves fast convergence by using conjugate gradient. Hence, LARS and sPLS algorithms use the same criterion to select active variables in the univariate case. However, the sPLS algorithm differs from LARS in that sPLS selects more than one variable at a time and utilizes the conjugate gradient method to compute coefficients at each step. The computational cost for computing coefficients at each step of the sPLS algorithm is less than or equal to the computational cost of computing step size in LARS since conjugate gradient methods avoid matrix inversion.

\subsection{{
Extensions of PLSR models to censored data}}
\subsubsection{{
PLS-Cox}}
\citealp{garth94}, showed that PLS regression could be obtained as a succession of simple and multiple linear regressions. \citealp{tenen99}, proposed a fairly similar approach but one which could cope with missing data by using the principles of the Nipals algorithm \citep{wold66}. As a result, Tenenhaus suggested that PLS regression be extended to logistic regression (PLS-LR) by replacing the succession of simple and multiple regressions by a succession of simple and multiple logistic regressions in an approach much simpler than that developed by \citealp{marx96}. By using this alternative formulation of the PLS regression, \citealp{bast01}, extended the PLS regression to any generalized linear regression model (PLS-GLR) and to the Cox model ({PLS-Cox}) as a special case. Further improvements have then been described (\citealp{bast04}) in the case of categorical descriptors with variable selection using hard thresholding and model validation by bootstrap resampling. Since then many developments in the framework of PLS and Cox regressions have appeared in the literature. \citealp{nguyen02}, directly applied PLS regression to survival data and used the resulting PLS components in the Cox model for predicting survival time. However such a direct application did not really generalize PLS regression to censored survival data since it did not take into account the failure time in the dimension reduction step. 
Based on a straightforward generalization of \citealp{garth94}, \citealp{ligu04}, presented a solution, Partial Cox Regression, quite similar to the one proposed by Bastien and Tenenhaus, using different weights to derive the PLS components but not coping with missing data. 

\subsubsection{{
(DK)(S)PLS(DR)}}
\paragraph{{
The PLSDR algorithm \citep{bast08}}}
Following \citealp{seg06}, who suggested initially computing the null deviance residuals and then using these as outcomes for the LARS-Lasso algorithm, \citealp{bast08}, proposed PLSDR, an alternative in high-dimensional settings using deviance residuals based PLS regression. This approach is advantageous both by its simplicity and its efficiency because it only needs to carry out null deviance residuals using a simple Cox model without covariates and use these as outcome in a standard PLS regression. The final Cox model is then carried out on the $m$ retained PLSDR components.

Moreover, following the principles of the Nipals algorithm, weights, loadings, and PLS components are computed as regression slopes. These slopes may be computed even when there are missing data:
let $t_{hi}={x_{h-1,i}w_h}/{w'_hw_h}$ the value of the  PLS component for individual $i$, with $x_{h-1,i}$ the vector of residual descriptors and $w_h$ the vector of weights at step $h$. $t_{hi}$ represents the slope of the OLS line without constant term related to the cloud of points $(w_h,x_{h-1,i})$. In such case, in computing the $h^{\textup{th}}$ PLS component, the denominator is computed only on the data available also for the denominator.

\IncMargin{1em}
\begin{algorithm}
  \LinesNumbered
  \SetKwData{Left}{left}\SetKwData{This}{this}\SetKwData{Up}{up}
  \SetKwFunction{Union}{Union}\SetKwFunction{FindCompress}{FindCompress}
  \SetKwInOut{Input}{input}\SetKwInOut{Output}{output}
  \DontPrintSemicolon
  \BlankLine
  \Begin{
	$d \leftarrow $ null deviance residuals of the Cox model without covariates.\;
  \Begin{\tcc{Computation of the sPLS components by using the sPLS regression with the null deviance residuals $d$ as outcome.}
	\Indp  
		$\hat \beta^{PLS} \leftarrow 0$.\tcc*[r]{Initialisation}
		$\Omega \leftarrow \{\}$.\;
		$k \leftarrow 1$.\;
		$y_1 \leftarrow d$.\;
	\Indm  
		\While(\tcc*[f]{Component derivation loop}){$(h \leqslant m)$}{
			$z \leftarrow X'y_1/\|X'y_1\|_2$.
			$w \leftarrow (\lvert z\rvert-\lambda/2)_+\sign{(z)}$.\;
			$\Omega \leftarrow \{i: \hat w_i\neq 0\}\cup\{i: \hat\beta_i^{PLS}\neq 0\}$.\;
			Fit PLS with $X_\Omega$ by using the $k$ number of latent components.\;
			$\hat\beta^{PLS} \leftarrow$ new PLS estimates of the direction vectors.\; 
			 $y_1 \leftarrow y_1-X\hat\beta^{PLS}$.\;
			 $h \leftarrow h+1$.\;
		}
		}
	\KwRet{Cox model on the $m$-retained sPLSDR components.}\;
  }
  \caption{The sPLSDR algorithm \protect\citep{Bastien2015}}\label{algo_slpsdr}
\end{algorithm}
\DecMargin{1em}

\paragraph{{
The DKsPLSDR algorithm \citep{Bastien2015}}}
In the case of very many descriptors, PLS regression being
invariant by orthogonal transformation (\citealp{jong94}), an even faster procedure could be derived by replacing the $X$ matrix by the matrix of principal components $Z$ $(XX'=ZZ')$. This could be viewed as the simple form of linear kernel PLS regression algorithms which have been proposed in the nineties (\citealp{lin93,ran94}) to solve computational problems posed by very large matrices in chemometrics. The objective of these methods was to obtain PLS components by working on a condensed matrix of a considerably smaller size than the original one. Moreover, in addition to dramatically reducing the size of the problem, non-linear pattern in the data could also be analyzed using non-linear kernel.

\citealp{ros01}, proposed a nonlinear extension of PLS regression using kernels. Assuming a nonlinear transformation of the input variables $\{x_i\}_{i=1}^n$ into a feature space $F$, \textit{i.e.} a mapping $\Phi: x_i\in \mathbb{R}^N \mapsto \Phi(x_i)\in F$, their goal was to construct a linear PLS regression model in $F$. They derived an algorithm named KPLS for Kernel PLS by performing the PLS regression on $\Phi(X)$. It amounts to replacing, in the expression of PLS components, the product $XX'$ by $\Phi(X)\Phi(X)'$ using the so-called kernel trick which allows the computation of dot products in high-dimensional feature spaces using simple functions defined on pairs of input patterns: $\Phi(x_i)\Phi(x_j)'=K(x_i,x_j)$. This avoids having to explicitly calculate the coordinates in the feature space which could be difficult for a highly dimensional feature space. By using the kernel functions corresponding to the canonical dot product in the feature space, non-linear optimization can be avoided and simple linear algebra used.

\citealp{ben03}, proposed to perform PLS regression directly on the kernel matrix $K$ instead of $\Phi(X)$. DKPLS corresponds to a low rank approximation of the kernel matrix. Moreover, \citealp{tena07} demonstrated that, for one dimensional output response, PLS of $\Phi(X)$ (KPLS) is equivalent to PLS on $K^{1/2}$ (DKPLS).

Using previous works, it becomes straightforward to derive a non-linear Kernel sPLSDR algorithm by replacing in the sPLSDR algorithm the $X$ matrix by a kernel matrix $K$. The main kernel functions are the linear kernel ($K(u,v)= <u,v>$) and the Gaussian kernel ($K(u,v)=\exp{(-\|u-v\|_2^2/2\sigma^2)}$).

However, non-linear kernel (sparse) PLS regression loses the explanation with the original descriptors unlike linear kernel PLS regression, which could limit the interpretation of the results.

\IncMargin{1em}
\begin{algorithm}
  \LinesNumbered
  \SetKwData{Left}{left}\SetKwData{This}{this}\SetKwData{Up}{up}
  \SetKwFunction{Union}{Union}\SetKwFunction{FindCompress}{FindCompress}
  \SetKwInOut{Input}{input}\SetKwInOut{Output}{output}
  \DontPrintSemicolon
  \BlankLine
  \Begin{
  	Computation of the kernel matrix.\;
	$d \leftarrow $ null deviance residuals of the Cox model without covariates.\;
  	Computation of the sPLS components by using the DKsPLS algorithm with the null deviance residuals $d$ as outcome.\;
	\KwRet{Cox model on the $m$-retained DKsPLSDR components.}\;
  }
  \caption{The DKsPLSDR algorithm \protect\citep{Bastien2015}}\label{algo_dkslpsdr}
\end{algorithm}
\DecMargin{1em}

\section{Simulation studies}
\subsection{Scheme of the studies}

\IncMargin{1em}
\begin{algorithm}
  \LinesNumbered
  \SetKwData{Left}{left}\SetKwData{This}{this}\SetKwData{Up}{up}
  \SetKwFunction{Union}{Union}\SetKwFunction{FindCompress}{FindCompress}
  \SetKwInOut{Input}{input}\SetKwInOut{Output}{output}
  \DontPrintSemicolon
  \BlankLine
  \Begin{
	\ForEach{simulation types $\in \{$eigengene, cluster, factorial$\}$}{
		\ForEach{link types $\in \{$none, linear, quadratic$\}$}{
					\For{i=1 to 100}{
						Simulate a dataset with exponential survival distribution and 40\% censored rate (100 observations $\times$ 1000 genes).\;
						Randomly split the dataset into a training set (7/10, 70 observations) and a test set (3/10, 30 observations)\;
						\ForEach{of the 7 (S)PLS based methods}{
							\ForEach{of the 12 cross-validation criteria}{Find the optimal number of components by $K$-fold cross-validation of the training data set, see Section \ref{sechypcv}.
								}
								}
						\ForEach{of the 14 prediction methods}{
							\ForEach{of the 12 cross-validation criteria}{
								Find the optimal tuning parameter $\hat\lambda_{train}$ by $K$-fold cross-validation of the training data set, see Section \ref{sechypcv}.\;
								Given $\hat\lambda_{train}$, estimate the vector of regression coefficients $\hat\beta_{train}$ on the whole training data set.\;
								Calculate the values of the 22 performance criteria on the test data set as described in Section \ref{perfmeas}.\;
							}
						}
			}
		}
  }
  }
  \caption{Summary of the procedure for evaluating the accuracy of the cross validation methods and revisit the performance of the component based methods.\label{sumpredme}}
\end{algorithm}
\DecMargin{1em}

The aim of our two in silico studies is twofold: evaluating the accuracy of the cross validation methods, see Section~\ref{cvcritchoice}, and revisit the performance of the component based methods, see Section~\ref{perfevalbench}.

We performed a simulation study (Algorithm \ref{sumpredme}) in order to evaluate the methods by simulating 100 datasets with exponential survival distribution and 40\% censored rate (100 observations $\times$ 1000 genes) according to three different simulation types (cluster by \cite{bair06}, factorial by \cite{kais62} and \cite{fan2002} or eigengene by \cite{lang13}), using either no link or a linear one between the response and the predictors.

We divided each of these 600 datasets into a training set, of 7/10 (70) of the observations, used for estimation, and a test set, of 3/10 (30) of the observations, used for evaluation or testing of the prediction capability of the estimated model. This choice was made to stay between the 2:1 scheme of \cite{bove07,vwie09,lll11} and the 9:1 scheme of \cite{li2006}. The division between training and test sets was balanced using the \verb+caret+ package, \cite{kuhn14}, both according to the response value and censor rate.


\subsection{Data generation}
\subsubsection{Eigengene: \cite{lang13}}
Given module seeds and a desired size for the genes modules around the seeds of $n_I$ genes, module genes expression profiles are generated such that the $k$-th rank correlated gene from module $I$ with its module seed $seed_I$  is :
\begin{equation}
\operatorname{cor}(x_{k,I},seed_{I})=1-k/n_I(1-r_{\textup{min}})=r_{k,I}\label{eqcor1}
\end{equation}
that is, the first gene has correlation $r_{i,I}\approx 1$ with the seed while the last ($n_I$-th) gene has correlation $r_{{n_i},I}\approx r_{\textup{min}}$.

The required correlation (\ref{eqcor1}) is achieved by calculating the $k$-th gene profile as the sum of the seed vector $seed_I$  and a noise term $a_k\varepsilon_k$
\begin{equation}
x_{k,I}=seed_{I}+{a}_{k}\varepsilon _{k}
\quad\textup{where}\quad
{a}_{k}=\sqrt{\frac{\operatorname{var}(seed_{I}^{{}})}{\operatorname{var}({{\varepsilon }_{k}})}\left( \frac{1}{r_{k,I}^{2}}-1 \right)}
\end{equation}

This technique produces modules consisting of genes distributed symmetrically around the module seed; in this sense, the simulated modules are spherical clusters whose centers coincide (on average) with the module seed.

In the simulations the parameters have been let as follow $I=4$, $r_{\textup{min}}=0.5$, $n_{I}=25$ with $seed_{I}$ and $\varepsilon_{k}\sim \mathcal{N}(0,1)$.

Survival and censoring times, with $0.4$ censoring probability, are generated from exponential survival distributions. When linked to survival (linear or quadratic case), only expressions from genes from the first two modules ($N=50$) are related to survival time.

Each simulated data set consists of $1000$ genes and $100$ samples. Only the first hundred genes are structured. The last $900$ are random noise generated from $\mathcal{N}(0,1)$.

\subsubsection{Cluster \cite{bair06}}
The gene expression data is distributed as:
\begin{equation}{{X}_{ij}}=\left\{ \begin{matrix}
   3+\varepsilon_{ij}^{{}}\text{ if }i\leq 50,j\leq 50  \\
   4+\varepsilon_{ij}^{{}}\text{ if }i>50,j\leq 50  \\
3.5+\varepsilon_{ij}^{{}}\text{ if }j>50.\\
\end{matrix} \right.\end{equation}
Where the ${{\varepsilon }_{ij}}$are drawn from a $\mathcal{N}(0,1)$.

Each simulated data set consists of $1000$ genes and $100$ samples. Survival and censoring times, with $0.4$ censoring probability, are generated from exponential survival distributions. When linked to survival (linear or quadratic case), only expressions from genes from the first $50$ genes are related to survival. 

\subsubsection{Factorial \cite{kais62}, \cite{fan2002}}
We have supposed that genes expressions are related to $4$ latent variables associated each to a specific biological function. Let for each group a specified population inter-correlation pattern matrix $R$. By applying principal component factorization (PCA) to the matrix $R$ and following Kaiser and Dickman, we can generate $4$ multivariate normally distributed sample data with a specific correlation pattern.

${{Z}_{(k\times N)}}={{F}_{(k\times k)}}X_{(k\times N)}^{{}}$ 

Where: 

$k$ is the number of descriptors (genes)

$N$ is the number of observations

$X$ is a matrix of uncorrelated random normal variables $\mathcal{N}(0,1)$

$F$ is a matrix containing principal component factor pattern coefficients obtained by applying Principal Components Analysis (PCA) to the given population correlation matrix $R$

$Z$ is the resultant sample data matrix, as if sampled from a population with the given population correlation matrix $R$

We have chosen a compound symmetry structure for the correlation matrix $R$ with a same correlation ($0.7$) between two descriptors of a same group, descriptors between different groups being independent.

Moreover the choice of the correlation coefficient allows specifying the percentage of variance explained by the first factorial axes. Given four groups with inter-genes correlation coefficient of $0.7$ corresponds to expend $70\%$ of the inertia in $4$ principal directions.

Survival and censoring times, with $0.4$ censoring probability, are generated from exponential survival distributions. When linked to survival (linear or quadratic case), only expressions from genes from the first two groups ($N=50$) are related to survival time. 

Each simulated data set consists of $1000$ genes and $100$ samples. Only the first hundred genes are structured. The last $900$ are random noise generated from $\mathcal{N}(0,1)$.

\begin{figure}[!tpb]
\centerline{\includegraphics[width=.96\columnwidth]{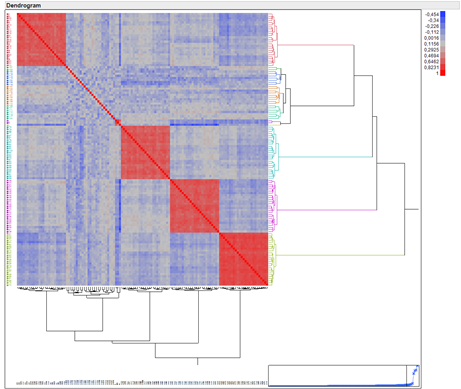}};;
\vspace{-.5cm}
\caption{Hierarchical clustering of the variables in a factorial-type simulated dataset.}\label{heatmap}
\end{figure}

Figure~\ref{heatmap} displays the pattern of correlation for the first 150 descriptors with the four groups of $25$ genes each well defined.

\subsection{Hyperparameters and cross-validation}\label{sechypcv}
In standard $K$-fold cross-validation of a dataset of size $n$, $K$ folds of size $\text{Floor}(n/K)$ are created by sampling from the data without replacement and each of the remaining $n\ \text{mod}\ K$ data points is assigned randomly to a different fold. In stratified or balanced crossvalidation \cite[p. 246]{brei84}, the data are first ordered by the response value or class. This list is broken up into $c$ bins each containing $K$ points with many similar response values. Any remaining points at the end of the list are assigned to an additional bin. A fold is formed by sampling one point without replacement from each of the bins. Except for the ordering of the data, this is equivalent to standard cross-validation. We used balanced cross-validation with respect to the response value and censor rate. The folds were design using the \verb+caret+ package, \cite{kuhn14}.

In traditional cross-validation, \textit{i.e.} with a dataset without censored events, each fold would yield a test set and a value of a prediction error measure (for instance log partial likelihood, integrated area under the curve, integrated area under the prediction error curve). When dealing with censored events and using the CV partial likelihood (CVLL, \cite{vevh93}) criterion, it is possible to make more efficient use of risk sets: \cite{vhou06} recommended to derive the CV log partial likelihood for the $j$th fold by subtraction; by subtracting the log partial likelihood evaluated on the full dataset from that evaluated on the full dataset minus the $j$th fold, called the $(K-1)/K$ dataset. This yields the van Houwelingen CV partial likelihood (vHCVLL) .

Hyperparameters were tuned using 7-fold cross-validation on the training set. The number of folds was chosen following the recommandation of \cite{wold01}, \cite{brei92} and \cite{koha95}. As in, \cite{bove07}, \cite{vwie09} and \cite{lll11}, mean values were then used to summarize these cross validation criteria over the 7 runs and the hyperparameters were chosen according to the best values of these measures.

\section{Highlighting relevant cross validation criteria}\label{cvcritchoice}
\subsection{The failure of the two usual criteria}
\begin{figure}[!tpb]
\centerline{{\includegraphics[width=.75\columnwidth]{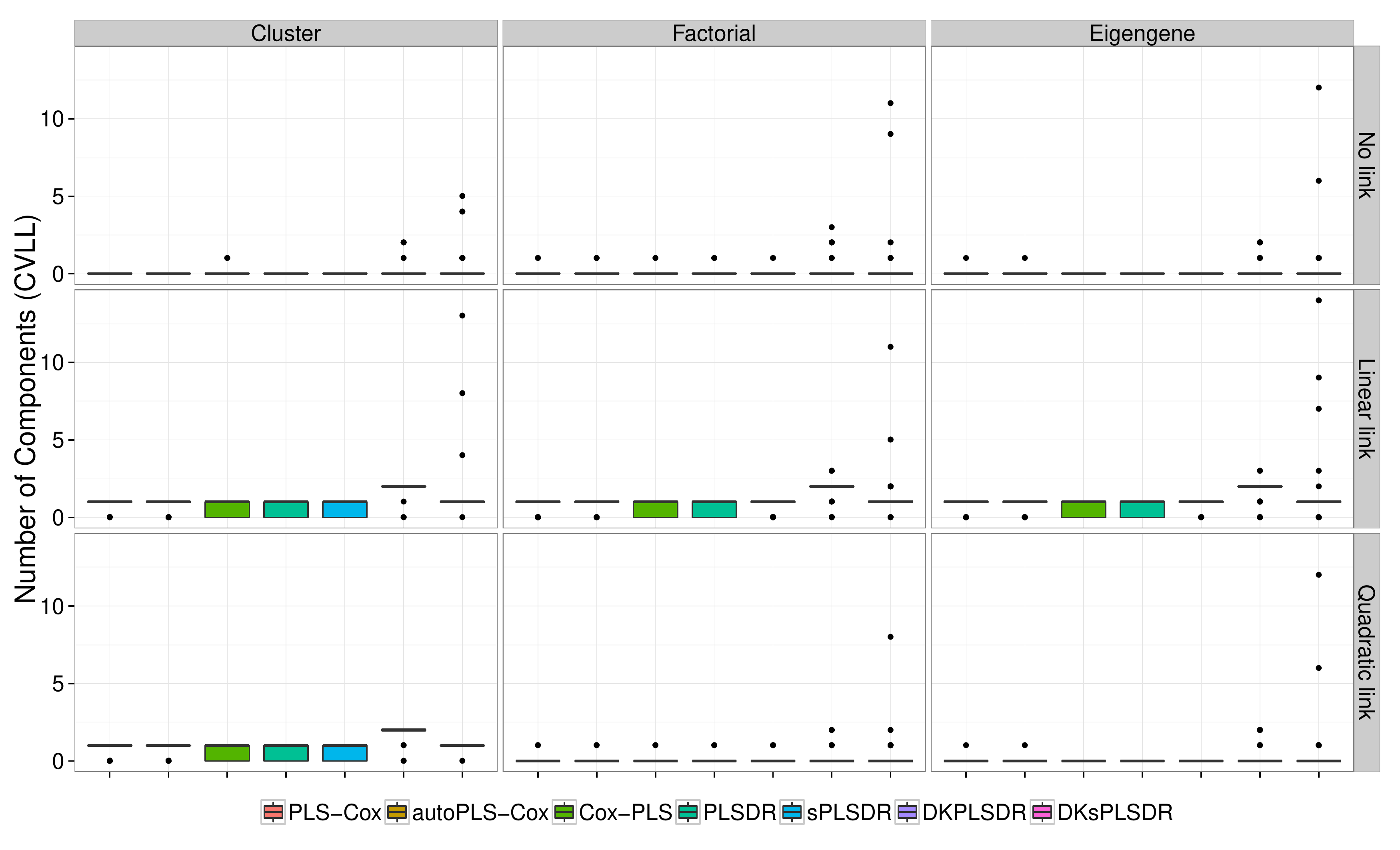}\phantomcaption\label{NbrComp_cvll}}\qquad{\includegraphics[width=.75\columnwidth]{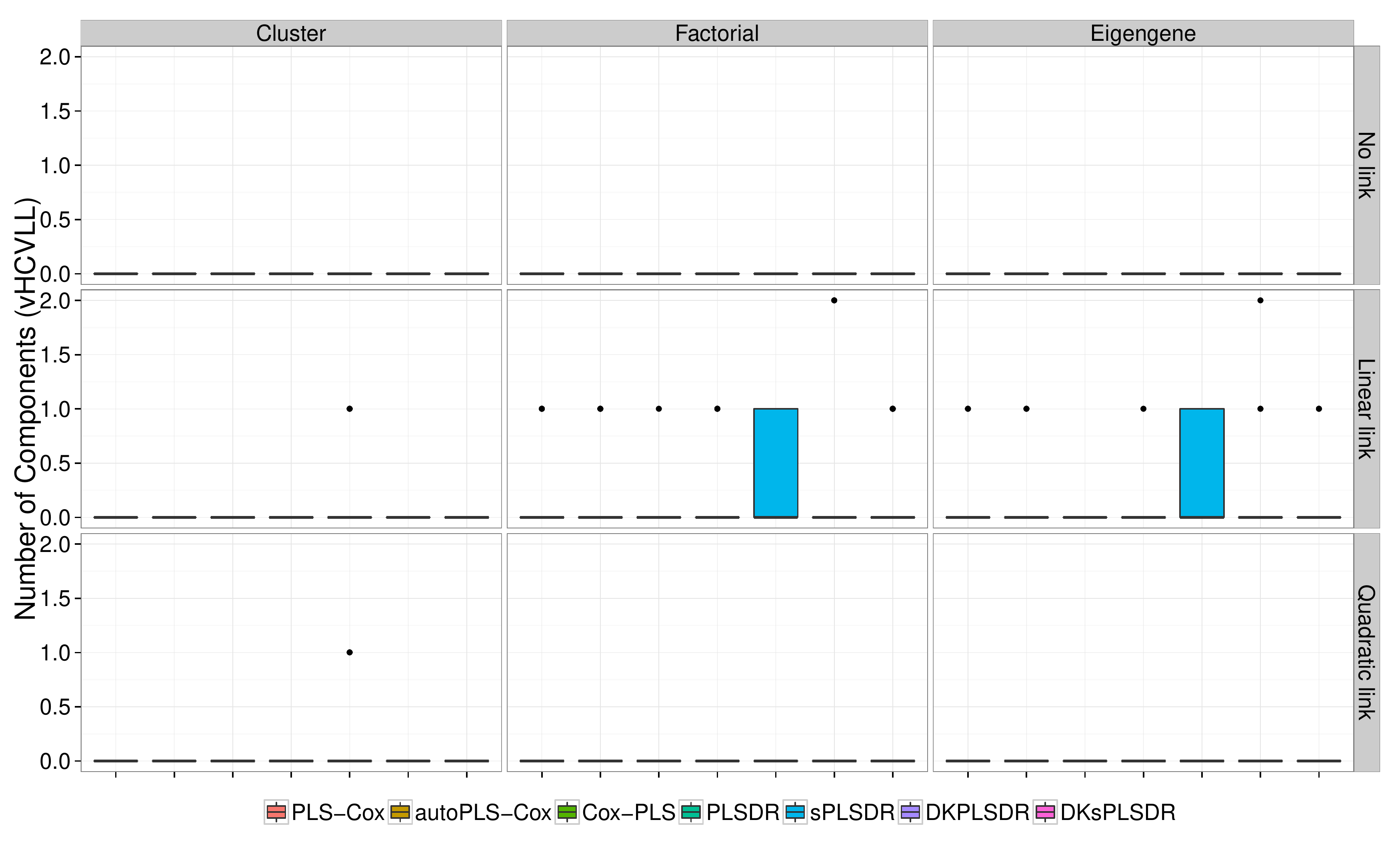}\phantomcaption\label{NbrComp_vanHcvll}}}
\vspace{-.5cm}
\caption*{\hspace{-.5cm}\mbox{Figure~\ref{NbrComp_cvll}:  Nbr of comp, LL criterion. \quad\quad\quad Figure~\ref{NbrComp_vanHcvll}:  Nbr of comp, vHLL criterion.}}
\end{figure}

The van Houwelingen CV partial likelihood (vHCVLL,  see Figure \ref{NbrComp_vanHcvll}) criterion behave poorly for all the PLS or sPLS based methods by selecting zero components where, according to our simulation types, the PLS-Cox, autoPLS-Cox, Cox-PLS, PLSDR, sPLSDR, DKPLSDR and DKsPLSDR methods were expected to select, for the factor or eigengene schemes, about two components and slightly more for the cluster scheme. As with the the classic CV partial likelihood (CVLL), it almost always selects at most one component and hence systematically underestimates the number of components. The simulations results for the selection of the number of components using CVLL are plotted on Figure \ref{NbrComp_cvll}. We confirmed this poor property by performing cross-validation on a simpler simulation scheme designed by \cite{coxnet11}.

\subsection{Proposal of new criteria}
\begin{figure}[!tpb]
\centerline{{\includegraphics[width=.75\columnwidth]{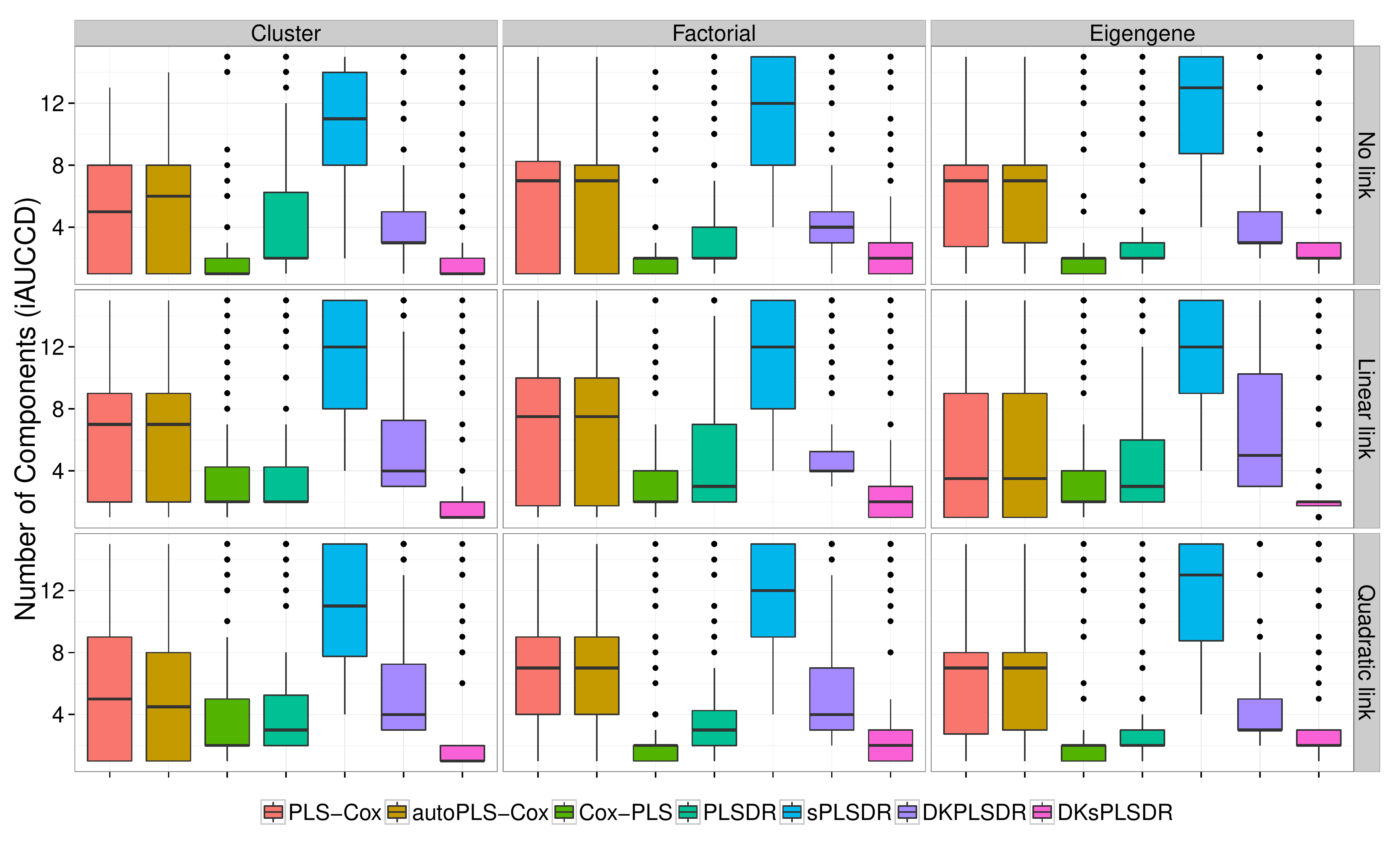}\phantomcaption\label{NbrComp_AUCcd}}\qquad{\includegraphics[width=.75\columnwidth]{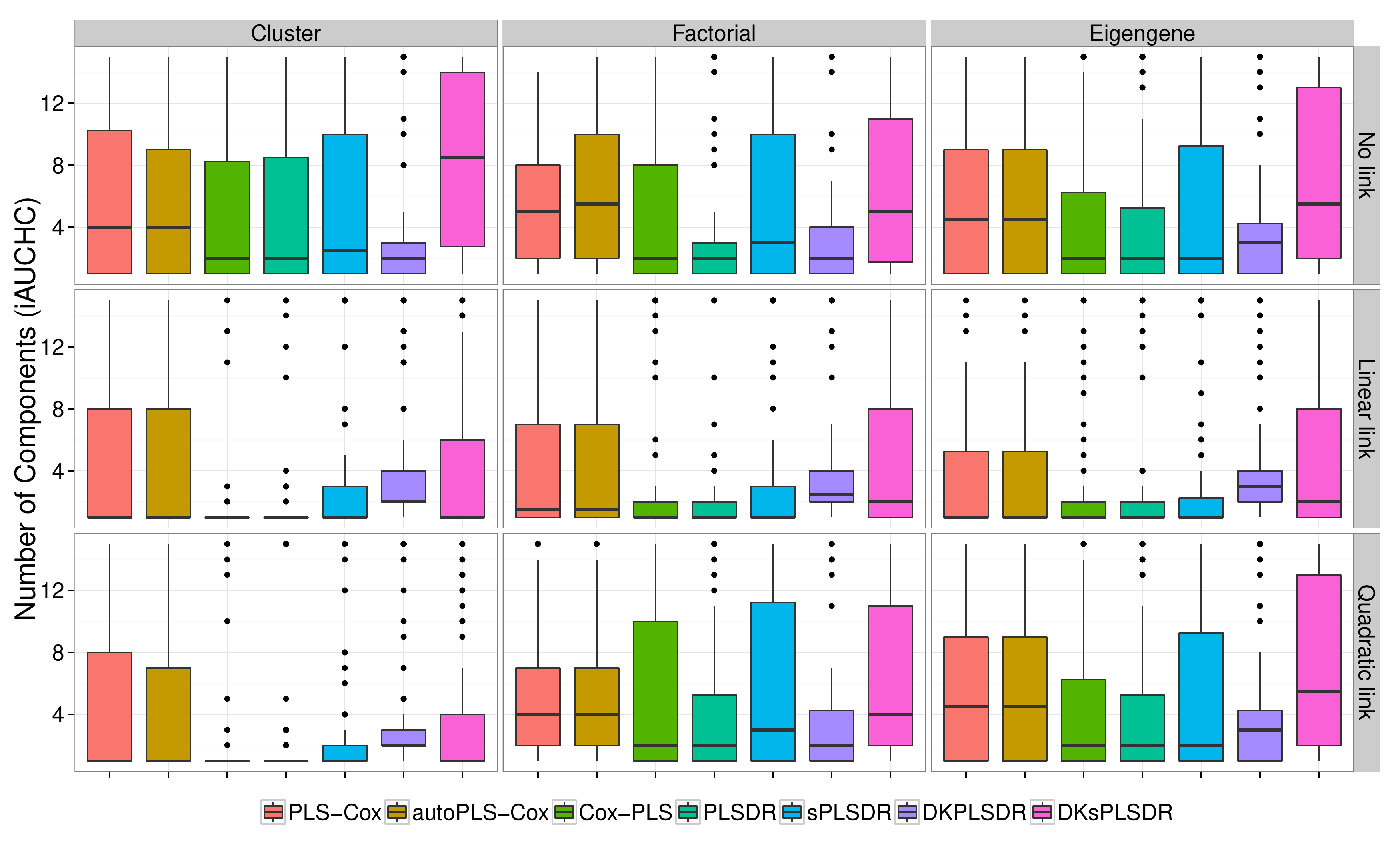}\phantomcaption\label{NbrComp_AUChc}}}
\vspace{-.5cm}
\caption*{\hspace{-.5cm}\mbox{Figure~\ref{NbrComp_AUCcd}:  Nbr of comp, iAUCCD criterion. \qquad Figure~\ref{NbrComp_AUChc}:  Nbr of comp, iAUCHC criterion.}}
\end{figure}

\begin{figure}[!tpb]
\centerline{{\includegraphics[width=.75\columnwidth]{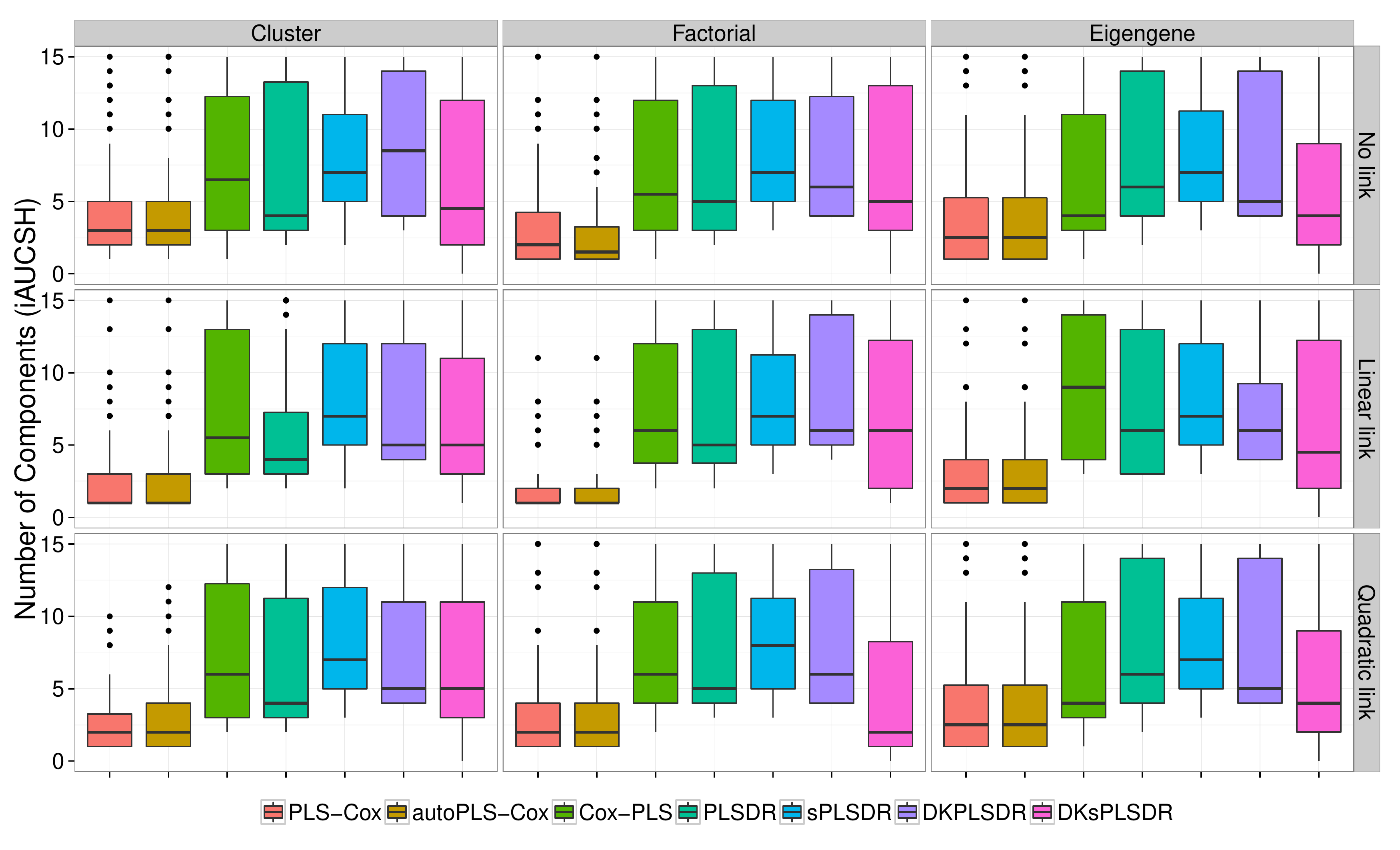}\phantomcaption\label{NbrComp_AUCsh}}\qquad{\includegraphics[width=.75\columnwidth]{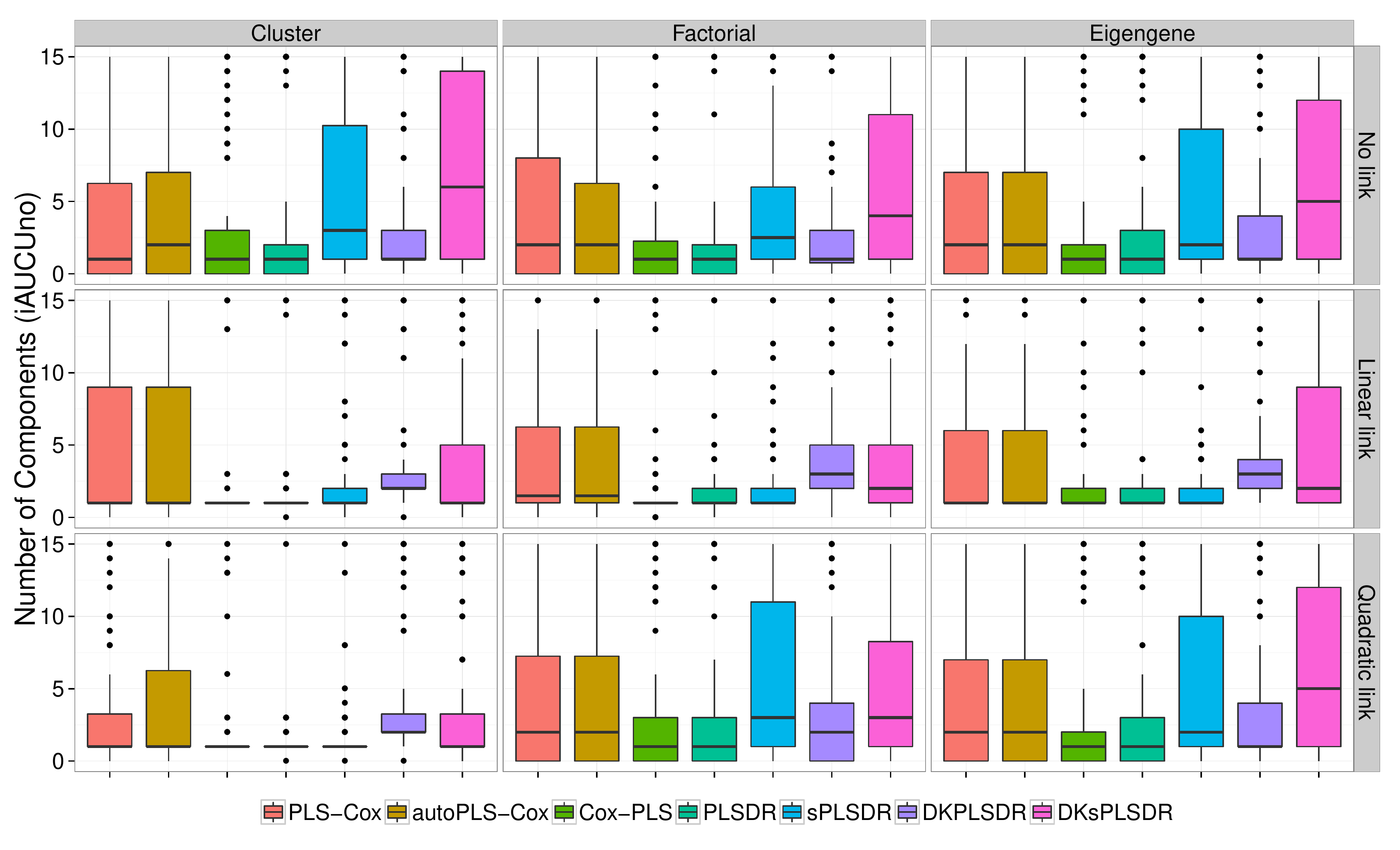}\phantomcaption\label{NbrComp_AUCUno}}}
\vspace{-.5cm}
\caption*{\hspace{-.5cm}\mbox{Figure~\ref{NbrComp_AUCsh}:  Nbr of comp, iAUCSH criterion. \qquad Figure~\ref{NbrComp_AUCUno}:  Nbr of comp, iAUCUno criterion.}}
\end{figure}

\begin{figure}[!tpb]
\centerline{{\includegraphics[width=.75\columnwidth]{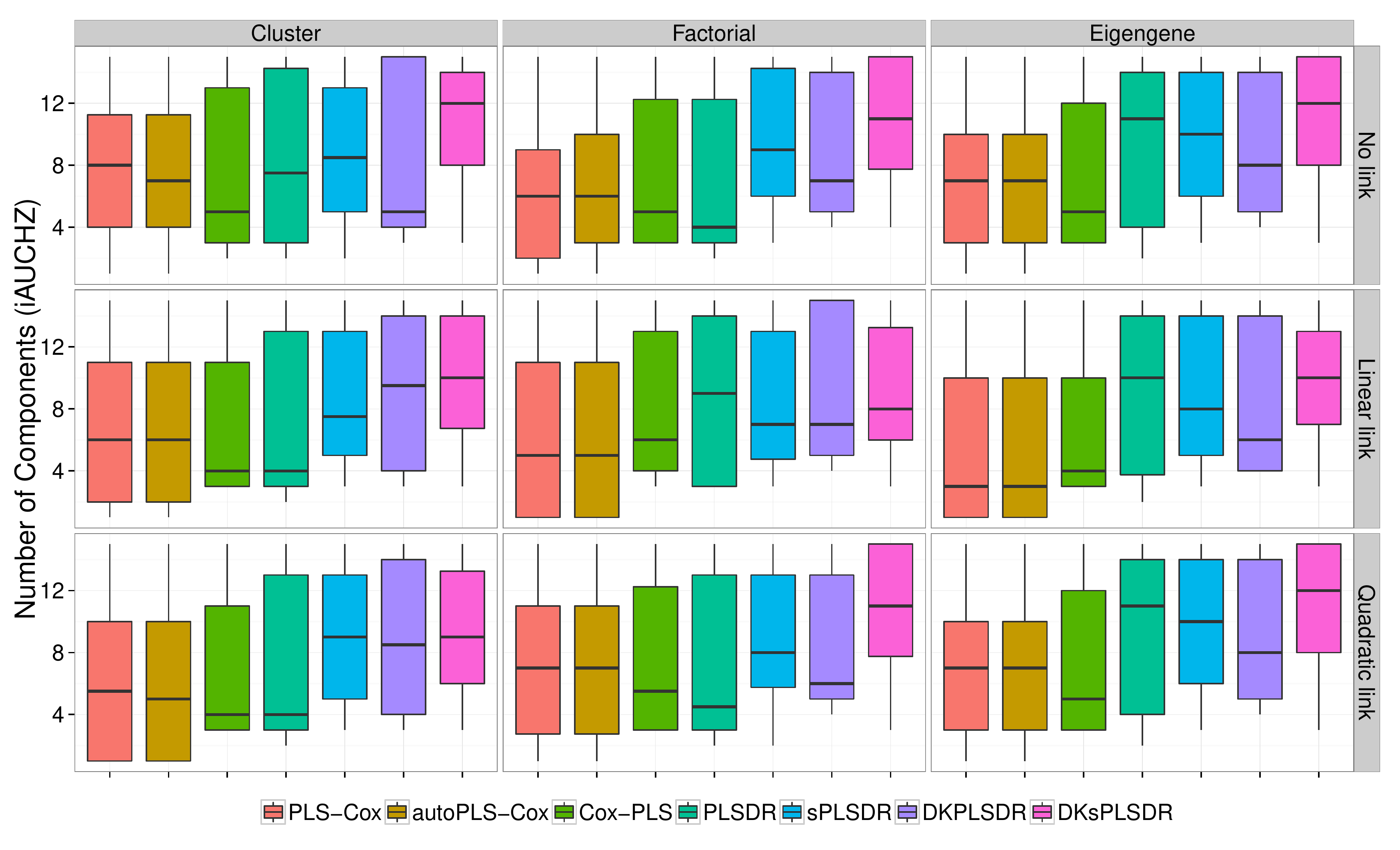}\phantomcaption\label{NbrComp_AUChztest}}\qquad{\includegraphics[width=.75\columnwidth]{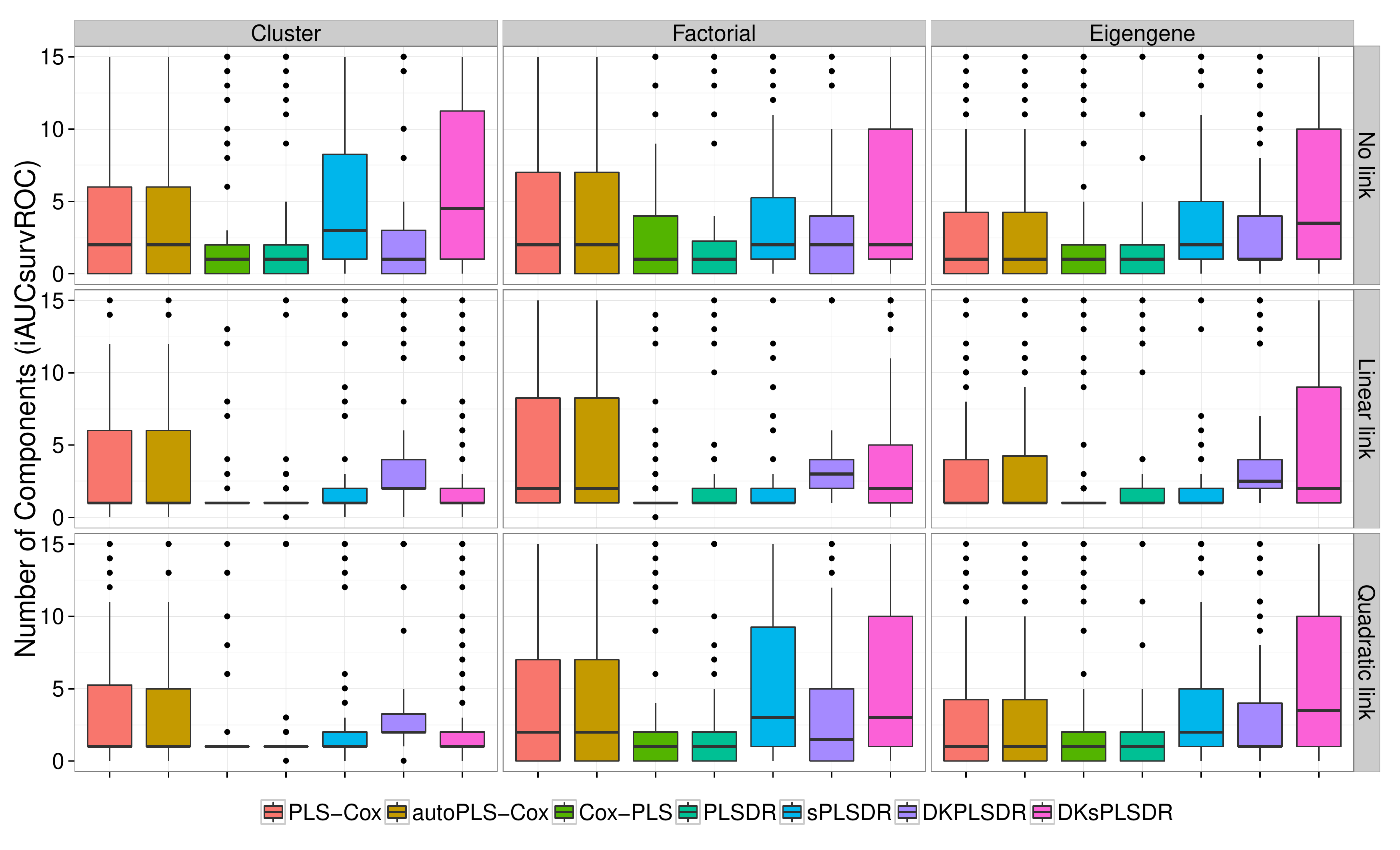}\phantomcaption\label{NbrComp_AUCsurvROCtest}}}
\vspace{-.5cm}
\caption*{\hspace{-.5cm}\mbox{Figure~\ref{NbrComp_AUChztest}:  Nbr of comp, iAUCHZ criterion. \qquad Figure~\ref{NbrComp_AUCsurvROCtest}:  Nbr of comp, iAUCSurvROC criterion.}}
\end{figure}

\begin{figure}[!tpb]
\centerline{{\includegraphics[width=.75\columnwidth]{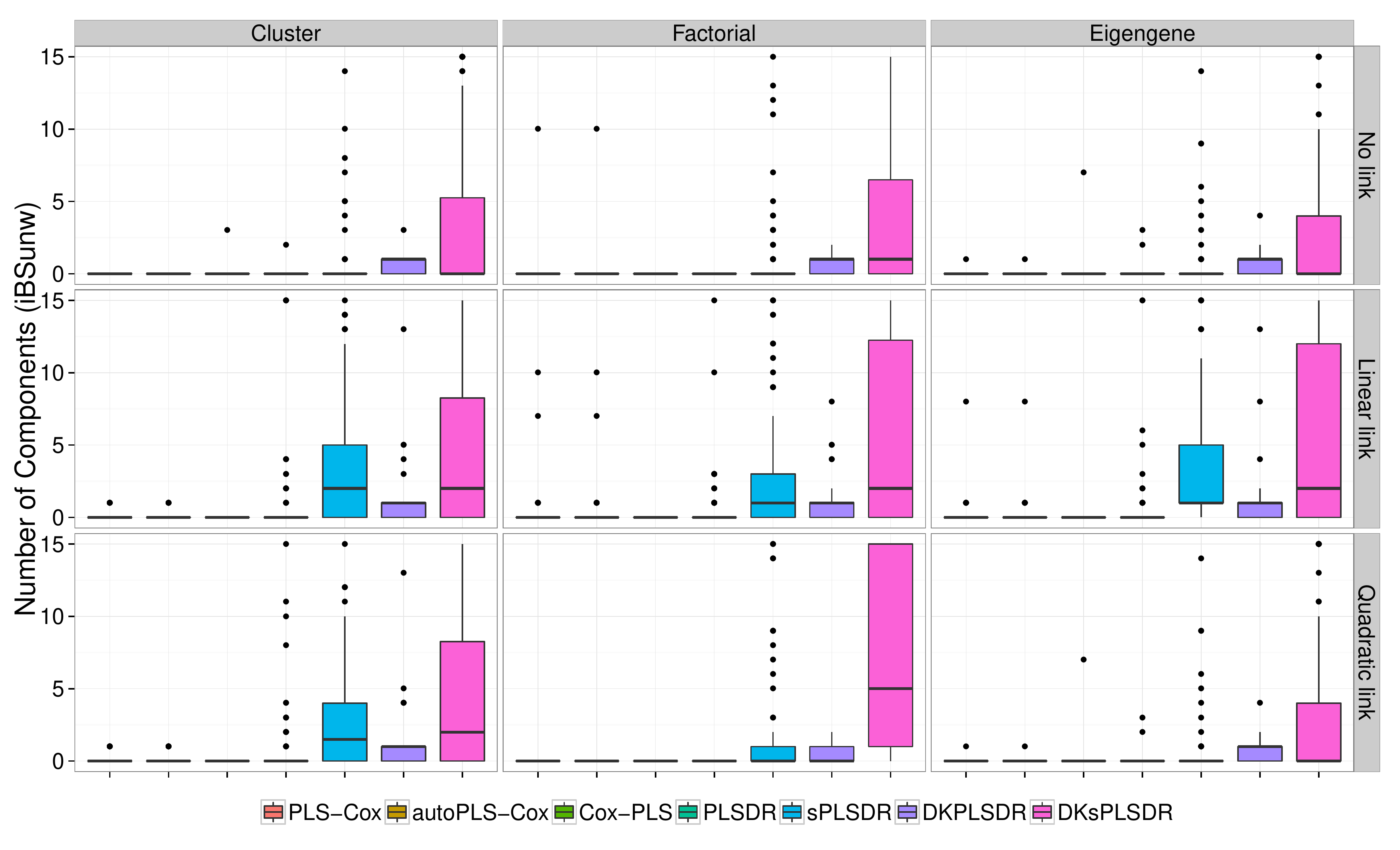}\phantomcaption\label{NbrComp_iBSunw}}\qquad{\includegraphics[width=.75\columnwidth]{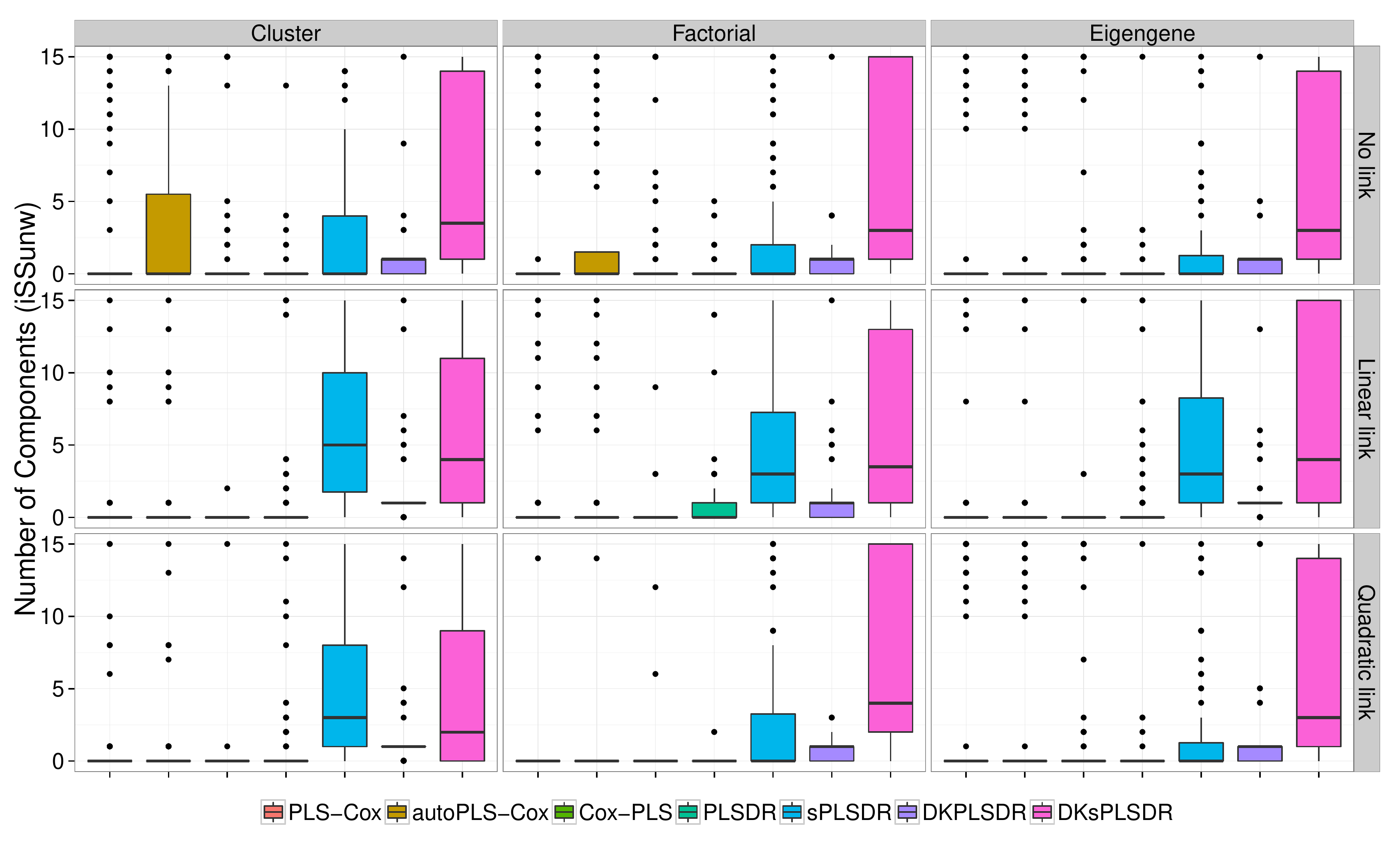}\phantomcaption\label{NbrComp_iSchmidSunw}}}
\vspace{-.5cm}
\caption*{\hspace{-.5cm}\mbox{Figure~\ref{NbrComp_iBSunw}:  Nbr of comp, iBSunw criterion. \qquad\qquad Figure~\ref{NbrComp_iSchmidSunw}:  Nbr of comp, iSSunw criterion.}}
\end{figure}

\begin{figure}[!tpb]
\centerline{{\includegraphics[width=.75\columnwidth]{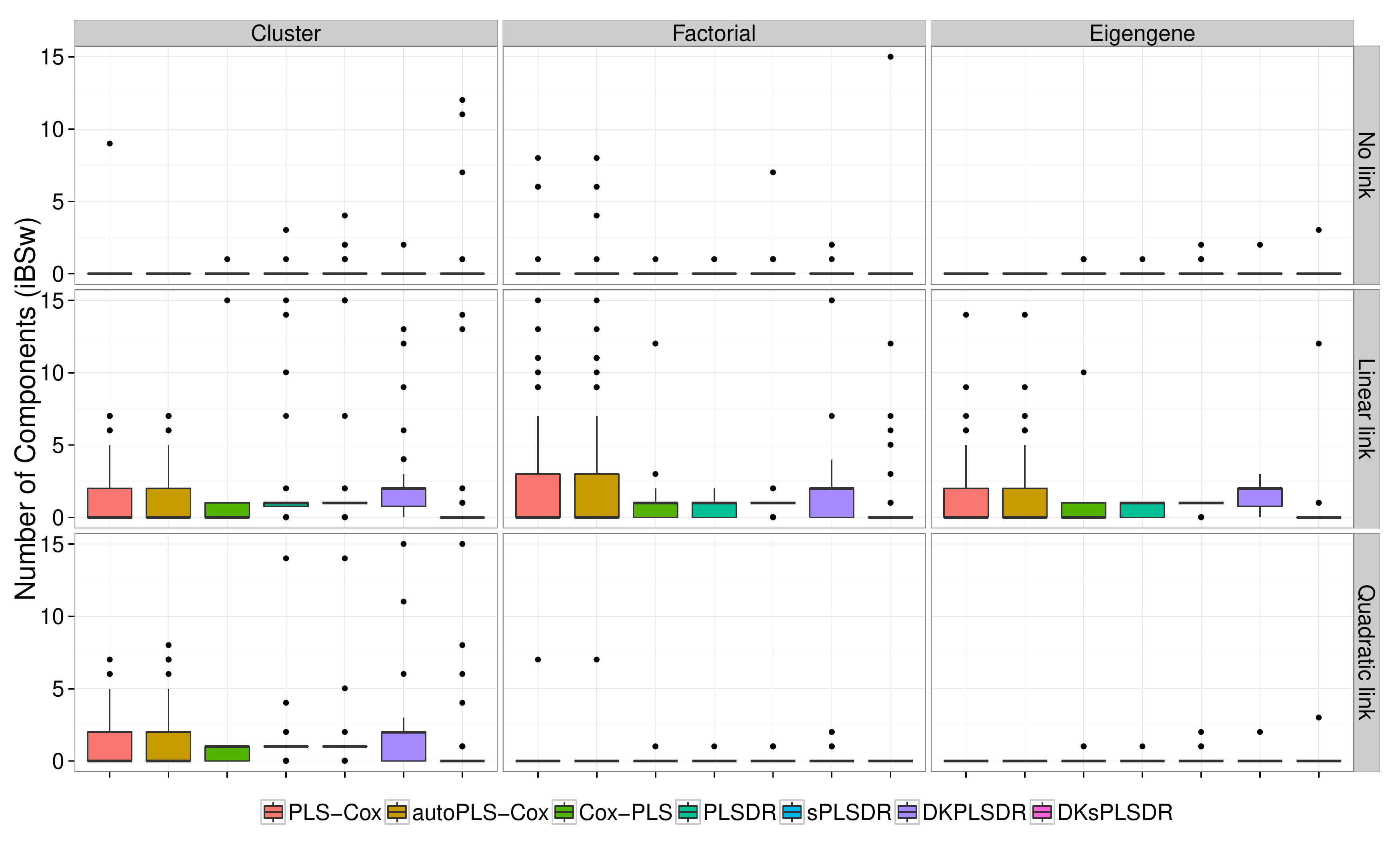}\phantomcaption\label{NbrComp_iBSw}}\qquad{\includegraphics[width=.75\columnwidth]{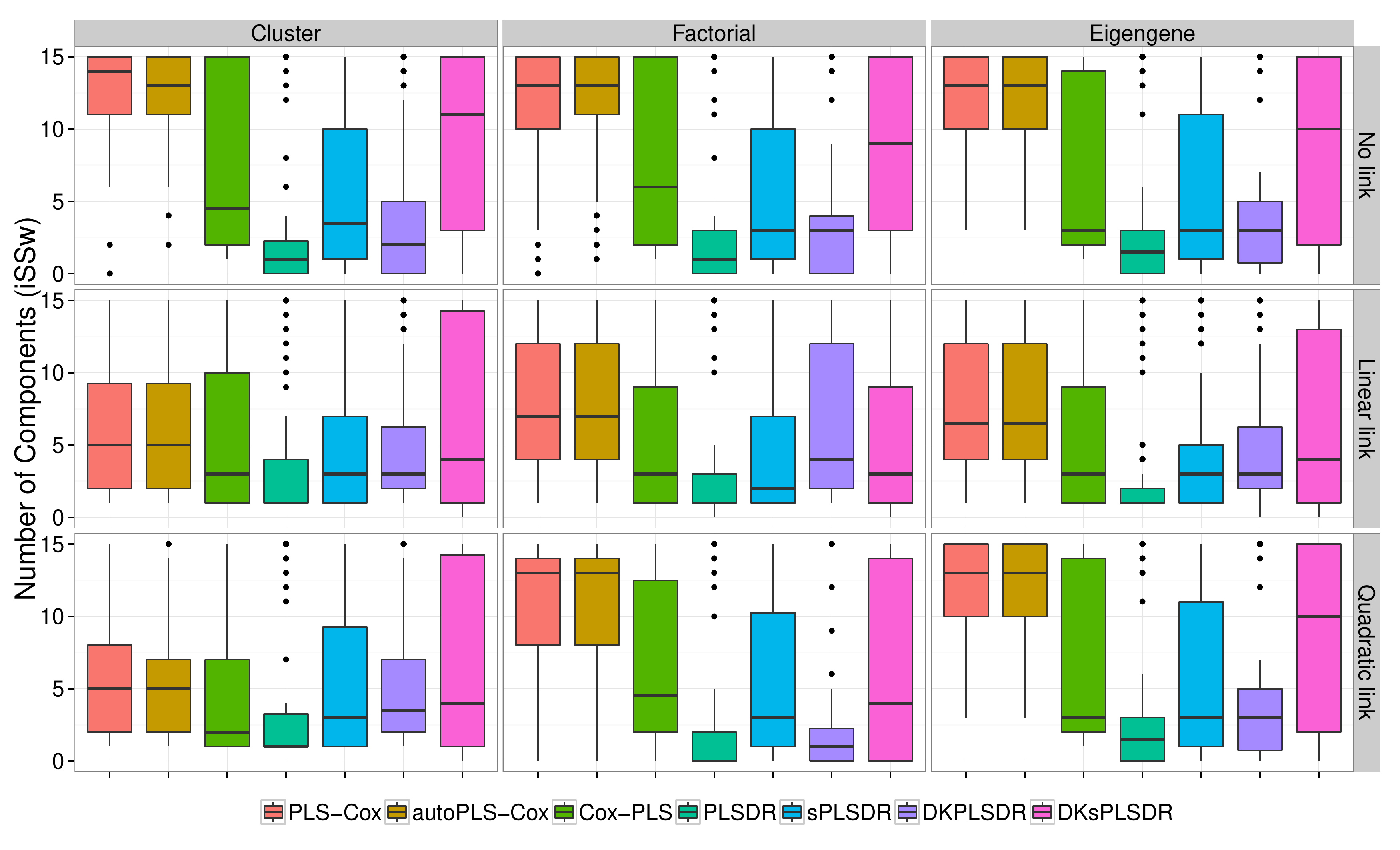}\phantomcaption\label{NbrComp_iSchmidSw}}}
\vspace{-.5cm}
\caption*{\hspace{-.5cm}\mbox{Figure~\ref{NbrComp_iBSw}:  Nbr of comp, iBSw criterion. \qquad\qquad Figure~\ref{NbrComp_iSchmidSw}:  Nbr of comp, iSSw criterion.}}
\end{figure}

As a consequence, we had to search for other CV criteria (CVC) for the models featuring components. \cite{li2006} used the integrated area under the curves of time-dependent ROC curves (iAUCsurvROC, \cite{hea00}) to carry out his cross-validations, implemented in the \verb+survcomp+ package, \citep{schr11}. Apart from that criterion (Figure \ref{NbrComp_AUCsurvROCtest}) we added five other integrated AUC measures: integrated \citeauthor{chdi06}'s (\citeyear{chdi06}) estimator (iAUCCD, Figure \ref{NbrComp_AUCcd}), integrated \citeauthor{huch10}'s (\citeyear{huch10}) estimator (iAUCHC, Figure \ref{NbrComp_AUChc}), integrated \citeauthor{sozh08}'s (\citeyear{sozh08}) estimator (iAUCSH, Figure \ref{NbrComp_AUCsh}), integrated \citeauthor{uno2007}'s (\citeyear{uno2007}) estimator (iAUCUno, Figure \ref{NbrComp_AUCUno}) and integrated \citeauthor{heager05}'s (\citeyear{heager05}) estimator (iAUCHZ, Figure \ref{NbrComp_AUChztest}) of cumulative/dynamic AUC for right-censored time-to-event data, implemented in the \verb+survAUC+ package, \cite{pota12}, and the \verb+risksetROC+ package, \cite{heag2012}. We also studied two versions of two prediction error criteria, the integrated (un)weighted Brier Score (\cite{graf99}, \cite{gesc06}, iBS(un)w, integrated (un)weighted squared deviation between predicted and observed survival, Figures \ref{NbrComp_iBSunw} and \ref{NbrComp_iBSw}) and the integrated (un)weighted Schmid Score (\cite{schm11}, iSS(un)w, integrated (un)weighted absolute deviation between predicted and observed survival, Figure \ref{NbrComp_iSchmidSunw} and \ref{NbrComp_iSchmidSw}), also implemented in the \verb+survAUC+ package, \cite{pota12}.

\subsection{Analysis of the results}\label{cvtechniques}
The simulation results highlighted the integrated \citeauthor{sozh08}'s estimator of cumulative/dynamic AUC for right-censored time-to-event data (iAUCSH),  implemented in the \verb+survAUC+ package, \cite{pota12}, as the best CV criterion for the PLS-Cox and the autoPLS-Cox methods even though it behaves poorly in all the other cases.

As for the other models featuring components, the iAUCsurvROC, iAUCUno criterion exhibited the best performances.

The two unweighted criteria iBSunw and iSSunw uniformly fail for all the models.

The iBSw criterion is too conservative and wrongly selects null models in more than half of the cases in the linear link scheme and in almost every times in the quadratic scheme.

The iSSw provides very poor results for Cox-PLS, sPLSDR and DKsPLSDR and average for PLSDR and DKPLSDR methods.

The two models SPLSDR and DKSPLSDR use an additional parameter: the thresholding parameter $\eta$. 
The same figures were produced for all the criteria (results not shown): both iAUCUno criterion and iAUCsurvROC criterion provided a reasonable spread for the $\eta$ parameter. 

\subsection{Recommendation}\label{recochanges}
In a word, this simulation campaign enables us to state the following recommendations to firmly improve the selection of the right number of components: use iAUCSH to cross-validate PLS-Cox or autoPLS-Cox models and either iAUCUno or iAUCsurvROC to cross-validate Cox-PLS, PLSDR, sPLSDR, DKPLSDR and DKsPLSDR. We implemented these recommendations (iAUCSH for PLS-Cox or autoPLS-Cox models and iAUCsurvROC for Cox-PLS, PLSDR, sPLSDR, DKPLSDR and DKsPLSDR) as the default cross validation techniques in the \verb+plsRcox+ package. We will apply them in the remaining of the article to assess goodness of fit of the model.

\clearpage

\clearpage

\section[Reassessing performance of (s)PLS based models]{Reassessing performance of (s)PLS \\ based models}\label{perfevalbench}
We will now provide evidence that the changes of the cross-validation criteria recommended in Section~\ref{recochanges} actually lead to performance improvements for the fitted models.
\subsection{Introduction to performance criteria analysis}\label{perfmeas}
We followed the methodological recommendations of \cite{vwie09} to design a simulation plan that ensures a good evaluation of the predictive performance of the models.
\begin{quote}
``The true evaluation of a predictor's performance is to be done on independent data. In the absence of independent data (the situation considered here) the predictive accuracy can be estimated as follows \cite{Dupuy2007}. The samples are split into mutually exclusive training and test sets. The gene expression and survival data of the samples in the training set are used to build the predictor. No data from the test set are used in predictor construction (including variable selection) by any of the methods compared. This predictor is considered to be representative of the predictor built on all samples (of which the training set is a subset). The test set is used to evaluate the performance of the predictor built from the training set: for each sample in the test set, survival is predicted from gene expression data. The predicted survival is then compared to the observed survival and summarized into an evaluation measure. To avoid dependency on the choice of training and test set, this procedure is repeated for multiple splits. The average of the evaluation measures resulting from each split is our estimate of the performance of the predictor built using the data from all samples.''
\end{quote}

As to the performance criteria themselves, \cite{schm11} made several points that we will take into account to carry out our performance comparison analysis.
\begin{quote}
``Evaluating the prognostic performance of prediction rules for continuous survival outcomes is an important topic of recent methodological discussion in survival analysis. The derivation of measures of prediction accuracy for survival data is not straightforward in the presence of censored observations (\cite{keoq88}, \citet{scst96}, \cite{rost04}). This is mainly due to the fact that traditional performance measures for continuous outcomes (such as the mean squared error or the $R^2$ fraction of explained variation) lead to biased predictions if applied to censored data (\citet{scst96}).

To overcome this problem, a variety of new approaches has been suggested in the literature. These developments can be classified into three groups: ``likelihood-based approaches'' (\citet{nage91}, \citet{xuoq99}, \citet{oqxu05}), ``ROC-based approa\-ches'' (\citet{hea00}, \citet{heager05}, \citet{cai2006}, \citet{uno2007}, \citet{pepe08}), and ``distance-based approa\-ches'' (\citet{korn90}, \citet{graf99}, \citet{sche00}, \citet{gesc06}, \citeyear{gesc07}, \cite{scho08}).

When using likelihood-based approaches, the log likelihood of a prediction model is related to the corresponding log likelihood obtained from a ``null model'' with no covariate information. ROC-based approaches use the idea that survival outcomes can be considered as time-dependent binary variables with levels -event- and -no event,- so that time-dependent misclassification rates and ROC curves can be computed for each threshold of a predictor variable of interest. If distance-based approaches are applied, a measure of prediction error is given by the distance between predicted and observed survival functions of the observations in a sample. None of these approaches has been adopted as a standard for evaluating survival predictions so far.''
\end{quote}

To assess the goodness of fit and prediction accuracy of all the methods, we found 23 performance measures (PM) that are commonly used (LRT, VarM, R2Nag, R2XO, R2OXS, iR2BSunw, iR2BSw, iRSSunw, iRSSw, iAUCCD, iAUCHC, iAUCSH, iAUCUno, IAUCHZ, iAUCSurvROC, C, UnoC, GHCI, SchemperV, iBSunw, iBSw, iSSunw, iSSw). We chose, on statistical grounds, 14 among them (LRT, R2XO, iR2BSw, iRSSw, iAUCCD, iAUCHC, iAUCSH, iAUCUno, IAUCHZ, iAUCSurvROC, GHCI, SchemperV, iBSw, iSSw) and reported the results of six indices of several kind: two $R^2$-like measure (a likelihood-based approach (LBA), R2XO, and a distance-based approach (DBA), iRSSw), a $C$ index (GHCI), two $iAUC$ (ROC-based approaches (ROCBA), iAUCCD and iAUCSurvROC), and an integrated robust prediction error (distance-based approach, iSSw), see Table~\ref{measselected}. The results for the remaining eight indices are similar to those shown. We now explain our process of selection of the performance criteria.


\begin{table}[!b]
\small
\centerfloat
\begin{tabular}{|l|c|ccc|ccc|}
\hline
Criterion&Type&\multicolumn{3}{c|}{As a Cross Validation Criterion}&\multicolumn{3}{c|}{As a Performance Measure}\\
Criterion&Type&Tested&Results&Recom. for&Is a&Selected on&Results\\
&&&&&PM ?&statistical&\\
&&&&&&grounds&\\
\hline
CVLL&LBA&{\bf Yes}&{\bf Yes}&&No&No&No\\
vHCVLL&LBA&{\bf Yes}&{\bf Yes}&&No&No&No\\
LRT $p$-value&LBA&No&No&&{\bf Yes}&{\bf Yes}&No\\
VarM&LBA&No&No&&{\bf Yes}&No&No\\
R2Nag&LBA&No&No&&{\bf Yes}&No&No\\
R2XO&LBA&No&No&&{\bf Yes}&{\bf Yes}&{\bf Yes}\\
R2OXS&LBA&No&No&&{\bf Yes}&No&No\\
iR2BSunw&DBA&No&No&&{\bf Yes}&No&No\\
iR2BSw&DBA&No&No&&{\bf Yes}&{\bf Yes}&No\\
\textit{iRSSunw}&DBA&No&No&&\textit{\textbf{New}}&No&No\\
\textit{iRSSw}&DBA&No&No&&\textit{\textbf{New}}&{\bf Yes}&{\bf Yes}\\
iAUCCD&ROCBA&{\bf Yes}&{\bf Yes}&&{\bf Yes}&{\bf Yes}&{\bf Yes}\\
iAUCHC&ROCBA&{\bf Yes}&{\bf Yes}&&{\bf Yes}&{\bf Yes}&No\\
iAUCSH&ROCBA&{\bf Yes}&{\bf Yes}&PLS$-$Cox,&{\bf Yes}&{\bf Yes}&No\\
&&&&autoPLS$-$Cox&&&\\
iAUCUno&ROCBA&{\bf Yes}&{\bf Yes}&(DK)(s)PLSDR&{\bf Yes}&{\bf Yes}&No\\
&&&&Cox$-$PLS&&&\\
iAUCHZ&ROCBA&{\bf Yes}&{\bf Yes}&&{\bf Yes}&{\bf Yes}&No\\
iAUCSurvROC&ROCBA&{\bf Yes}&{\bf Yes}&(DK)(s)PLSDR&{\bf Yes}&{\bf Yes}&{\bf Yes}\\
&&&&Cox$-$PLS&&&\\
C&ROCBA&No&No&&{\bf Yes}&No&No\\
UnoC&ROCBA&No&No&&{\bf Yes}&No&Sup. Info.\\
GHCI&ROCBA&No&No&&{\bf Yes}&{\bf Yes}&{\bf Yes}\\
SchemperV&DBA&No&No&&{\bf Yes}&{\bf Yes}&No\\
iBSunw&DBA&{\bf Yes}&{\bf Yes}&&{\bf Yes}&No&No\\
iBSw&DBA&{\bf Yes}&{\bf Yes}&&{\bf Yes}&{\bf Yes}&Sup. Info.\\
iSSunw&DBA&{\bf Yes}&{\bf Yes}&&{\bf Yes}&No&No\\
iSSw&DBA&{\bf Yes}&{\bf Yes}&&{\bf Yes}&{\bf Yes}&{\bf Yes}\\
\hline
Total Number&25&12&&12&23&14&6 (+2 SI)\\
\hline
\end{tabular}
\caption{Criteria and their use in the cross validation step or as a performance measures for assessing the quality of the model.\label{measselected}}
\end{table}


\eject

\subsection{Selection of performance criteria}

The likelihood ratio test (LRT, \cite{Lehmann2005}) evaluates the null hypothesis $\mathcal{H}_0 \ : \ \beta = 0$, i.e. the built predictor has no effect on survival. The null hypothesis is evaluated using the likelihood ratio test statistic $LLR(\hat\beta)=-2(l(0)-l(\hat\beta))$, with $l(.)$ denoting the value of the log-likelihood function. Under the null hypothesis this test statistic has a $\chi^2$ distribution, which is used to calculate the $p$-value. The $p$-value summarizes the evidence against $\mathcal{H}_0$: the lower the $p$-value the more probable that $\mathcal{H}_0$ is not true.The $p$-value of the likelihood ratio test has been used as an evaluation measure for predictive performance of gene expression based predictors of survival by many others \cite{Bair2004,bove07,Park2002,seg06}.

\bigskip

In the Cox model, an alternative measure of predictive performance is the variance of the martingale residuals (VarM, \textit{cf.} section~\ref{matingaleresidualsdef}). 
As in \cite{vwie09}, we found that this measure is not able to discriminate very well between good and poor predictors in the considered setting (data not shown). It is therefore omitted here.

\bigskip

To quantify the proportion of variability in survival data of the test set that can be explained by the predictor, we use the coefficient of determination (henceforth called $R^2$). A predictor with good predictive performance explains a high proportion of variability in the survival data of the test set, and vice versa a poor predictor explains little variability in the test set. However, the traditional definition of the $R^2$ cannot be used in the context of censored data and modified criteria have been proposed in the past. Three types of likelihood-based $R^2$ coefficients for right-censored time-to-event data are were put forward (R2NAG, R2XO and R2OXS). 
\begin{itemize}
\item The coefficient (R2Nag) proposed by \cite{nage91}:
\begin{equation}
R^2_{Nag}=1-\exp{\left(-\frac2n(l(\hat\beta)-l(0))\right)}
\end{equation}
where $l(.)$ denotes the log-likelihood function.
\item The coefficient (R2XO) proposed by \cite{xuoq99} that is restricted to proportional hazards regression models, because here the means of squared residuals $MSE$ in the $R^2_{adj}$ measure for linear regression are replaced by the (weighted) sums of squared \textit{Schoenfeld} residuals, denoted by $J(\beta)$:
\begin{equation}
R^2_{XO}=1-\frac{J(\hat\beta)}{J(0)}\cdot
\end{equation}
\item The coefficient (R2OXS) proposed by \cite{oqxu05} who replaced the number of observations $n$ by the number of events $e$:
\begin{equation}
R^2_{OXS}(\hat\beta)=1-\exp{\left(-\frac2e(l(\hat\beta)-l(0))\right)}=1-\left(\frac{L(\hat\beta)}{L(0)}\right)^{-2/e}\cdot
\end{equation}
\end{itemize}
All three were implemented in the \verb+survAUC+ package, \cite{pota12}. Others have also used these modified $R^2$ statistics to assess predictive performance of gene expression based predictors on survival \cite{Bair2004,seg06}.

\cite{hiel10} carried out a comparison of the properties of these three coefficients. In a word, R2Nag is strongly influenced by censoring (negative correlation with censoring); R2OXS is less influenced by censoring and exhibits a positive correlation with censoring. From those three R2XO is the less influenced by censoring. As a consequence, we selected the R2XO as the $R^2$-like measure to compare the models. 

\bigskip

The weighted Brier score $BSw(t)$ (\cite{Brier1950, Hothorn2004, Radespiel-Troger2003}) is a distance-based measure of prediction error that is based on the squared deviation between survival functions. It is defined as a function of time $t>0$ by
\begin{equation}
BSw(t)=\frac1n\sum_{i=1}^n\left[\frac{\hat S(t \mid \mathbf{X}_i)^2I(t_i\leqslant t\wedge \delta_i=1)}{\hat G(t_i)}+\frac{(1-\hat S(t \mid \mathbf{X}_i))^2I(t_i>t)}{\hat G(t_i)}\right]\label{eqBSdef}
\end{equation}
where $\hat G(.)$ denotes the Kaplan-Meier estimate of the censoring distribution, that is the Kaplan–Meier estimate based on the observations $(t_i,1-\delta_i)$ and $I$ stands for the indicator function. The expected Brier score of a prediction model which ignores all predictor variables corresponds to the KM estimate. To derive the unweighted Brier score, $BSunw(t)$, clear the $\hat G(t_i)$ value of the denominators.

The Schmid score $SS(t)$ (\cite{schm11}) is a distance-based measure of prediction error that is based on the absolute deviation between survival functions, instead of the squared one for the Brier-Score. It is a robust improvement over the following empirical measure of absolute deviation between survival functions that was suggested by \cite{sche00} as a function of time $t>0$ by:
\begin{equation}
SH(t)=\frac1n\sum_{i=1}^n\left[\frac{\hat S(t \mid \mathbf{X}_i) I(t_i\leqslant t\wedge \delta_i=1)}{\hat G(t_i)}+\frac{(1-\hat S(t \mid \mathbf{X}_i))I(t_i>t)}{\hat G(t_i)}\right]\label{eqSHdef}
\end{equation}
where $\hat G(.)$ denotes the Kaplan-Meier estimate of the censoring distribution which is based on the observations $(t_i,1-\delta_i)$ and $I$ stands for the indicator function. With the same notations, the Schmid score is defined as a function of time $t>0$ by:
\begin{equation}
SS(t)=\frac1n\sum_{i=1}^n\lvert I(t_i> t) - \hat S(t \mid \mathbf{X}_i) \rvert \left[\frac{I(t_i\leqslant t\wedge \delta_i=1)}{ \hat G(t_i^-)}+\frac{I(t_i>t)}{\hat G(t_i)}\right]\label{eqSSdef}
\end{equation}
where $t_i^-$ is a survival time that is marginally smaller than $t_i$. To derive the unweighted Schmid score, $SSunw(t)$, clear the $\hat G(t_i^-)$ and $\hat G(t_i)$ values of the denominators.

The values of the Brier-Score range between 0 and 1. Good predictions at time $t$ result in small Brier-Scores. The numerator of the first summand is the squared predicted probability that individual $i$ survives until time $t$ if he actually died (uncensored) before $t$, or zero otherwise. The better the survival function is estimated, the smaller is this probability. Analogously, the numerator of the second summand is the squared probability that individual $i$ dies before time $t$ if he was observed at least until $t$, or zero otherwise. Censored observations with survival times smaller than $t$ are weighted with 0. The Brier-score as defined in Eq. \ref{eqBSdef} depends on $t$. It makes sense to use the integrated Brier-Score ($IBS$) given by
\begin{equation}
IBS = \frac1{\max{(t_i)}}\int_0^{\max(t_i)}BS(t)dt.
\end{equation}
as a score to assess the goodness of the predicted survival functions of all observations at every time $t$ between 0 and $\max{(t_i)}$, $i=1,\ldots,N$. Note that the $IBS$ is also appropriate for prediction methods that do not involve Cox regression models: it is more general than the $R^2$ and the $p$-value criteria associated to the log likelihood test and has thus become a standard evaluation measure for survival prediction methods (\cite{Hothorn2006,Schumacher2007}).

Denoting by $BS^0$, the Kaplan-Meier estimator based on the $t_i$, $\delta_i$, which corresponds to a prediction without covariates, we first define $R^2_{BS}$ for all $t>0$: 
\begin{equation}
R^2_{BS}(t)=1-\frac{BS(t)}{BS^0(t)}\cdot
\end{equation}
Then the integrated iR2BSw, \cite{graf99}, is defined by:
\begin{equation}
iR2BSw=\frac{1}{\max(t_i)}\int_0^{\max(t_i)}R^2_{BS}(t)dt.
\end{equation}
This criterion has already been used in \cite{bove07} and \cite{lll11}. The integrated iR2BSw is slightly influenced by censoring, \cite{hiel10}, and, as a measure based on the the quadratic norm, not robust. 

As a consequence, we propose and use a similar measure based on the Schmid score, the integrated R Schmid Score weighted (iRSSw), by turning the traditional $R^2$, derived from the quadratic norm, into the R coefficient of determination for least absolute deviation, introduced by \cite{mcke87}. Denoting by $SS^0$ the Schmid score which corresponds to a prediction without covariates, we first define $R_{SS}$ for all $t>0$: 
\begin{equation}
R_{SS}(t)=1-\frac{SS(t)}{SS^0(t)}\cdot
\end{equation}
Then the integrated iRSSw, is defined by:
\begin{equation}
iRSSw=\frac{1}{\max(t_i)}\int_0^{\max(t_i)}R_{SS}(t)dt.
\end{equation}

The C-index provides a global assessment of a fitted survival model for the continuous event time rather than focuses on the prediction of t-year survival for a fixed time and is arguably the most widely used measure of predictive accuracy for censored data regression models. It is a rank-correlation measure motivated to quantify the correlation between the ranked predicted and observed survival times. The C index estimates the probability of concordance between predicted and observed responses. A value of 0.5 indicates no predictive discrimination and a value of 1.0 indicates perfect separation of patients with different outcomes. A popular nonparametric C-statistic for estimating was proposed by \cite{harr96}. It is  computed by forming all pairs $\{(y_i,x_i,\delta_i), (y_j,x_j,\delta_j)\}$ of the observed data, where the smaller follow-up time is a failure time and defined as:
\begin{equation}
c=\frac{\sum_{1\leqslant i<j \leqslant n}I(y_i<y_j)I(\hat\beta'\mathbf{X_i}>\hat\beta'\mathbf{X_j})I(\delta_i=1)+I(y_j<y_i)I(\hat\beta'\mathbf{X_j}>\hat\beta'\mathbf{X_i})I(\delta_j=1)}{\sum_{1\leqslant i<j \leqslant n}I(y_i<y_j)I(\delta_i=1)+I(y_j<y_i)I(\delta_j=1)}
\end{equation}
We used the improved version (GHCI) by \cite{gohe05} for the Cox proportional hazards models as a performance comparison criterion. Their estimator $K_n(\hat\beta)$ is a function of the regression parameters and the covariate distribution only and does not use the observed event and censoring times. For this reason it is asymptotically unbiased, unlike Harrell's C-index based on informative pairs. It focuses on the concordance probability as a measure of discriminatory power within the framework of the Cox model. The appeal of this formulation is that it provides a stable estimator of predictive accuracy that is easily computed:
\begin{equation}
K_n(\hat\beta)=\frac2{n(n-1)}\sum_{1\leqslant i<j \leqslant n}\left\{\frac{I(\hat\beta'(\mathbf{X_j}-\mathbf{X_i})<0)}{1+\exp{(\hat\beta'(\mathbf{X_j}-\mathbf{X_i}))}}+\frac{I(\hat\beta'(\mathbf{X_i}-\mathbf{X_j})<0)}{1+\exp{(\hat\beta'(\mathbf{X_i}-\mathbf{X_j}))}}\right\}\cdot
\end{equation}
In contrast to Harrell's C-index, the effect of the observed times on $K_n(\hat\beta)$ is mediated through
the partial likelihood estimator $\hat\beta$, and, since the effect of censoring on the bias of $\hat\beta$ is negligible, the measure is robust to censoring. In addition, $K_n(\hat\beta)$ remains invariant under monotone transformations of the survival times.

\subsection{Ranking the performance of the CV criteria}
We stated several recommendations, in Section~\ref{cvcritchoice} based of the accuracy of the selection of the number of components. Selecting the right number of components is a goal \textit{per se}. 

\begin{figure}[!tpb]
\centerline{{\includegraphics[width=.75\columnwidth]{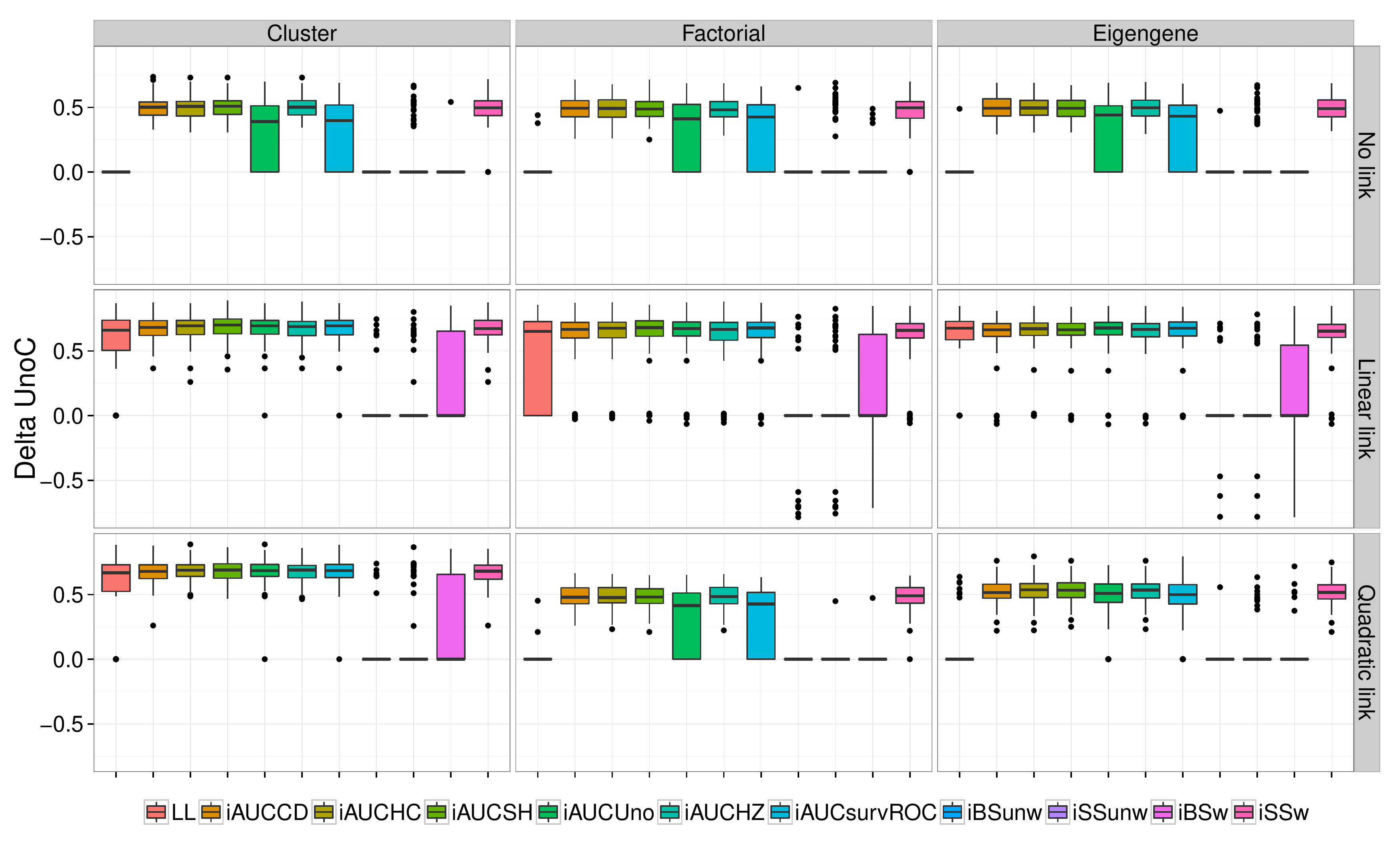}\phantomcaption\label{incplsRcox_UnoCstat}}\qquad{\includegraphics[width=.75\columnwidth]{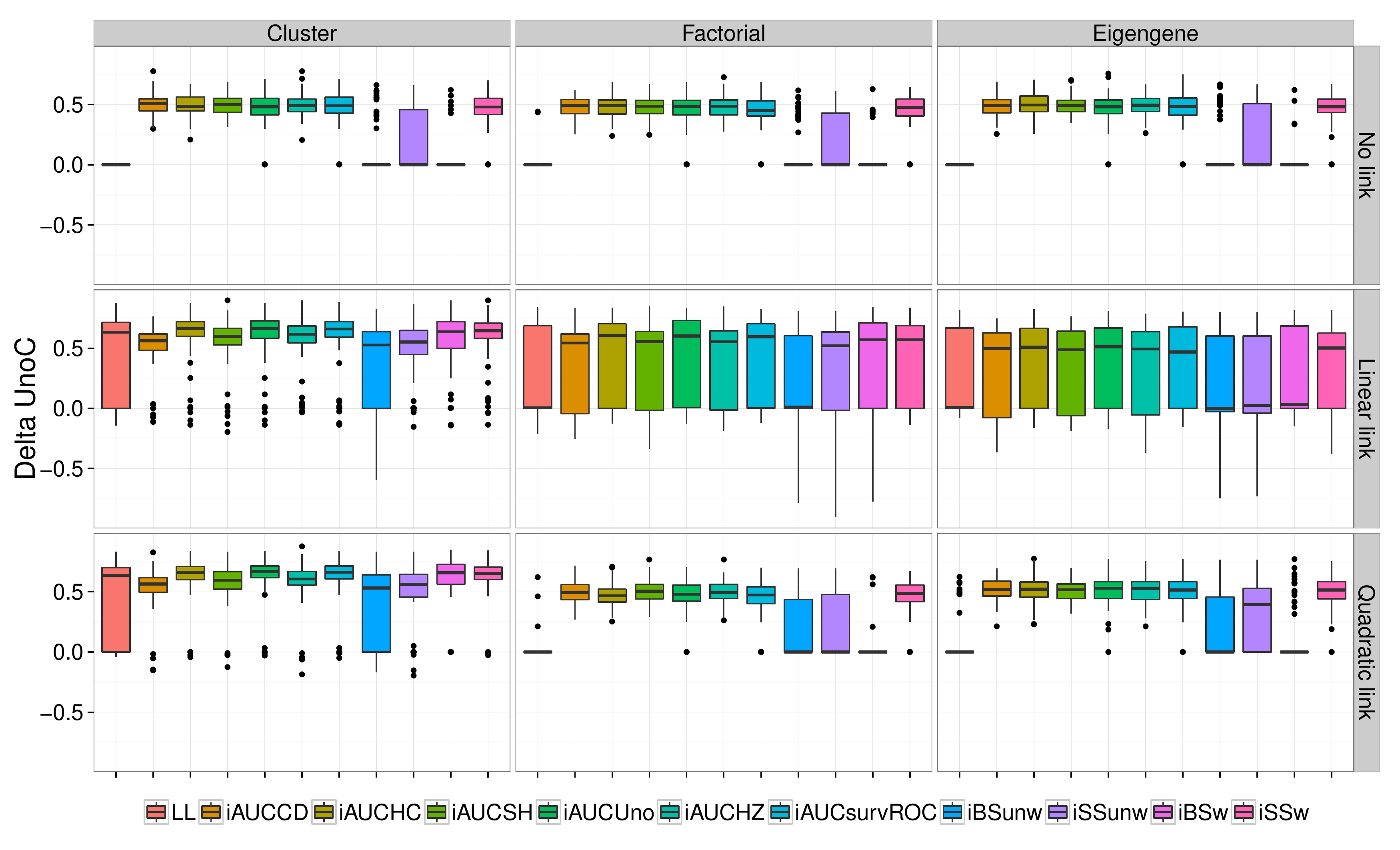}\phantomcaption\label{inccoxsplsDR_UnoCstat}}}
\vspace{-.5cm}
\caption*{\hspace{-2cm}\mbox{Delta of UnoC (CV criteria $-$ vHCVLL value). \quad Figure~\ref{incplsRcox_UnoCstat} (left): PLS$-$Cox. Figure~\ref{inccoxsplsDR_UnoCstat} (right): sPLSDR.}}
\end{figure}

\begin{figure}[!tpb]
\centerline{{\includegraphics[width=.75\columnwidth]{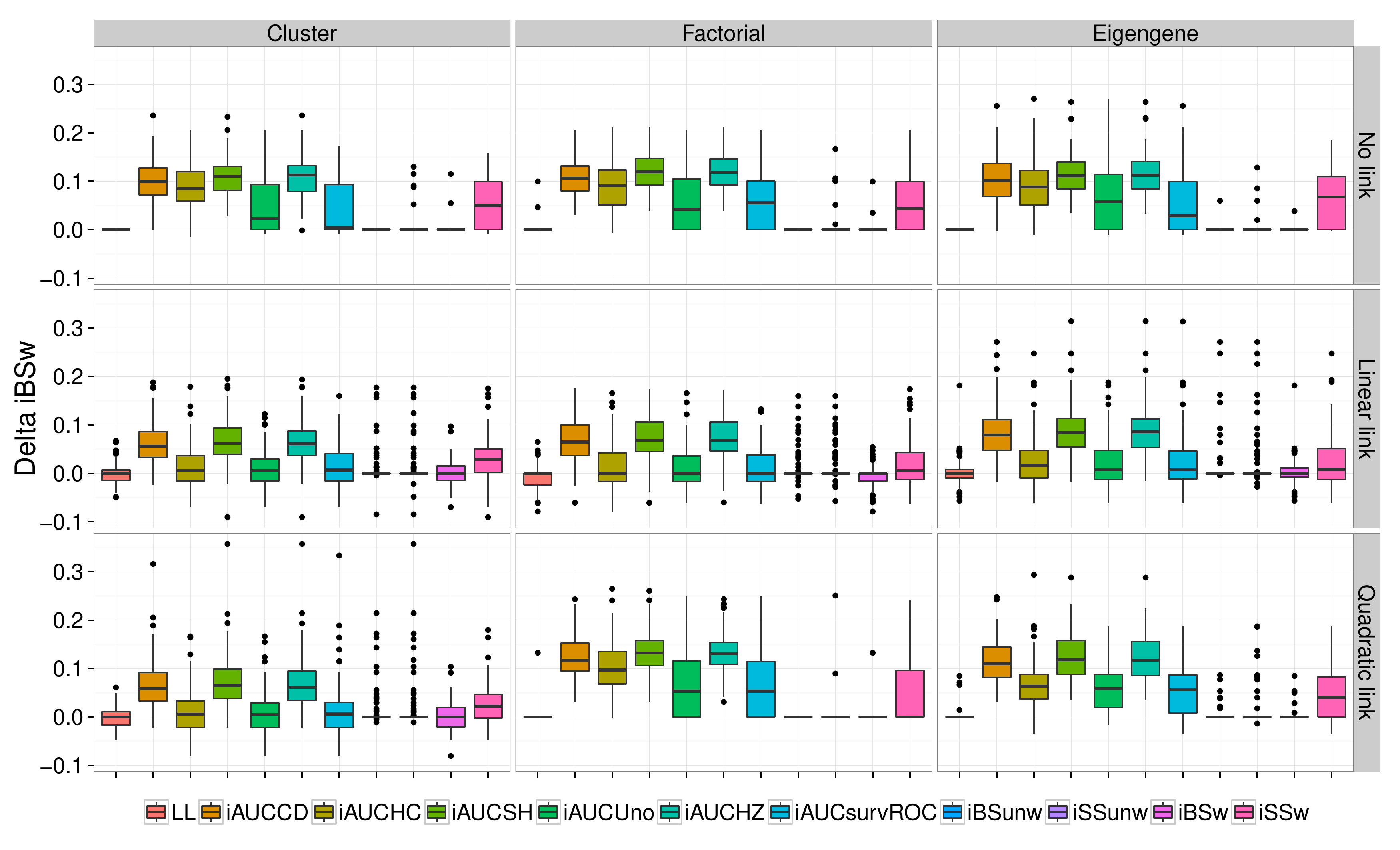}\phantomcaption\label{inciBSWcoxplsDR}}\qquad{\includegraphics[width=.75\columnwidth]{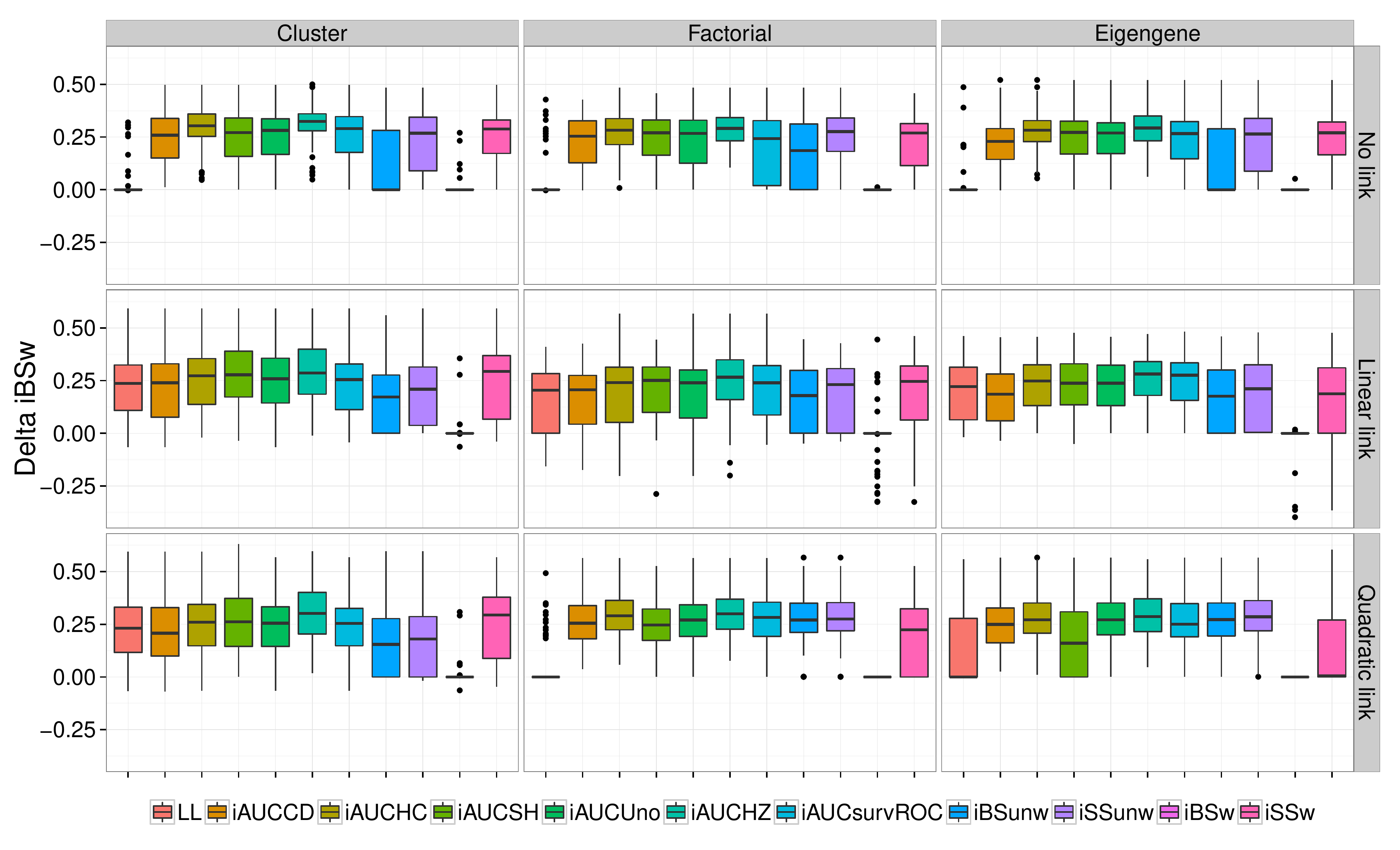}\phantomcaption\label{inciBSWcoxDKsplsDR}}}
\vspace{-.5cm}
\caption*{\hspace{-2cm}\mbox{Delta of iBSW (CV criteria $-$ vHCVLL value). \quad Figure~\ref{inciBSWcoxplsDR} (left): PLSDR. Figure~\ref{inciBSWcoxDKsplsDR} (right): DKsPLSDR.}}
\end{figure}

\begin{figure}[!tpb]
\centerline{{\includegraphics[width=.75\columnwidth]{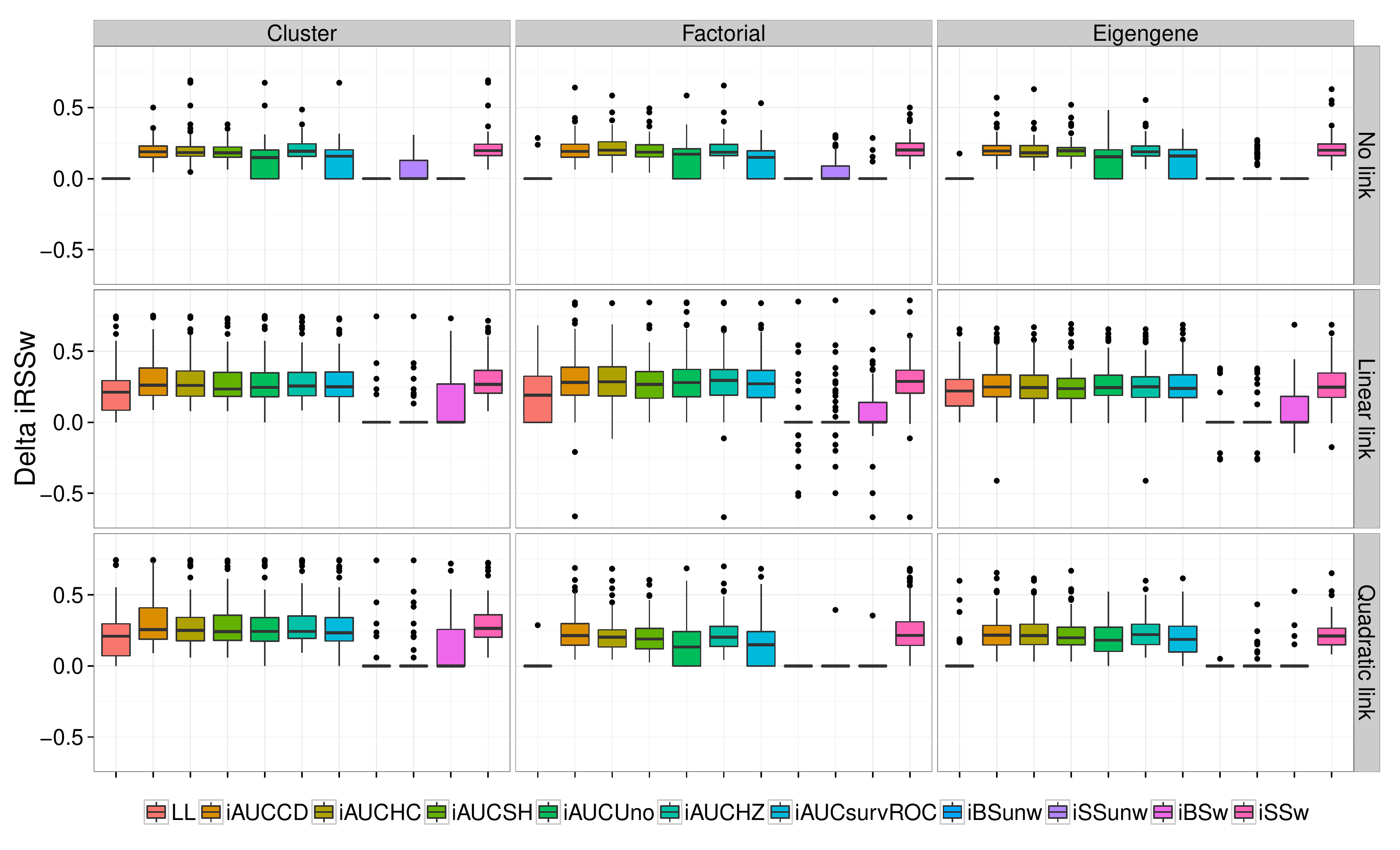}\phantomcaption\label{inciRSSWautoplsRcox}}\qquad{\includegraphics[width=.75\columnwidth]{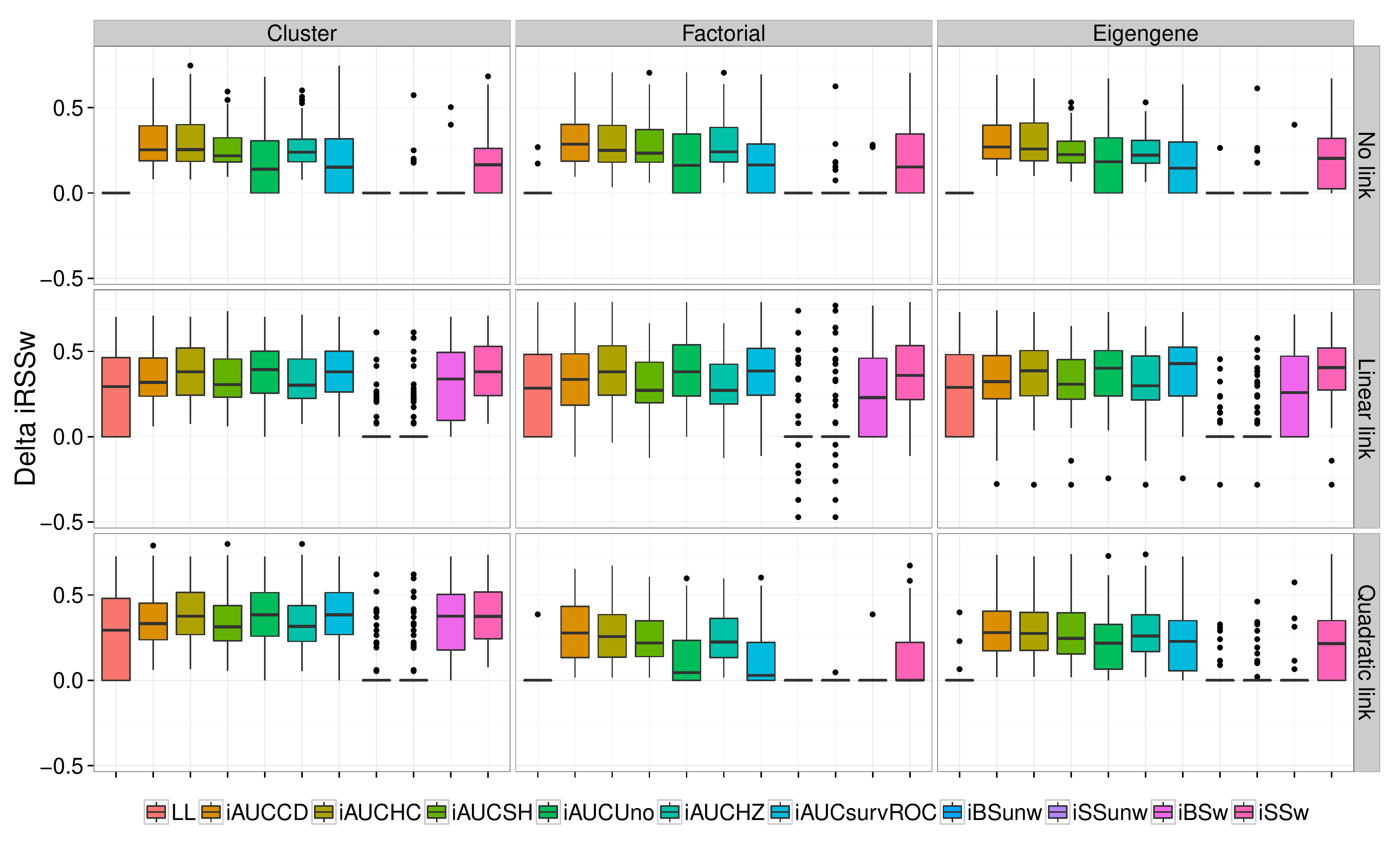}\phantomcaption\label{inciRSSWplsDR}}}
\vspace{-.5cm}
\caption*{\hspace{-2cm}\mbox{Delta of iRSSW (CV criteria $-$ vHCVLL value). \quad Figure~\ref{inciRSSWautoplsRcox} (left): autoPLS$-$Cox. Figure~\ref{inciRSSWplsDR} (right): PLSDR.}}
\end{figure}

\begin{figure}[!tpb]
\centerline{{\includegraphics[width=.75\columnwidth]{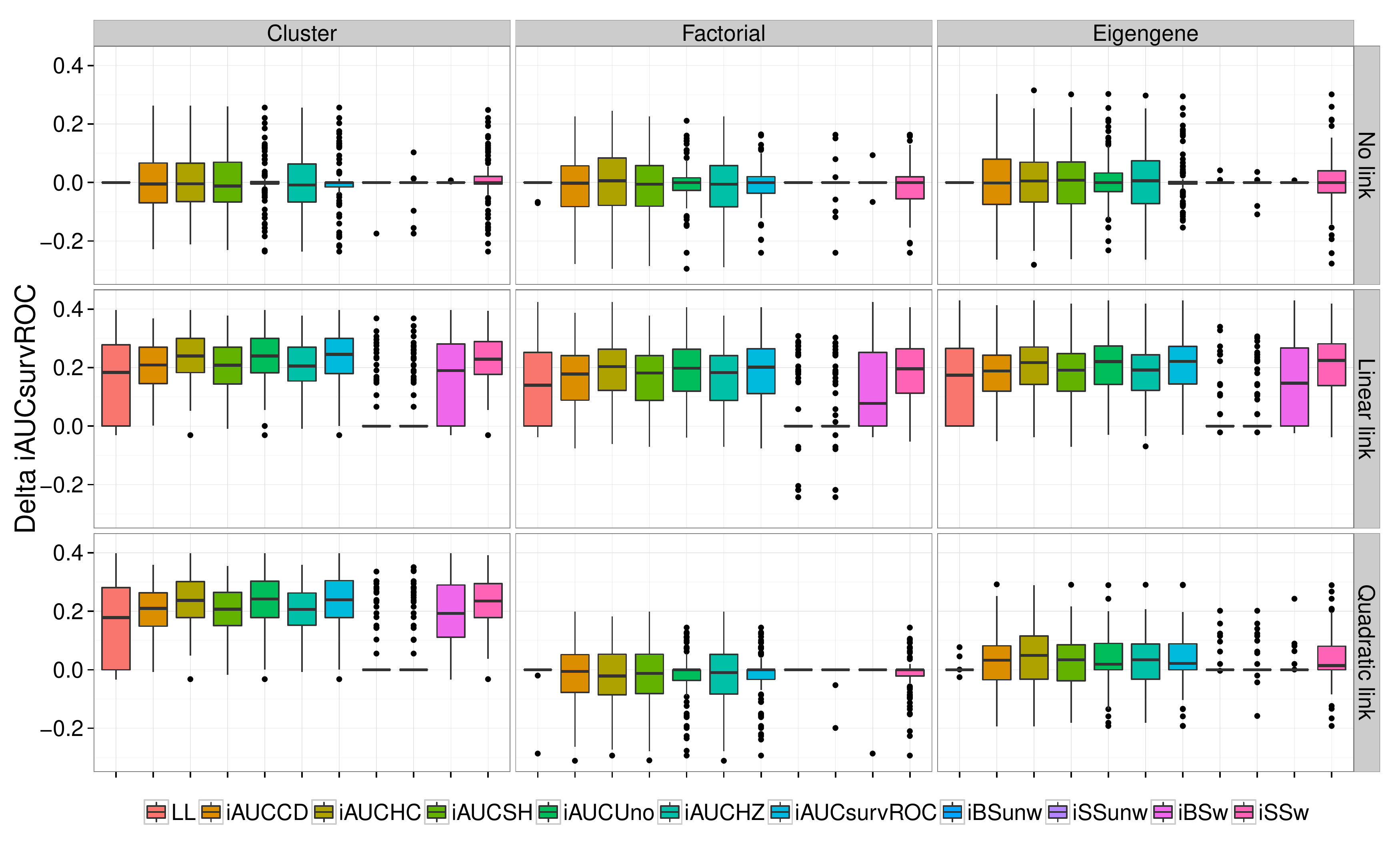}\phantomcaption\label{inciAUCSurvROCplsDR}}\qquad{\includegraphics[width=.75\columnwidth]{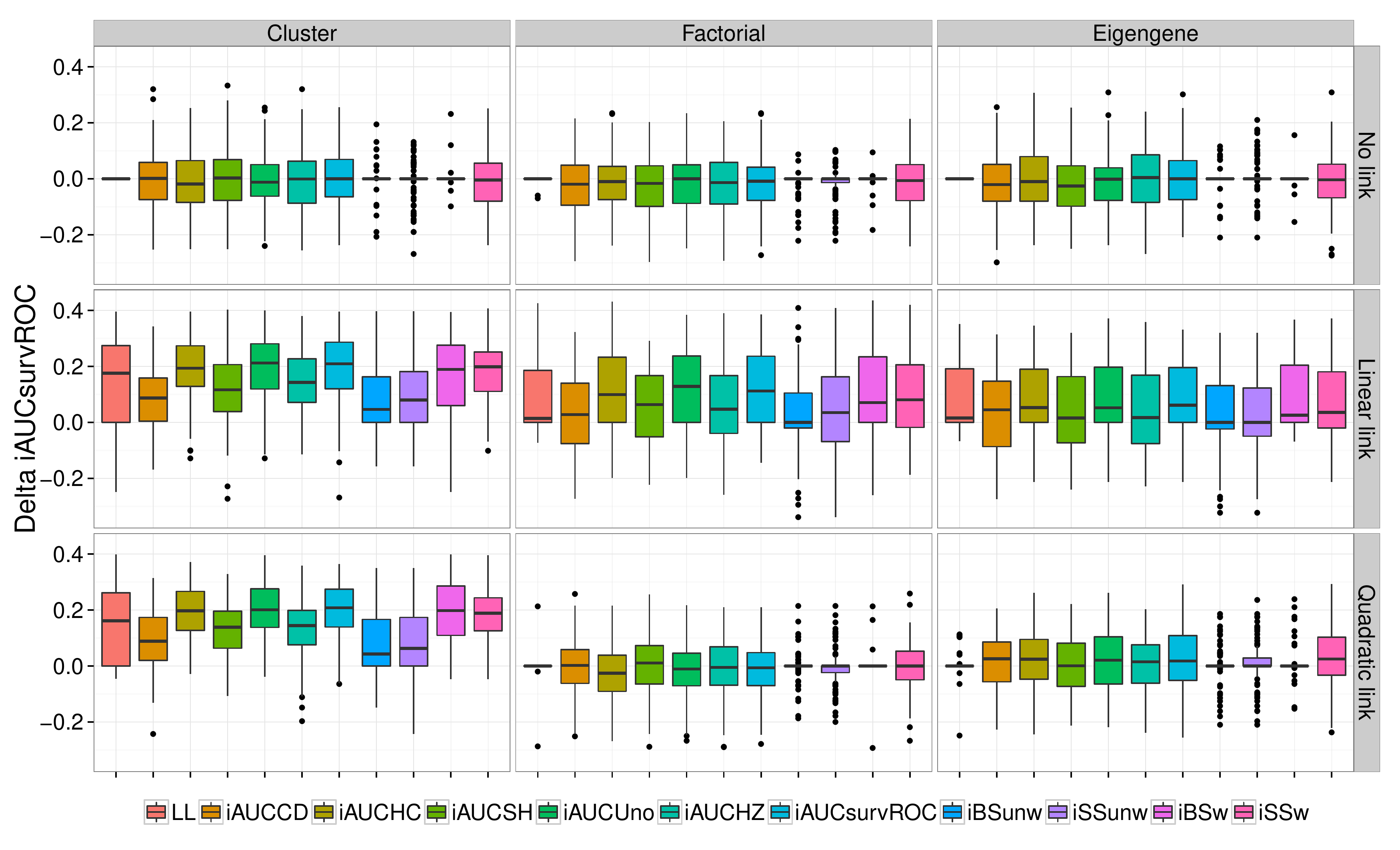}\phantomcaption\label{inciAUCSurvROCsplsDR}}}
\vspace{-.5cm}
\caption*{\hspace{-2cm}\mbox{Delta of iAUCSurvROC (CV criteria $-$ vHCVLL value). \quad Figure~\ref{inciAUCSurvROCplsDR} (left): PLSDR. Figure~\ref{inciAUCSurvROCsplsDR} (right): sPLSDR.}}
\end{figure}

Moreover, these recommendations are also relevant from a performance criteria point of view (see Section \ref{perfmeas}) as the following analysis showed.
\begin{enumerate}
\item For all the models and simulation types, we carried out the cross-validation according to all of the 12 criteria and, for each of these criteria, we derived the value of all the 14 the performance measures.
\item In order to lay the stress on the improvements of performance made when switching from the classic and the van Houwelingen log likelihood cross validation techniques to the recommended ones, we computed, for every datasets and models, all the paired differences between CVLL or vHCVLL and the eleven other CV techniques.
\begin{itemize}
\item Paired comparison with CVLL. For every simulated dataset we evaluated: Delta = Performance Measure(with any CV criteria $\neq$ CVLL) $-$ Performance Measure(with CVLL)
\item Paired comparison with vHCVLL. For every simulated dataset we evaluated: Delta = Performance Measure(with any CV criteria $\neq$ vHCVLL) $-$ Performance Measure(with vHCVLL)
\end{itemize}
An analysis of these results showed a steady advantage of the recommended criteria versus either CVLL or vHCVLL especially in the linear and quadratic cases.\\
In the case of paired comparison with vHCVLL and for some couples of the type (performance measure, model), namely (UnoC, PLS$-$Cox), (UnoC, sPLSDR), (iBSW, PLSDR), (iBSW, DKsPLSDR), (iRSSW, autoPLS$-$Cox), (iRSSW, PLSDR), (iAUCSurvROC, PLSDR) and (iAUCSurvROC, sPLSDR), those deltas are plotted on Figures \ref{incplsRcox_UnoCstat}, \ref{inccoxsplsDR_UnoCstat}, \ref{inciBSWcoxplsDR}, \ref{inciBSWcoxDKsplsDR}, \ref{inciRSSWautoplsRcox}, \ref{inciRSSWplsDR}, \ref{inciAUCSurvROCplsDR} and \ref{inciAUCSurvROCsplsDR}.
\end{enumerate}

\vfill\eject

\subsection{Performance comparison revisited}\label{perfmeasrev}
\subsubsection{Selection of competing benchmark methods}
\cite{coxnet11}, introduced the \verb+coxnet+ procedure, which is an elastic net-type procedure for the Cox model, in a similar but not equivalent way than two competing ones: \verb+coxpath+ (\verb+glmpath+ package, \citealp{park07}) and \verb+penalized+ (\verb+penalized+ package, \citealp{goeman10}). In Section 3 of the same article, these authors extensively compared \verb+coxnet+ to \verb+coxpath+ and to \verb+penalized+ for the lasso penalty that is the only one relevant for these comparisons since the three procedures use different elastic net penalties. Their results show tremendous timing advantage for \verb+coxnet+ over the two other procedures. The \verb+coxnet+ procedure was integrated in the \verb+glmnet+ package (\citealp{glmnet10}) and is called in the \verb+R+ language by applying the \verb+glmnet+ function with the option \verb+family=cox+: \verb+coxnet+ is \verb+glmnet+ for the Cox model. The timing results of \cite{coxnet11} on both simulated and real datasets show some advantage to \verb+coxpath+ over \verb+penalized+.

As to pure lasso-type penalty algorithms, we selected two of them: ``Univariate Shrinkage in the Cox Model for High Dimensional data'' (\verb+uniCox+, \citealp{uniCox}) and `Gradient Lasso for Cox Proportional Hazards Model'' (\verb+glcoxph+, \citealp{sohn09}). 

The \verb+uniCox+ package implements ``Univariate Shrinkage in the Cox Model for High Dimensional data'' (\citealp{uniCox}). Being ``essentially univariate'', it differs from applying a classical lasso penalty when fitting the Cox model and hence from both coxnet/glmnet and coxpath/glmpath. It can be used on highly correlated and even rectangular datatsets. 

In their article, \cite{sohn09}, show that the \verb+glcoxph+ package is very competitive compared with popular existing methods \verb+coxpath+ by \cite{park07} and \verb+penalized+ by \cite{goeman10} in its computational time, prediction and selectivity. As a very competitive procedure to \verb+coxpath+, that we included in our benchmarks, and since no comparisons were carried out with \verb+coxnet+, we selected \verb+glcoxph+ as well.

Cross validation criteria were recommended for several of our benchmark methods by their authors. We followed these recommendations -classic CV partial likelihood for coxpath, glcoxph and uniCox; van Houwelingen CV partial likelihood for coxnet with both the $\lambda_{min}$, the value of $\lambda$ that gives minimum of the mean cross-validated error, or $\lambda_{1se}$, the largest value of $\lambda$ such that the cross-validated error is within 1 standard error of the minimum of the mean cross-validated error, criteria- and used the same 7 folds fo the training set as those described in Section~\ref{sechypcv} for the other models.

It seemed unfair to compare the methods using a performance measure that is recommended as a cross-validation criterion for some, but not all, of them. Hence, we decided not to use any of the three recommended cross-validation criteria iAUCSH, iAUCUno or iAUCsurvROC -even if it has already been used by \cite{li2006}- as a performance measure, in order to strive to perform fair comparisons with the methods that are recommended to be cross validated using partial likelihood with either the classic or van Houwelingen technique.  

As a consequence and in order to still provide results for a ROC-based performance measure on a fair basis, we selected the \citeauthor{chdi06}'s (\citeyear{chdi06}) estimator of cumulative/dynamic AUC for right-censored time-to-event data in a form restricted to Cox regression. The iAUCCD summary measure is given by the integral of AUC on $[0, \max(\text{times})]$ (weighted by the estimated probability density of the time-to-event outcome).

\subsubsection{Results}
For coxnet, coxlars or ridgecox with both the $\lambda_{min}$ or $\lambda_{1se}$ CV criteria, the $\lambda_{min}$ criterion yield similar yet superior results than the $\lambda_{1se}$ one whose main default is to select too often no explanatory variable (a null model) for the linear or quadratic links. As a consequence, we only reported results for the former one.

\begin{figure}[!tpb]
\centerline{{\includegraphics[width=.75\columnwidth]{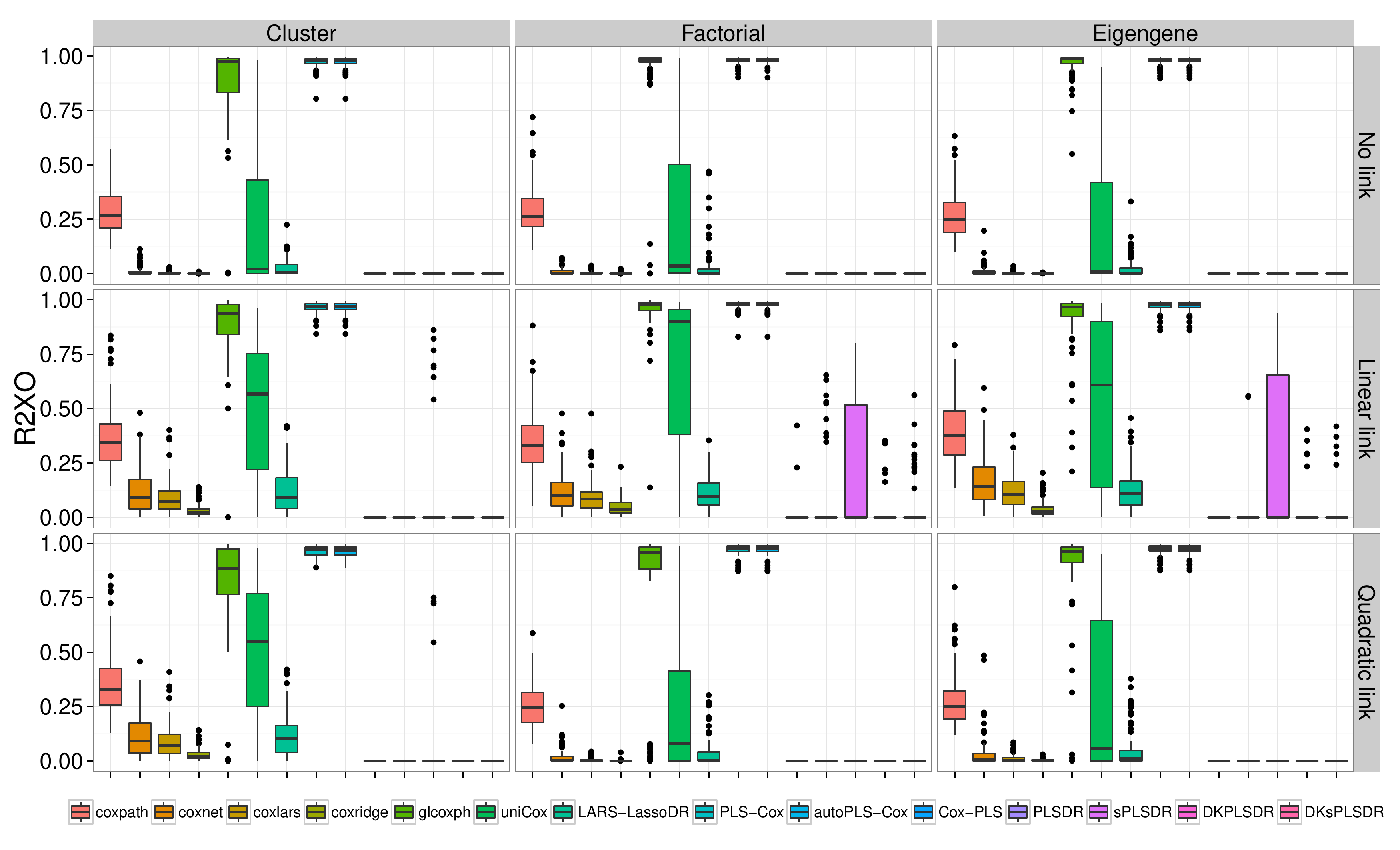}\phantomcaption\label{modmins_vanHcvll_R2XO}}\qquad{\includegraphics[width=.75\columnwidth]{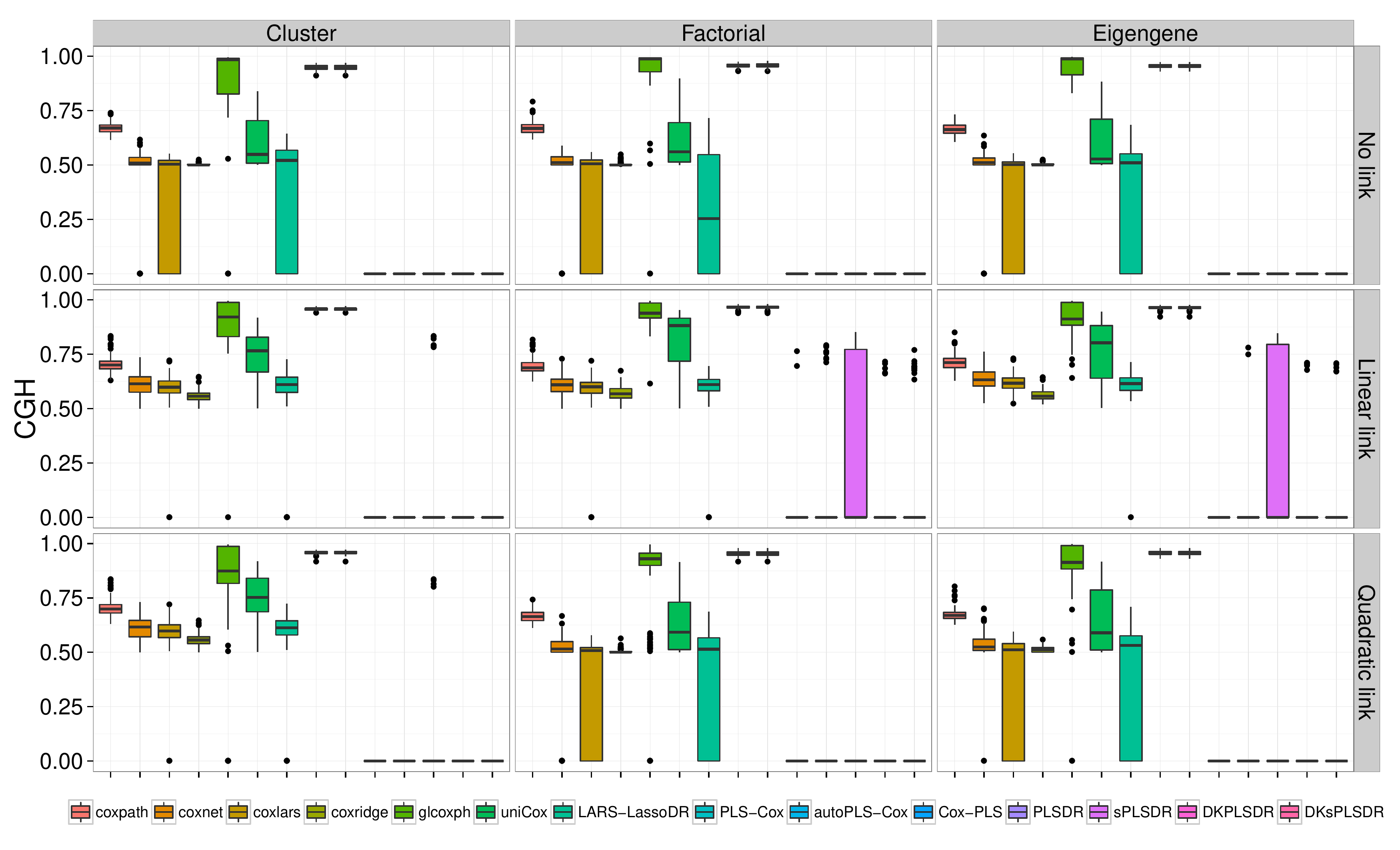}\phantomcaption\label{modmins_vanHcvll_GonenHellerCI}}}
\vspace{-.5cm}
\caption*{\hspace{-2cm}\mbox{Performance, vHCVLL CV. \quad Figure~\ref{modmins_vanHcvll_R2XO} (left):  R2XO measure. Figure~\ref{modmins_vanHcvll_GonenHellerCI} (right): GHCI measure.}}
\end{figure}

\begin{figure}[!tpb]
\centerline{{\includegraphics[width=.75\columnwidth]{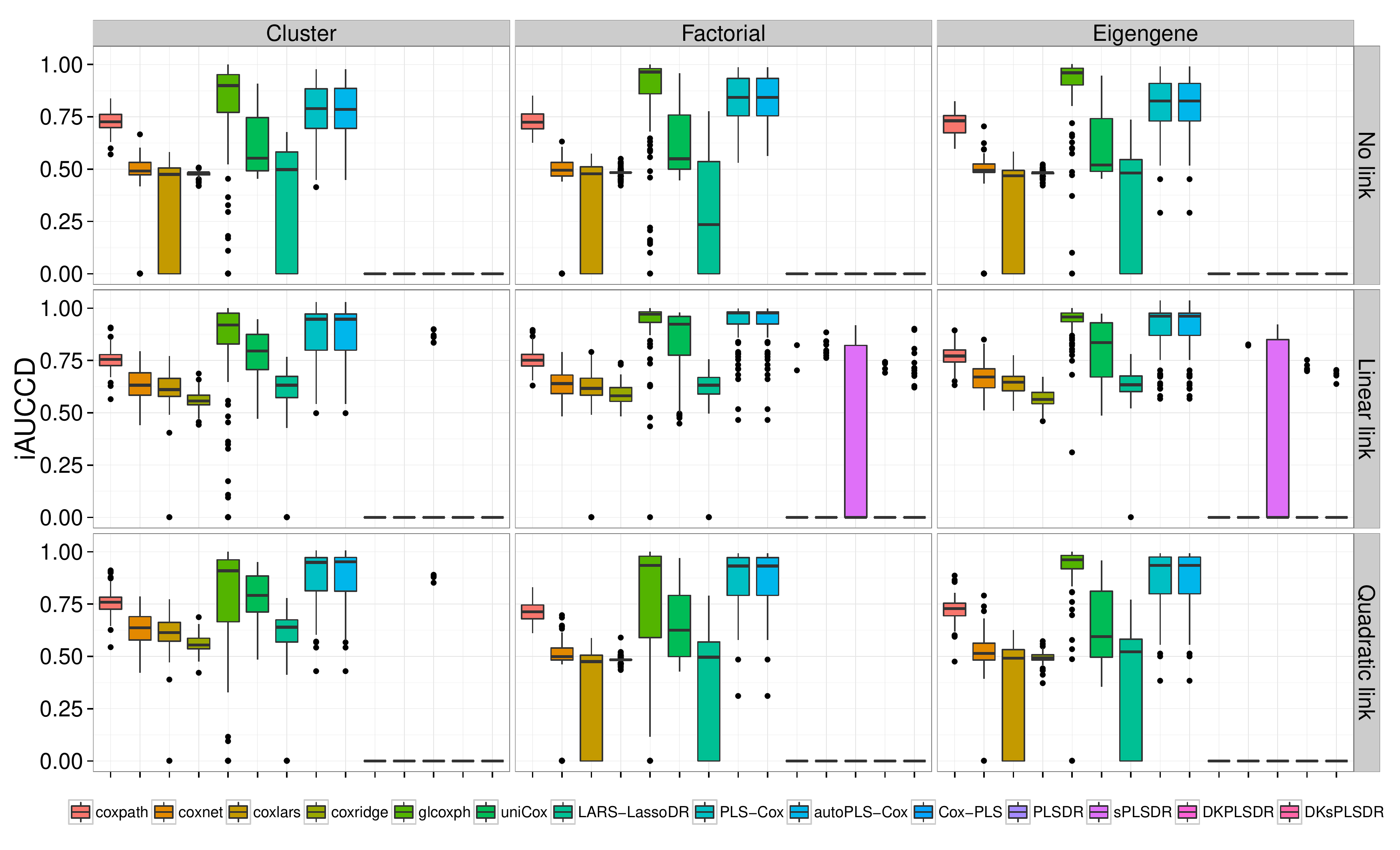}\phantomcaption\label{modmins_vanHcvll_iAUC_CD}}\qquad{\includegraphics[width=.75\columnwidth]{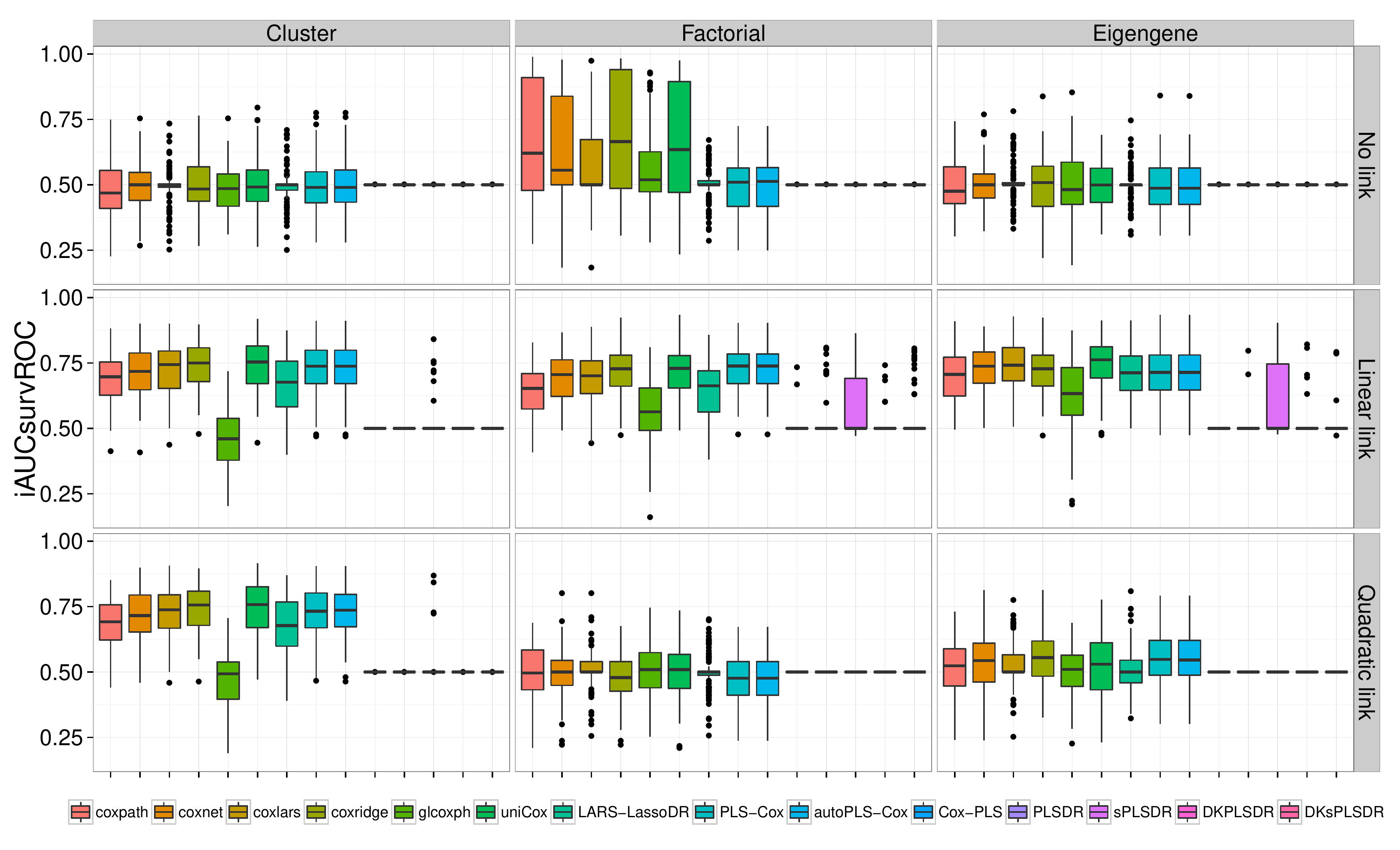}\phantomcaption\label{modmins_vanHcvll_iAUC_SurvROCtest}}}
\vspace{-.5cm}
\caption*{\hspace{-2cm}\mbox{Performance, vHCVLL CV. \quad Figure~\ref{modmins_vanHcvll_iAUC_CD} (left):  iAUCCD measure. Figure~\ref{modmins_vanHcvll_iAUC_SurvROCtest} (right): iAUCSurvROC measure.}}
\end{figure}

\begin{figure}[!tpb]
\centerline{{\includegraphics[width=.75\columnwidth]{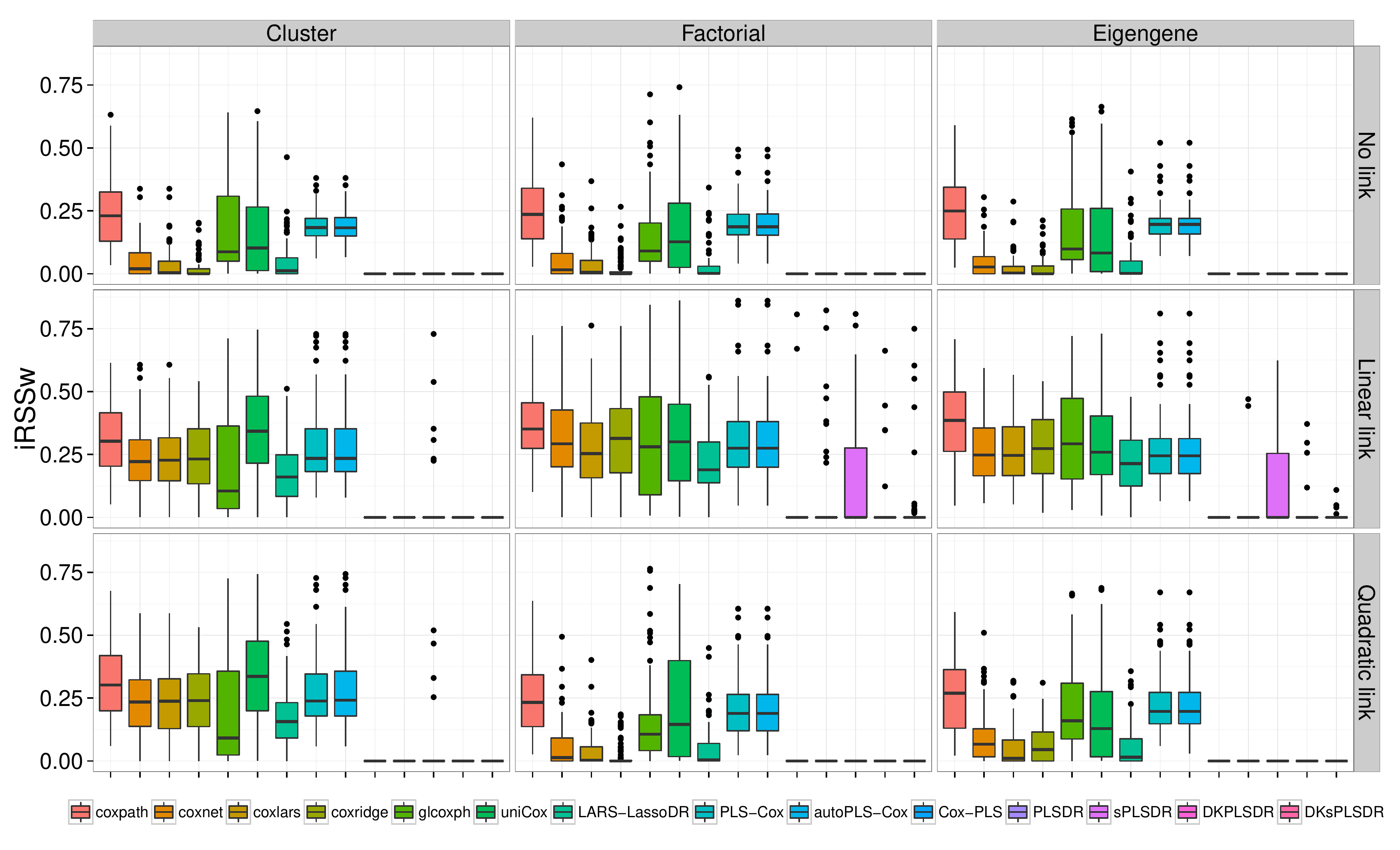}\phantomcaption\label{modmins_vanHcvll_iRSSw}}\qquad{\includegraphics[width=.75\columnwidth]{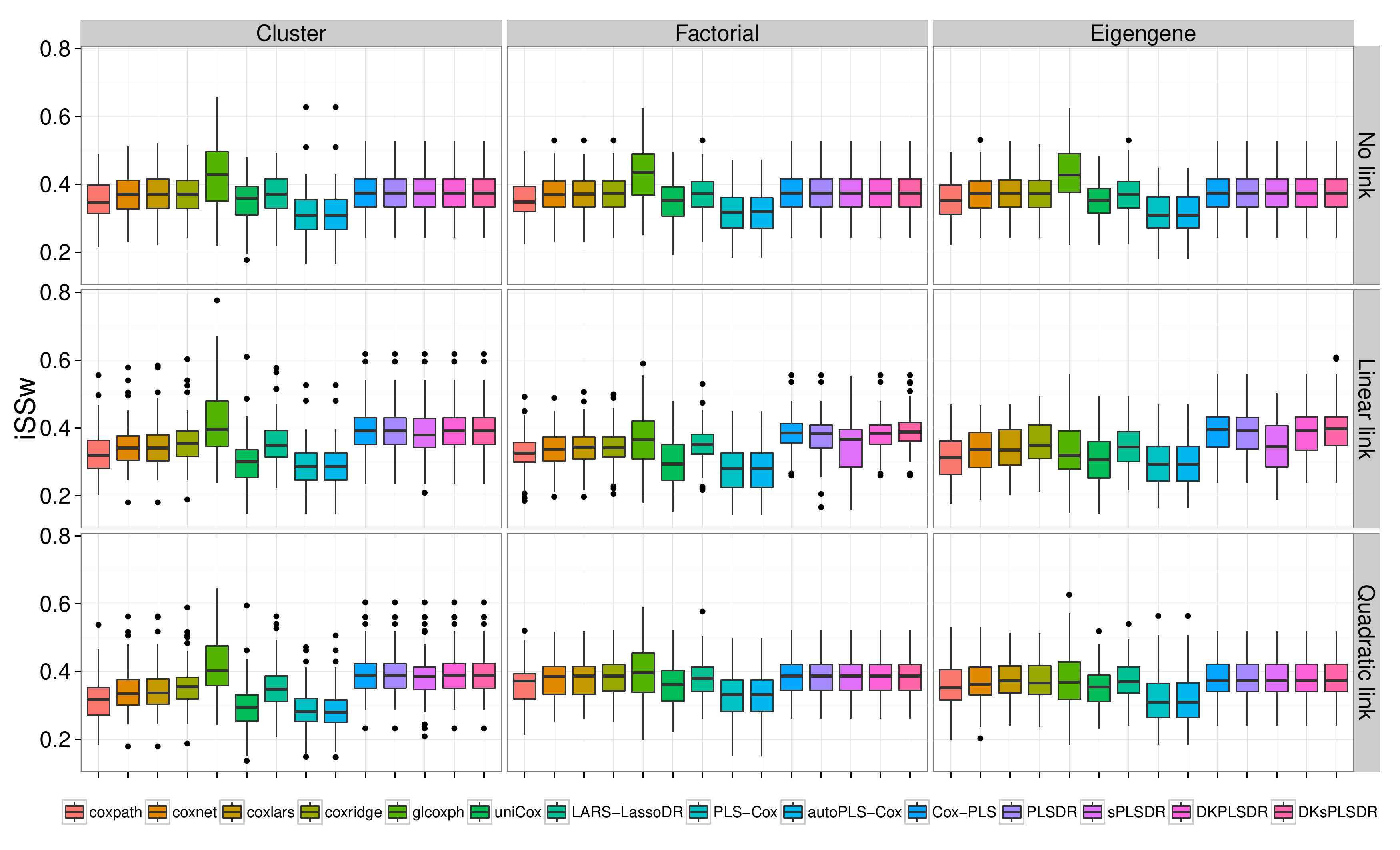}\phantomcaption\label{modmins_vanHcvll_irobustSw}}}
\vspace{-.5cm}
\caption*{\hspace{-2cm}\mbox{Performance, vHCVLL CV. \quad Figure~\ref{modmins_vanHcvll_iRSSw} (left):  iRSSw measure. Figure~\ref{modmins_vanHcvll_irobustSw} (right): iSSw measure.}}
\end{figure}

We plotted some of the performance measures when the cross-validation is done according to the vHCVLL criterion on Figures \ref{modmins_vanHcvll_R2XO}, \ref{modmins_vanHcvll_GonenHellerCI}, \ref{modmins_vanHcvll_iAUC_CD}, \ref{modmins_vanHcvll_iAUC_SurvROCtest}, \ref{modmins_vanHcvll_iRSSw} and \ref{modmins_vanHcvll_irobustSw}. The results are terrible for all the (s)PLS$-$like models apart from PLS$-$Cox and autoPLS$-$Cox.

We then provide, for each of the (s)PLS$-$like method, the increases in terms of performance measures when switching from the vHCVLL as a cross validation criterion to the recommended one in Section~\ref{recochanges}. Virtually, for PLS-Cox and autoPLS-Cox we switch to the iAUCSH cross-validation criterion and for other (s)PLS based models to either iAUCUno or iAUCSurvROC . 

\begin{figure}[!tpb]
\centerline{{\includegraphics[width=.75\columnwidth]{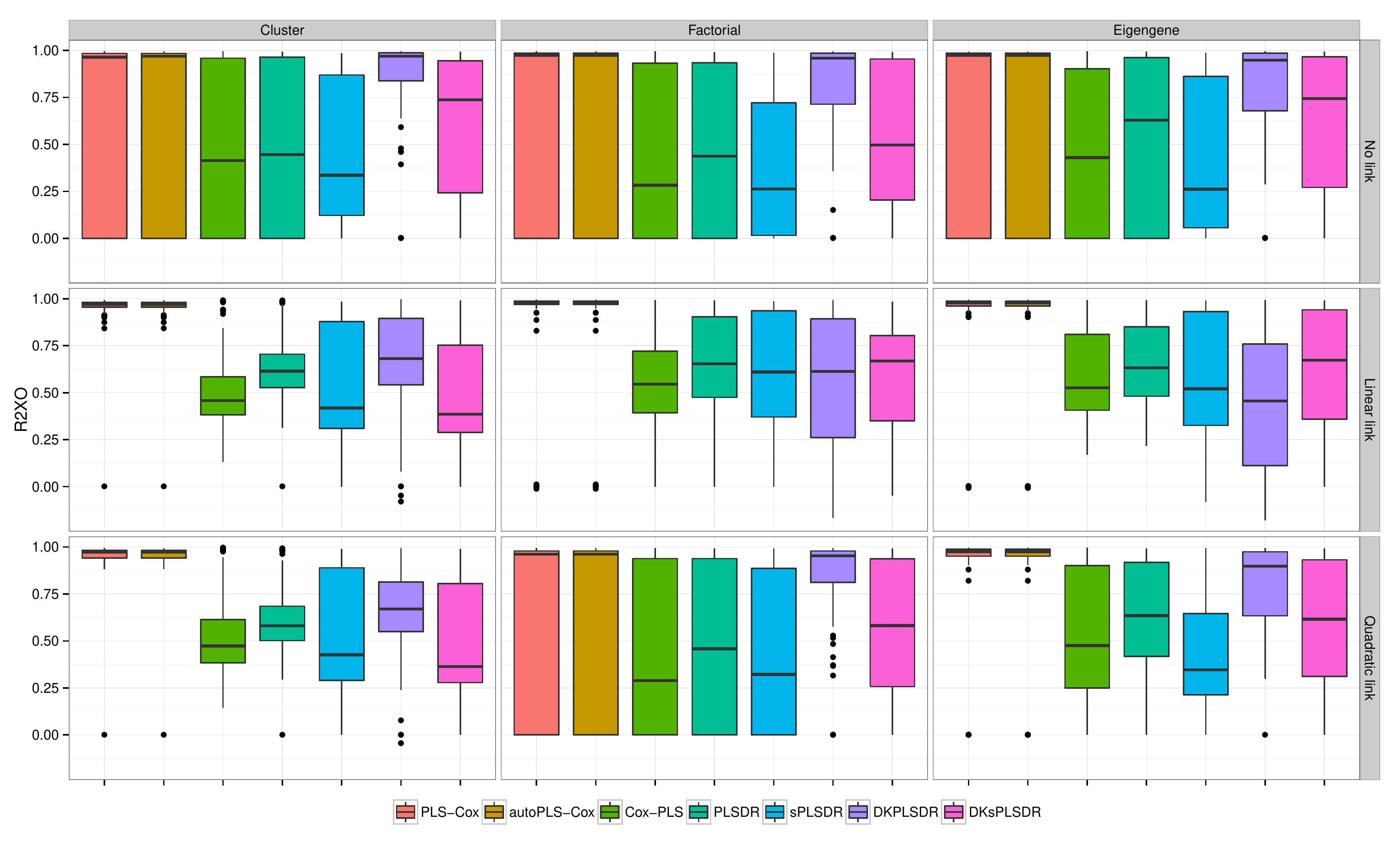}\phantomcaption\label{modmins_incAUCUno_R2XO}}\qquad{\includegraphics[width=.75\columnwidth]{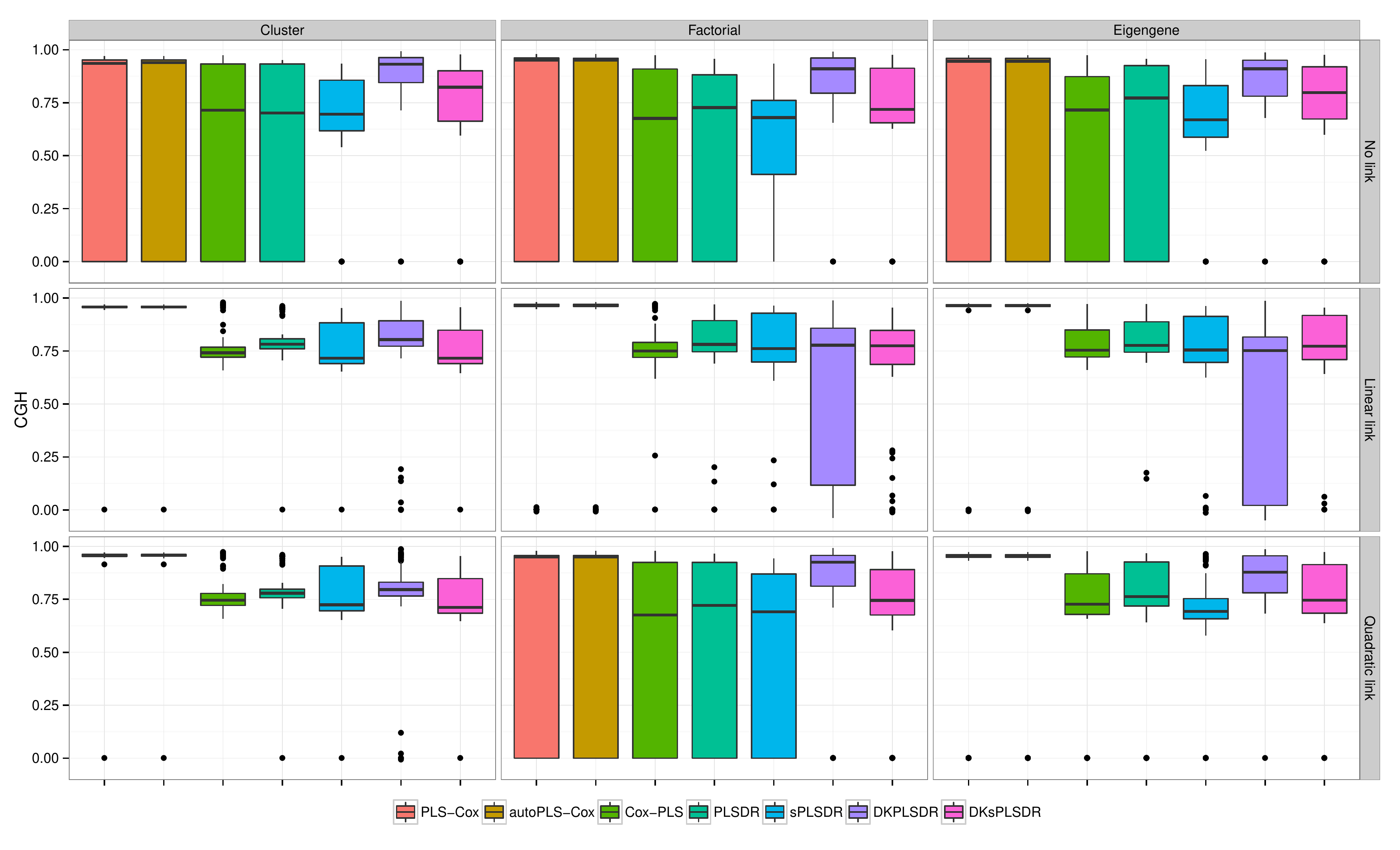}\phantomcaption\label{modmins_incAUCUno_GonenHellerCI}}}
\vspace{-.5cm}
\caption*{\hspace{-2cm}\mbox{Delta (iAUCUno CV  $-$ vHCVLL value). \quad Figure~\ref{modmins_incAUCUno_R2XO} (left): R2XO measure. Figure~\ref{modmins_incAUCUno_GonenHellerCI} (right):  GHCI measure.}}
\end{figure}

\begin{figure}[!tpb]
\centerline{{\includegraphics[width=.75\columnwidth]{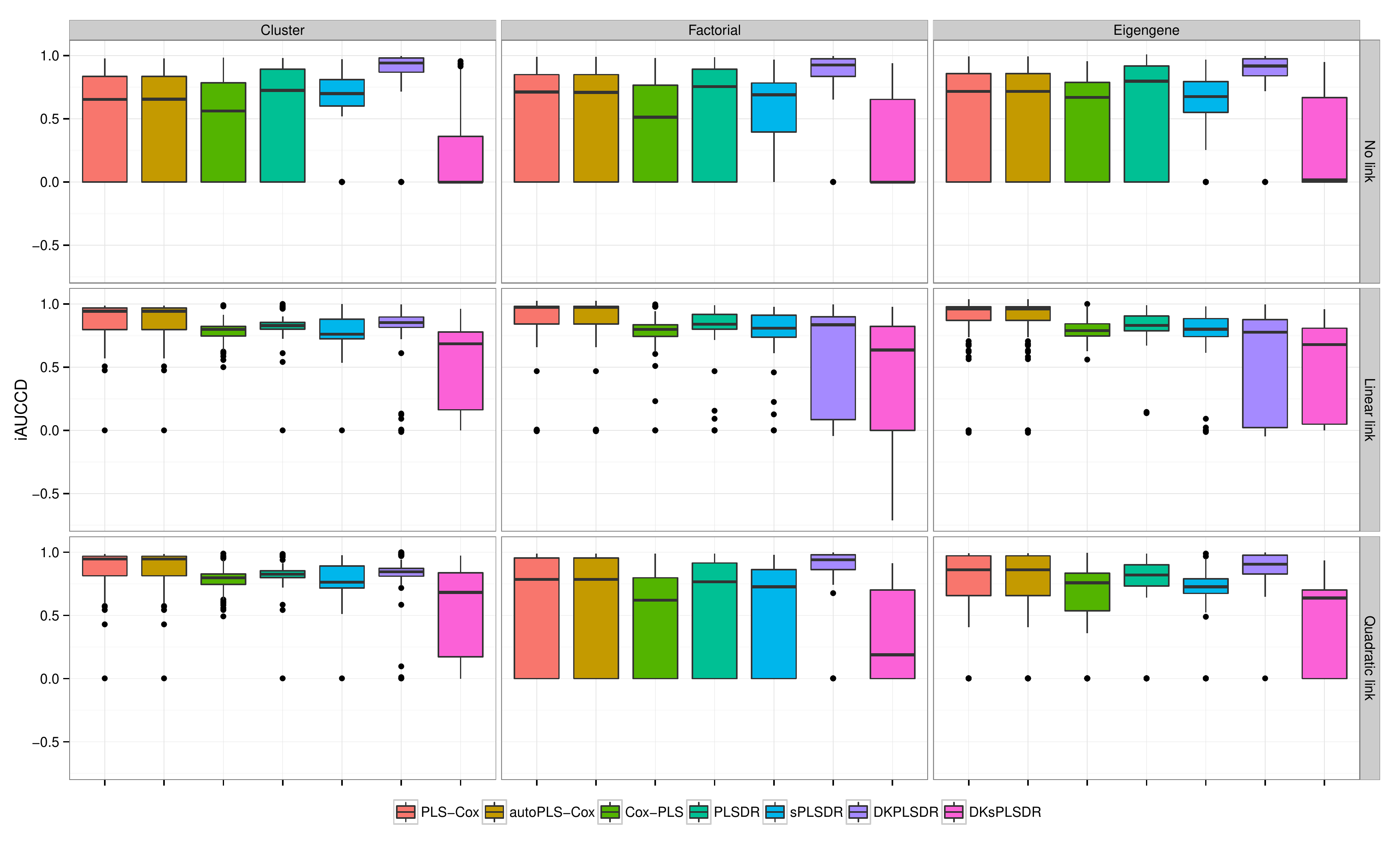}\phantomcaption\label{modmins_incAUCUno_iAUC_CD}}\qquad{\includegraphics[width=.75\columnwidth]{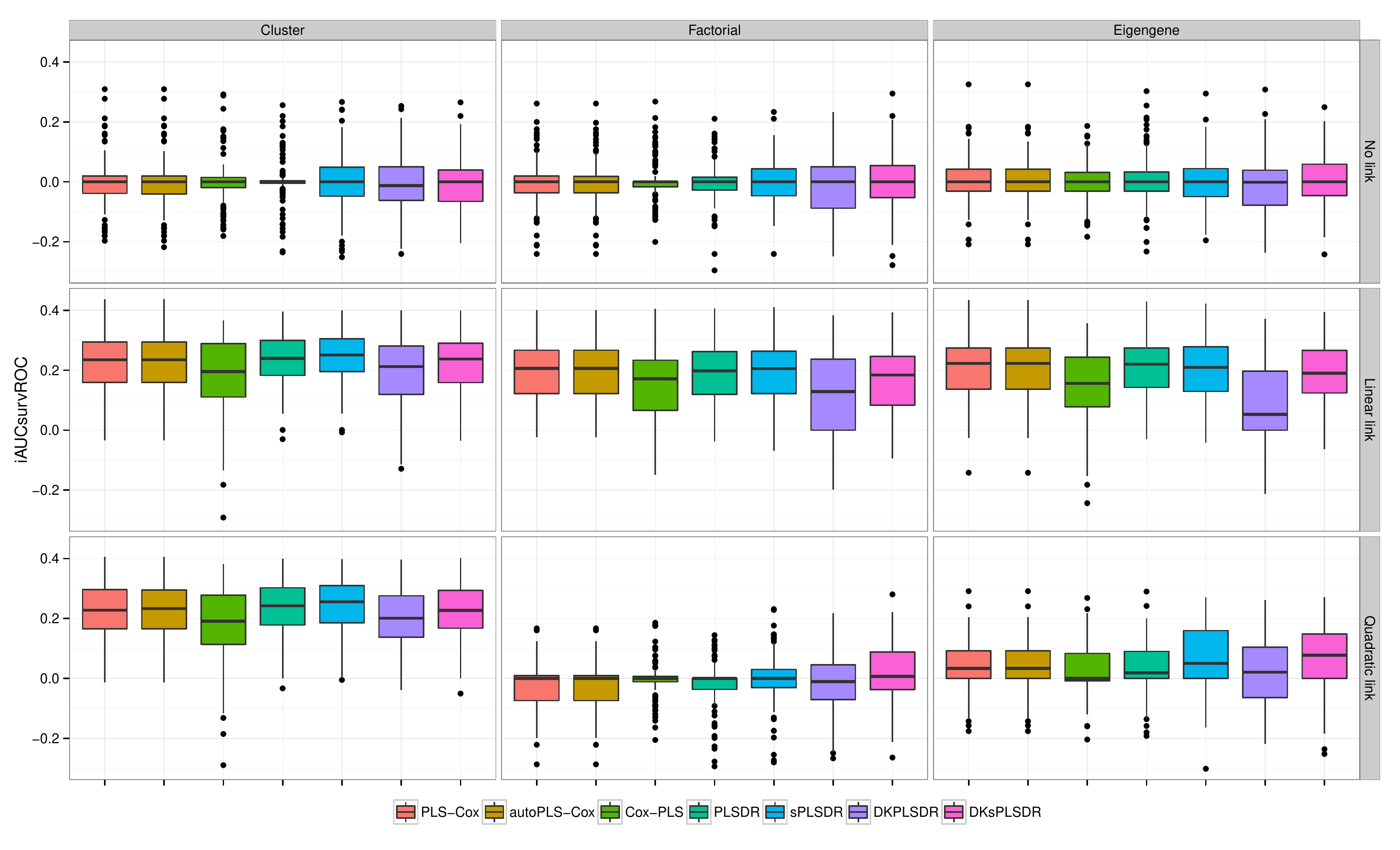}\phantomcaption\label{modmins_incAUCUno_iAUC_SurvROCtest}}}
\vspace{-.5cm}
\caption*{\hspace{-3.5cm}\mbox{Delta (iAUCUno CV  $-$ vHCVLL value). \quad Figure~\ref{modmins_incAUCUno_iAUC_CD} (left): iAUCCD measure. Figure~\ref{modmins_incAUCUno_iAUC_SurvROCtest} (right):  iAUCSurvROC measure.}}
\end{figure}

\begin{figure}[!tpb]
\centerline{{\includegraphics[width=.75\columnwidth]{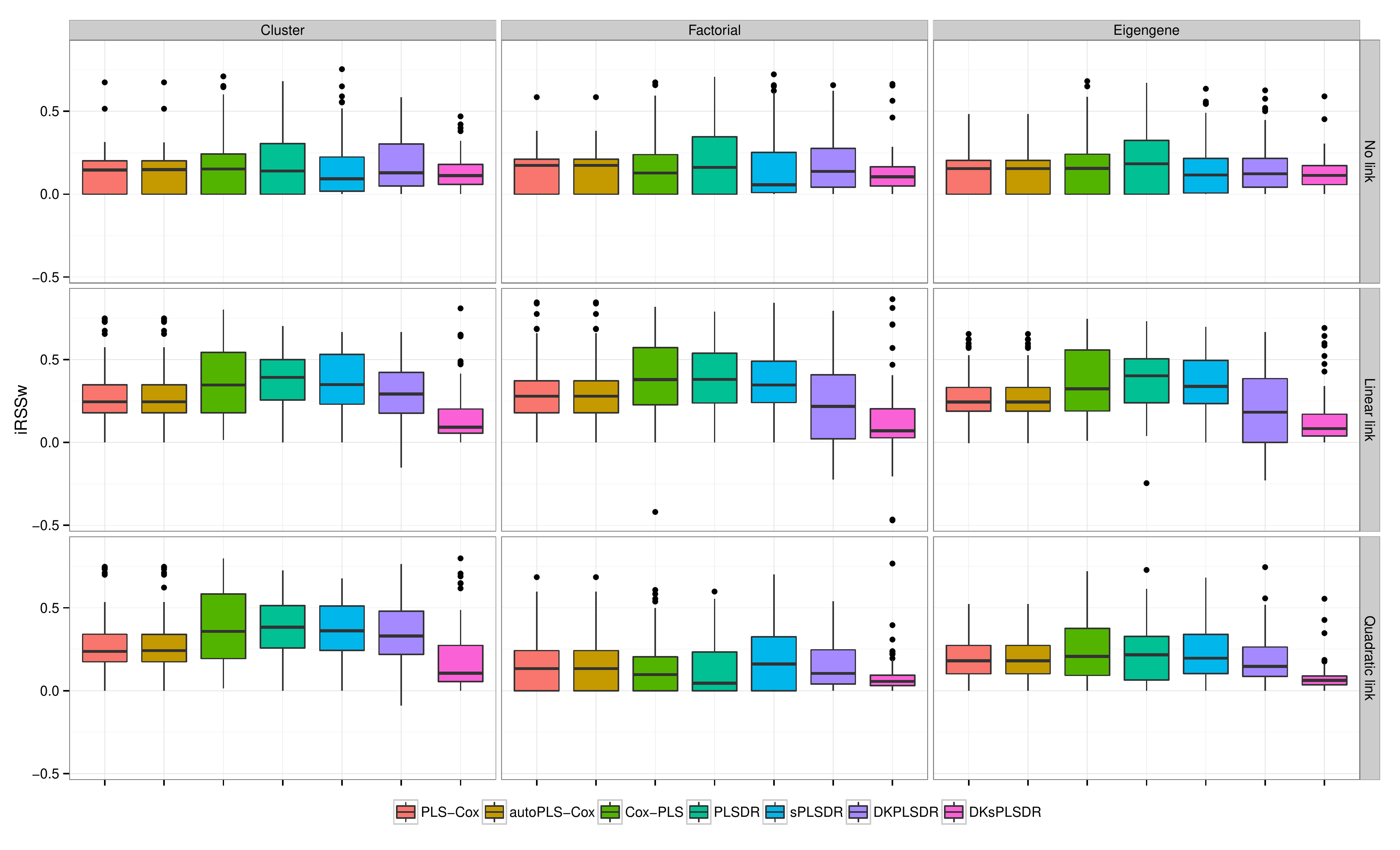}\phantomcaption\label{modmins_incAUCUno_iRSSw}}\qquad{\includegraphics[width=.75\columnwidth]{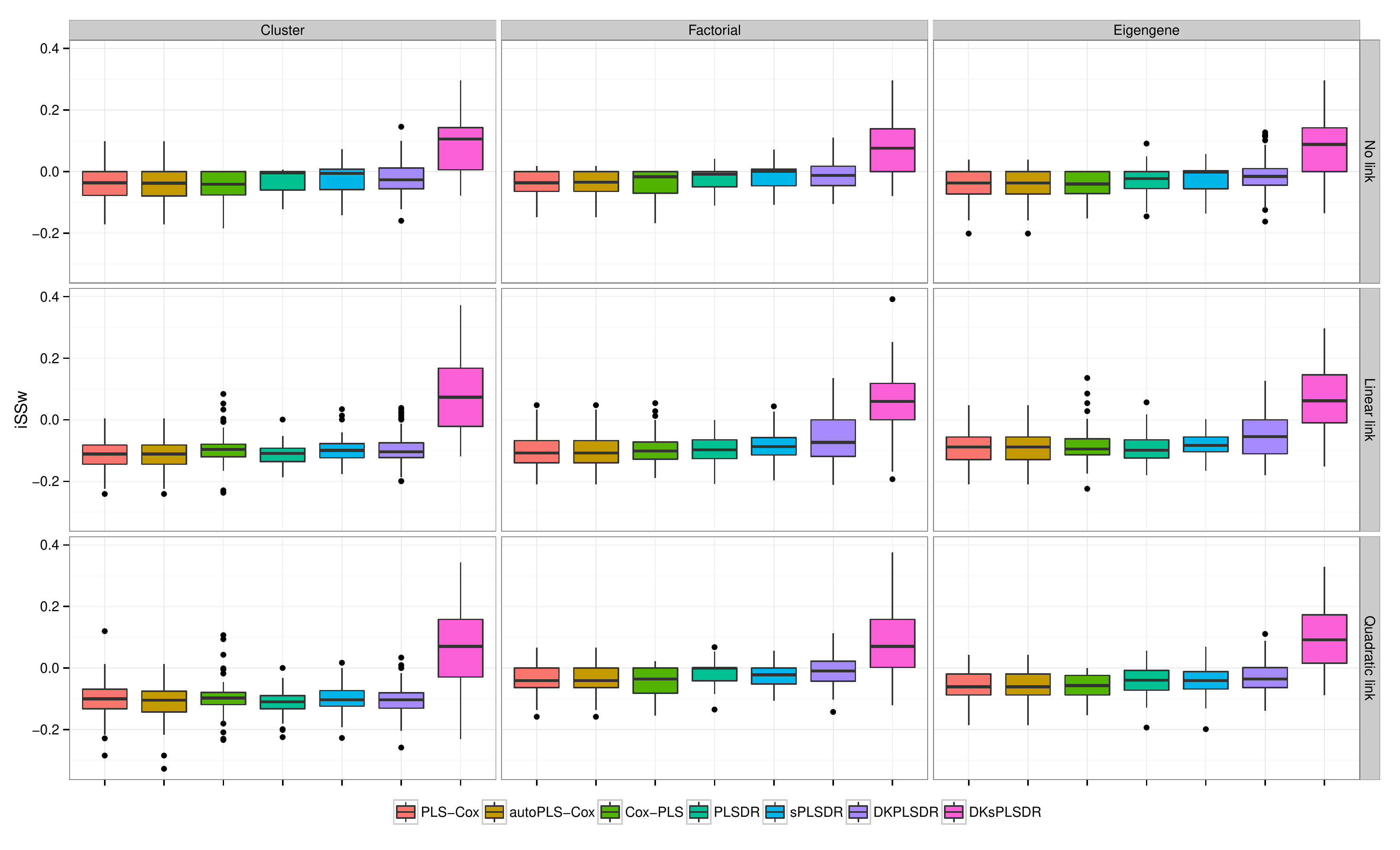}\phantomcaption\label{modmins_incAUCUno_irobustSw}}}
\vspace{-.5cm}
\caption*{\hspace{-2cm}\mbox{Delta (iAUCUno CV  $-$ vHCVLL value). \quad Figure~\ref{modmins_incAUCUno_iRSSw} (left): iRSSw measure. Figure~\ref{modmins_incAUCUno_irobustSw} (right):  iSSw measure.}}
\end{figure}

\begin{figure}[!tpb]
\centerline{{\includegraphics[width=.75\columnwidth]{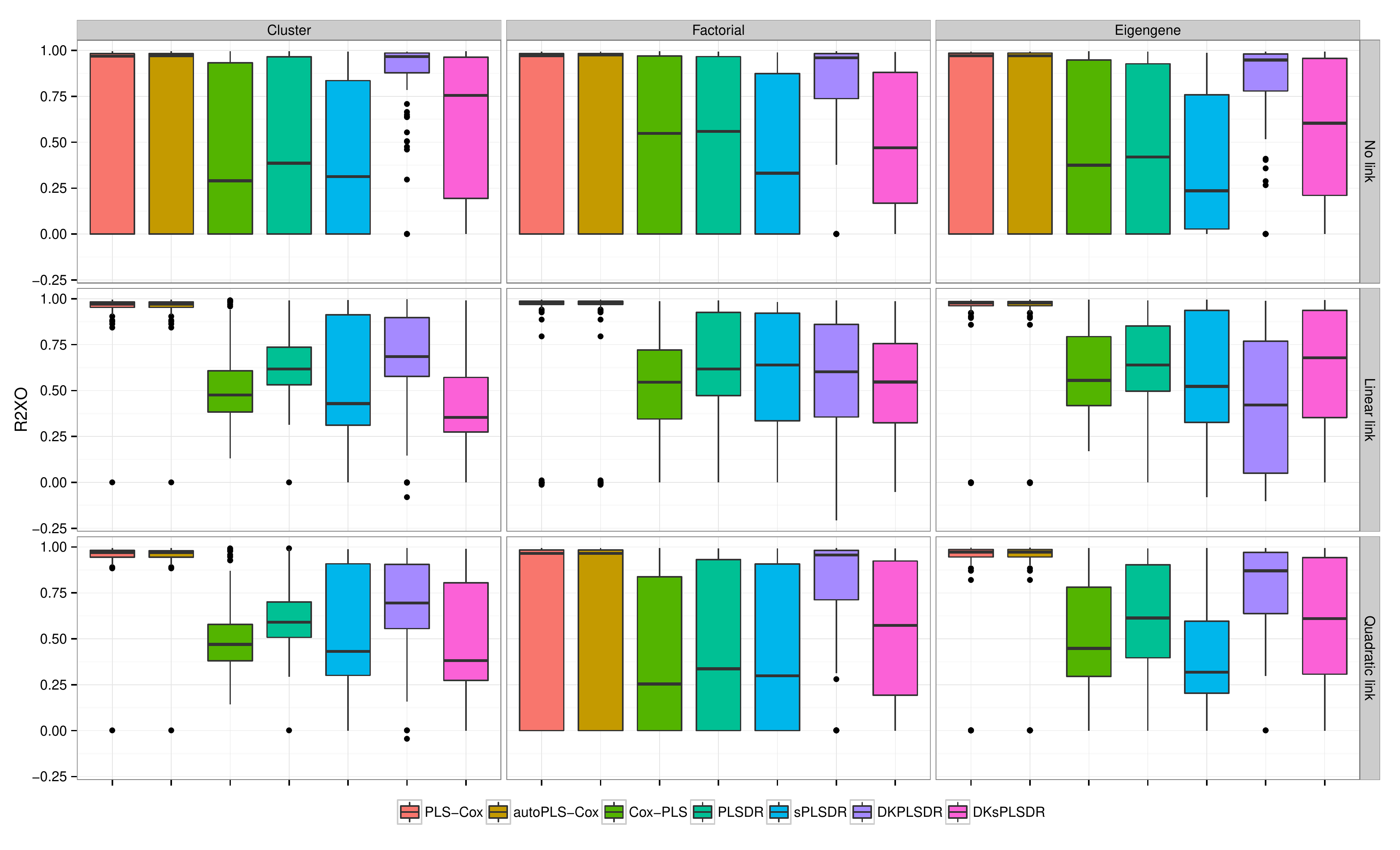}\phantomcaption\label{modmins_incAUCSurvROCtest_R2XO}}\qquad{\includegraphics[width=.75\columnwidth]{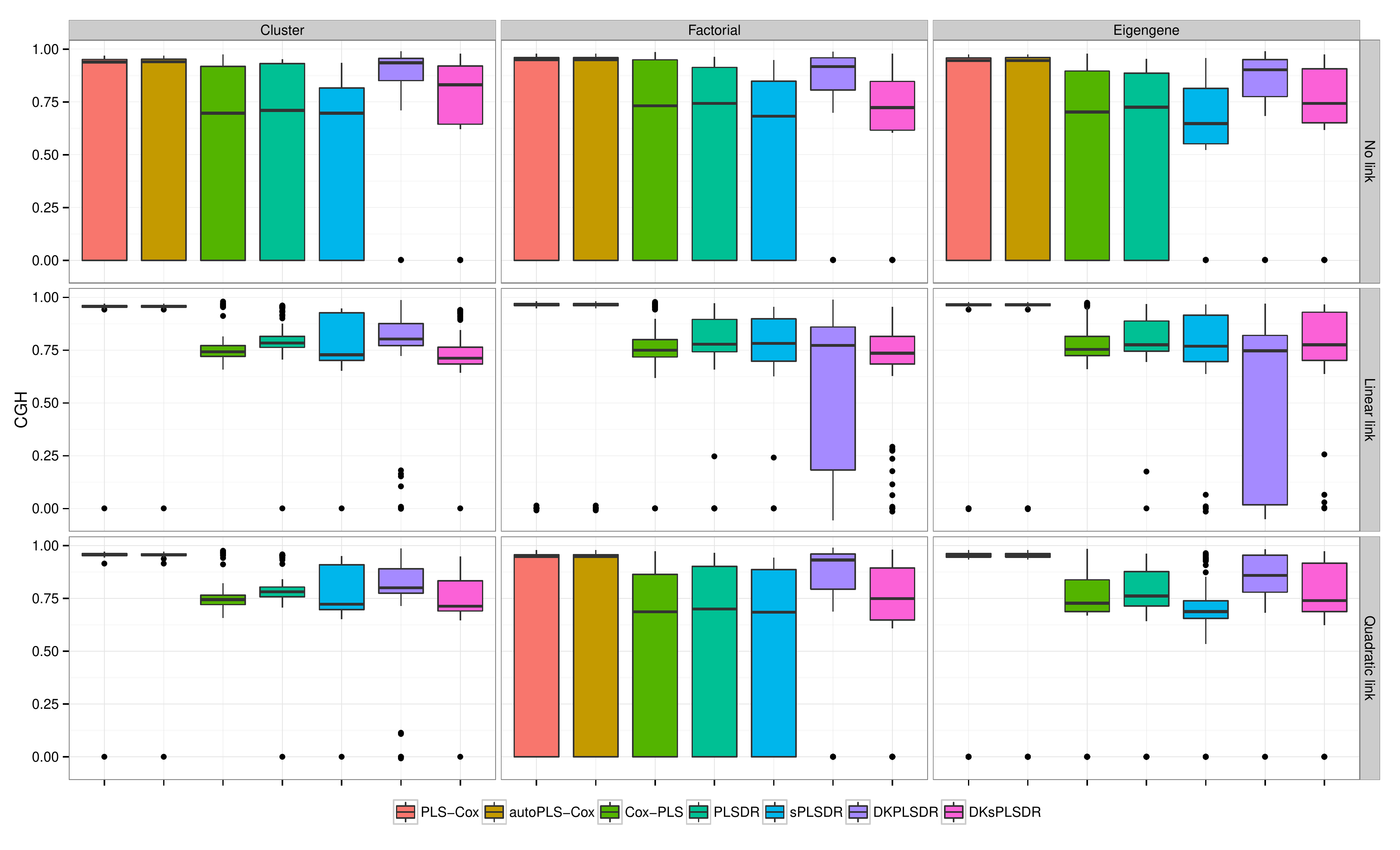}\phantomcaption\label{modmins_incAUCSurvROCtest_GonenHellerCI}}}
\vspace{-.5cm}
\caption*{\hspace{-2.5cm}\mbox{Delta (iAUCSurvROC CV  $-$ vHCVLL value). \quad Figure~\ref{modmins_incAUCSurvROCtest_R2XO} (left): R2XO measure. Figure~\ref{modmins_incAUCSurvROCtest_GonenHellerCI} (right):  GHCI measure.}}
\end{figure}

\begin{figure}[!tpb]
\centerline{{\includegraphics[width=.75\columnwidth]{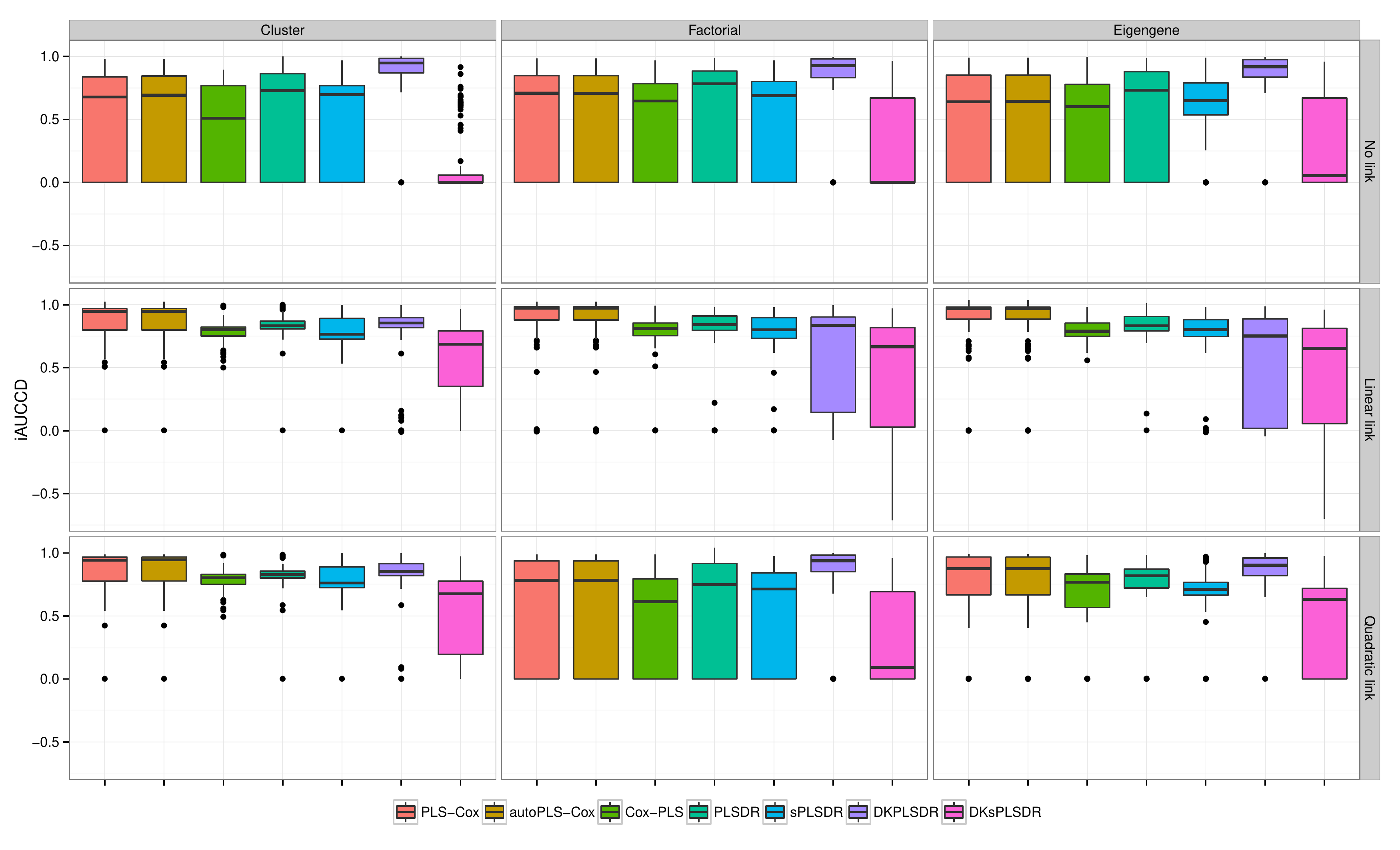}\phantomcaption\label{modmins_incAUCSurvROCtest_iAUC_CD}}\qquad{\includegraphics[width=.75\columnwidth]{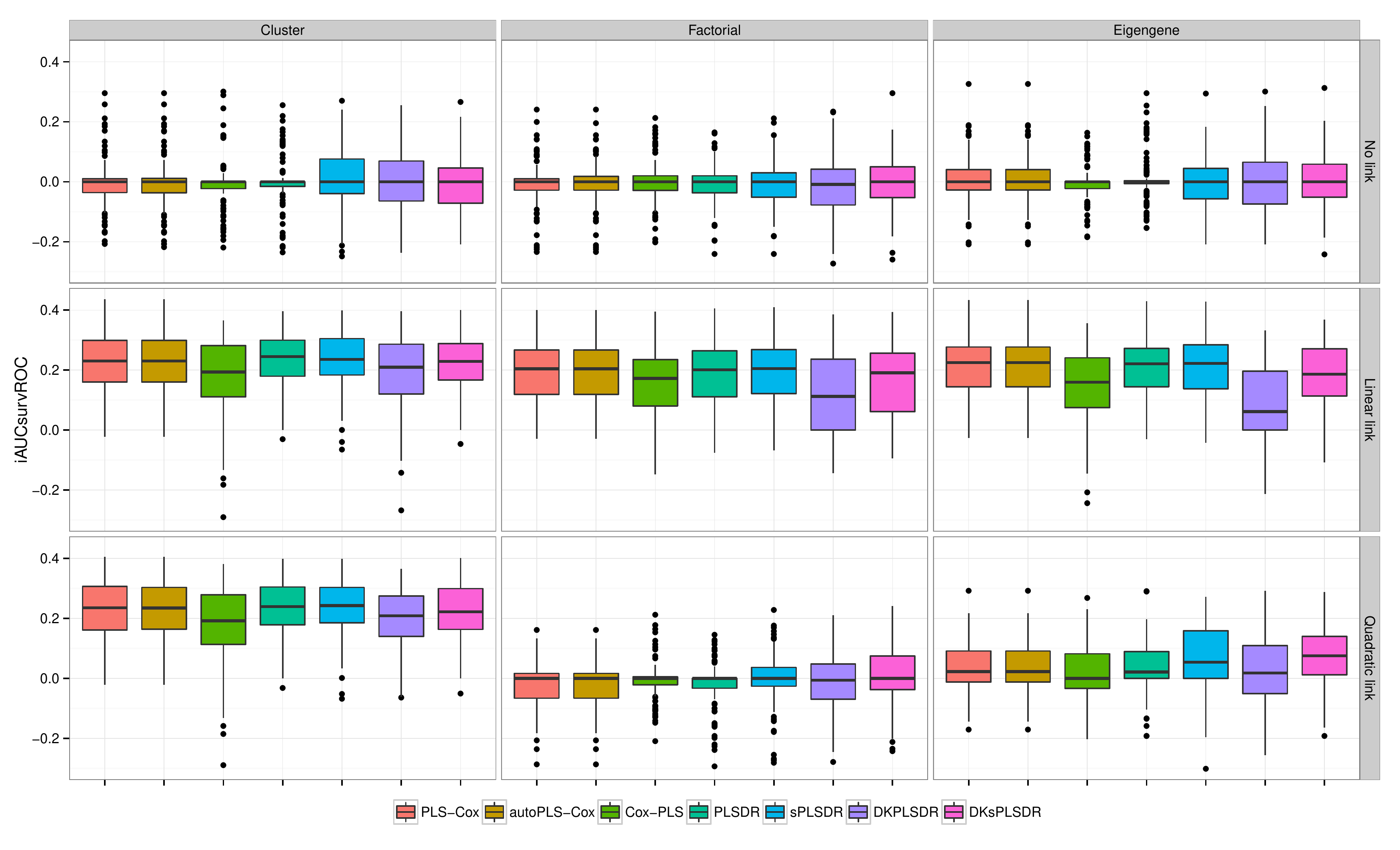}\phantomcaption\label{modmins_incAUCSurvROCtest_iAUC_SurvROCtest}}}
\vspace{-.5cm}
\caption*{\hspace{-2.5cm}\mbox{Delta (iAUCSurvROC CV  $-$ vHCVLL value). \quad Figure~\ref{modmins_incAUCSurvROCtest_iAUC_CD} (left): iAUCCD. Figure~\ref{modmins_incAUCSurvROCtest_iAUC_SurvROCtest} (right):  iAUCSurvROC.}}
\end{figure}

\begin{figure}[!tpb]
\centerline{{\includegraphics[width=.75\columnwidth]{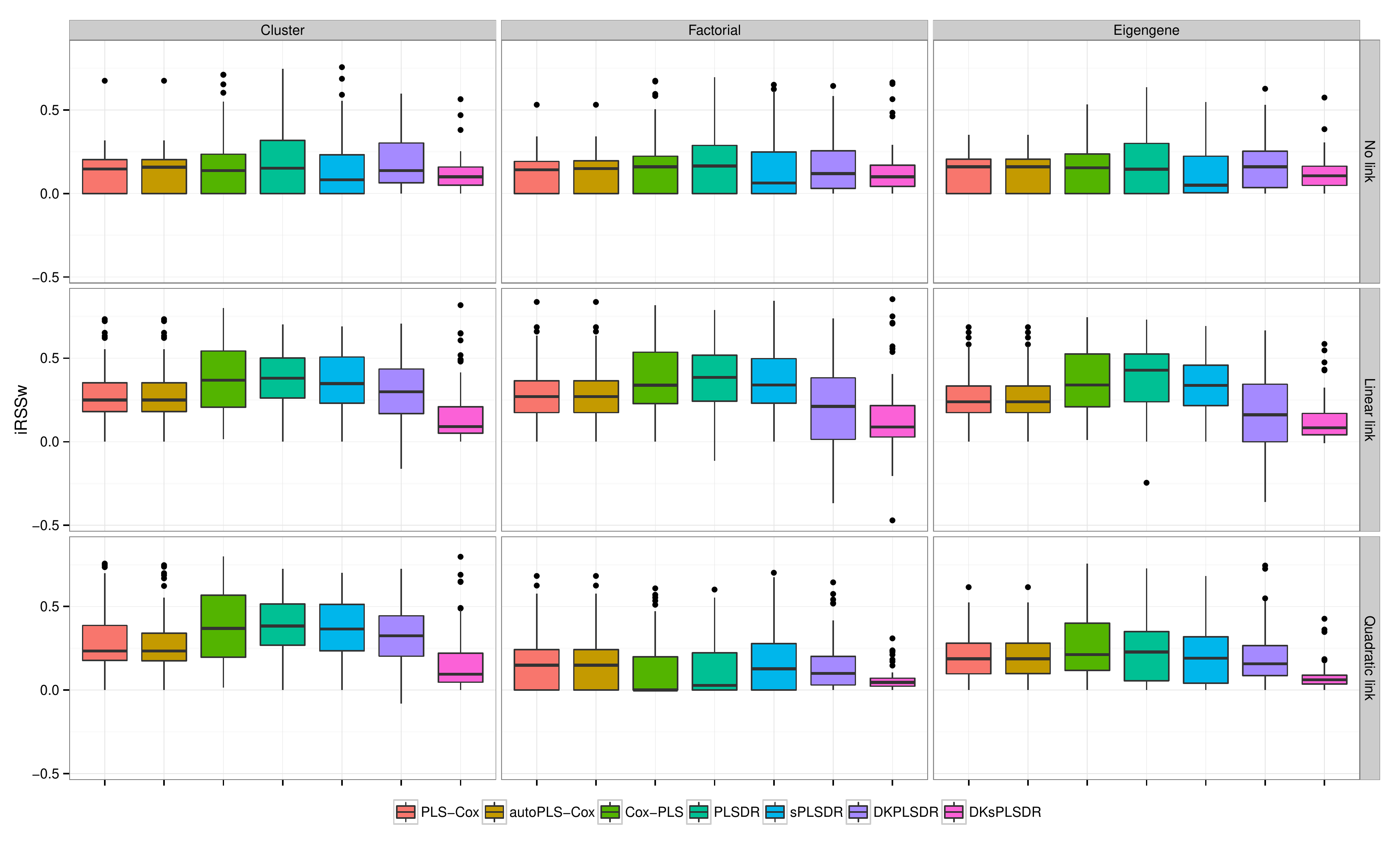}\phantomcaption\label{modmins_incAUCSurvROCtest_iRSSw}}\qquad{\includegraphics[width=.75\columnwidth]{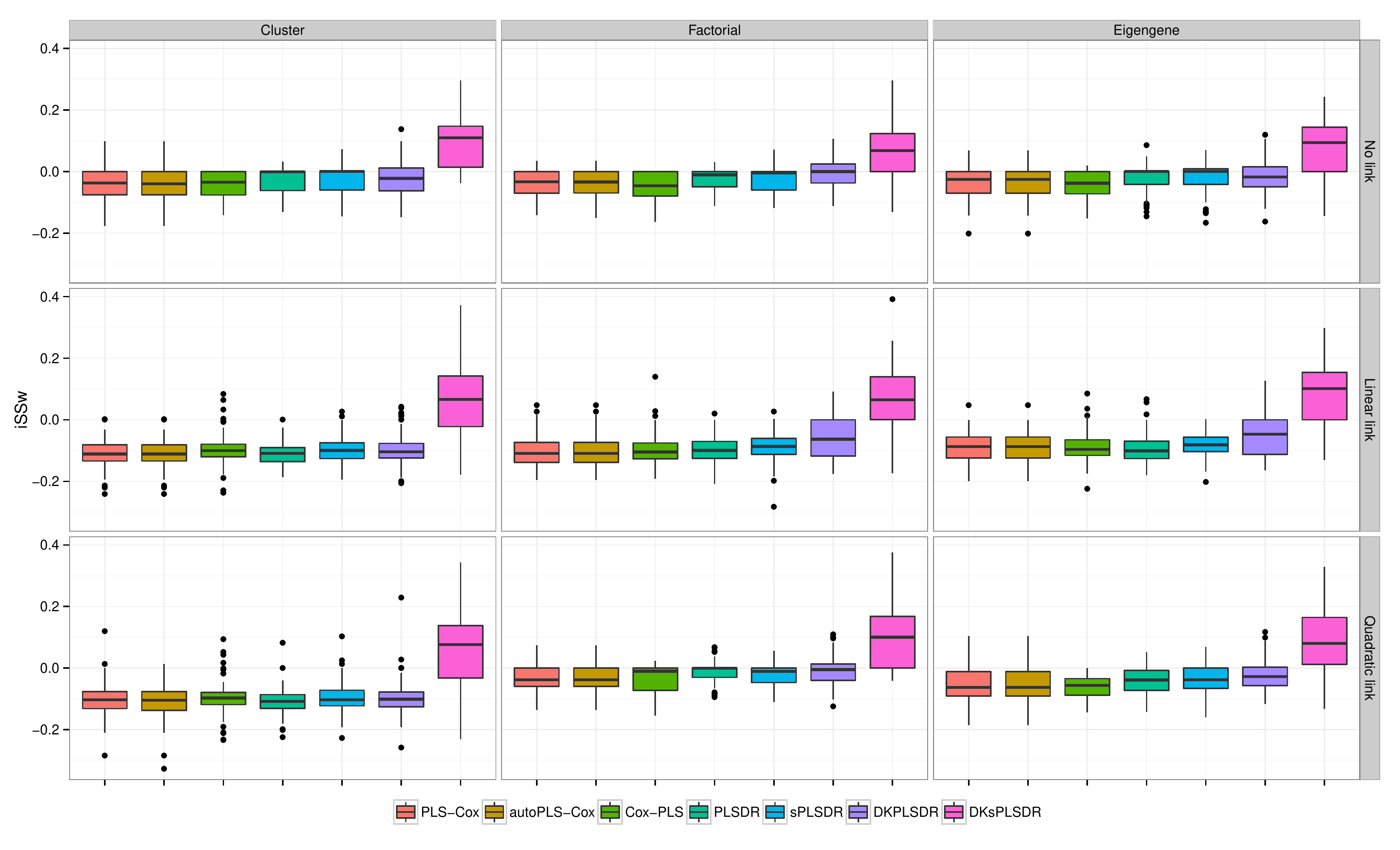}\phantomcaption\label{modmins_incAUCSurvROCtest_irobustSw}}}
\vspace{-.5cm}
\caption*{\hspace{-2.5cm}\mbox{Delta (iAUCSurvROC CV  $-$ vHCVLL value). \quad Figure~\ref{modmins_incAUCSurvROCtest_iRSSw} (left): iRSSw measure. Figure~\ref{modmins_incAUCSurvROCtest_irobustSw} (right):  iSSw measure.}}
\end{figure}

For iAUCUno, these results are plotted on Figures \ref{modmins_incAUCUno_R2XO}, \ref{modmins_incAUCUno_GonenHellerCI}, \ref{modmins_incAUCUno_iAUC_CD}, \ref{modmins_incAUCUno_iAUC_SurvROCtest}, \ref{modmins_incAUCUno_iRSSw} and \ref{modmins_incAUCUno_irobustSw} whereas for iAUCSurvROC they are displayed on Figures \ref{modmins_incAUCSurvROCtest_R2XO}, \ref{modmins_incAUCSurvROCtest_GonenHellerCI}, \ref{modmins_incAUCSurvROCtest_iAUC_CD}, \ref{modmins_incAUCSurvROCtest_iAUC_SurvROCtest}, \ref{modmins_incAUCSurvROCtest_iRSSw} and \ref{modmins_incAUCSurvROCtest_irobustSw}. These figures show a firm increase for the 6 criteria (R2XO, GHCI, iAUCCD, iAUCSurvROC, IRSSW, iSSW). 

\eject

\begin{figure}[!tpb]
\centerline{{\includegraphics[width=.75\columnwidth]{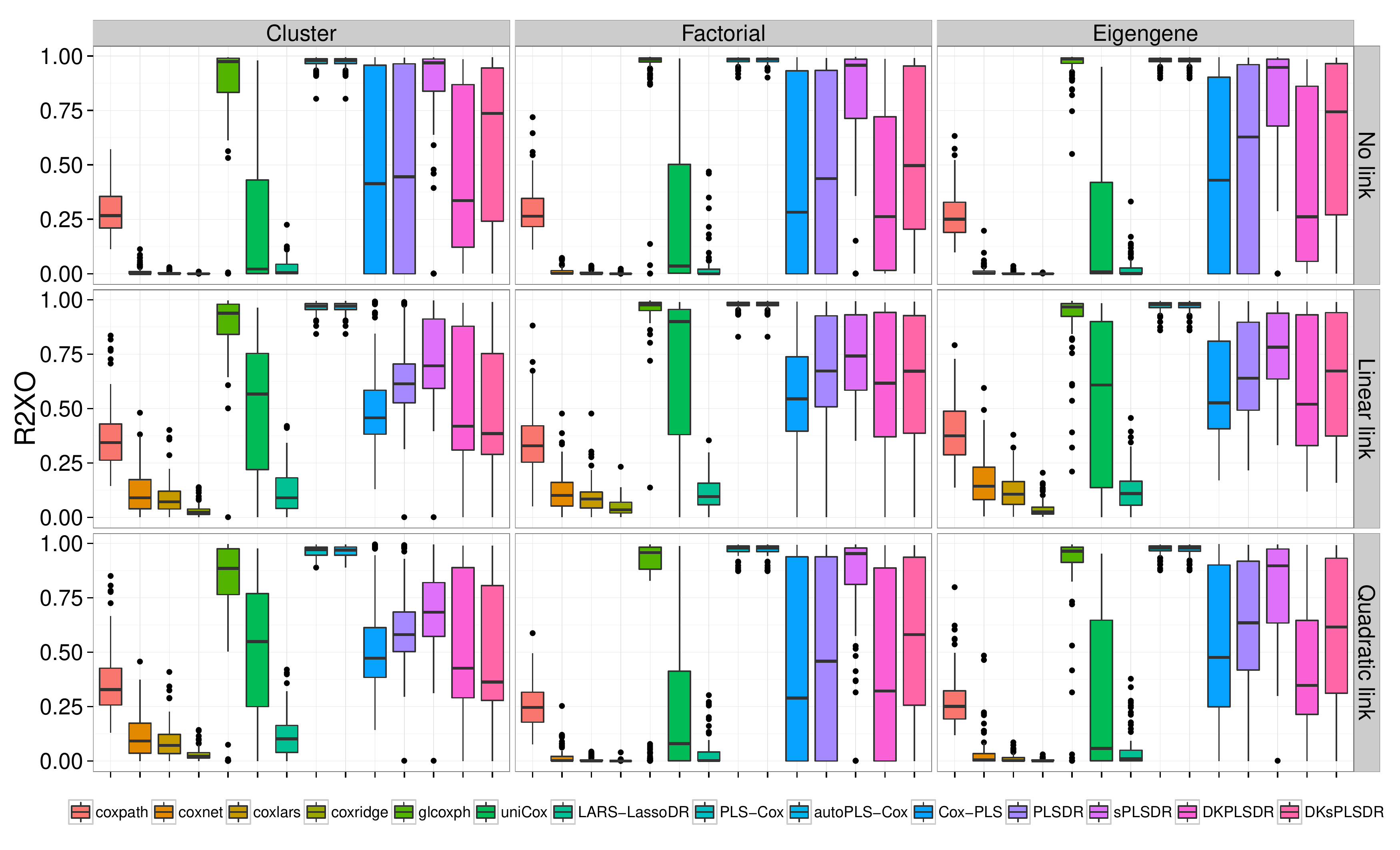}\phantomcaption\label{modmins_AUCUno_R2XO}}\qquad{\includegraphics[width=.75\columnwidth]{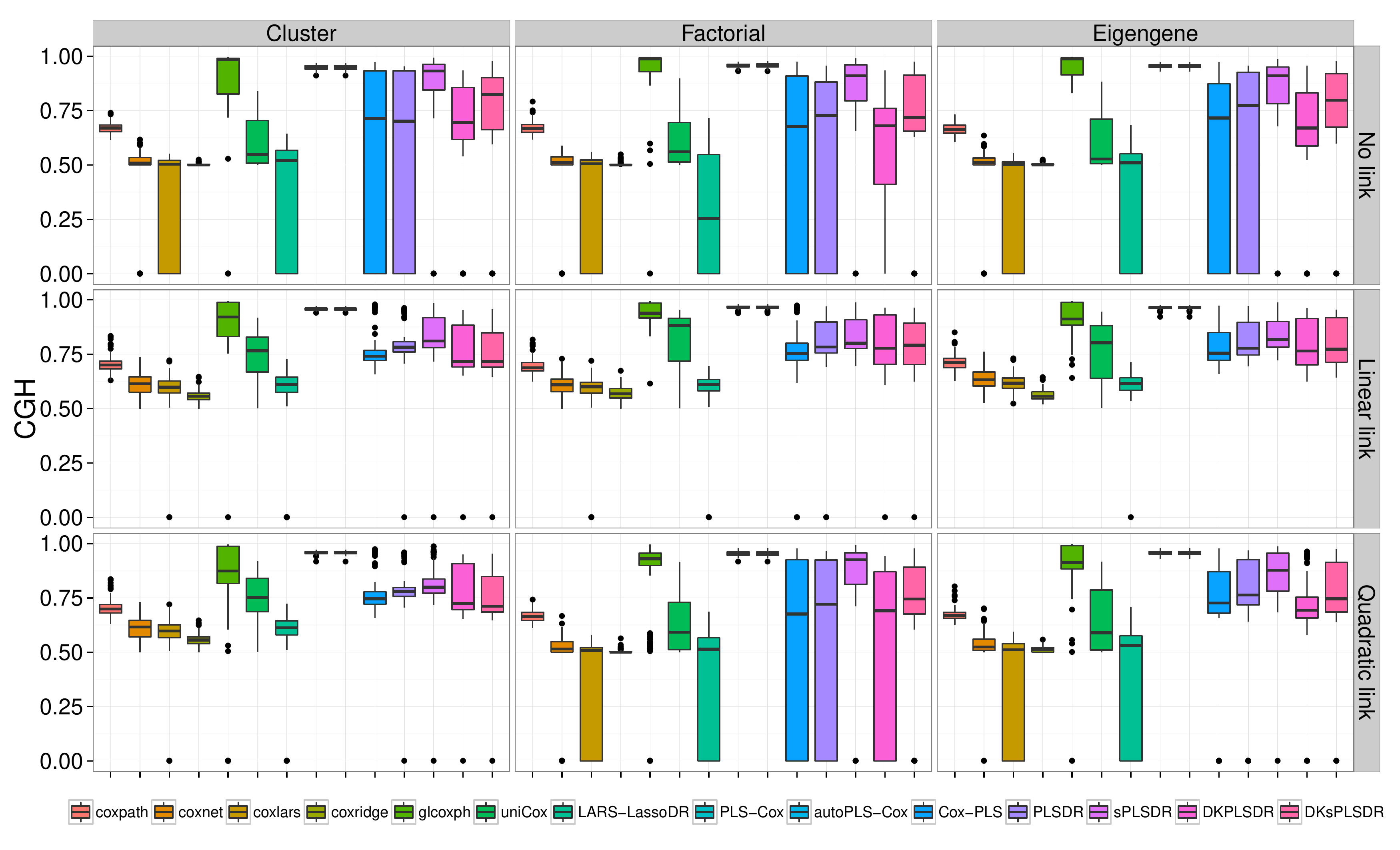}\phantomcaption\label{modmins_AUCUno_GonenHellerCI}}}
\vspace{-.5cm}
\caption*{\hspace{-2.5cm}\mbox{Performance, AUCUno CV. \quad Figure~\ref{modmins_AUCUno_R2XO} (left):  R2XO measure. Figure~\ref{modmins_AUCUno_GonenHellerCI} (right): GHCI measure.}}
\end{figure}

\begin{figure}[!tpb]
\centerline{{\includegraphics[width=.75\columnwidth]{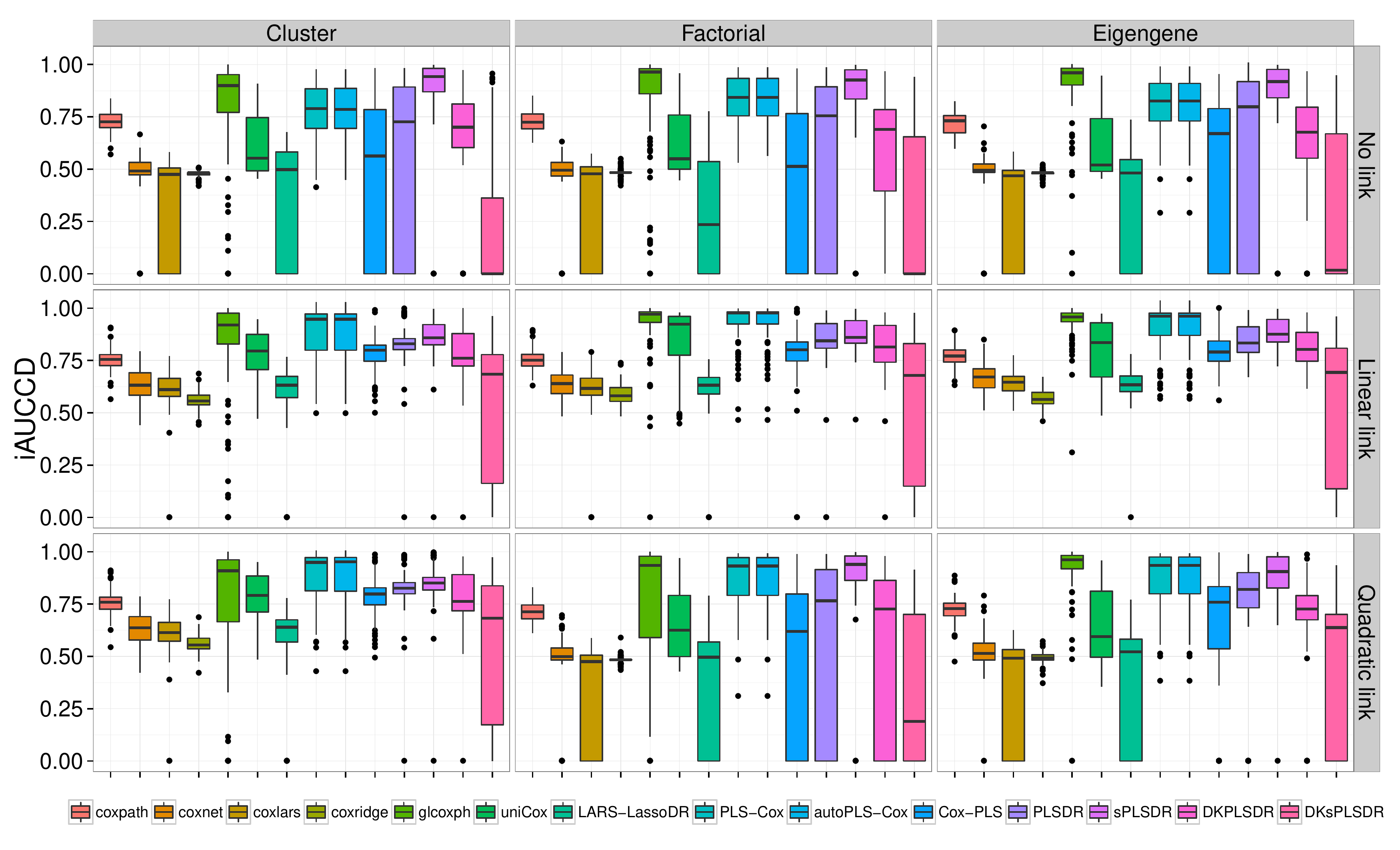}\phantomcaption\label{modmins_AUCUno_iAUC_CD}}\qquad{\includegraphics[width=.75\columnwidth]{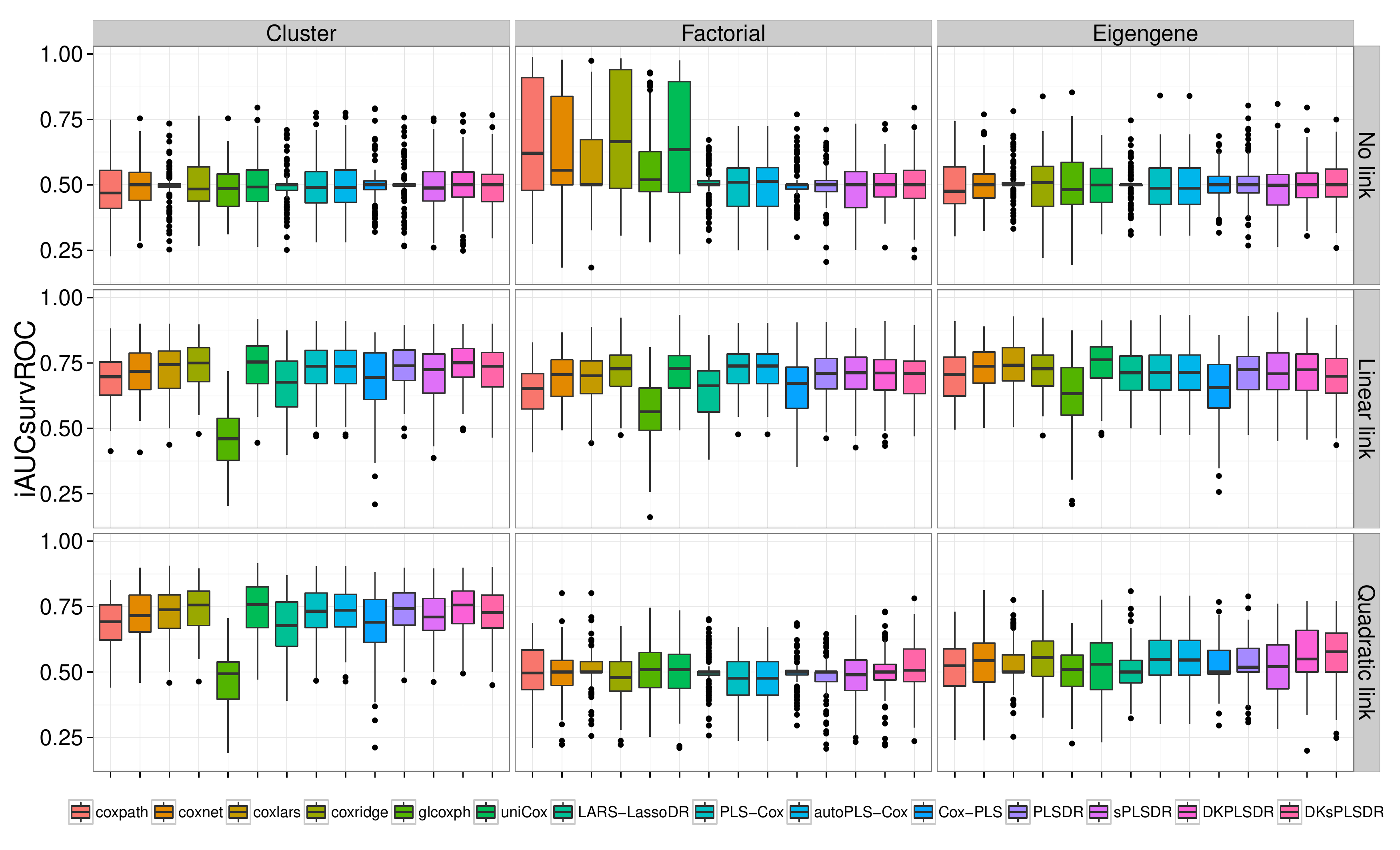}\phantomcaption\label{modmins_AUCUno_iAUC_SurvROCtest}}}
\vspace{-.5cm}
\caption*{\hspace{-2.5cm}\mbox{Performance, AUCUno CV. \quad Figure~\ref{modmins_AUCUno_iAUC_CD} (left):  iAUCCD measure. Figure~\ref{modmins_AUCUno_iAUC_SurvROCtest} (right): iAUCSurvROC measure.}}
\end{figure}

\begin{figure}[!tpb]
\centerline{{\includegraphics[width=.75\columnwidth]{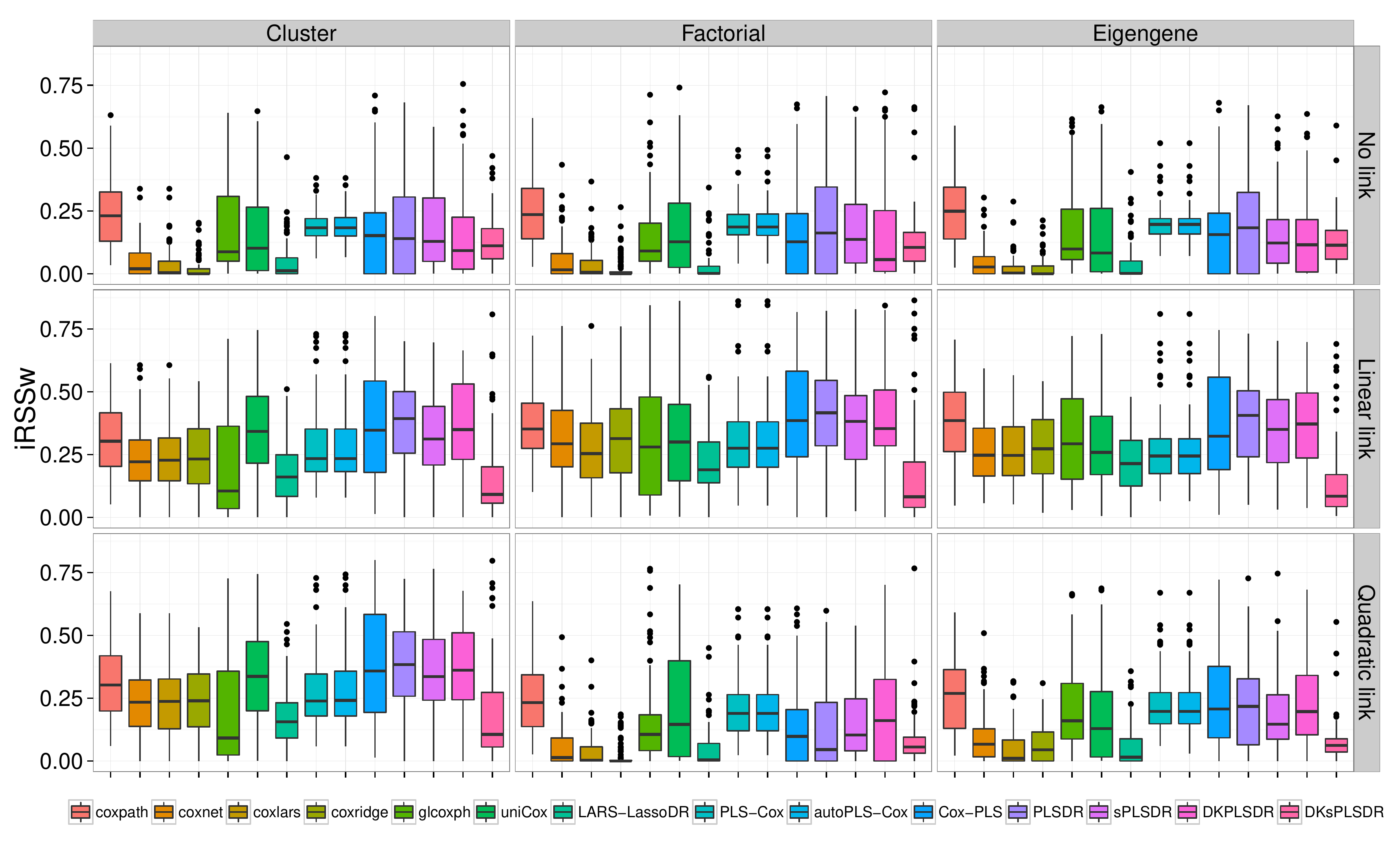}\phantomcaption\label{modmins_AUCUno_iRSSw}}\qquad{\includegraphics[width=.75\columnwidth]{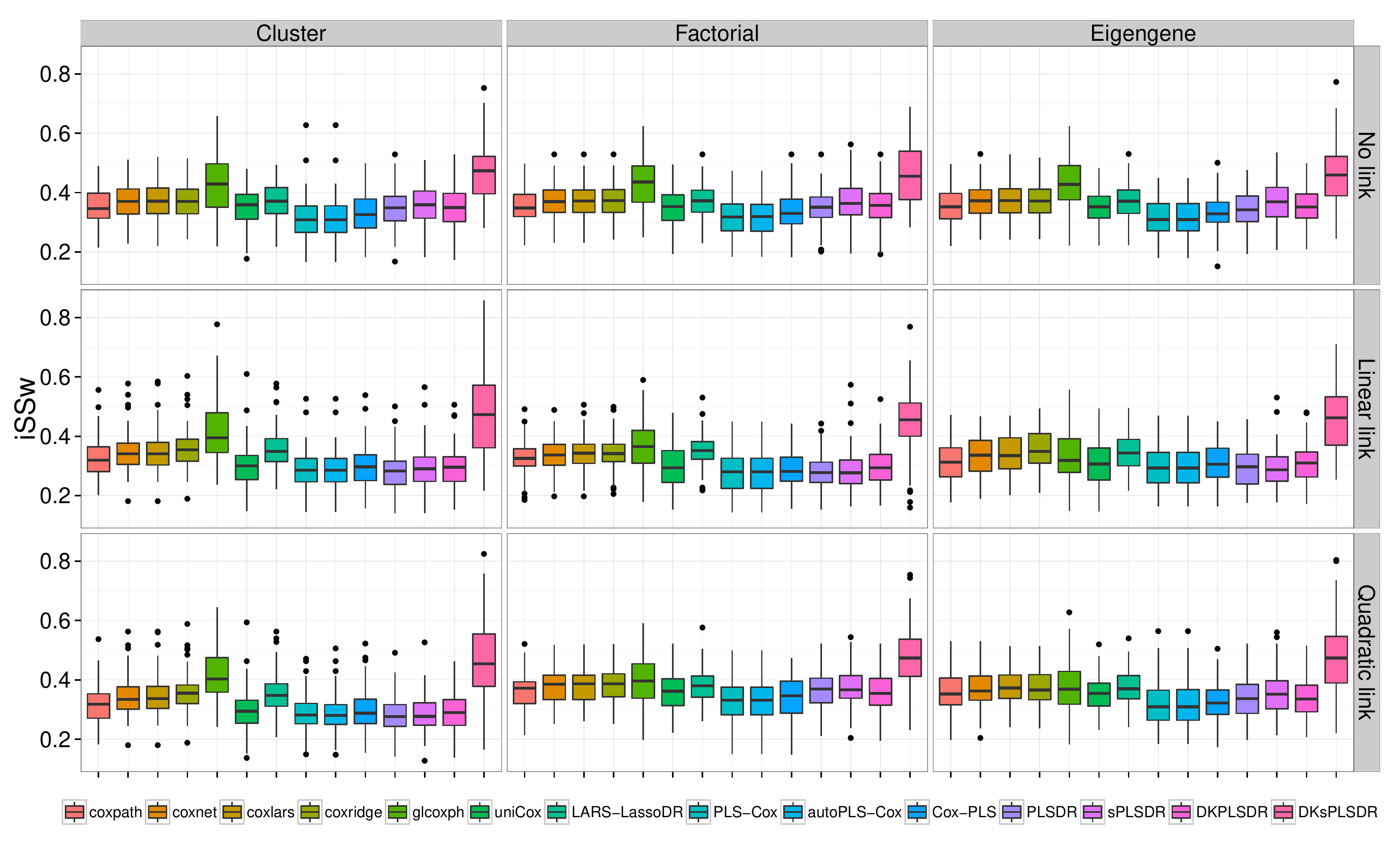}\phantomcaption\label{modmins_AUCUno_irobustSw}}}
\vspace{-.5cm}
\caption*{\hspace{-2.5cm}\mbox{Performance, AUCUno CV. \quad Figure~\ref{modmins_AUCUno_iRSSw} (left):  iRSSw measure. Figure~\ref{modmins_AUCUno_irobustSw} (right): iSSw measure.}}
\end{figure}

\begin{figure}[!tpb]
\centerline{{\includegraphics[width=.75\columnwidth]{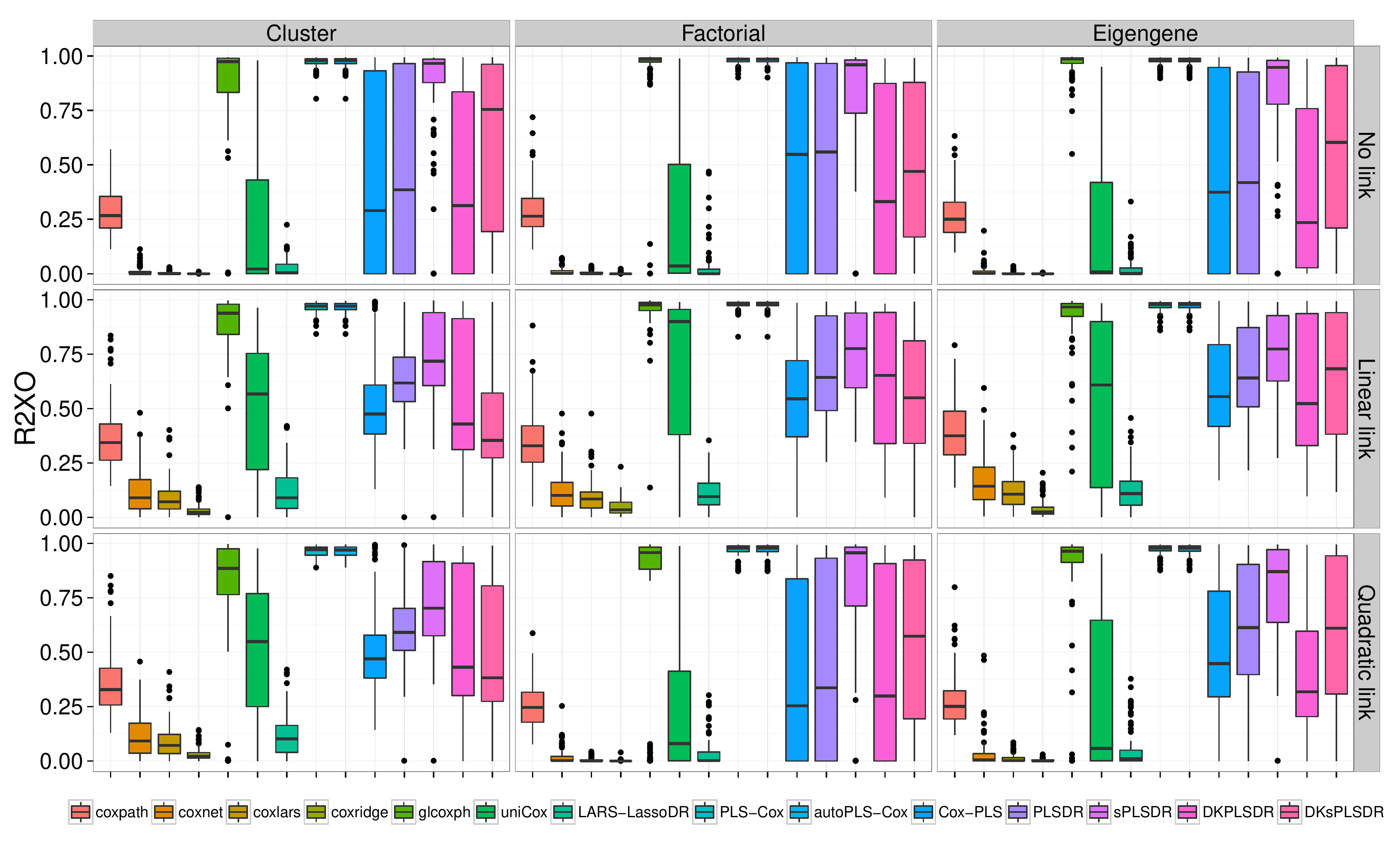}\phantomcaption\label{modmins_AUCsurvROCtest_R2XO}}\qquad{\includegraphics[width=.75\columnwidth]{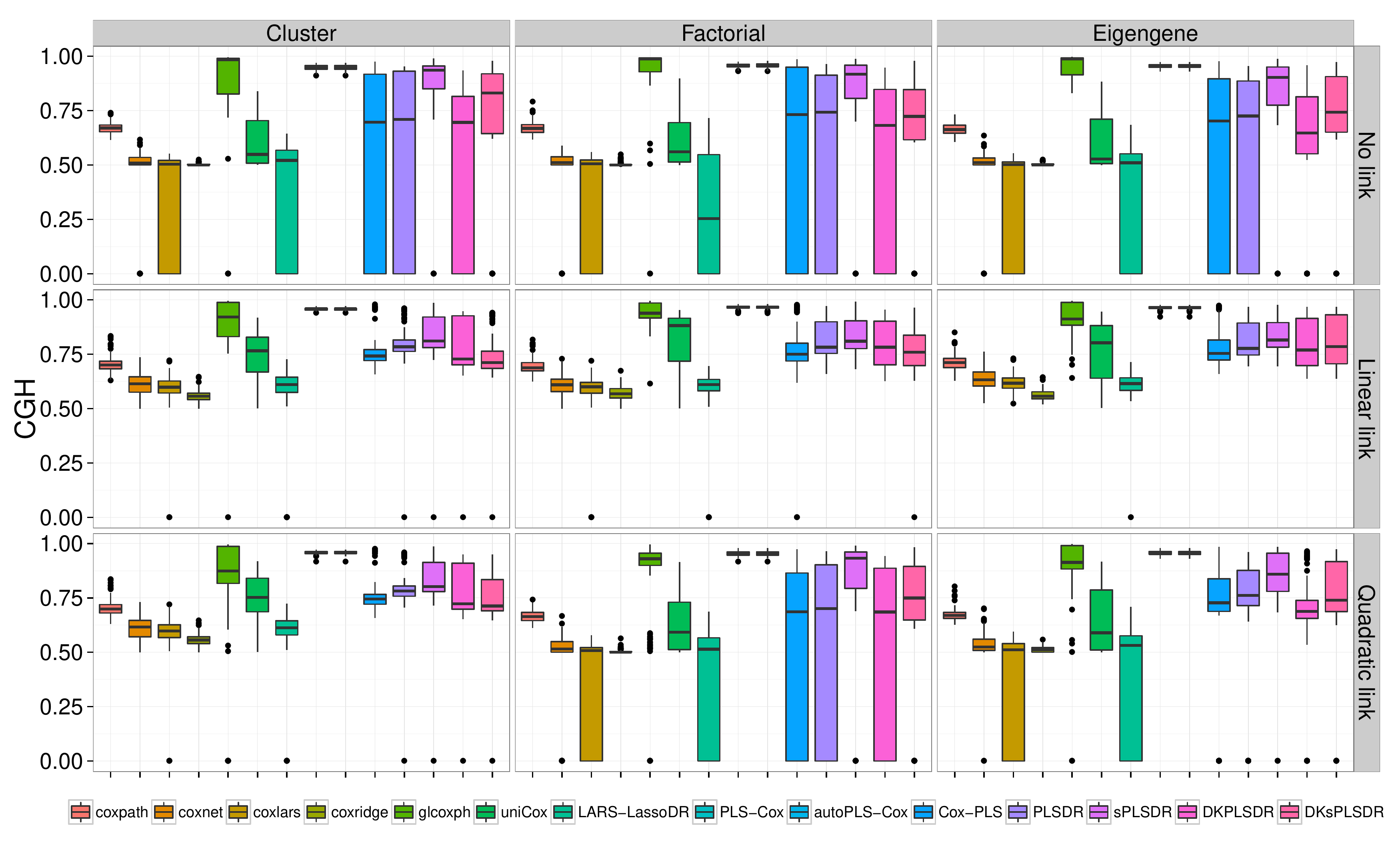}\phantomcaption\label{modmins_AUCsurvROCtest_GonenHellerCI}}}
\vspace{-.5cm}
\caption*{\hspace{-2.5cm}\mbox{Performance, iAUCsurvROC CV. \quad Figure~\ref{modmins_AUCsurvROCtest_R2XO} (left):  R2XO measure. Figure~\ref{modmins_AUCsurvROCtest_GonenHellerCI} (right): GHCI measure.}}
\end{figure}

\begin{figure}[!tpb]
\centerline{{\includegraphics[width=.75\columnwidth]{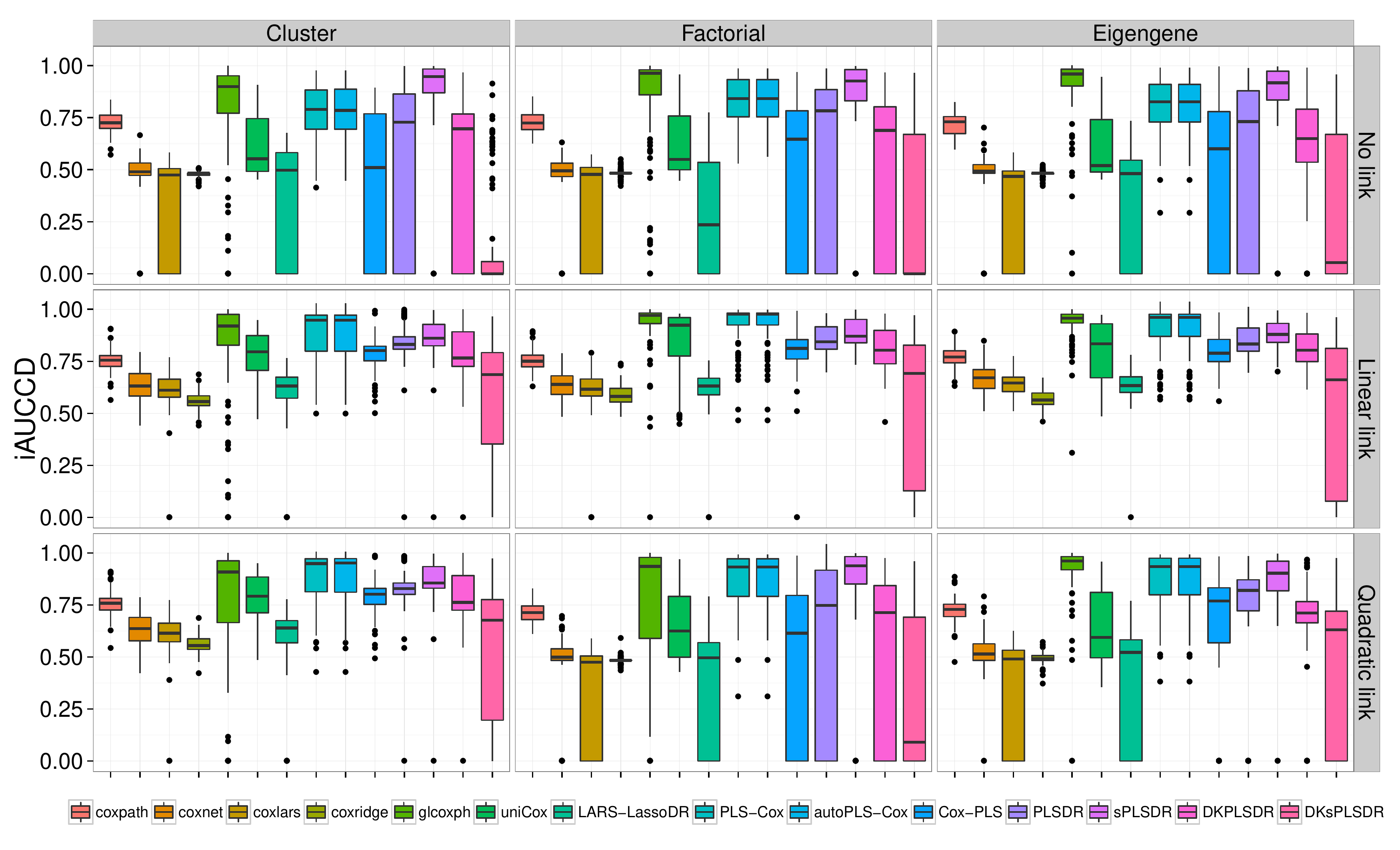}\phantomcaption\label{modmins_AUCsurvROCtest_iAUC_CD}}\qquad{\includegraphics[width=.75\columnwidth]{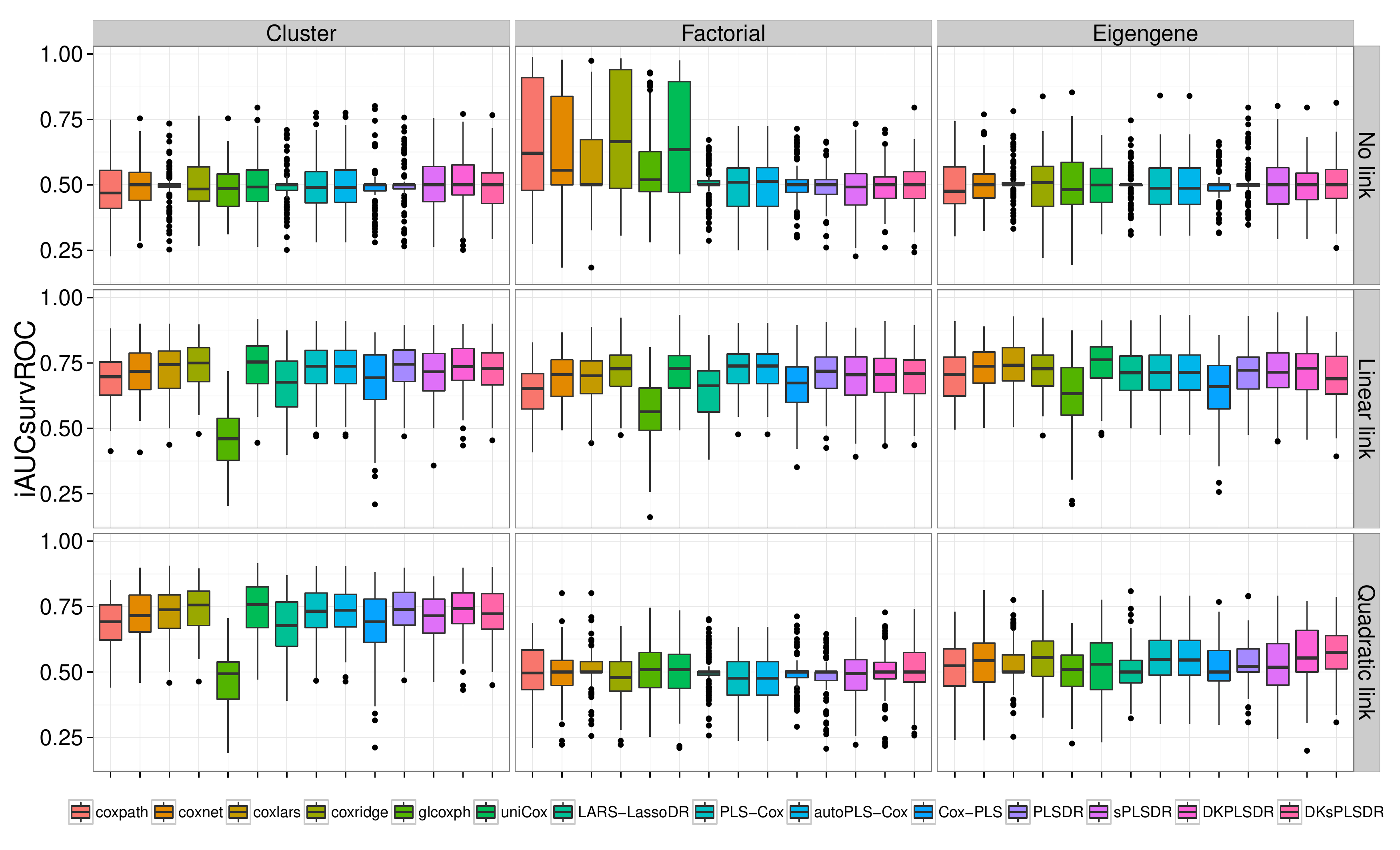}\phantomcaption\label{modmins_AUCsurvROCtest_iAUC_SurvROCtest}}}
\vspace{-.5cm}
\caption*{\hspace{-2.5cm}\mbox{Performance, iAUCsurvROC CV. \quad Figure~\ref{modmins_AUCsurvROCtest_iAUC_CD} (left):  iAUCCD measure. Figure~\ref{modmins_AUCsurvROCtest_iAUC_SurvROCtest} (right): iAUCSurvROC measure.}}
\end{figure}

\begin{figure}[!tpb]
\centerline{{\includegraphics[width=.75\columnwidth]{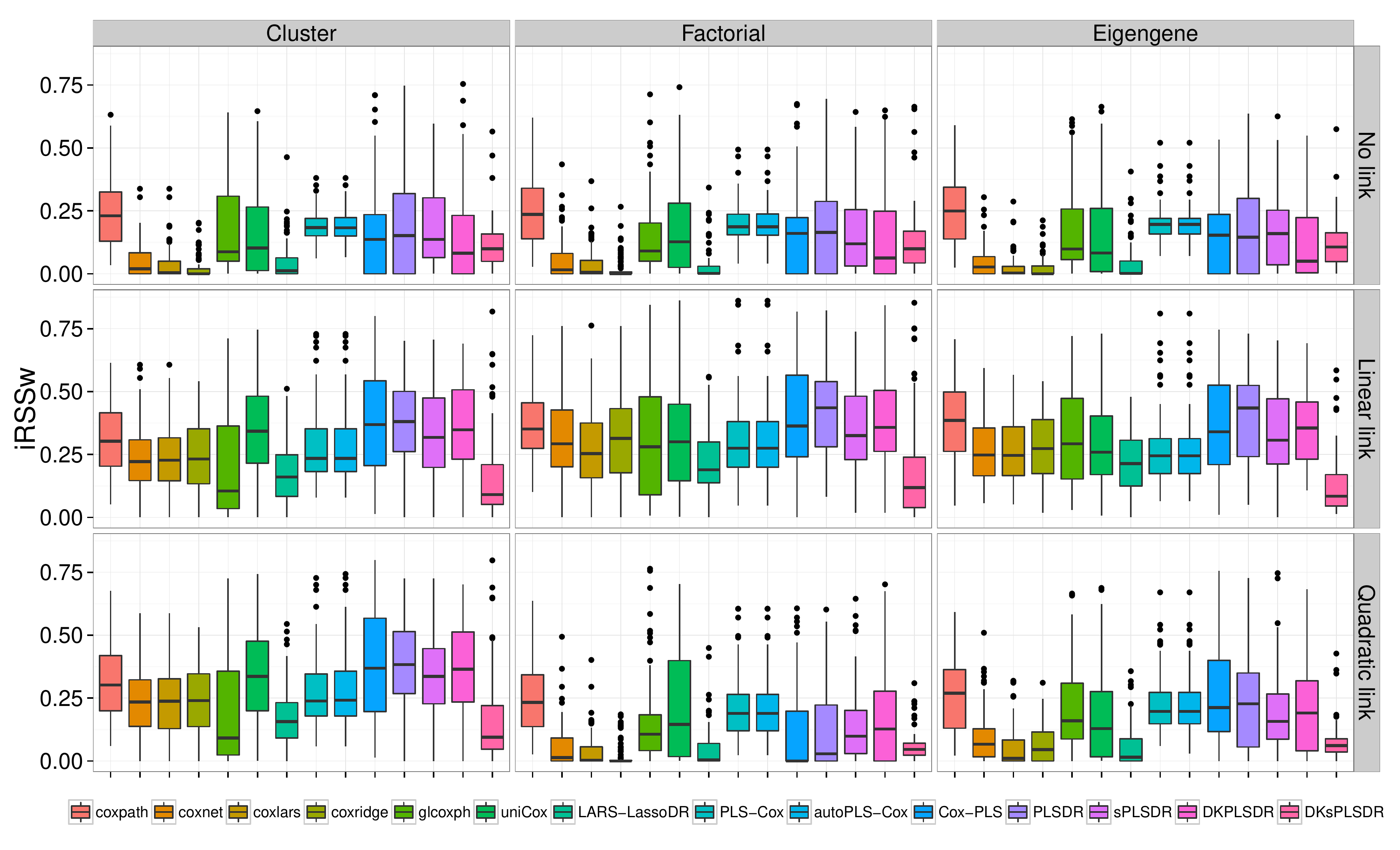}\phantomcaption\label{modmins_AUCsurvROCtest_iRSSw}}\qquad{\includegraphics[width=.75\columnwidth]{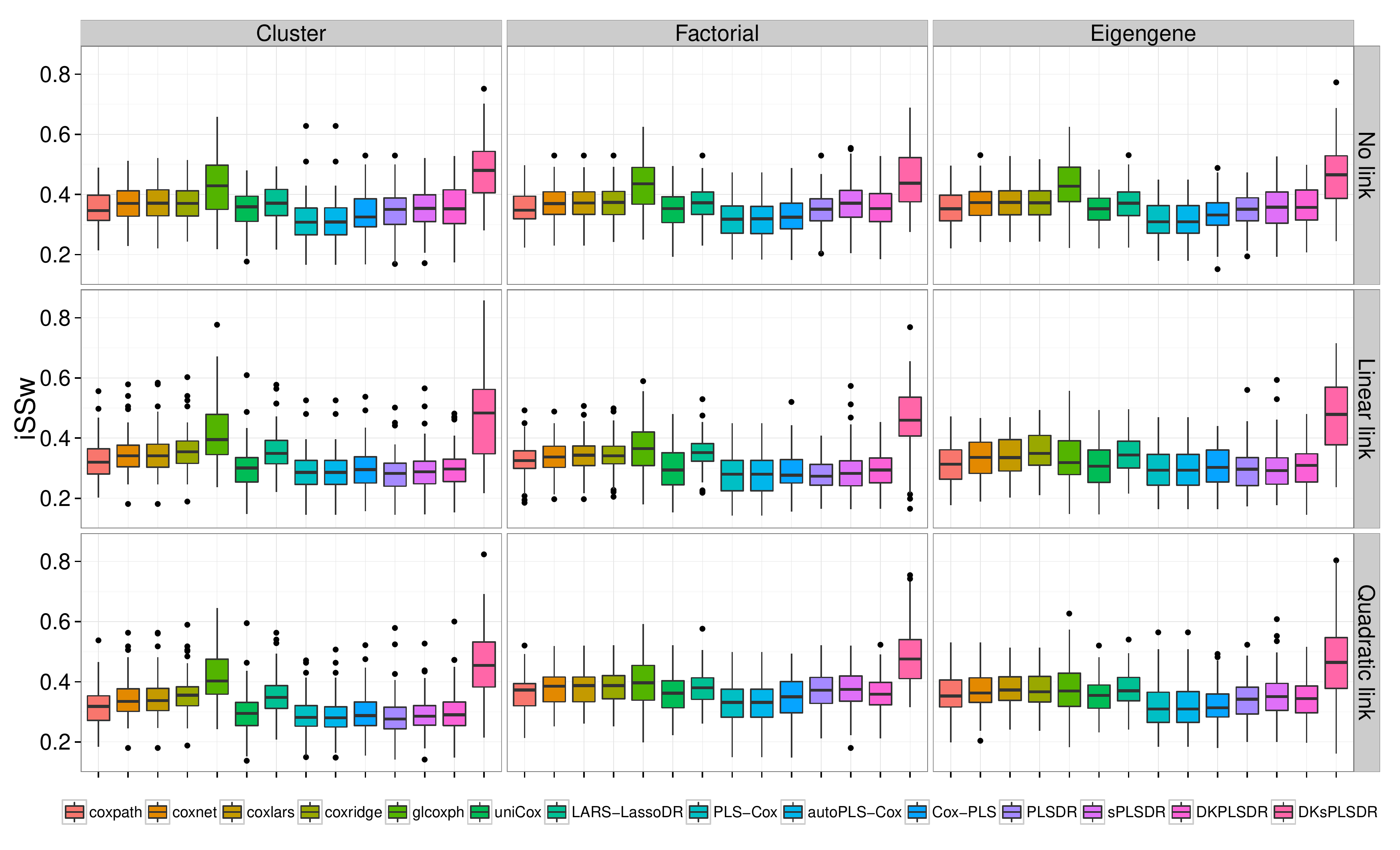}\phantomcaption\label{modmins_AUCsurvROCtest_irobustSw}}}
\vspace{-.5cm}
\caption*{\hspace{-2.5cm}\mbox{Performance, iAUCsurvROC CV. \quad Figure~\ref{modmins_AUCsurvROCtest_iRSSw} (left):  iRSSw measure. Figure~\ref{modmins_AUCsurvROCtest_irobustSw} (right): iSSw measure.}}
\end{figure}

As can be seen for iAUCUno (Figures \ref{modmins_AUCUno_R2XO}, \ref{modmins_AUCUno_GonenHellerCI}, \ref{modmins_AUCUno_iAUC_CD}, \ref{modmins_AUCUno_iAUC_SurvROCtest}, \ref{modmins_AUCUno_iRSSw} and \ref{modmins_AUCUno_irobustSw}) and iAUCSurvROC (Figures \ref{modmins_AUCsurvROCtest_R2XO}, \ref{modmins_AUCsurvROCtest_GonenHellerCI}, \ref{modmins_AUCsurvROCtest_iAUC_CD}, \ref{modmins_AUCsurvROCtest_iAUC_SurvROCtest}, \ref{modmins_AUCsurvROCtest_iRSSw} and \ref{modmins_AUCsurvROCtest_irobustSw}), the improvement of the performances due to switch to the recommended CV criteria is high enough to even have some (S)PLS based models, for instance SPLSDR, show some advantage over the other benchmark methods.

\section{Conclusion}
When cross-validating standard or extended Cox models, the commonly used criterion is the cross-validated partial loglikelihood using a naive or a van Houwelingen scheme. Quite astonishingly, these two cross-validation methods fail with all the 7 extensions of partial least squares regression to the Cox model, namely PLS-Cox, autoPLS-Cox, Cox-PLS, PLSDR, sPLSDR, DKPLSDR and DKsPLSDR.

In our simulation study, we introduced 12 cross validation criteria based on three different kind of model quality assessment:
\begin{itemize}
\item Likelihood (2): Verweij and Van Houwelingen (classic CVLL, 1993), van Houwelingen et al. (vHCVLL, 2006).
\item Integrated AUC measures (6): Chambless and Diao's (iAUCCD, 2006), Hung and Chiang's (iAUCHC, 2010), Song and Zhou's (iAUCSH, 2008), Uno et al.'s (iAUCUno,
2007), Heagerty and Zheng's (iAUCHZ, 2005), Heagerty et al.'s (iAUCsurvROC, 2000).
\item Prediction error criteria (4): integrated (un)weighted Brier Score (iBS(un)w, Gerds and Schumacher (2006)) or Schmid Score (iSS(un)w, Schmid et al. (2011))
\end{itemize}

Our simulation study was successful in finding good CV criterion for PLS or sPLS based extensions of the Cox model:
\begin{itemize}
\item iAUCsh for PLS-Cox and autoPLS-Cox.
\item iAUCSurvROC and iAUCUno ones for Cox-PLS, (DK)PLSDR and (DK)sPLSDR.
\end{itemize}

The derivation of measures of prediction accuracy for survival data is not straightforward in the presence of censored observations. To overcome this problem, a variety of new approaches has been suggested in the literature. We spotted 23 performance measures that can be classified into three groups:
\begin{itemize}
\item Likelihood-based approaches (llrt, varresmart, 3 R2-type).
\item ROC-based approaches such as integrated AUC (iAUCCD, iAUCHC, iAUCSH, iAUCUno, iAUCHZ,
iAUCsurvROC), 3 C-index (Harrell, GHCI, UnoC).
\item Distance-based approaches such as the V of Schemper and Henderson (2000) or derived from Brier or Schmid
Scores (iBS(un)w, iSS(un)w and 4 derived R2-type measures).
\end{itemize}

Using the newly found cross-validation, and these measures of prediction accuracy, we performed a benchmark reanalysis that showed enhanced performances of these techniques and a much better behaviour even against other well known competitors such as coxnet, coxpath, uniCox and glcoxph.

Hence the recommended criteria not only improve the accuracy of the choice of the number of components but also strongly raise the performances of the models, which enables some of them to overperform the other benchmark methods.

\vfill\eject

\clearpage

\section*{References}

\bibliographystyle{plainnat}
\bibliography{bibpls}

\clearpage

\section*{Supplemental Information}
\subsection*{Insights on the implementation of the methods}
We detail the implementation of the algorithms that we used in the article and start with some shared properties of these. Whenever the deviance residuals and survival models were to be derived, we used the \verb+survival+ R-package (\citealp{ther00,survival}).
As a PLS regression function in the \verb+plsRcox+ R-package, we made three wrappers using either the \verb+pls+ function of the \verb+pls+ R-package (\citealp{pls}), the \verb+plsR+ function of the \verb+plsRglm+ R-package (\citealp{plsRglm}) or the \verb+pls+ function of the \verb+mixOmics+ R-package (\citealp{mixOmics}). The last two are based on the NIPALS algorithm and hence automatically handle missing data (\citealp{tenen98}) in the explanatory variables. In addition, the \verb+pls+ function of the \verb+mixOmics+ R-package (\citealp{mixOmics}) can quickly handle big datasets such as Dataset~5 with 242 rows and 44754 variables. As a consequence, we had to make the \verb+spls+ function of the \verb+spls+ R-package use this function instead of the \verb+pls+ function of the \verb+pls+ R-package (\citealp{pls}).

\begin{enumerate}
\item The authors made a wrapper in the \verb+plsRcox+ R-package (\citealp{plsRcox}) to derive a sPLSDR implementation from the \verb+spls+ R-package (\citealp{spls}).
\item The authors made a wrapper in the \verb+plsRcox+ R-package (\citealp{plsRcox}) to derive a DKsPLSDR implementation from the \verb+spls+ (\citealp{spls}) and the \verb+kernlab+ (\citealp{kernlab}) R-packages.
\item The coxpath implementation was the \verb+coxpath+ function found in the \verb+glmpath+ R-package (\citealp{glmpath}). 
\item The coxnet implementation was the \verb+glmnet+ function, with the Cox family option, in the \verb+glmnet+ R-package (\citealp{glmnet10,coxnet11}).
\item The coxlars implementation was the \verb+glmnet+ function, with the Cox family option and the \verb+alpha+ option set to $0$, in the \verb+glmnet+ R-package (\citealp{glmnet10,coxnet11}).
\item The coxridge implementation was the \verb+glmnet+ function, with the Cox family option and the \verb+alpha+ option set to $1$, in the \verb+glmnet+ R-package (\citealp{glmnet10,coxnet11}).
\item PLS-Cox has not yet been implemented in \verb+R+ and the authors made it available as the function \verb+plsRcox+ of the \verb+plsRcox+ R-package (\citealp{plsRcox}).
\item autoPLS-Cox is PLS-Cox with sparse PLS components and automatic selection of their optimal number. The computed components are sparse since, to have a non zero coefficient, a variable must be significant at a given $\eta$ level in the cox regression of the response by this variable adjusted by the previously found components. The model stops adding a new component when there is no longer any of the explanatory variable that is significant at a given $\eta$ level. The authors made it available as the function \verb+plsRcox+ of the \verb+plsRcox+ R-package (\citealp{plsRcox}). The number of components can also be determined using cross-validation, the components being still sparse.
\item The authors made a wrapper in the \verb+plsRcox+ R-package (\citealp{plsRcox}) to derive a LARS-LassoDR implementation from the \verb+lars+ R-package (\citealp{lars}).
\item The authors made a wrapper in the \verb+plsRcox+ R-package (\citealp{plsRcox}) to derive a Cox-PLS implementation from either the \verb+pls+ or \verb+plsRglm+ R-packages.
\item The authors made a wrapper in the \verb+plsRcox+ R-package (\citealp{plsRcox}) to derive a PLSDR implementation from either the \verb+pls+ or \verb+plsRglm+ R-packages.
\item The authors made a wrapper in the \verb+plsRcox+ R-package (\citealp{plsRcox}) to derive a DKPLSDR implementation from the \verb+pls+ (\citealp{pls}) and the \verb+kernlab+ (\citealp{kernlab}) R-packages.
\item The uniCox implementation was the \verb+uniCox+ function found in the \verb+uniCox+ R-package (\citealp{uniCox}). 
\item The glcoxph implementation was the \verb+glcoxph+ function found in the \verb+glcoxph+ R-package (\citealp{glcoxph}). 
\end{enumerate}

\clearpage

\begin{figure}[!tpb]
\centerline{{\includegraphics[width=.75\columnwidth]{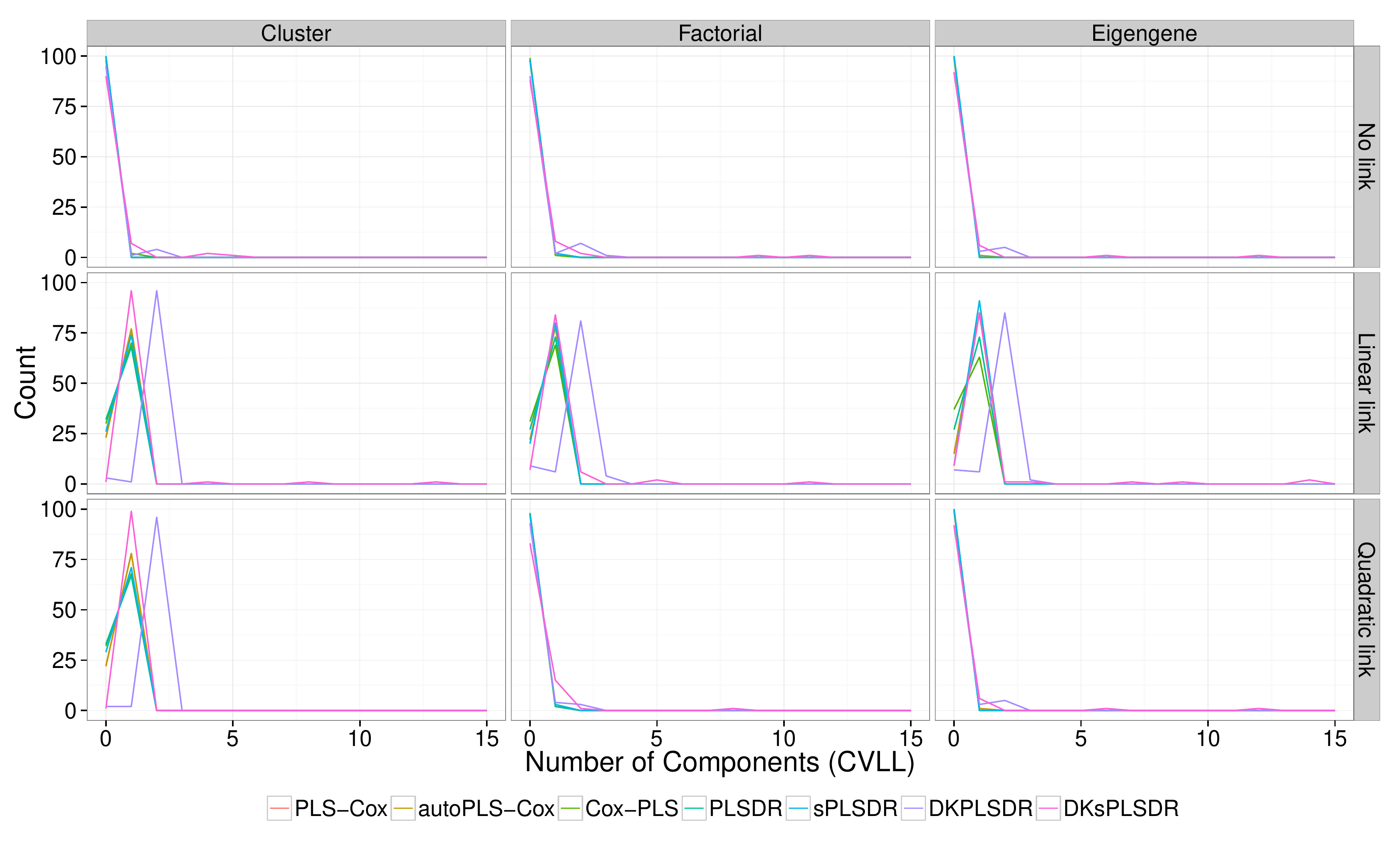}\phantomcaption\label{NbrComp_cvll_alt}}\qquad{\includegraphics[width=.75\columnwidth]{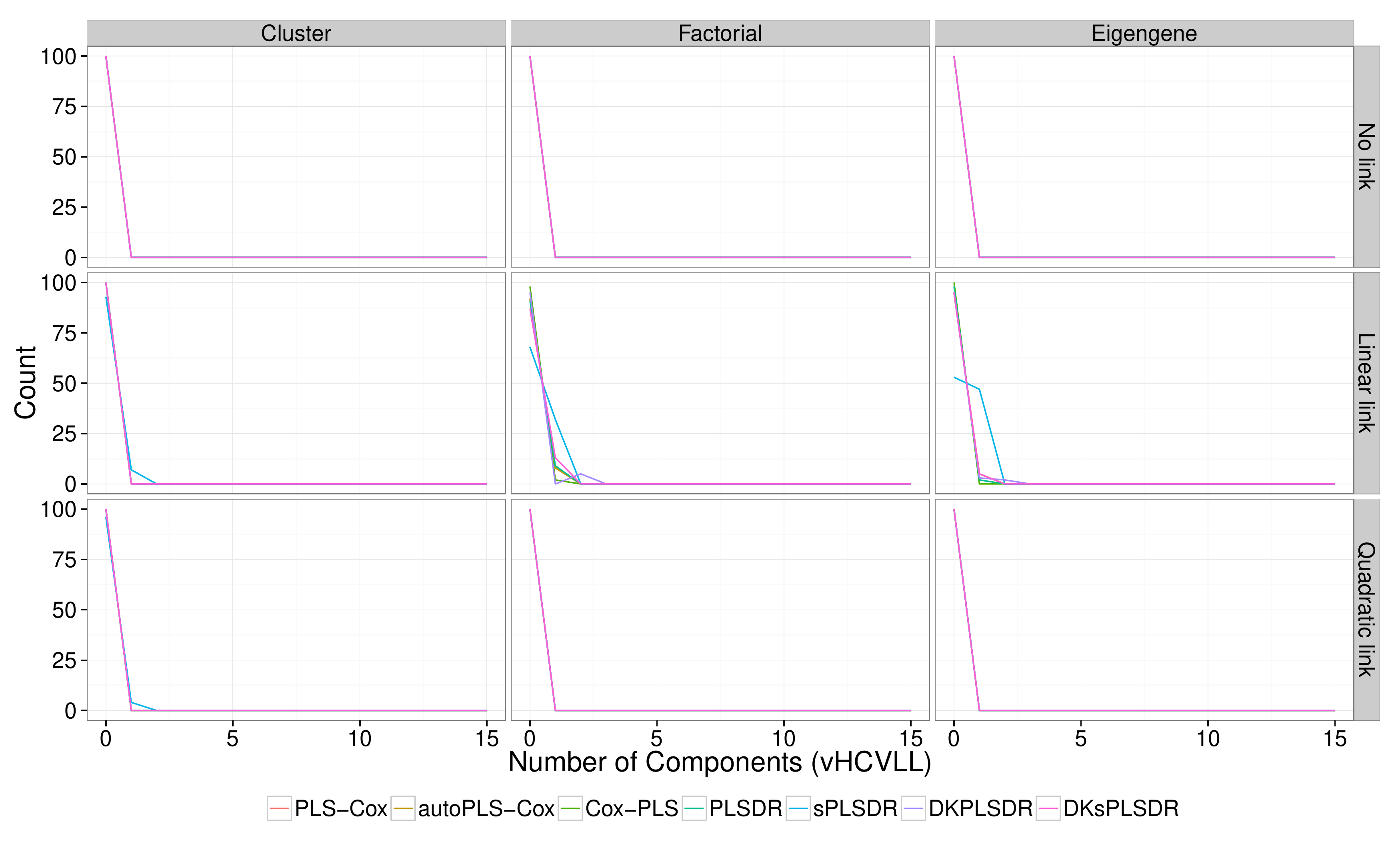}\phantomcaption\label{NbrComp_vanHcvll_alt}}}
\vspace{-.5cm}
\caption*{\hspace{-1.0cm}\mbox{Figure~\ref{NbrComp_cvll_alt}:  Nbr of comp, LL criterion. \qquad\qquad\qquad Figure~\ref{NbrComp_vanHcvll_alt}:  Nbr of comp, vHLL criterion.}}
\end{figure}

\begin{figure}[!tpb]
\centerline{{\includegraphics[width=.75\columnwidth]{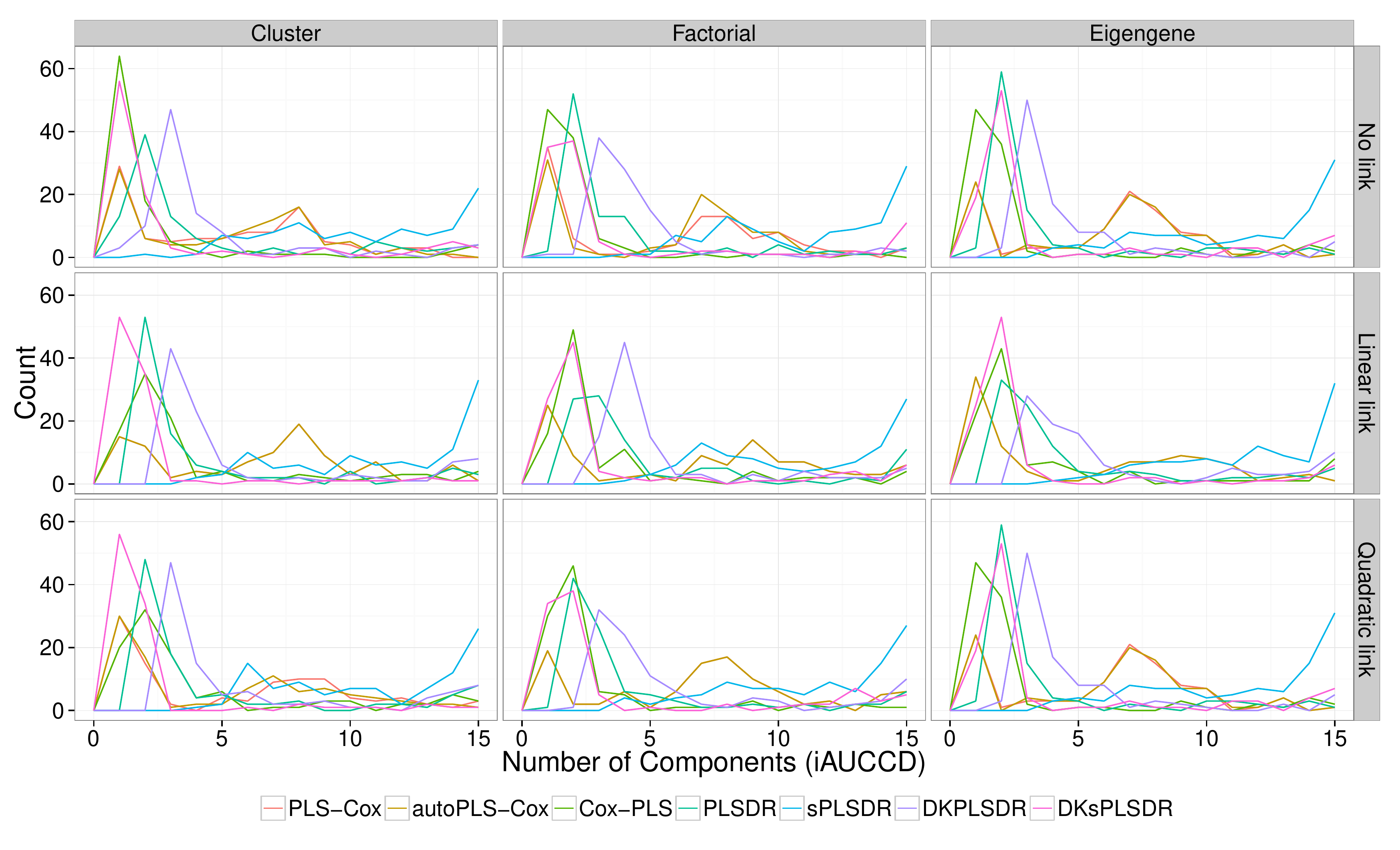}\phantomcaption\label{NbrComp_AUCcd_alt}}\qquad{\includegraphics[width=.75\columnwidth]{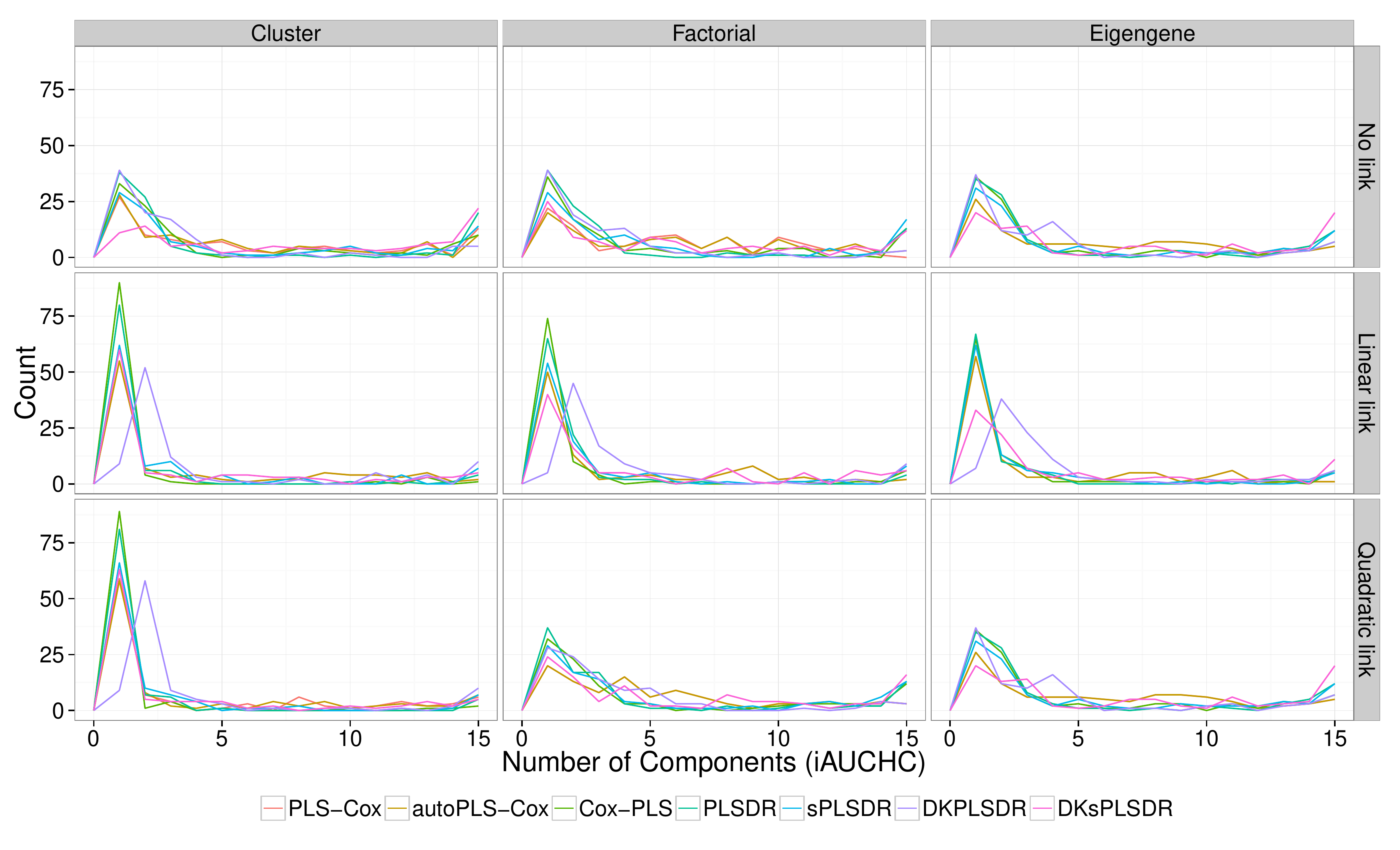}\phantomcaption\label{NbrComp_AUChc_alt}}}
\vspace{-.5cm}
\caption*{\hspace{-1.0cm}\mbox{Figure~\ref{NbrComp_AUCcd_alt}:  Nbr of comp, iAUCCD criterion. \qquad Figure~\ref{NbrComp_AUChc_alt}:  Nbr of comp, iAUCHC criterion.}}
\end{figure}

\begin{figure}[!tpb]
\centerline{{\includegraphics[width=.75\columnwidth]{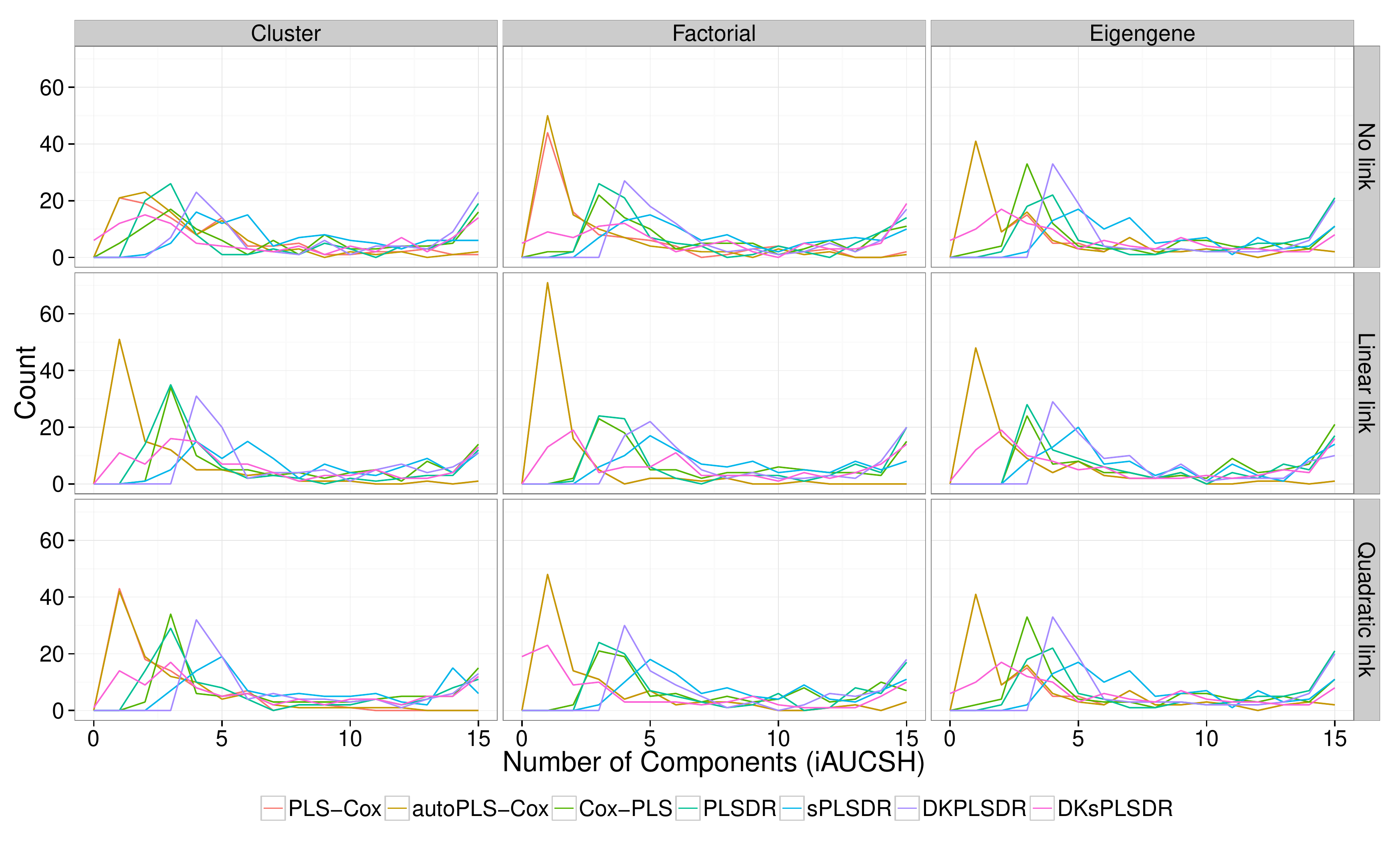}\phantomcaption\label{NbrComp_AUCsh_alt}}\qquad{\includegraphics[width=.75\columnwidth]{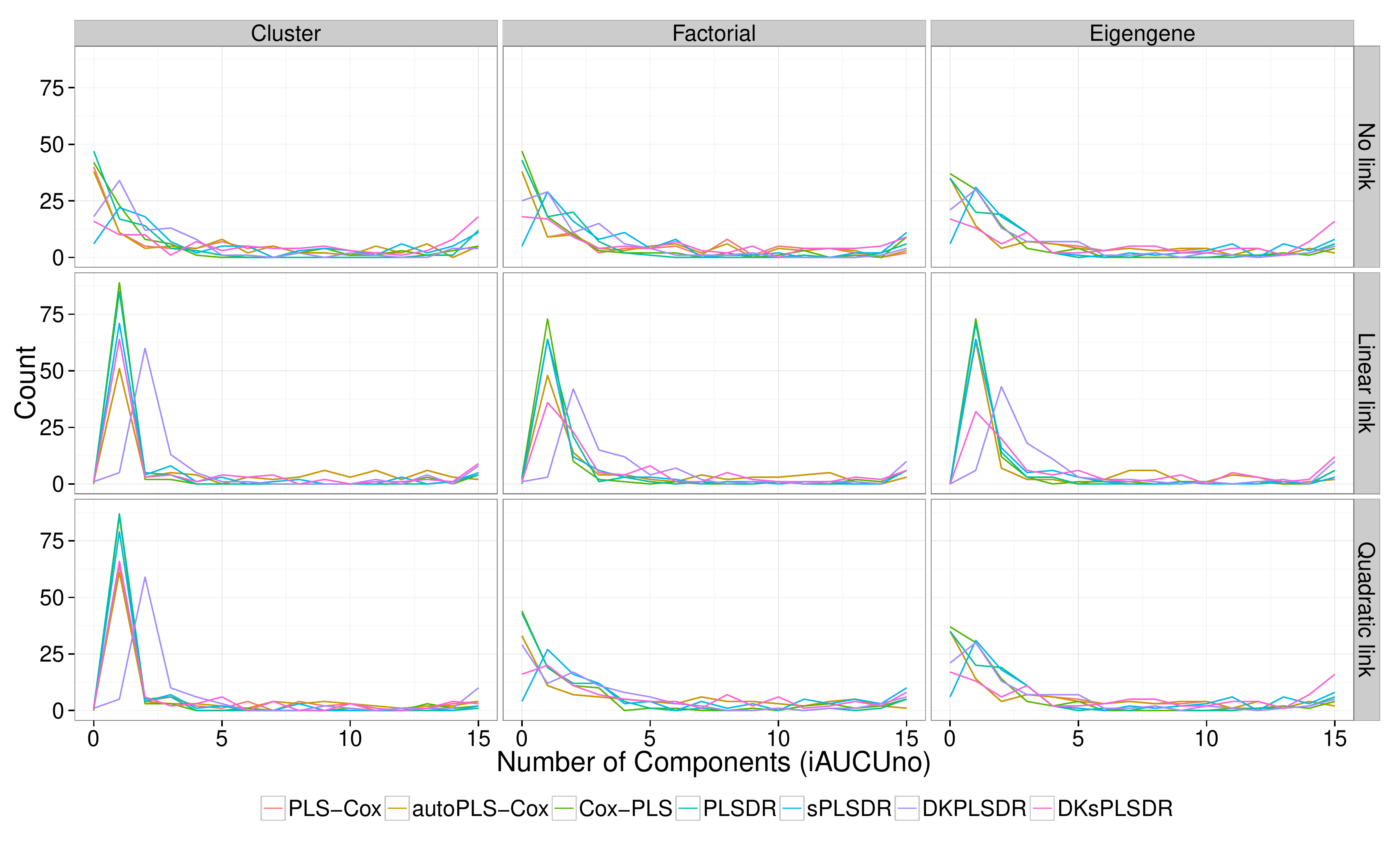}\phantomcaption\label{NbrComp_AUCUno_alt}}}
\vspace{-.5cm}
\caption*{\hspace{-1.0cm}\mbox{Figure~\ref{NbrComp_AUCsh_alt}:  Nbr of comp, iAUCSH criterion. \qquad Figure~\ref{NbrComp_AUCUno_alt}:  Nbr of comp, iAUCUno criterion.}}
\end{figure}

\clearpage

\begin{figure}[!tpb]
\centerline{{\includegraphics[width=.75\columnwidth]{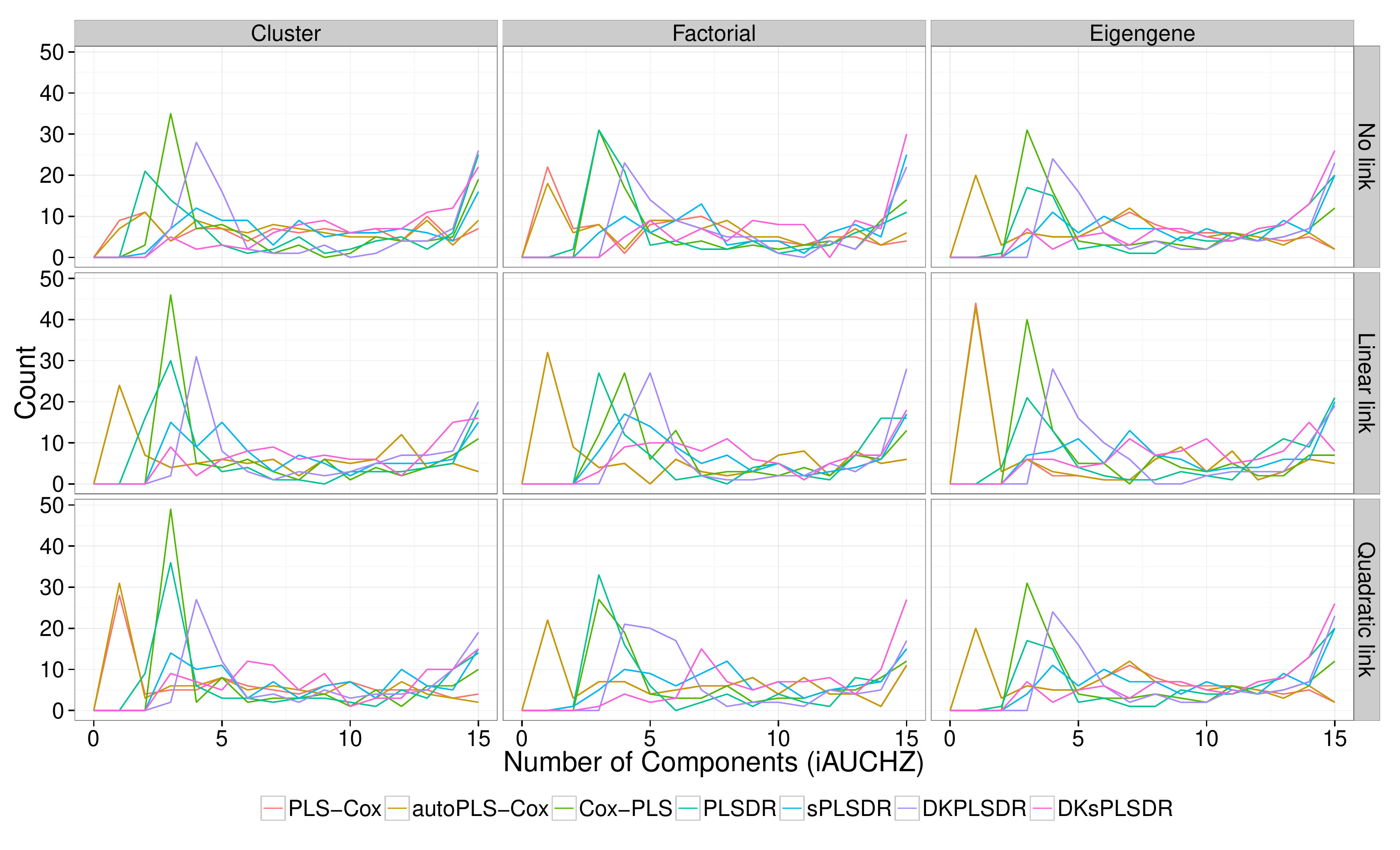}\phantomcaption\label{NbrComp_AUChztest_alt}}\qquad{\includegraphics[width=.75\columnwidth]{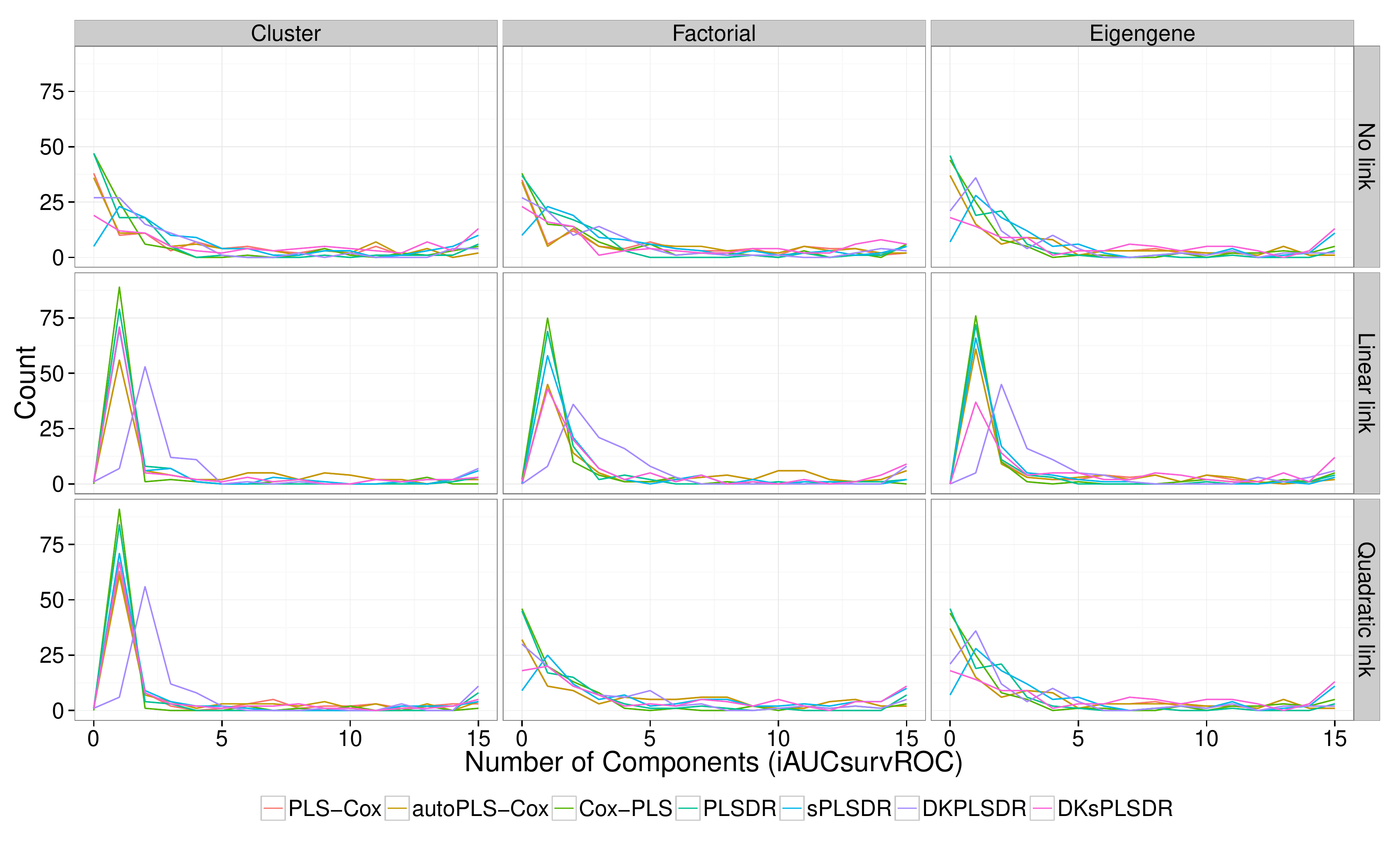}\phantomcaption\label{NbrComp_AUCsurvROCtest_alt}}}
\vspace{-.5cm}
\caption*{\hspace{-1.0cm}\mbox{Figure~\ref{NbrComp_AUChztest_alt}:  Nbr of comp, iAUCHZ criterion. \qquad Figure~\ref{NbrComp_AUCsurvROCtest_alt}:  Nbr of comp, iAUCSurvROC criterion.}}
\end{figure}

\begin{figure}[!tpb]
\centerline{{\includegraphics[width=.75\columnwidth]{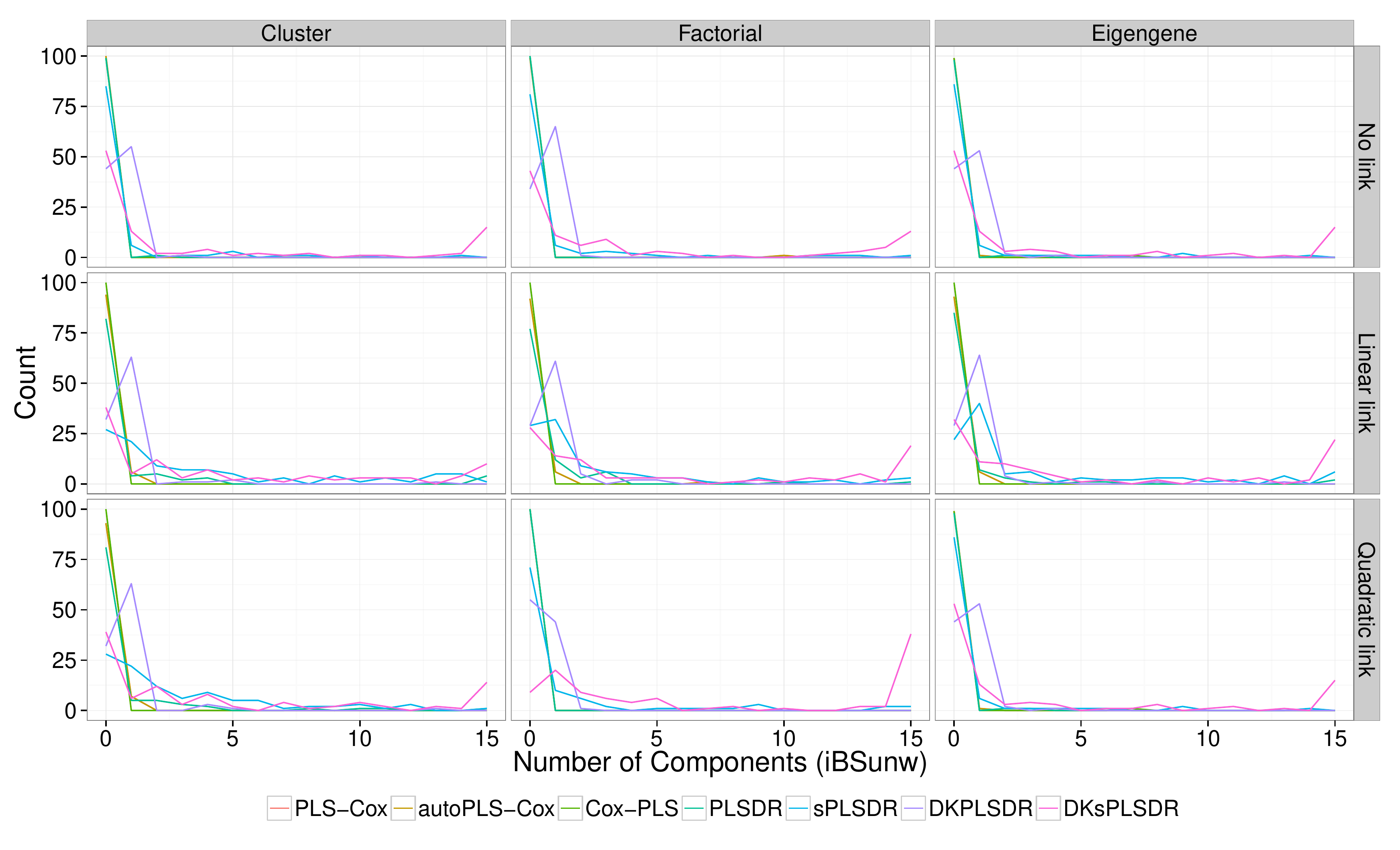}\phantomcaption\label{NbrComp_iBSunw_alt}}\qquad{\includegraphics[width=.75\columnwidth]{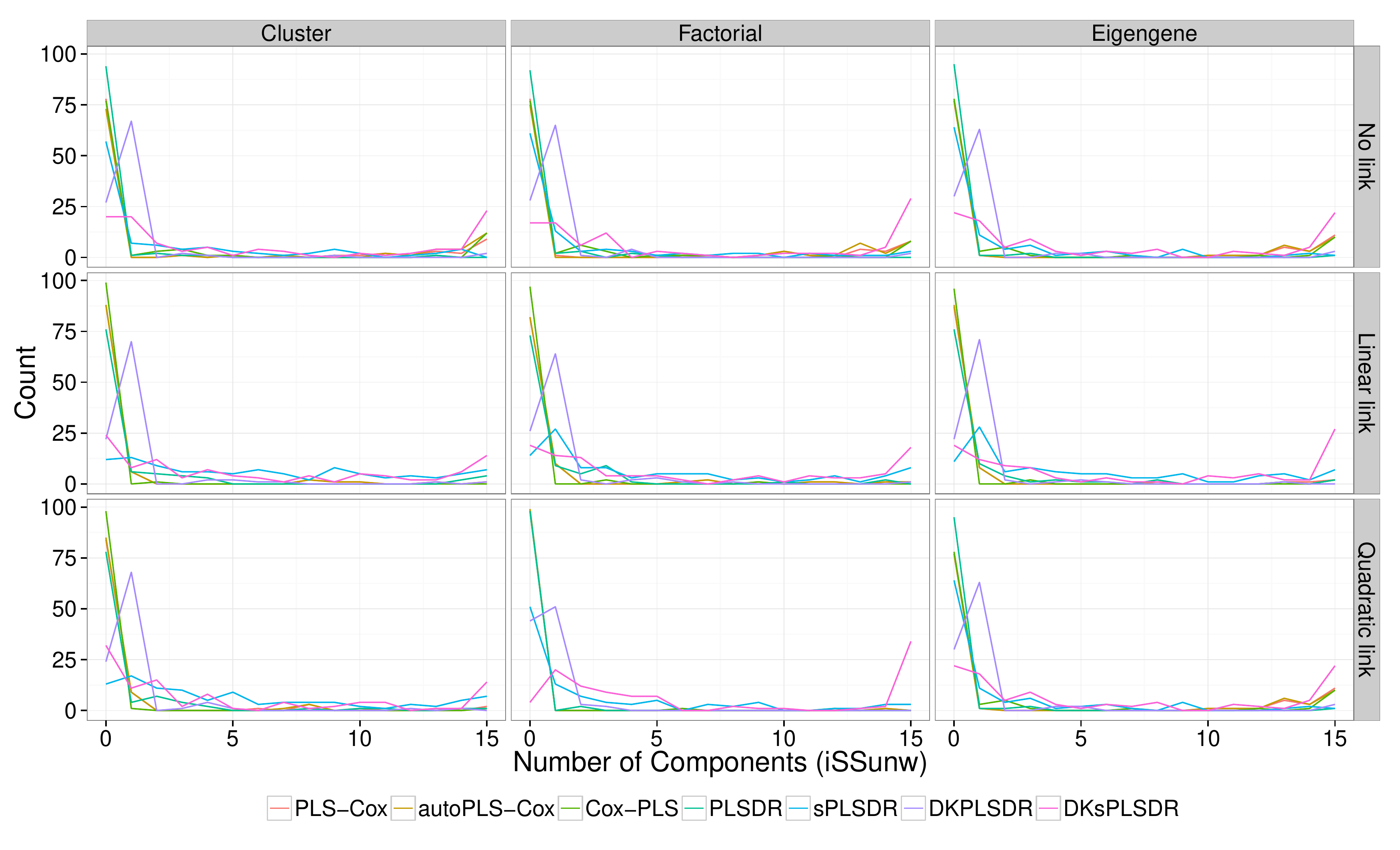}\phantomcaption\label{NbrComp_iSchmidSunw_alt}}}
\vspace{-.5cm}
\caption*{\hspace{-1.0cm}\mbox{Figure~\ref{NbrComp_iBSunw_alt}:  Nbr of comp, iBSunw criterion. \qquad\qquad Figure~\ref{NbrComp_iSchmidSunw_alt}:  Nbr of comp, iSSunw criterion.}}
\end{figure}

\begin{figure}[!tpb]
\centerline{{\includegraphics[width=.75\columnwidth]{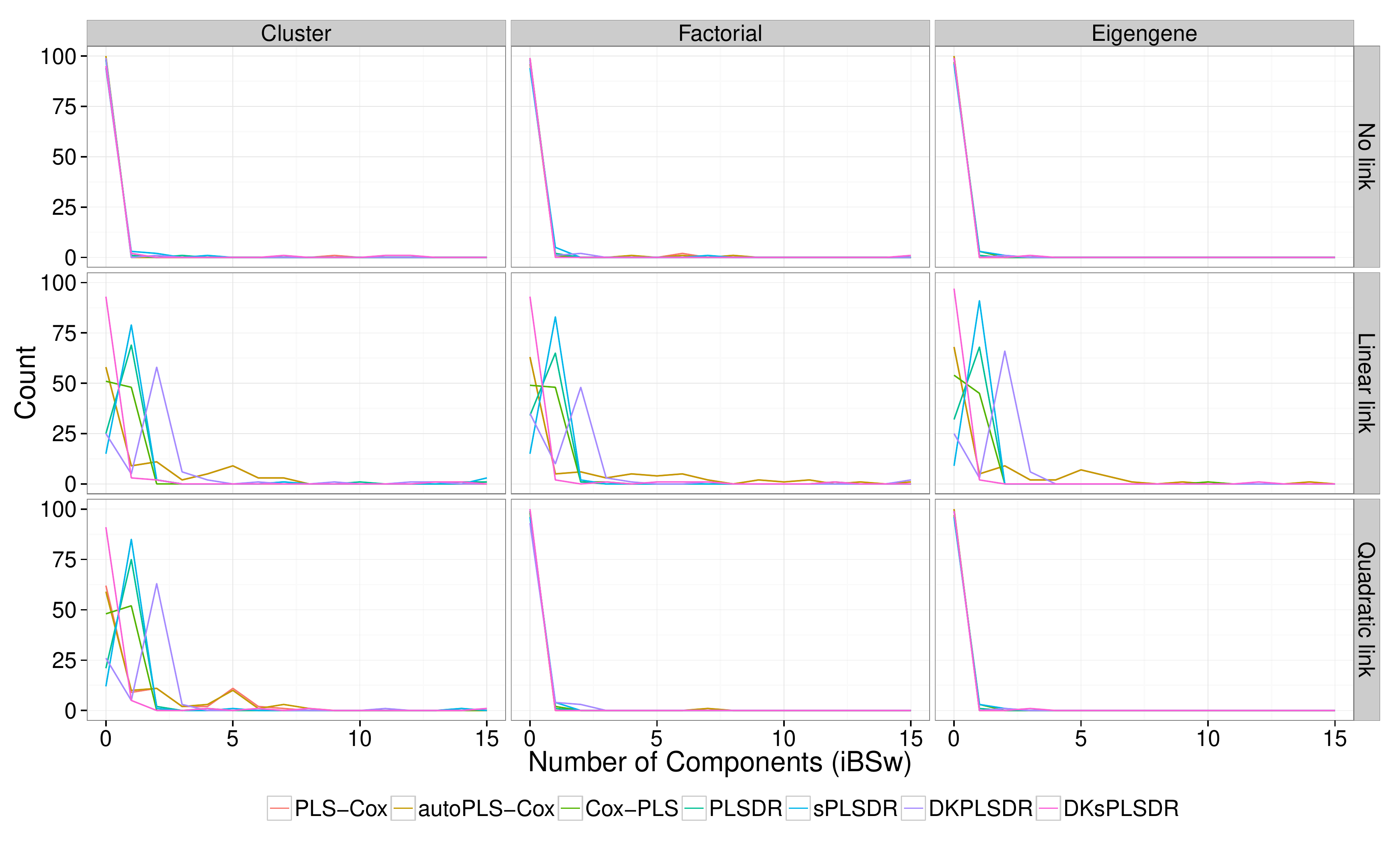}\phantomcaption\label{NbrComp_iBSw_alt}}\qquad{\includegraphics[width=.75\columnwidth]{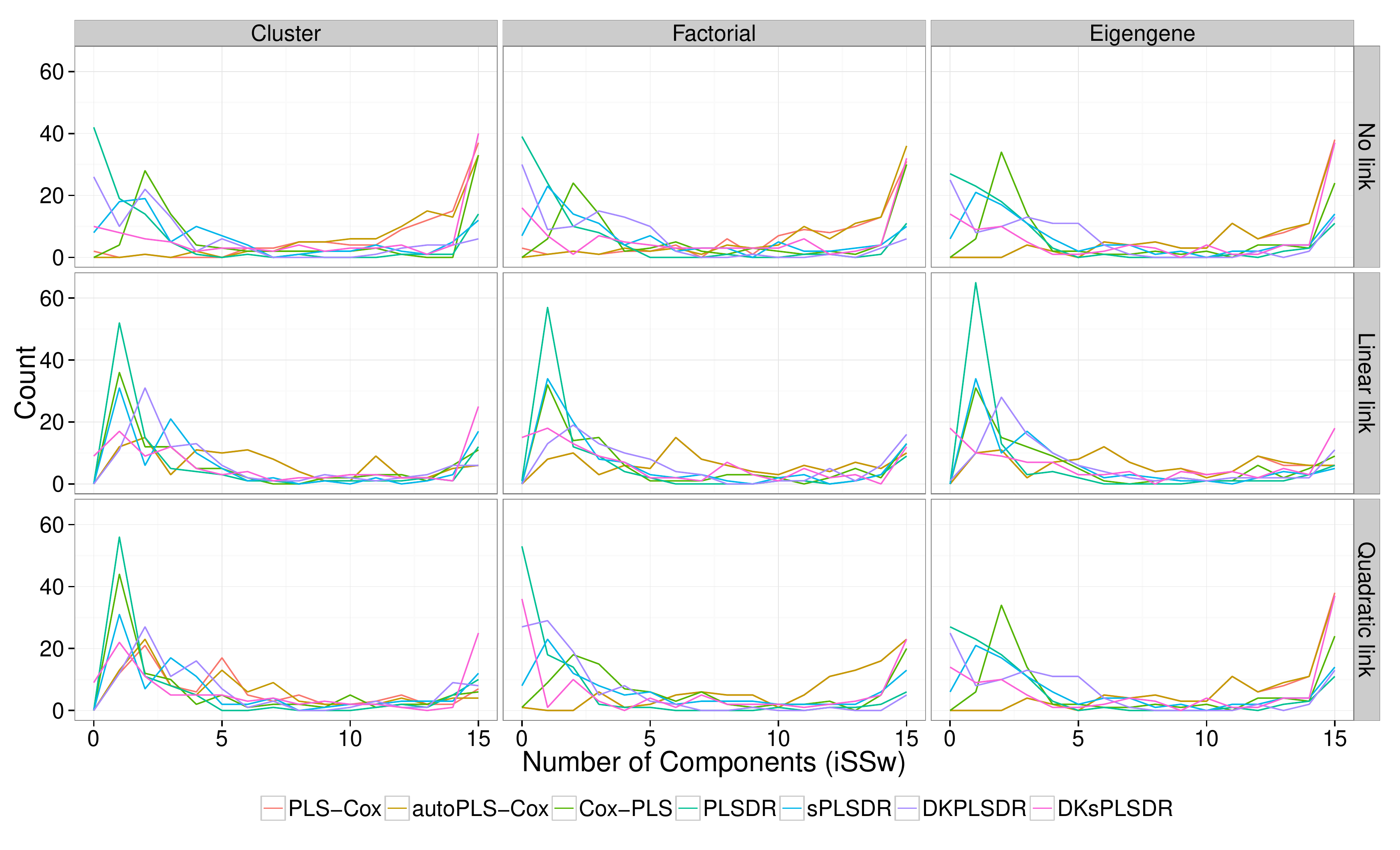}\phantomcaption\label{NbrComp_iSchmidSw_alt}}}
\vspace{-.5cm}
\caption*{\hspace{-1.0cm}\mbox{Figure~\ref{NbrComp_iBSw_alt}:  Nbr of comp, iBSw criterion. \qquad\qquad Figure~\ref{NbrComp_iSchmidSw_alt}:  Nbr of comp, iSSw criterion.}}
\end{figure}

\clearpage

\begin{figure}[!tpb]
\centerline{{\includegraphics[width=.75\columnwidth]{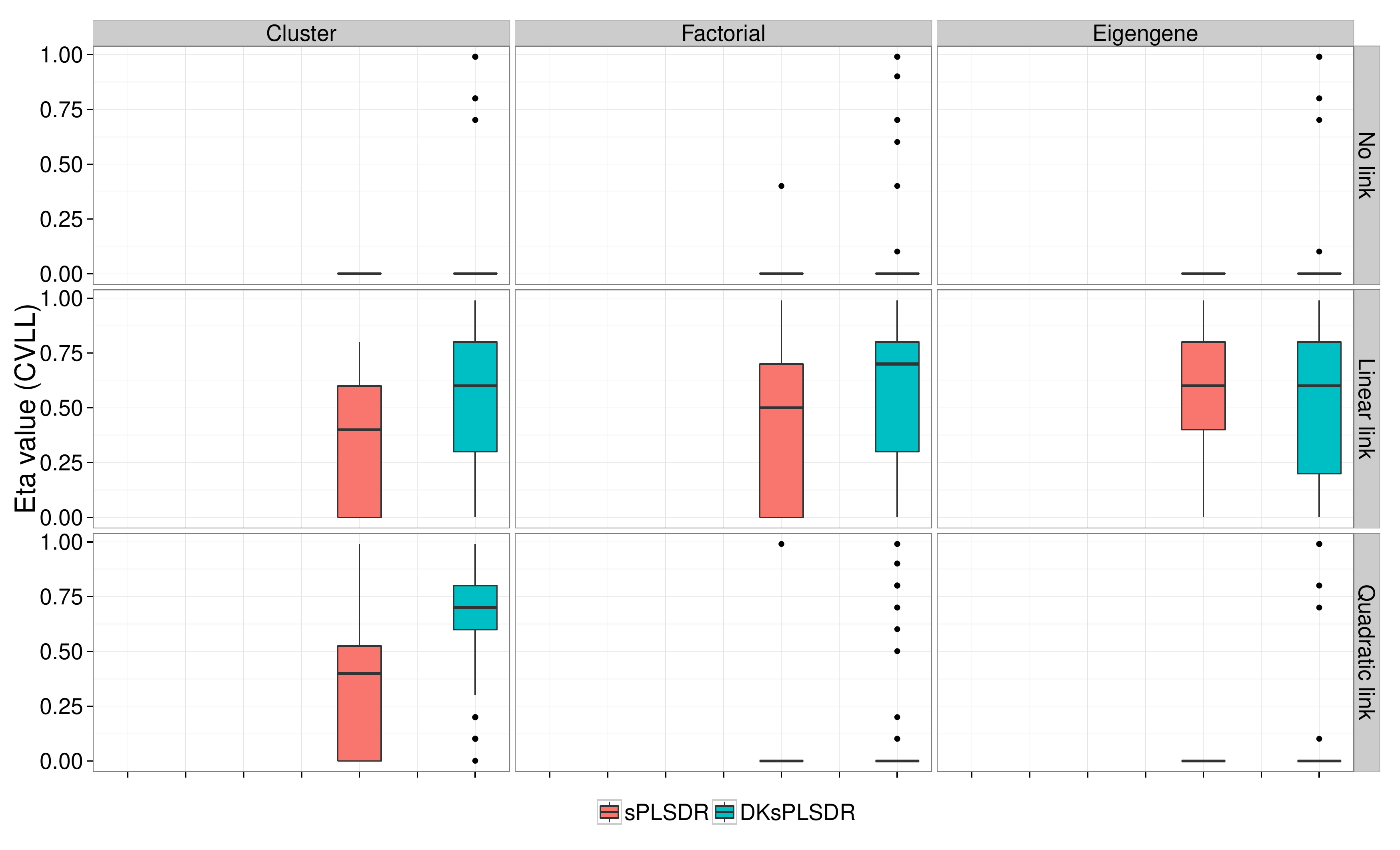}\phantomcaption\label{eta_cvll}}\qquad{\includegraphics[width=.75\columnwidth]{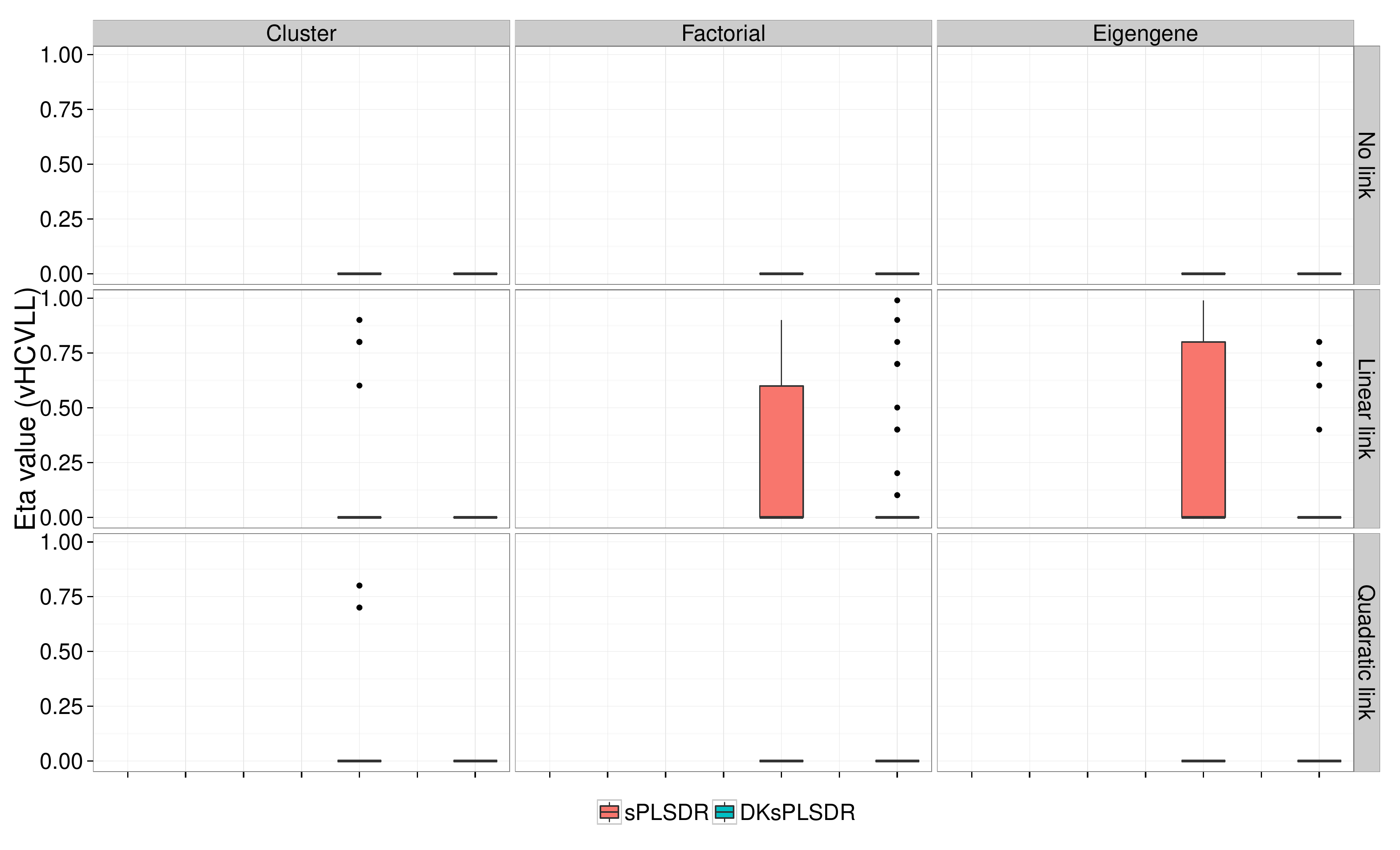}\phantomcaption\label{eta_vanHcvll}}}
\vspace{-.5cm}
\caption*{Figure~\ref{eta_cvll}:  $\eta$, LL criterion. \qquad\qquad\qquad Figure~\ref{eta_vanHcvll}:  $\eta$, vHLL criterion.}
\end{figure}

\begin{figure}[!tpb]
\centerline{{\includegraphics[width=.75\columnwidth]{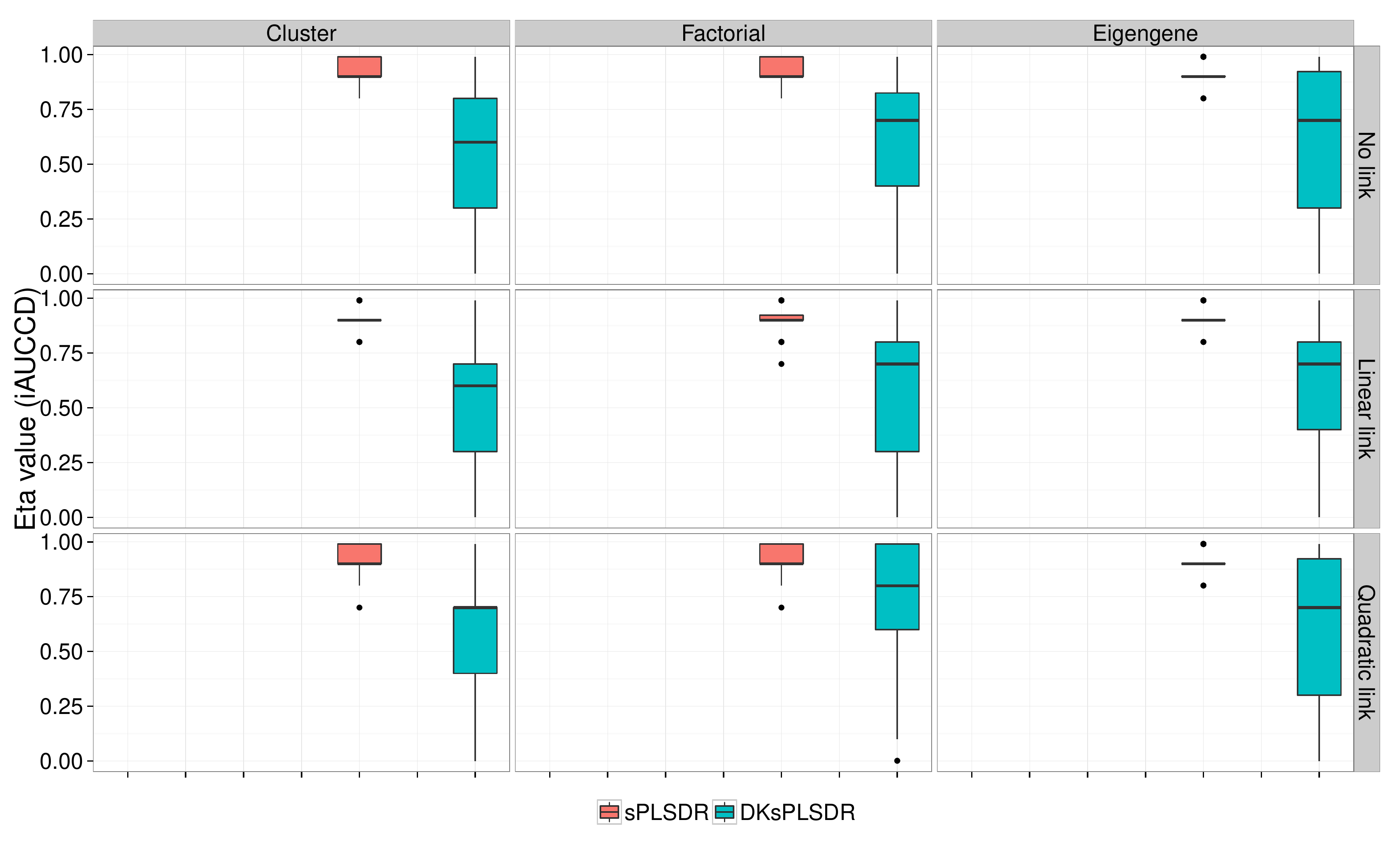}\phantomcaption\label{eta_AUCcd}}\qquad{\includegraphics[width=.75\columnwidth]{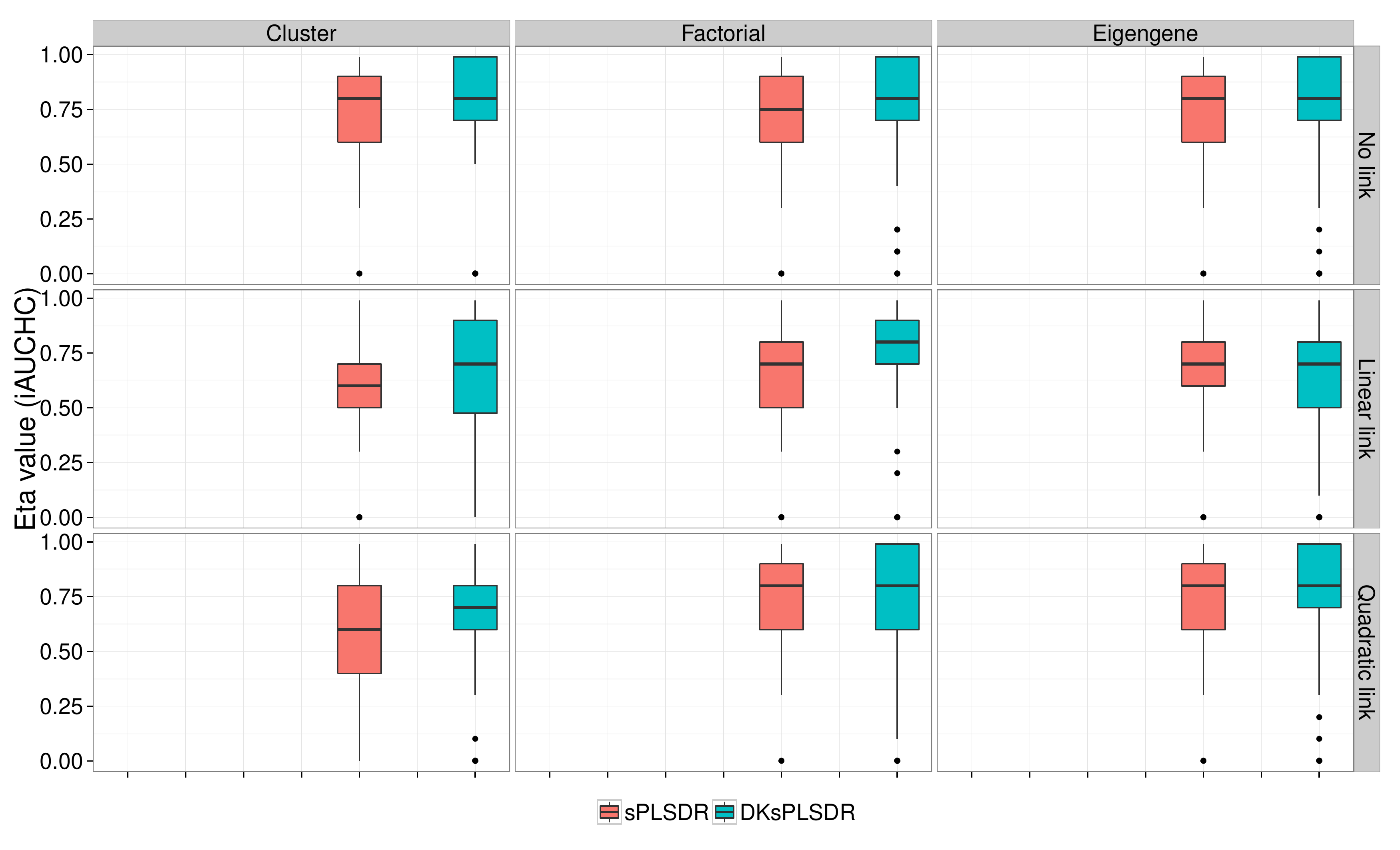}\phantomcaption\label{eta_AUChc}}}
\vspace{-.5cm}
\caption*{\mbox{Figure~\ref{eta_AUCcd}:  $\eta$, iAUCCD criterion. \qquad Figure~\ref{eta_AUChc}:  $\eta$, iAUCHC criterion.}}
\end{figure}

\begin{figure}[!tpb]
\centerline{{\includegraphics[width=.75\columnwidth]{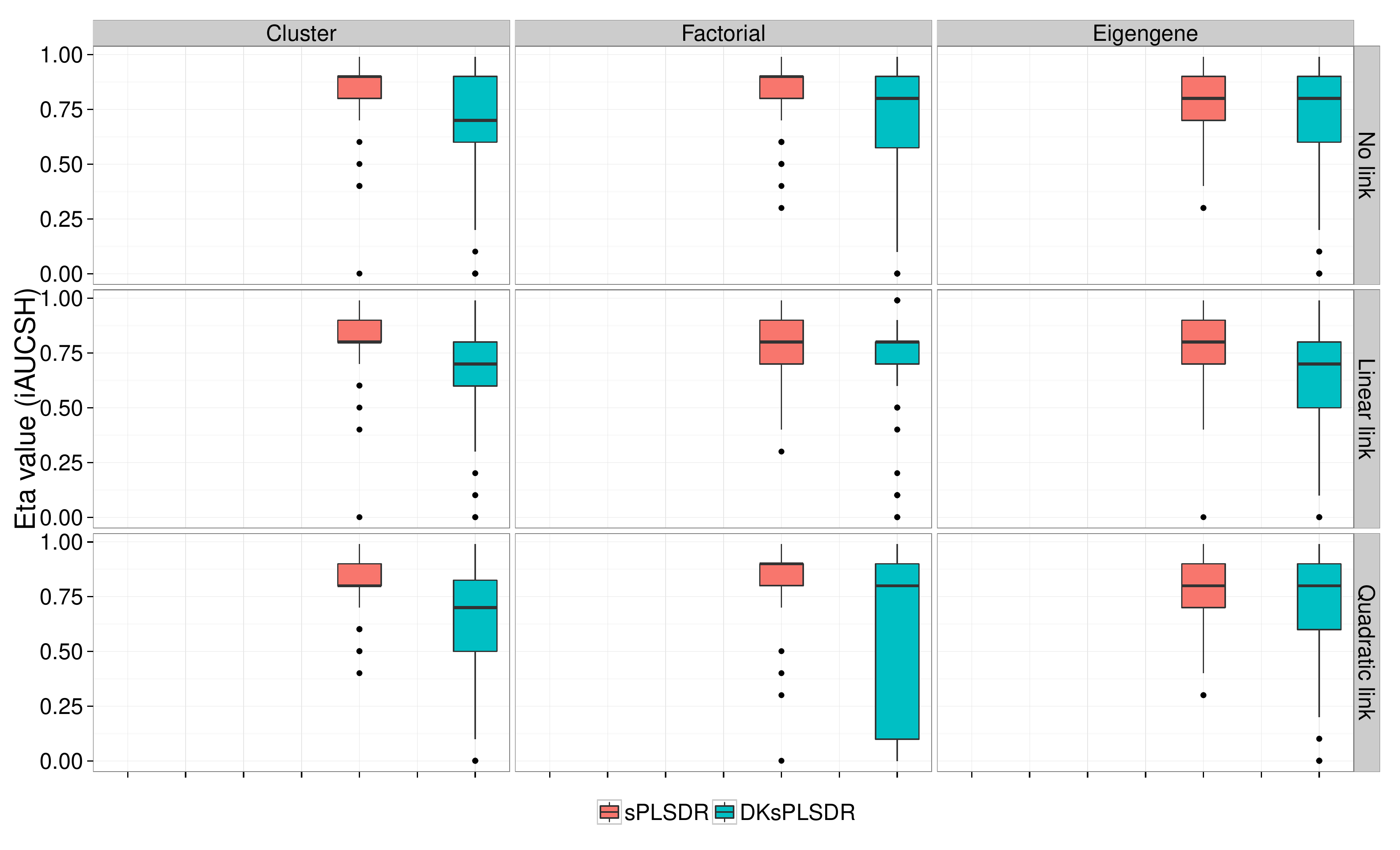}\phantomcaption\label{eta_AUCsh}}\qquad{\includegraphics[width=.75\columnwidth]{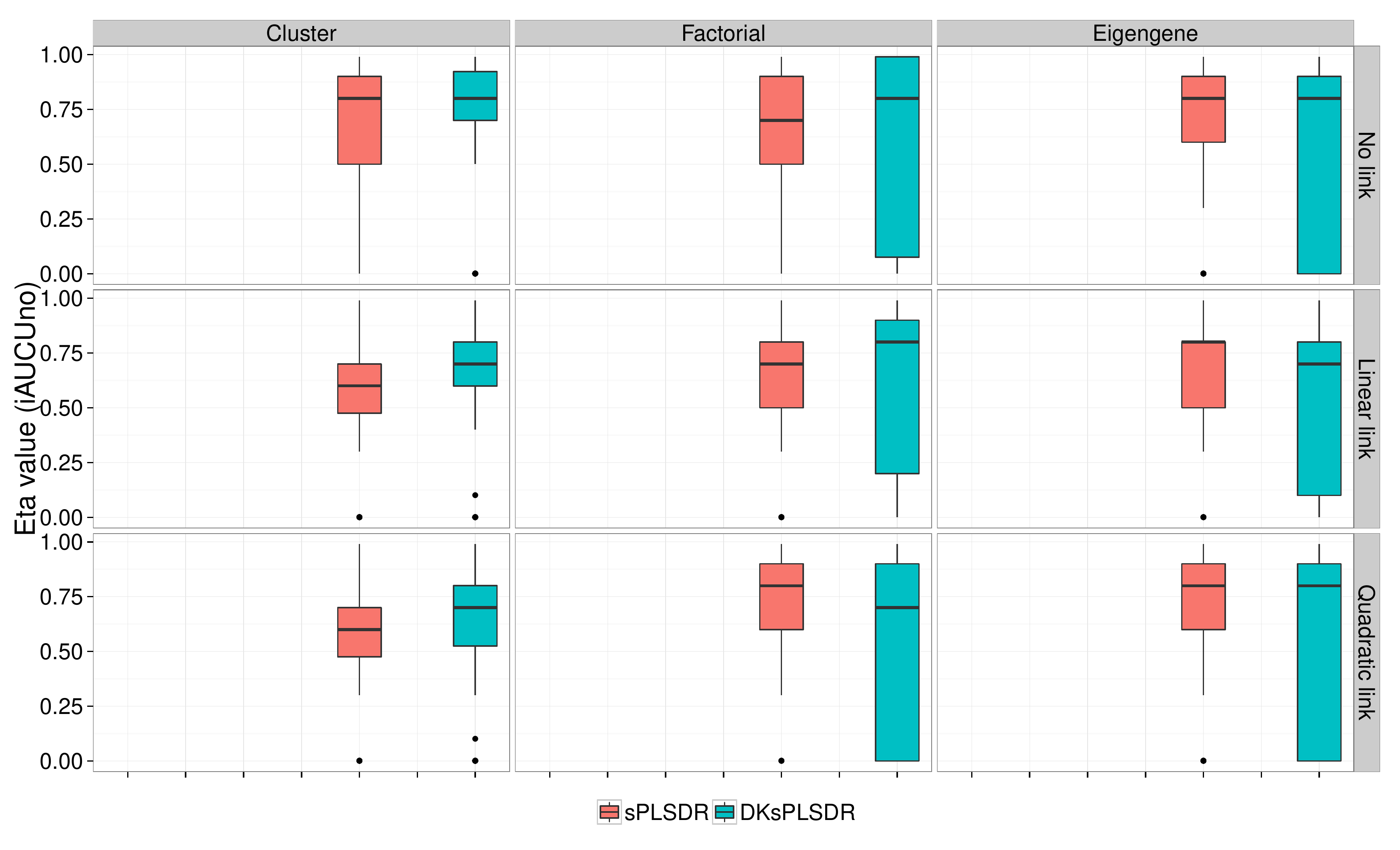}\phantomcaption\label{eta_AUCUno}}}
\vspace{-.5cm}
\caption*{Figure~\ref{eta_AUCsh}:  $\eta$, iAUCSH criterion. \qquad Figure~\ref{eta_AUCUno}:  $\eta$, iAUCUno criterion.}
\end{figure}

\clearpage

\begin{figure}[!tpb]
\centerline{{\includegraphics[width=.75\columnwidth]{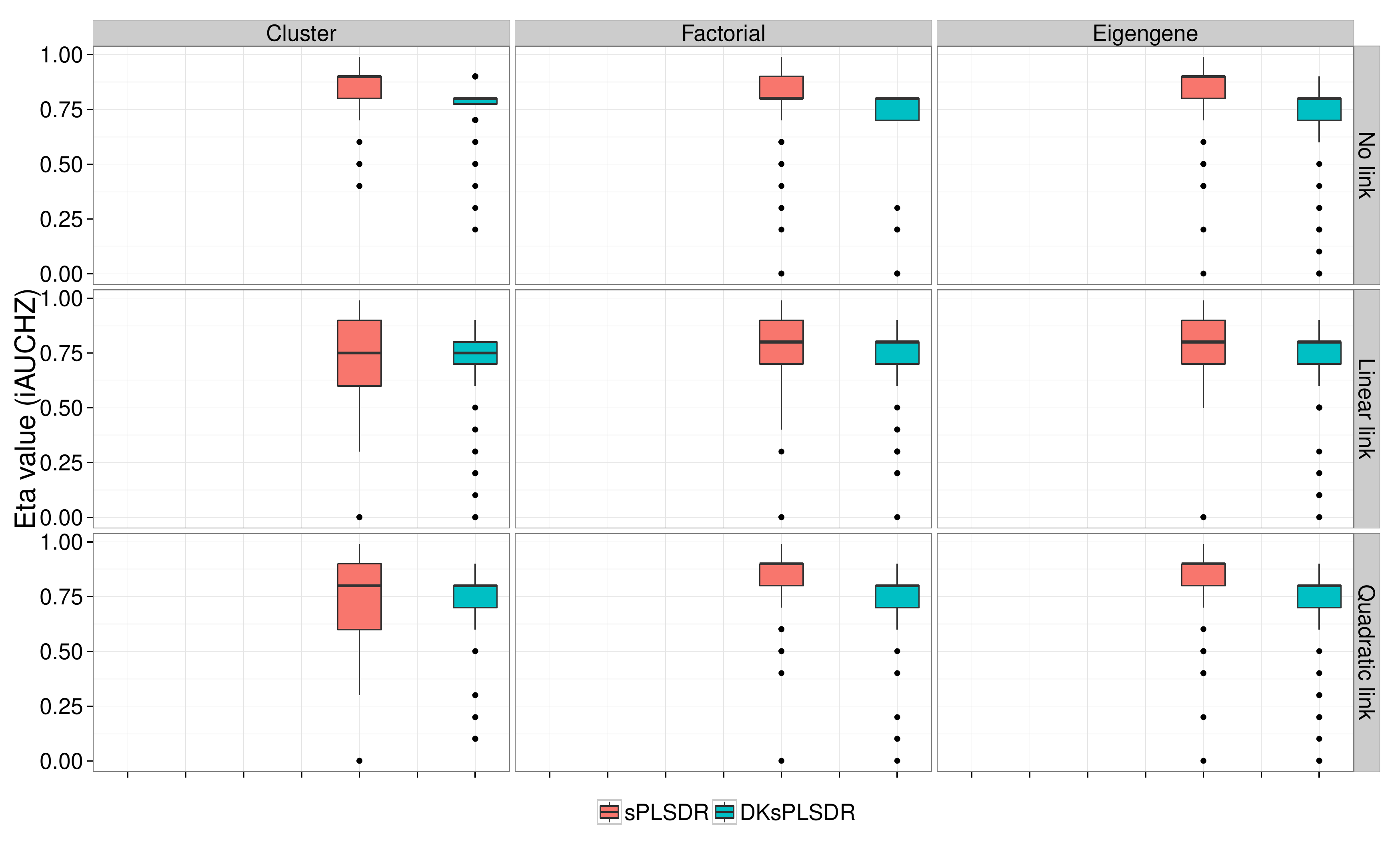}\phantomcaption\label{eta_AUChztest}}\qquad{\includegraphics[width=.75\columnwidth]{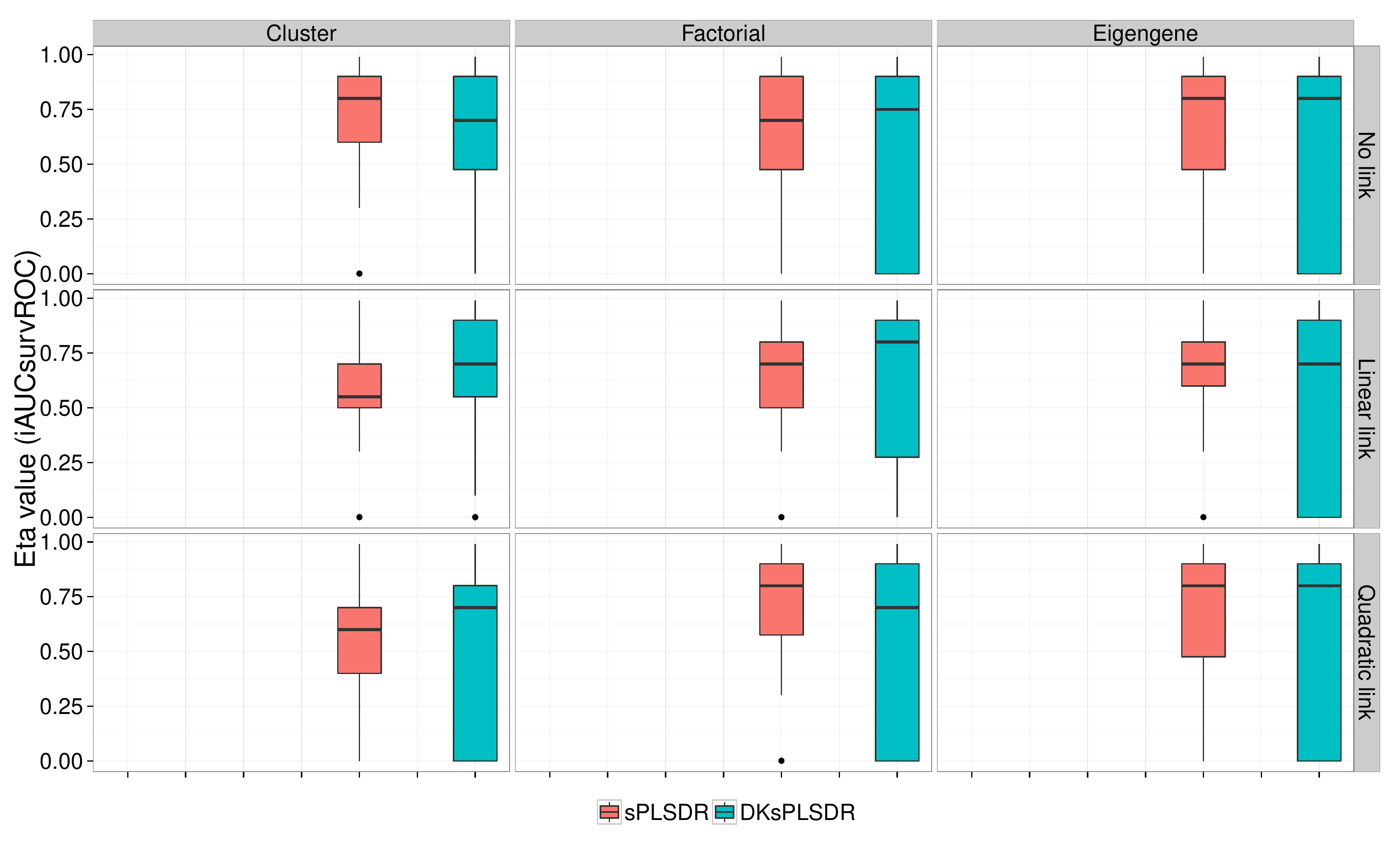}\phantomcaption\label{eta_AUCsurvROCtest}}}
\vspace{-.5cm}
\caption*{\mbox{Figure~\ref{eta_AUChztest}:  $\eta$, iAUCHZ criterion. \qquad Figure~\ref{eta_AUCsurvROCtest}:  $\eta$, iAUCSurvROC criterion.}}
\end{figure}

\begin{figure}[!tpb]
\centerline{{\includegraphics[width=.75\columnwidth]{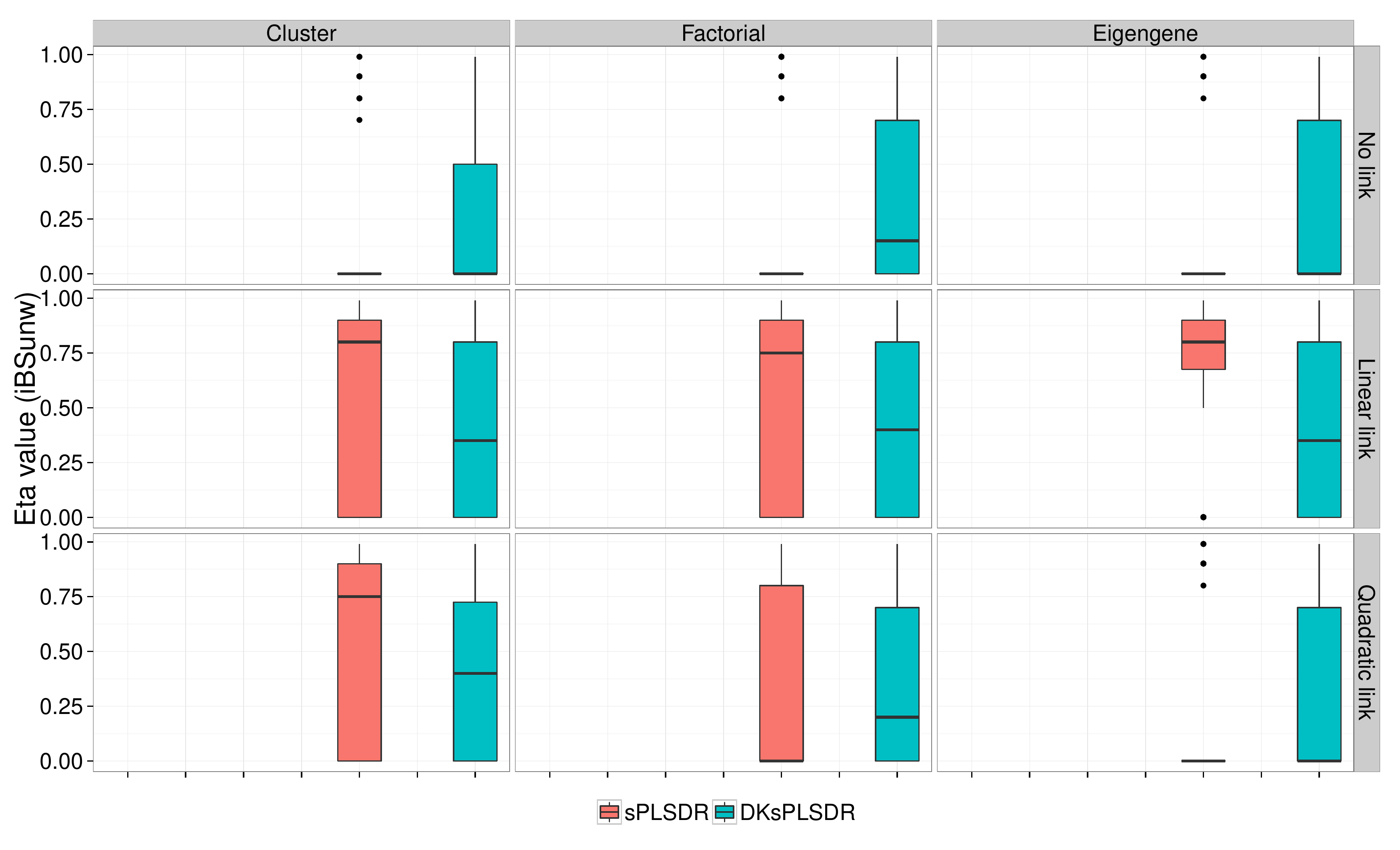}\phantomcaption\label{eta_iBSunw}}\qquad{\includegraphics[width=.75\columnwidth]{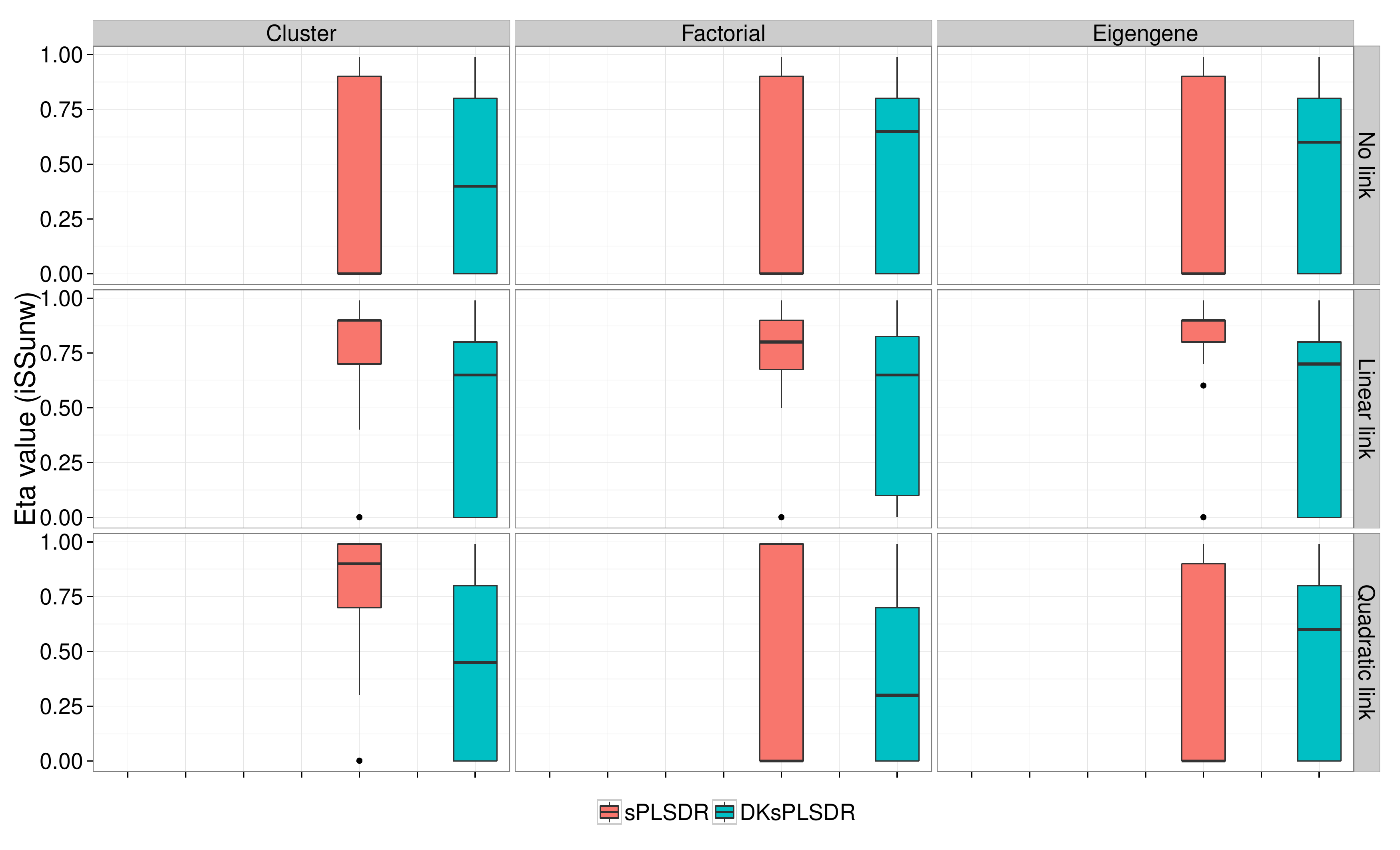}\phantomcaption\label{eta_iSchmidSunw}}}
\vspace{-.5cm}
\caption*{Figure~\ref{eta_iBSunw}:  $\eta$, iBSunw criterion. \qquad\qquad Figure~\ref{eta_iSchmidSunw}:  $\eta$, iSSunw criterion.}
\end{figure}

\begin{figure}[!tpb]
\centerline{{\includegraphics[width=.75\columnwidth]{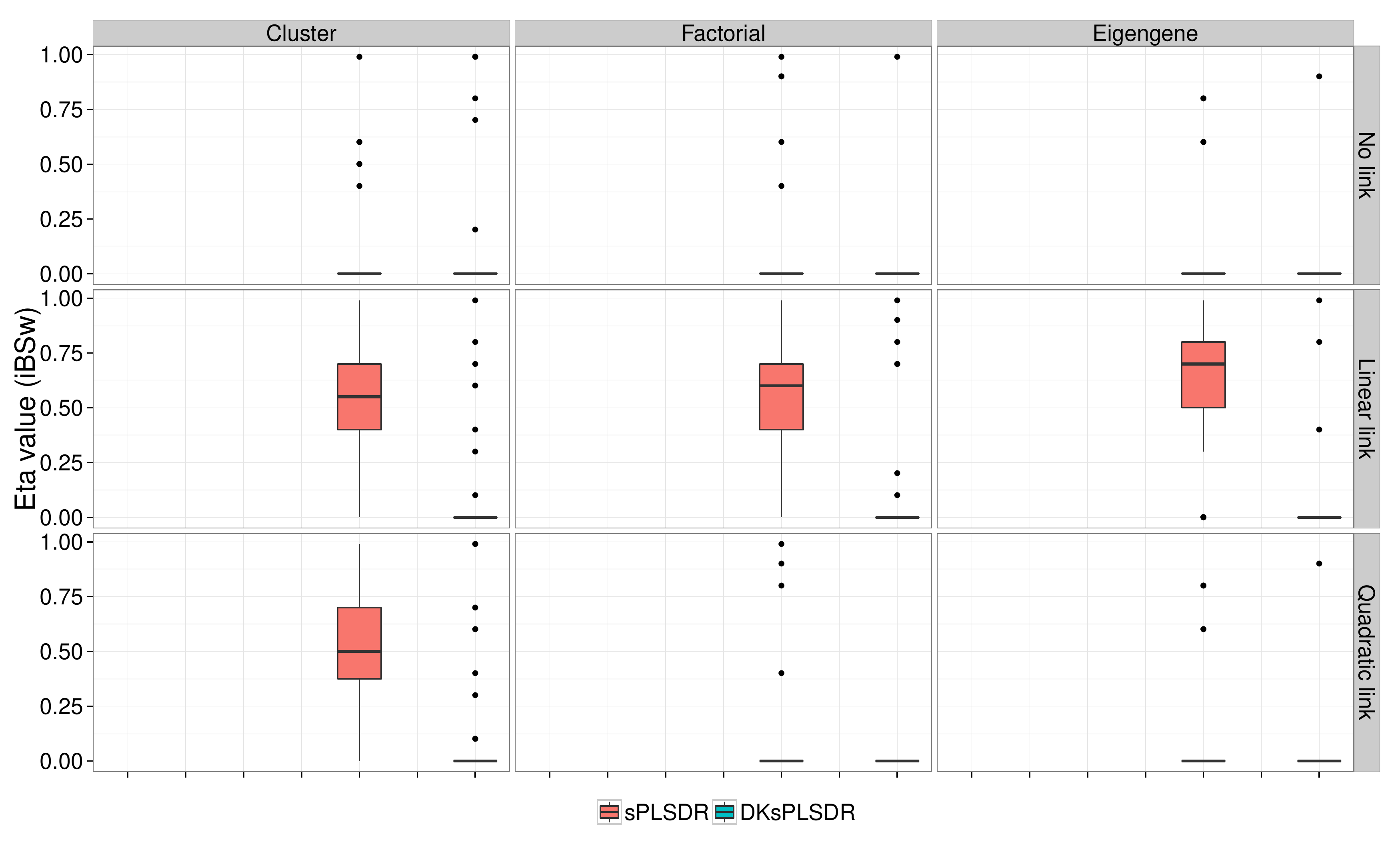}\phantomcaption\label{eta_iBSw}}\qquad{\includegraphics[width=.75\columnwidth]{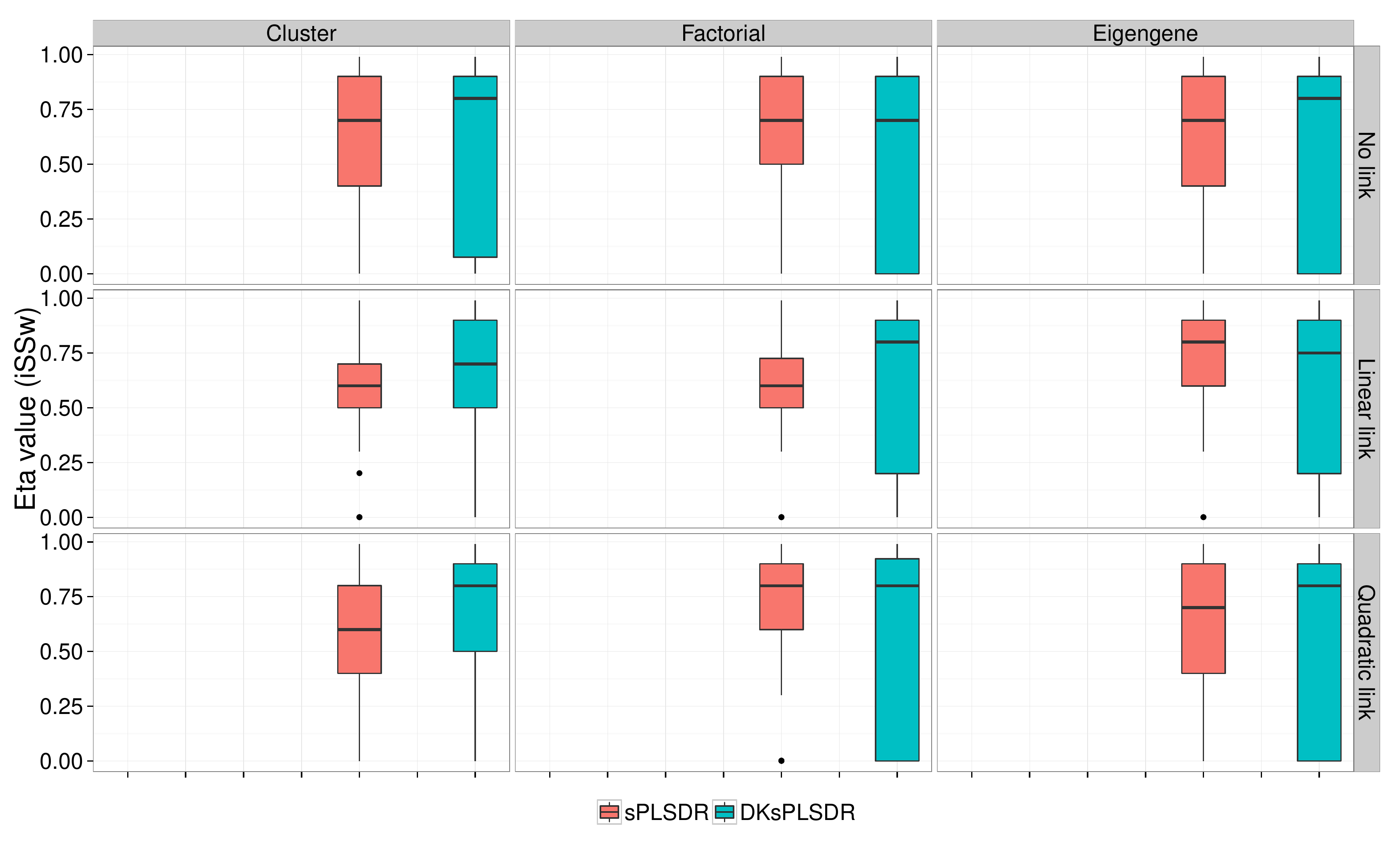}\phantomcaption\label{eta_iSchmidSw}}}
\vspace{-.5cm}
\caption*{Figure~\ref{eta_iBSw}:  $\eta$, iBSw criterion. \qquad\qquad Figure~\ref{eta_iSchmidSw}:  $\eta$, iSSw criterion.}
\end{figure}

\clearpage

\begin{figure}[!tpb]
\centerline{{\includegraphics[width=.75\columnwidth]{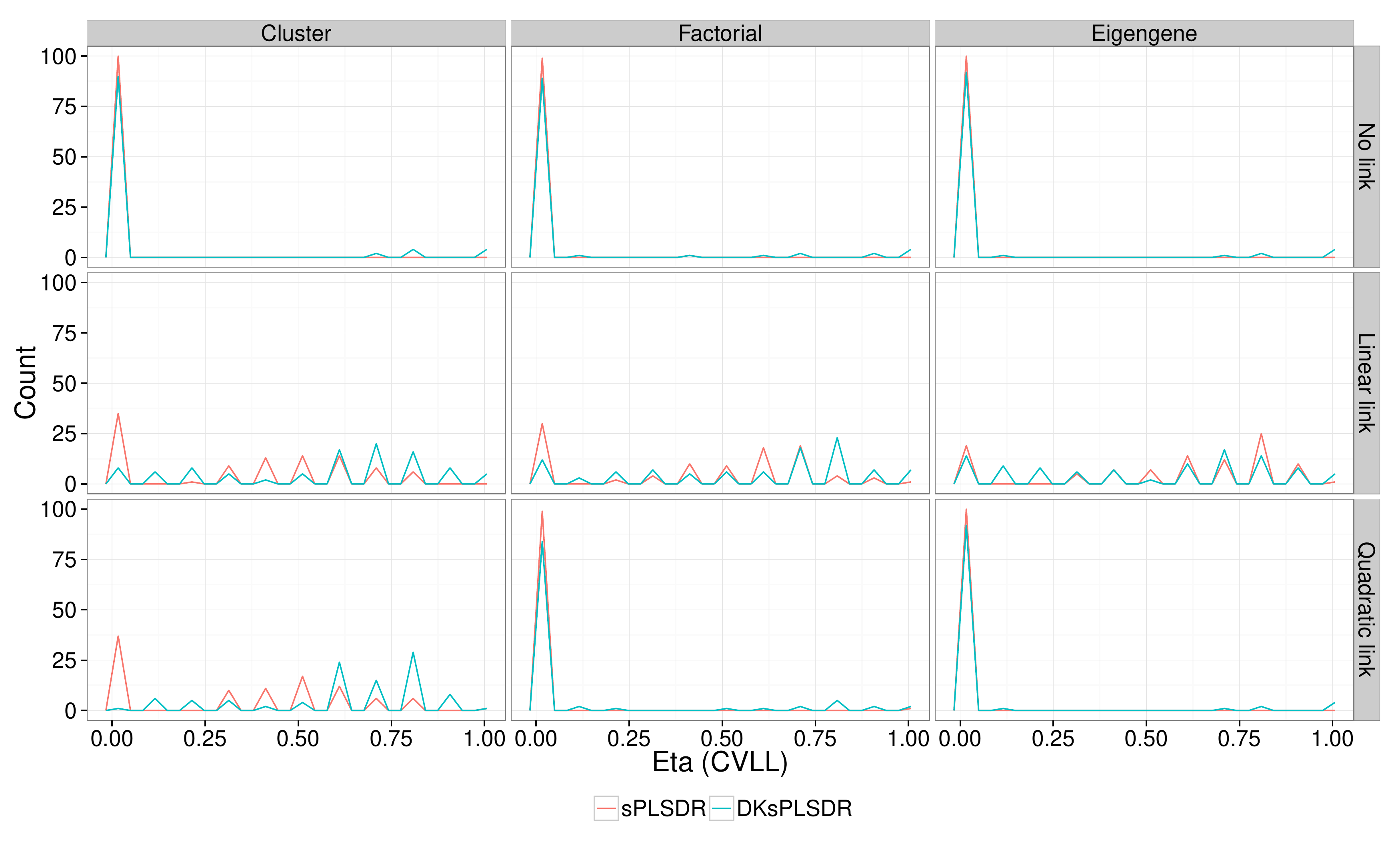}\phantomcaption\label{eta_cvll_alt}}\qquad{\includegraphics[width=.75\columnwidth]{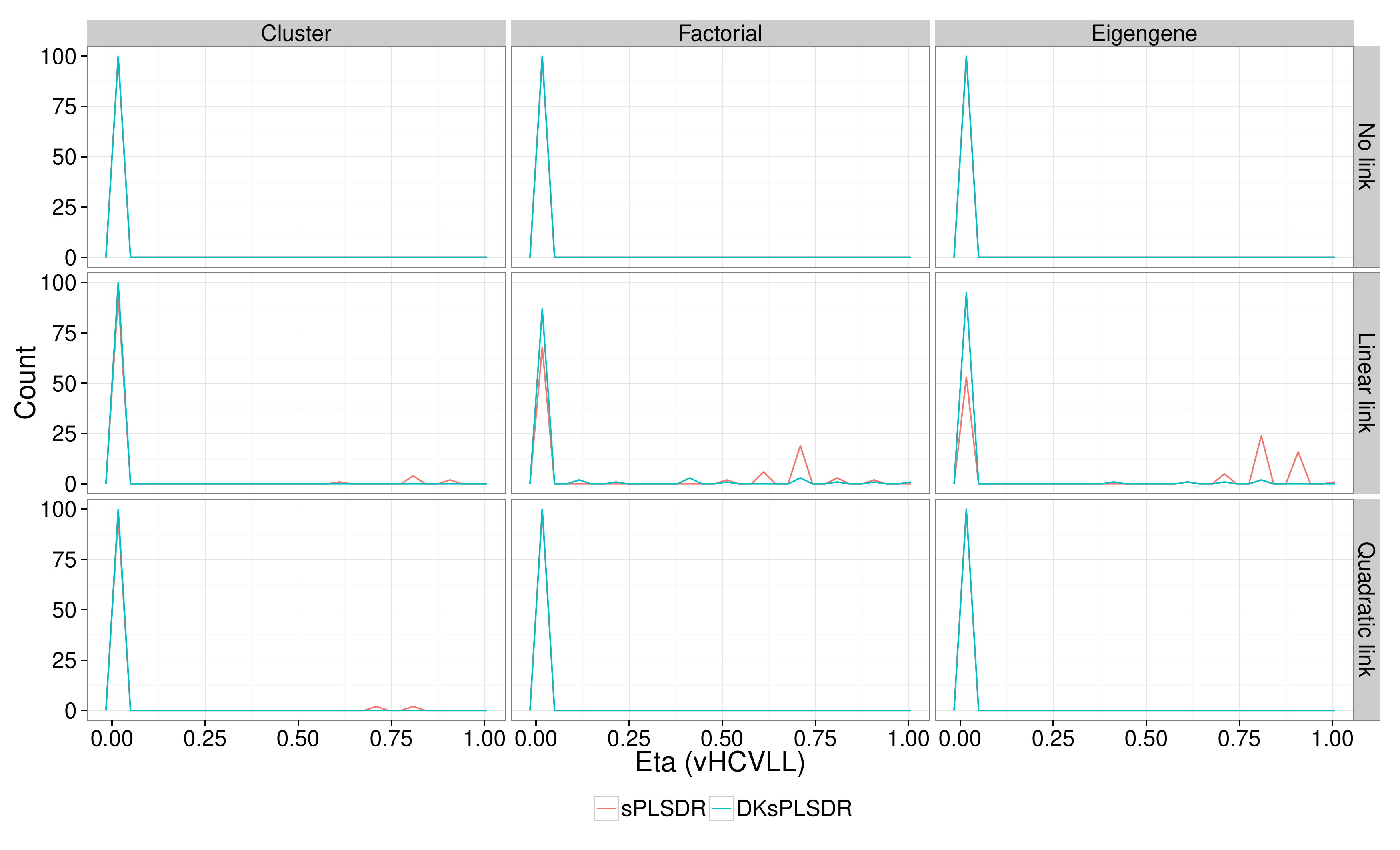}\phantomcaption\label{eta_vanHcvll_alt}}}
\vspace{-.5cm}
\caption*{Figure~\ref{eta_cvll_alt}:  $\eta$, LL criterion. \qquad\qquad\qquad Figure~\ref{eta_vanHcvll_alt}:  $\eta$, vHLL criterion.}
\end{figure}

\begin{figure}[!tpb]
\centerline{{\includegraphics[width=.75\columnwidth]{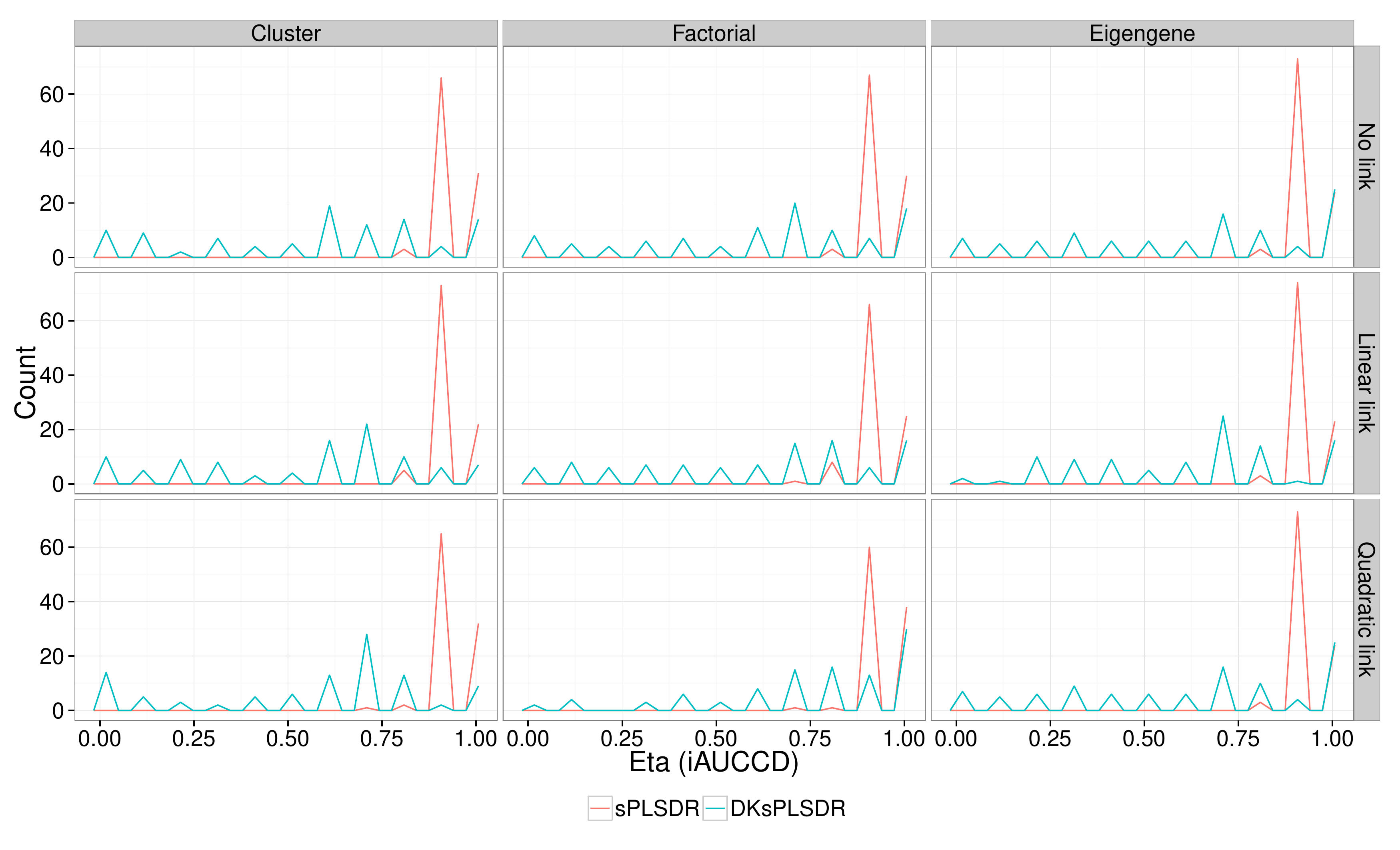}\phantomcaption\label{eta_AUCcd_alt}}\qquad{\includegraphics[width=.75\columnwidth]{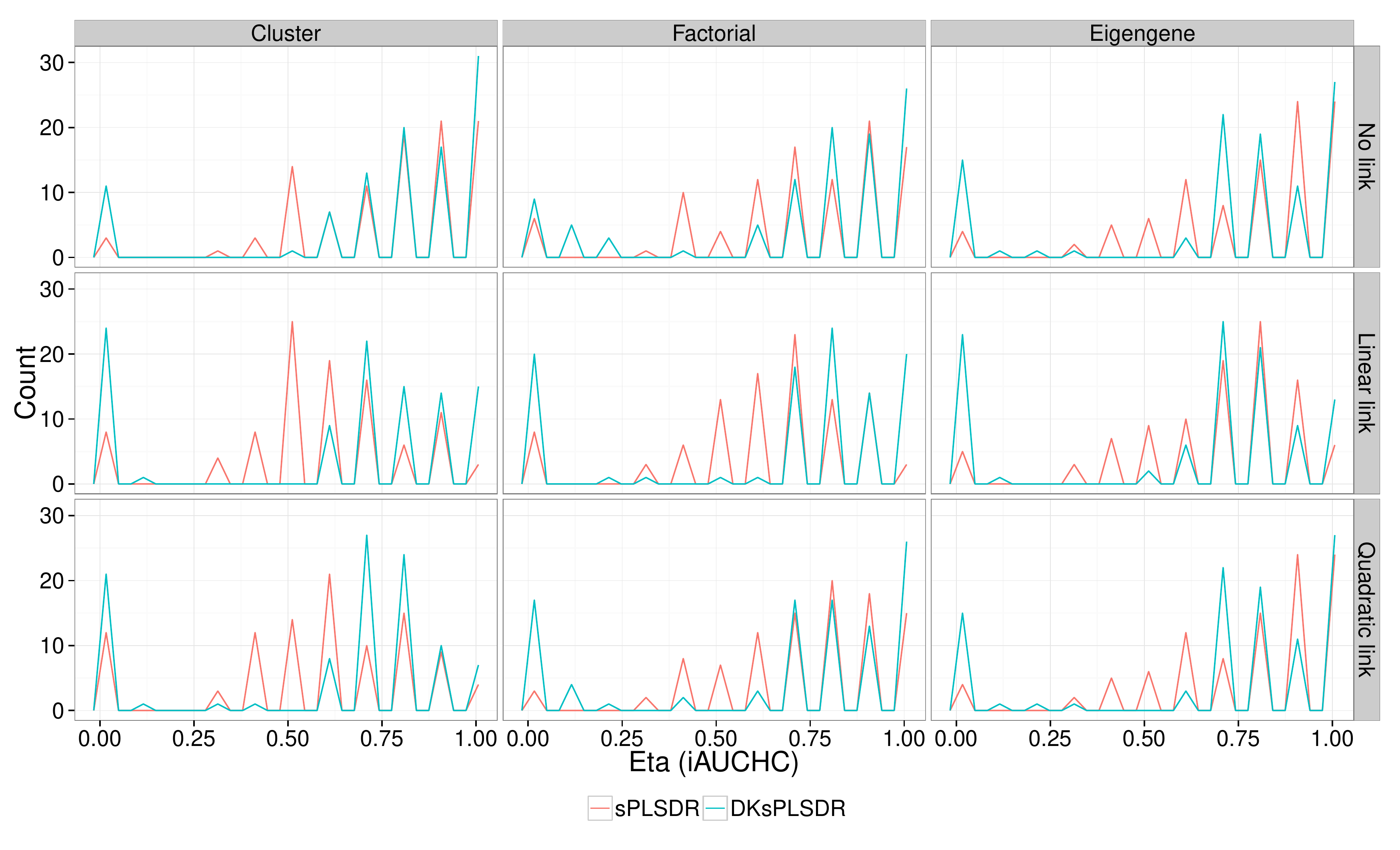}\phantomcaption\label{eta_AUChc_alt}}}
\vspace{-.5cm}
\caption*{\mbox{Figure~\ref{eta_AUCcd_alt}:  $\eta$, iAUCCD criterion. \qquad Figure~\ref{eta_AUChc_alt}:  $\eta$, iAUCHC criterion.}}
\end{figure}

\begin{figure}[!tpb]
\centerline{{\includegraphics[width=.75\columnwidth]{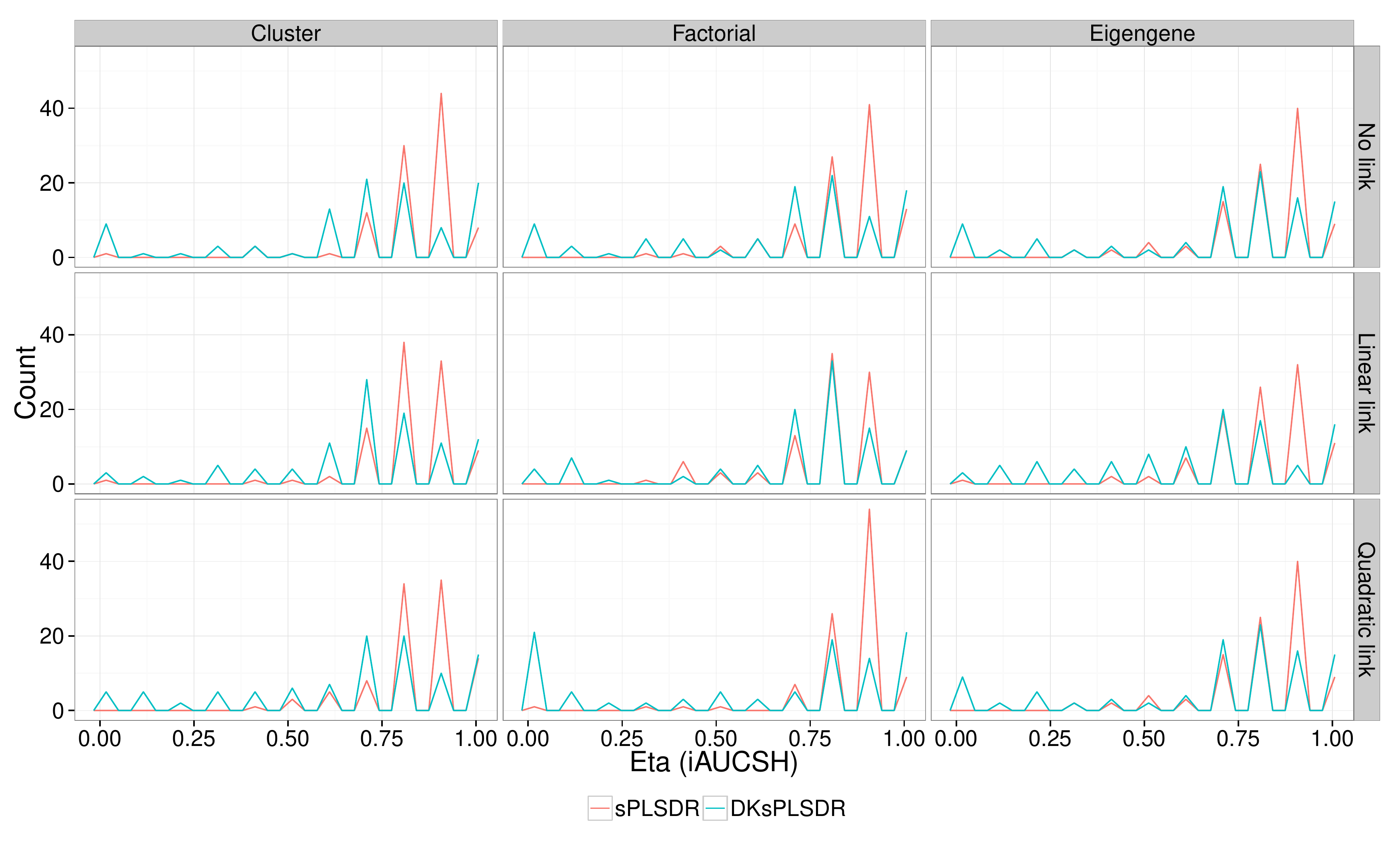}\phantomcaption\label{eta_AUCsh_alt}}\qquad{\includegraphics[width=.75\columnwidth]{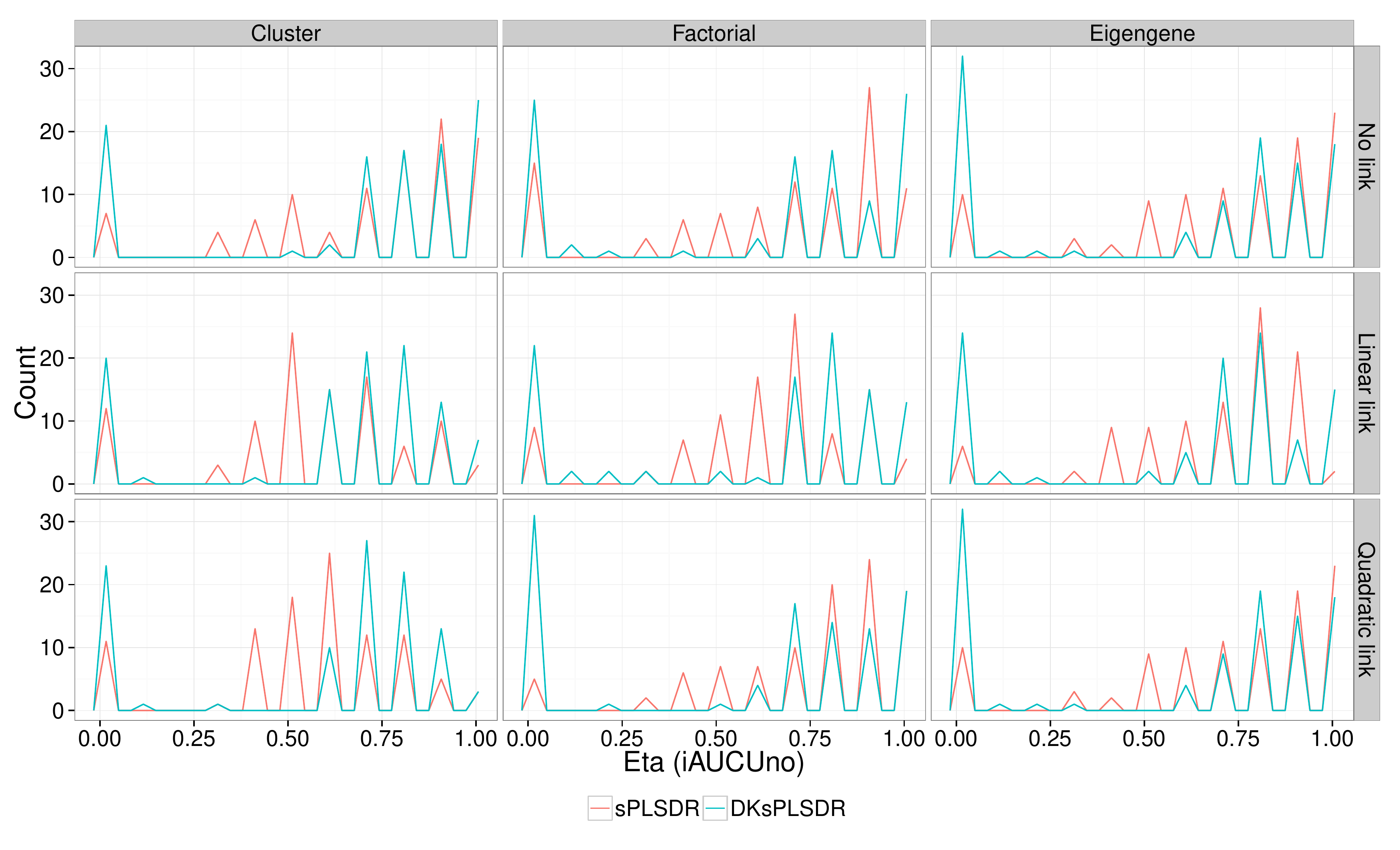}\phantomcaption\label{eta_AUCUno_alt}}}
\vspace{-.5cm}
\caption*{Figure~\ref{eta_AUCsh_alt}:  $\eta$, iAUCSH criterion. \qquad Figure~\ref{eta_AUCUno_alt}:  $\eta$, iAUCUno criterion.}
\end{figure}

\clearpage

\begin{figure}[!tpb]
\centerline{{\includegraphics[width=.75\columnwidth]{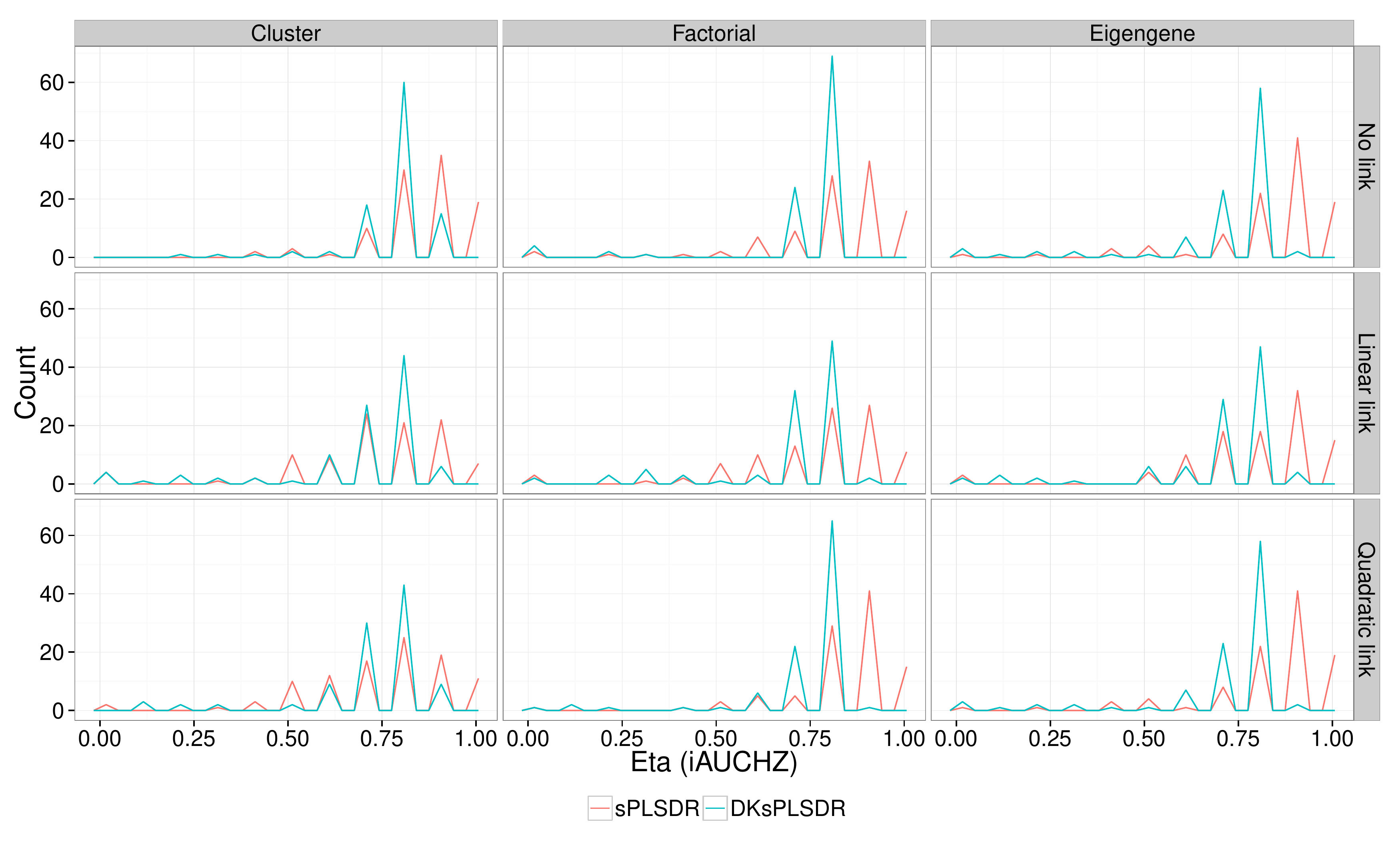}\phantomcaption\label{eta_AUChztest_alt}}\qquad{\includegraphics[width=.75\columnwidth]{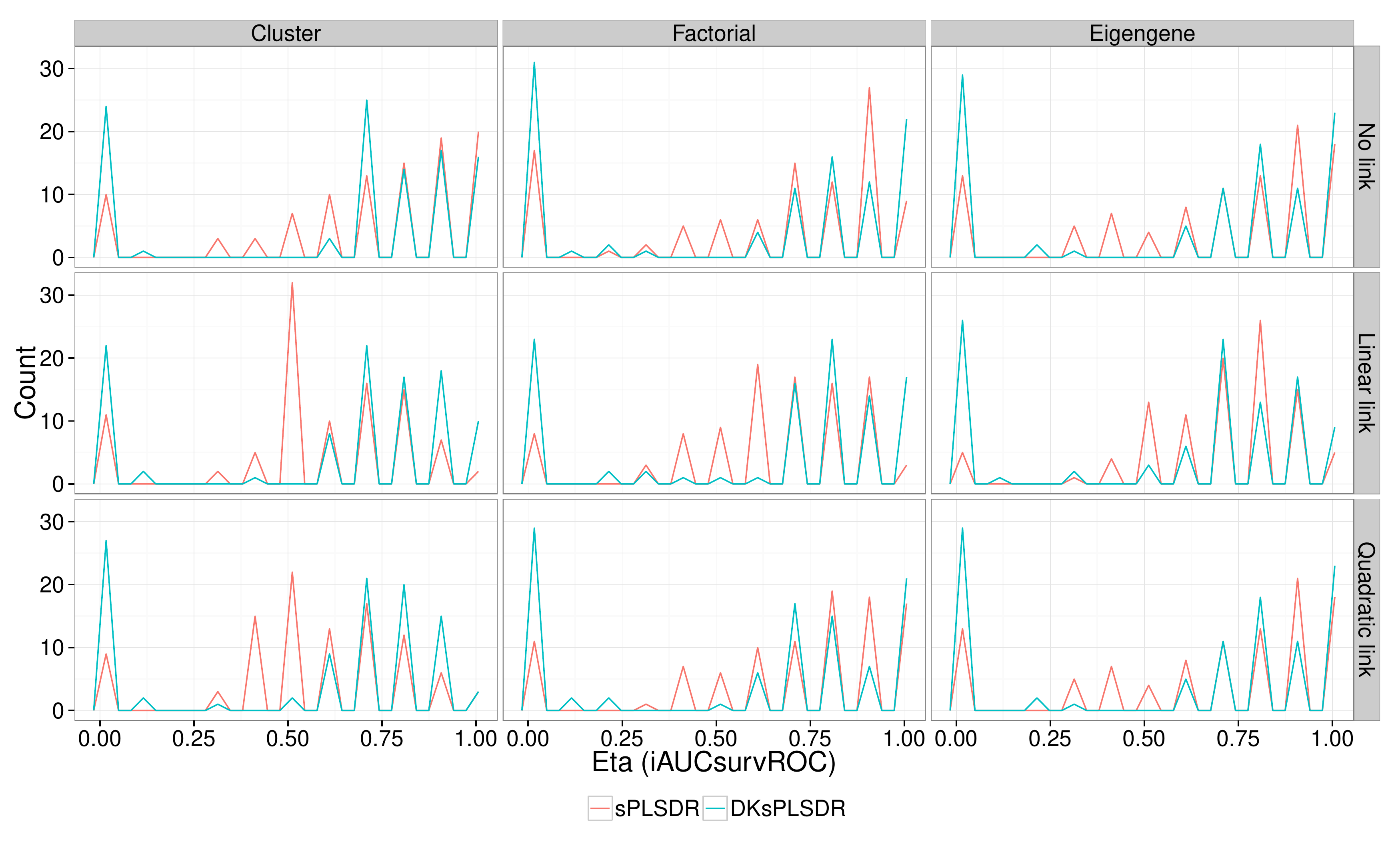}\phantomcaption\label{eta_AUCsurvROCtest_alt}}}
\vspace{-.5cm}
\caption*{\mbox{Figure~\ref{eta_AUChztest_alt}:  $\eta$, iAUCHZ criterion. \qquad Figure~\ref{eta_AUCsurvROCtest_alt}:  $\eta$, iAUCSurvROC criterion.}}
\end{figure}

\begin{figure}[!tpb]
\centerline{{\includegraphics[width=.75\columnwidth]{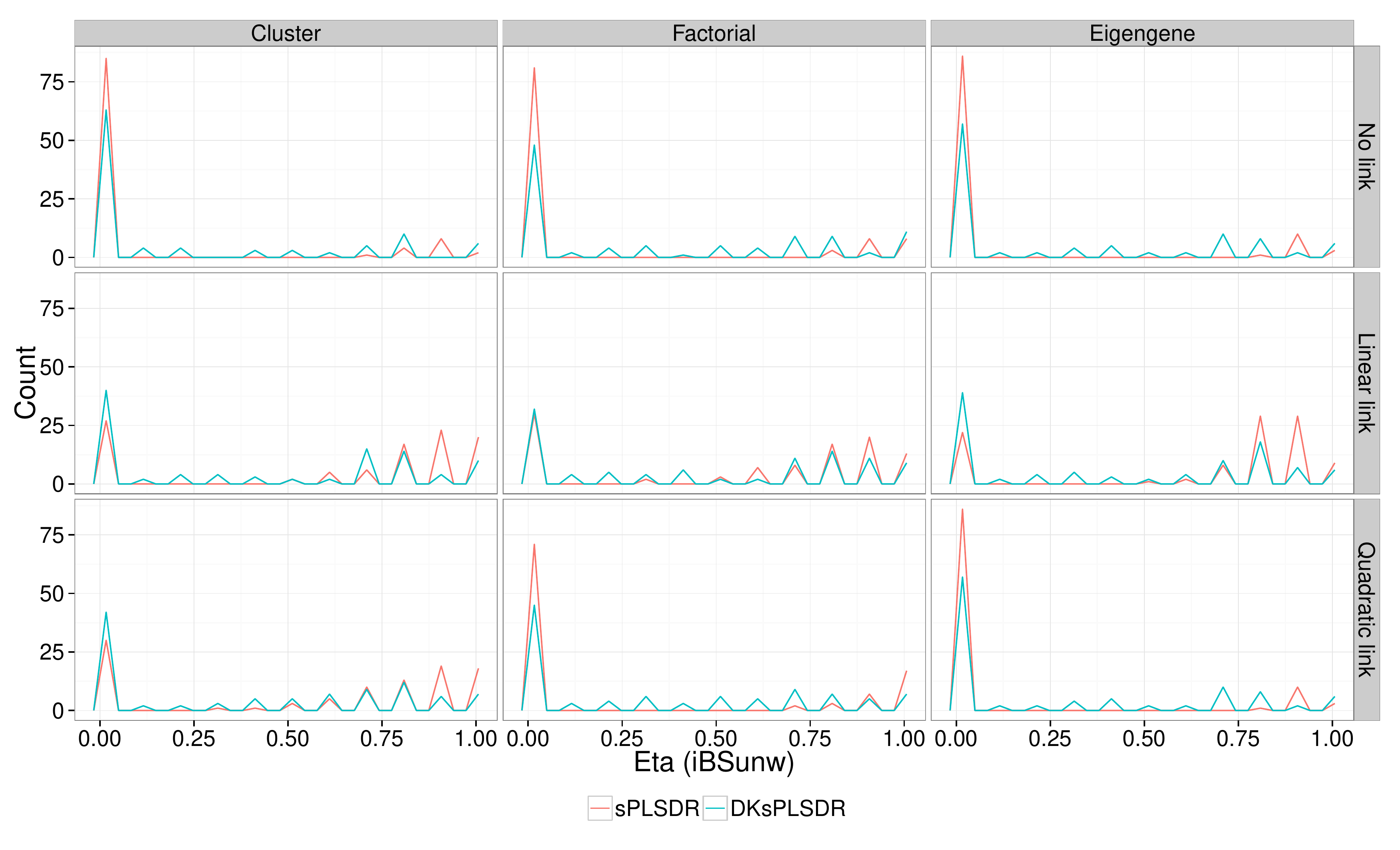}\phantomcaption\label{eta_iBSunw_alt}}\qquad{\includegraphics[width=.75\columnwidth]{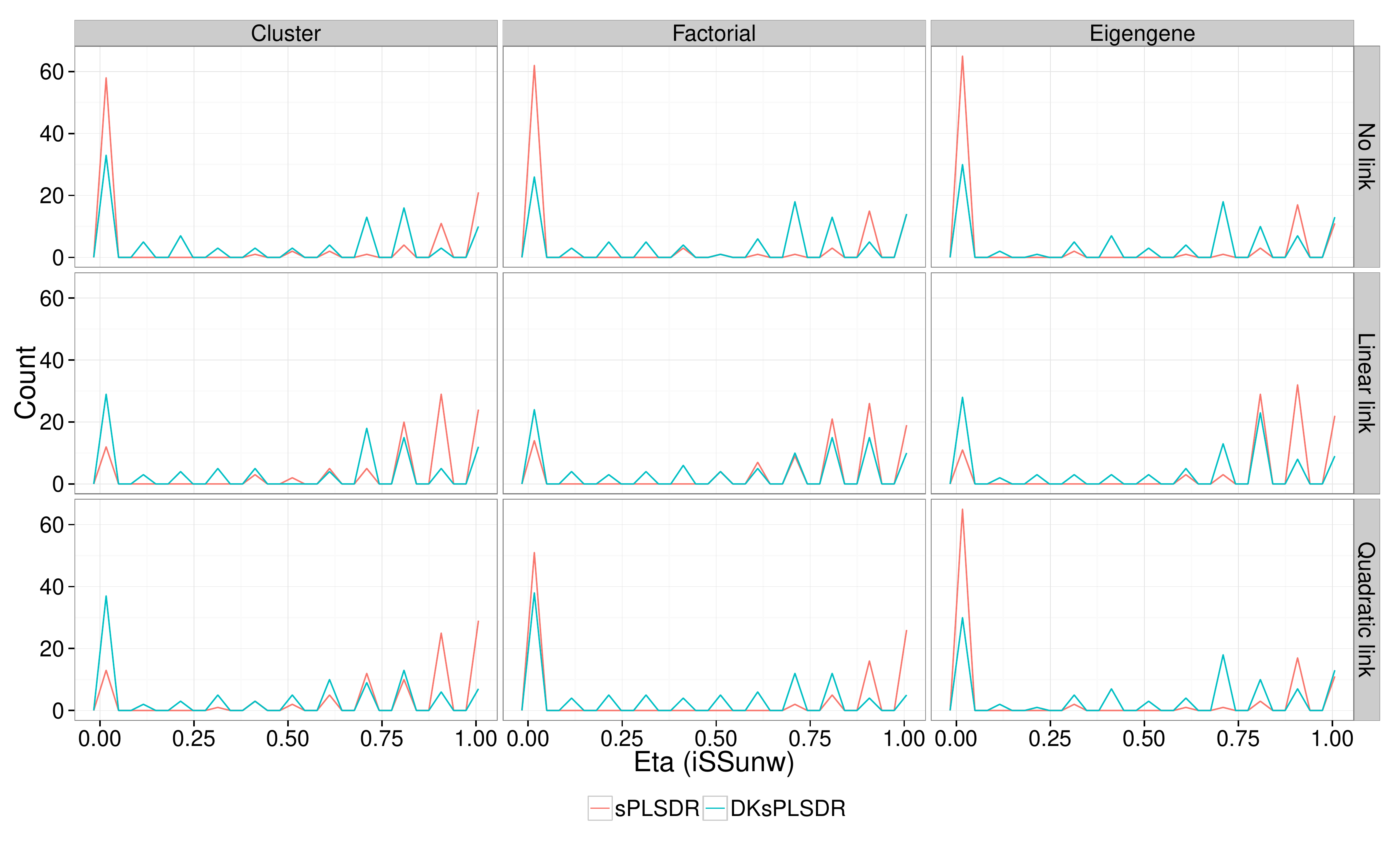}\phantomcaption\label{eta_iSchmidSunw_alt}}}
\vspace{-.5cm}
\caption*{Figure~\ref{eta_iBSunw_alt}:  $\eta$, iBSunw criterion. \qquad\qquad Figure~\ref{eta_iSchmidSunw_alt}:  $\eta$, iSSunw criterion.}
\end{figure}

\begin{figure}[!tpb]
\centerline{{\includegraphics[width=.75\columnwidth]{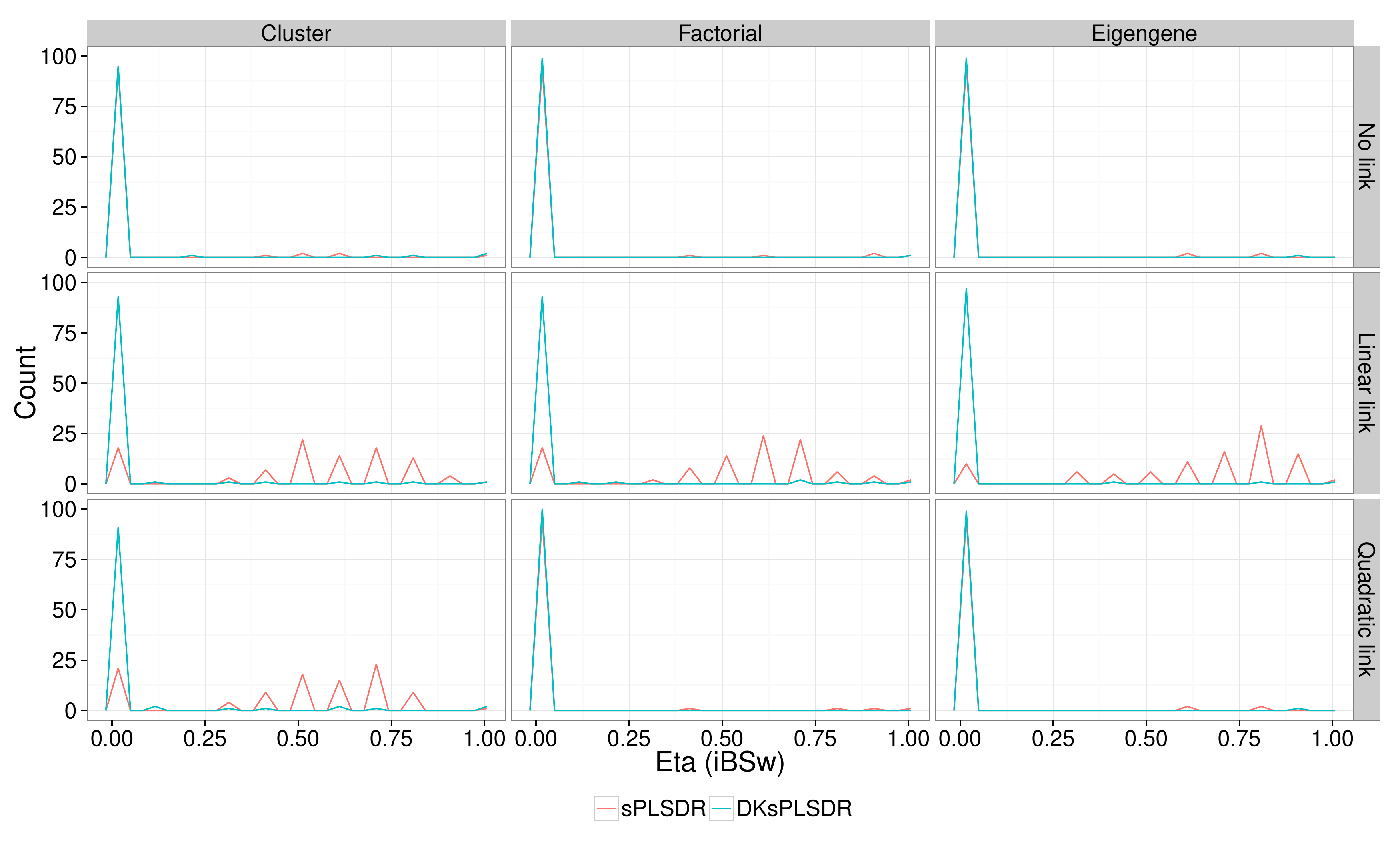}\phantomcaption\label{eta_iBSw_alt}}\qquad{\includegraphics[width=.75\columnwidth]{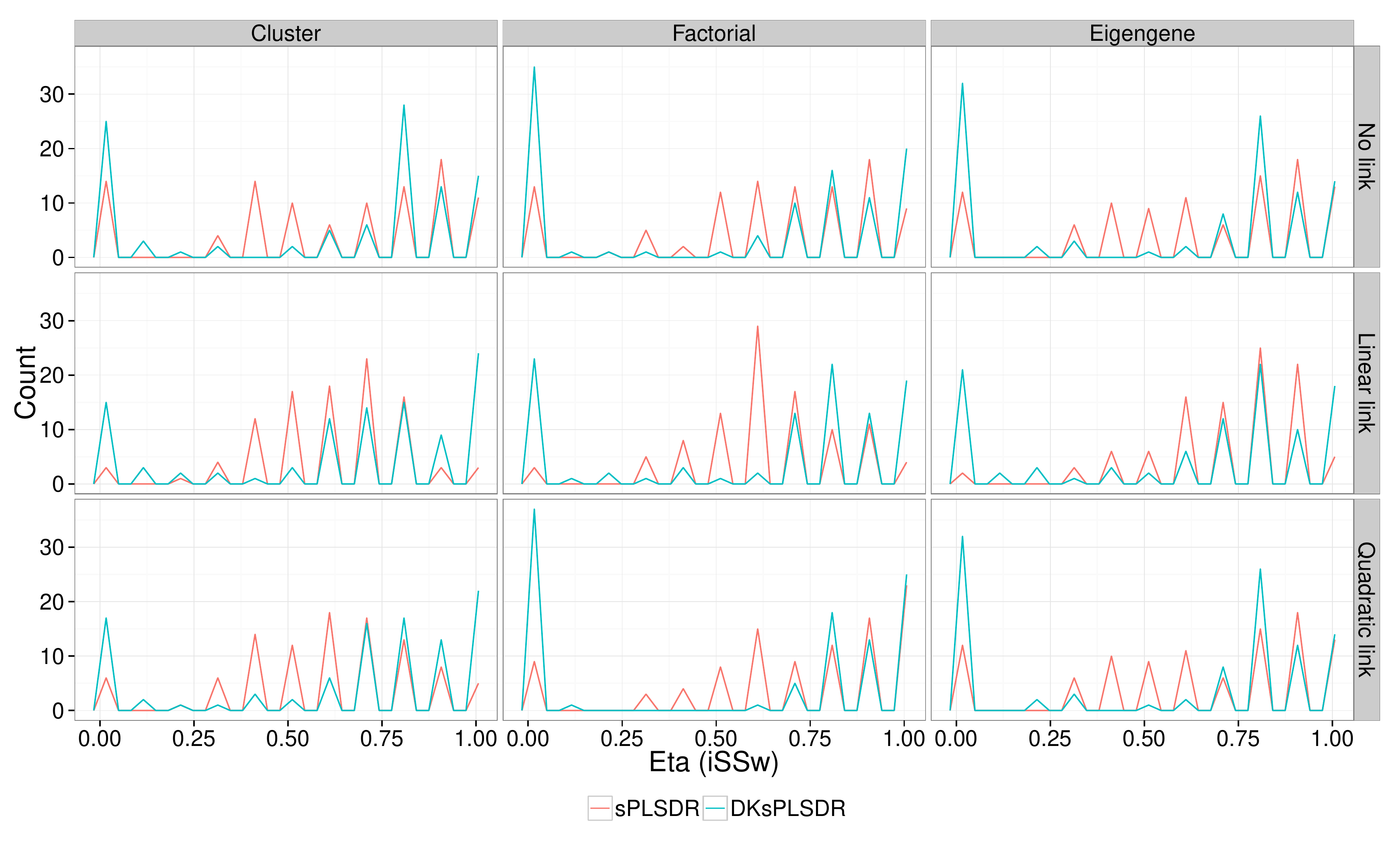}\phantomcaption\label{eta_iSchmidSw_alt}}}
\vspace{-.5cm}
\caption*{Figure~\ref{eta_iBSw_alt}:  $\eta$, iBSw criterion. \qquad\qquad Figure~\ref{eta_iSchmidSw_alt}:  $\eta$, iSSw criterion.}
\end{figure}

\clearpage


\begin{figure}[!tpb]
\centerline{{\includegraphics[width=.75\columnwidth]{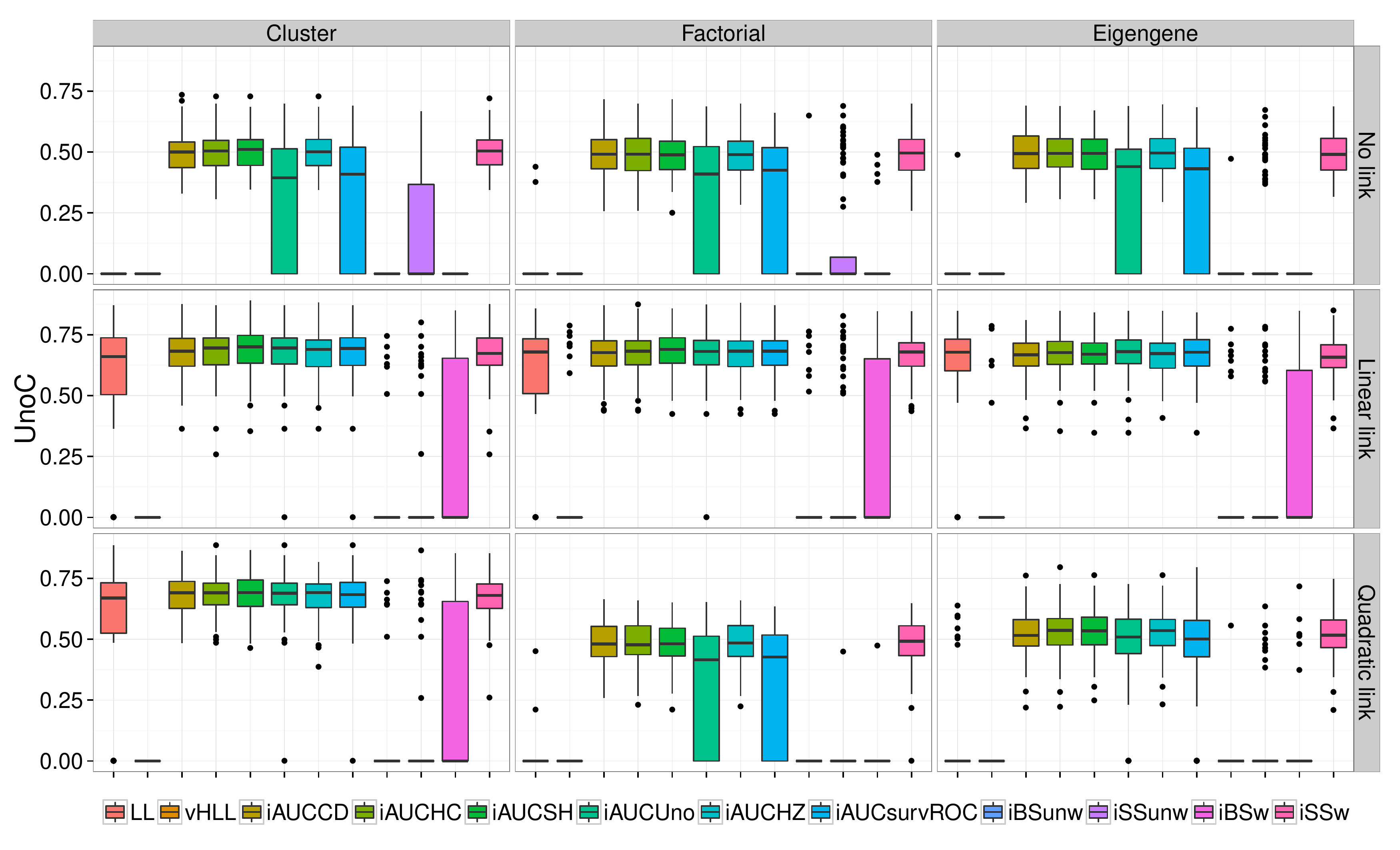}\phantomcaption\label{autoplsRcoxUnoCstat}}\qquad{\includegraphics[width=.75\columnwidth]{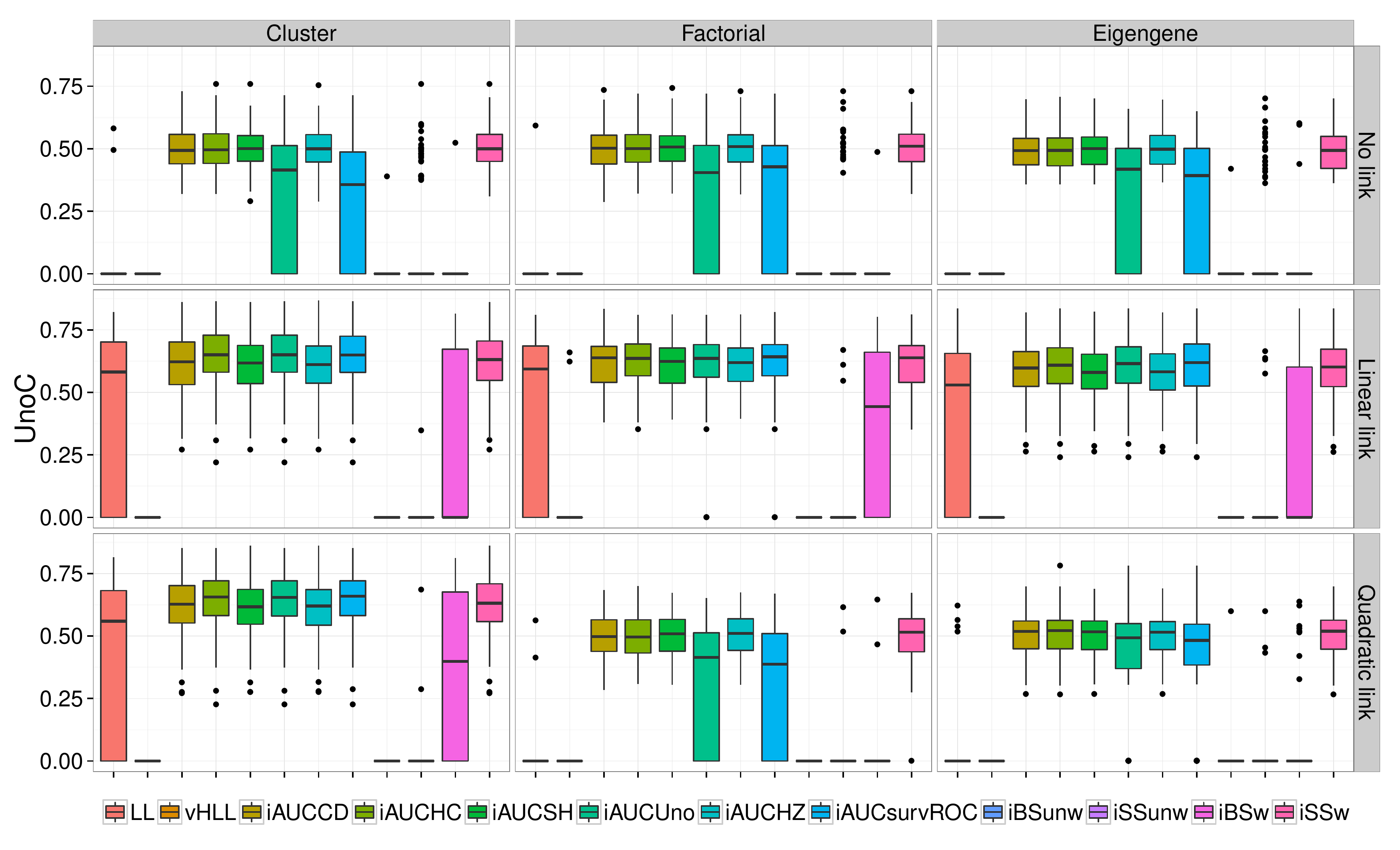}\phantomcaption\label{coxplsUnoCstat}}}
\vspace{-.5cm}
\caption*{\hspace{-2.0cm}\mbox{Figure~\ref{autoplsRcoxUnoCstat}:  UnoC vs CV criterion. autoplsRcox. \qquad\qquad Figure~\ref{coxplsUnoCstat}:  UnoC vs CV criterion. coxpls.}}
\end{figure}

\begin{figure}[!tpb]
\centerline{{\includegraphics[width=.75\columnwidth]{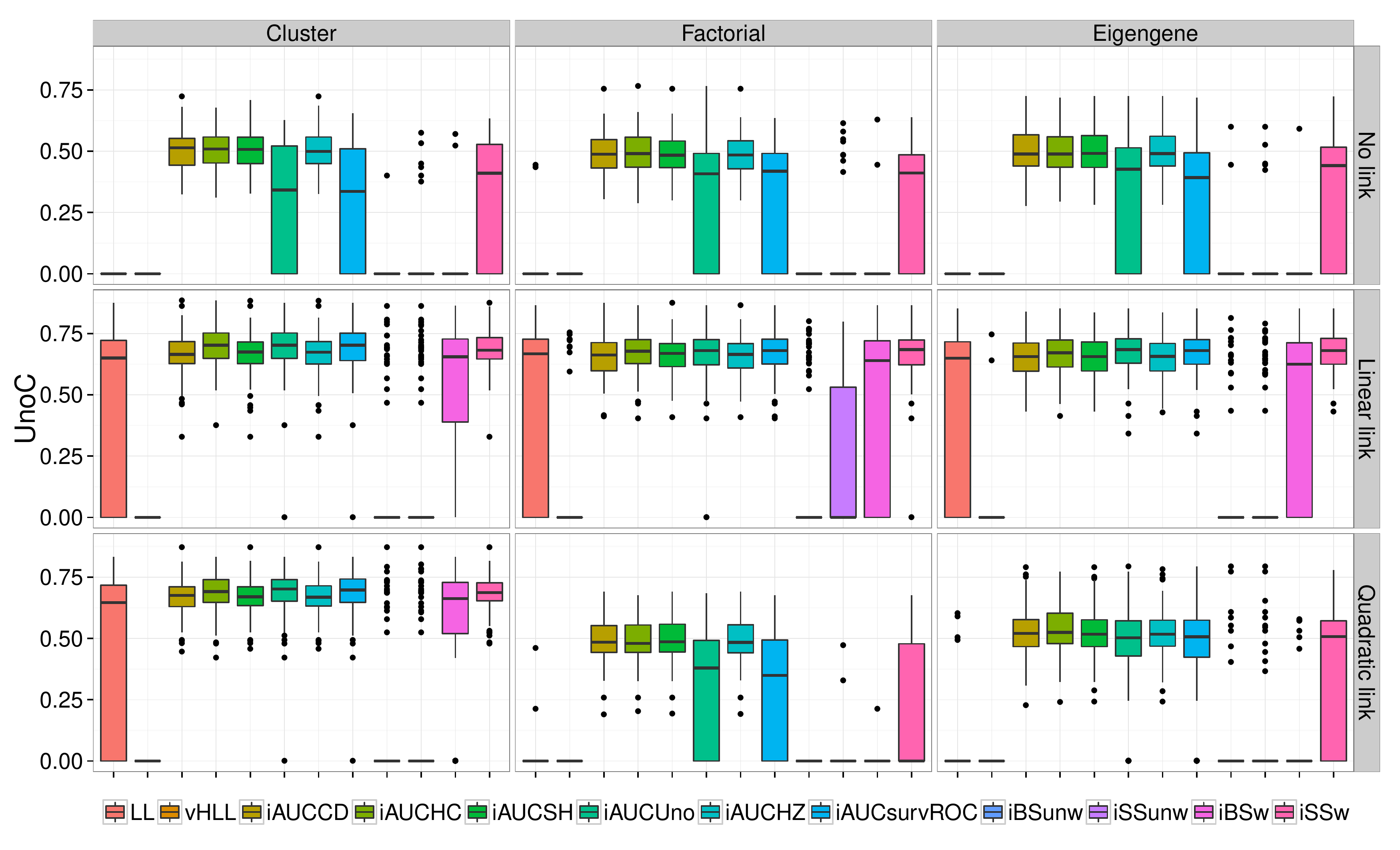}\phantomcaption\label{coxplsDRUnoCstat}}\qquad{\includegraphics[width=.75\columnwidth]{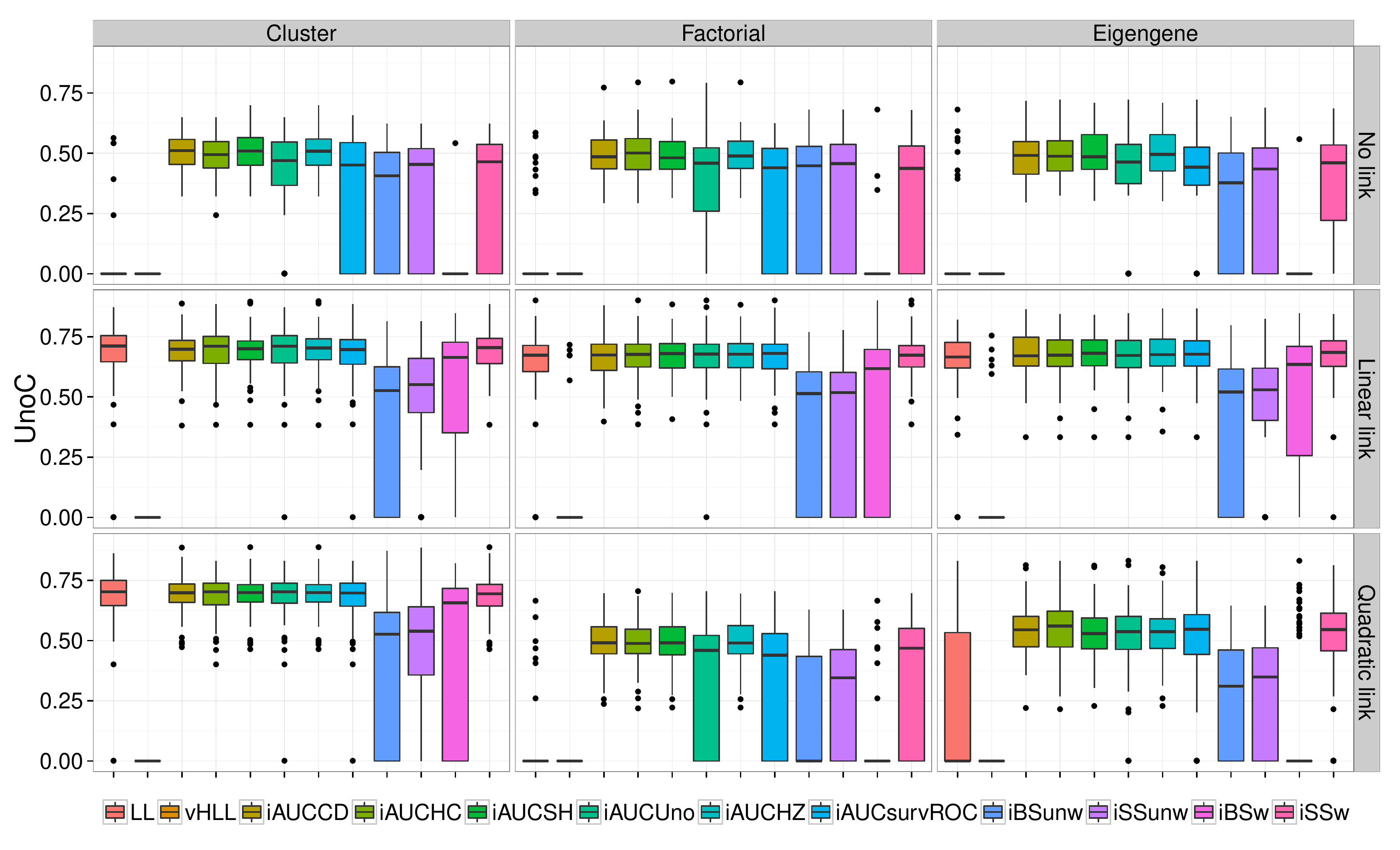}\phantomcaption\label{coxDKplsDRUnoCstat}}}
\vspace{-.5cm}
\caption*{\hspace{-1.5cm}\mbox{Figure~\ref{coxplsDRUnoCstat}:  UnoC vs CV criterion. coxplsDR. \qquad\qquad Figure~\ref{coxDKplsDRUnoCstat}:  UnoC vs CV criterion. coxDKplsDR.}}
\end{figure}



\begin{figure}[!tpb]
\centerline{\includegraphics[width=.75\columnwidth]{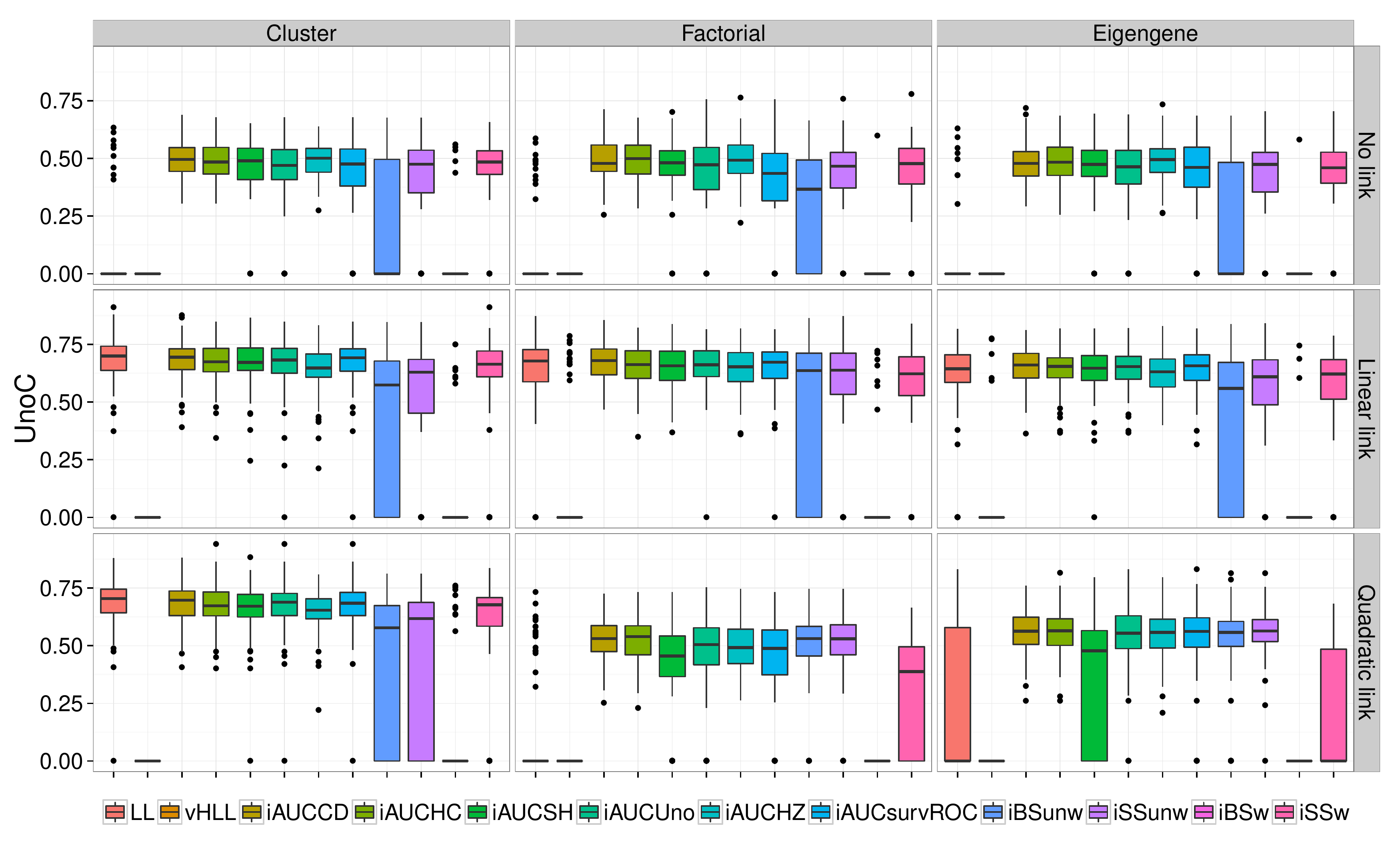}}
\vspace{-.5cm}
\caption{UnoC vs CV criterion. DKsplsDR.}\label{coxDKsplsDRUnoCstat}
\end{figure}

\clearpage

\begin{figure}[!tpb]
\centerline{{\includegraphics[width=.75\columnwidth]{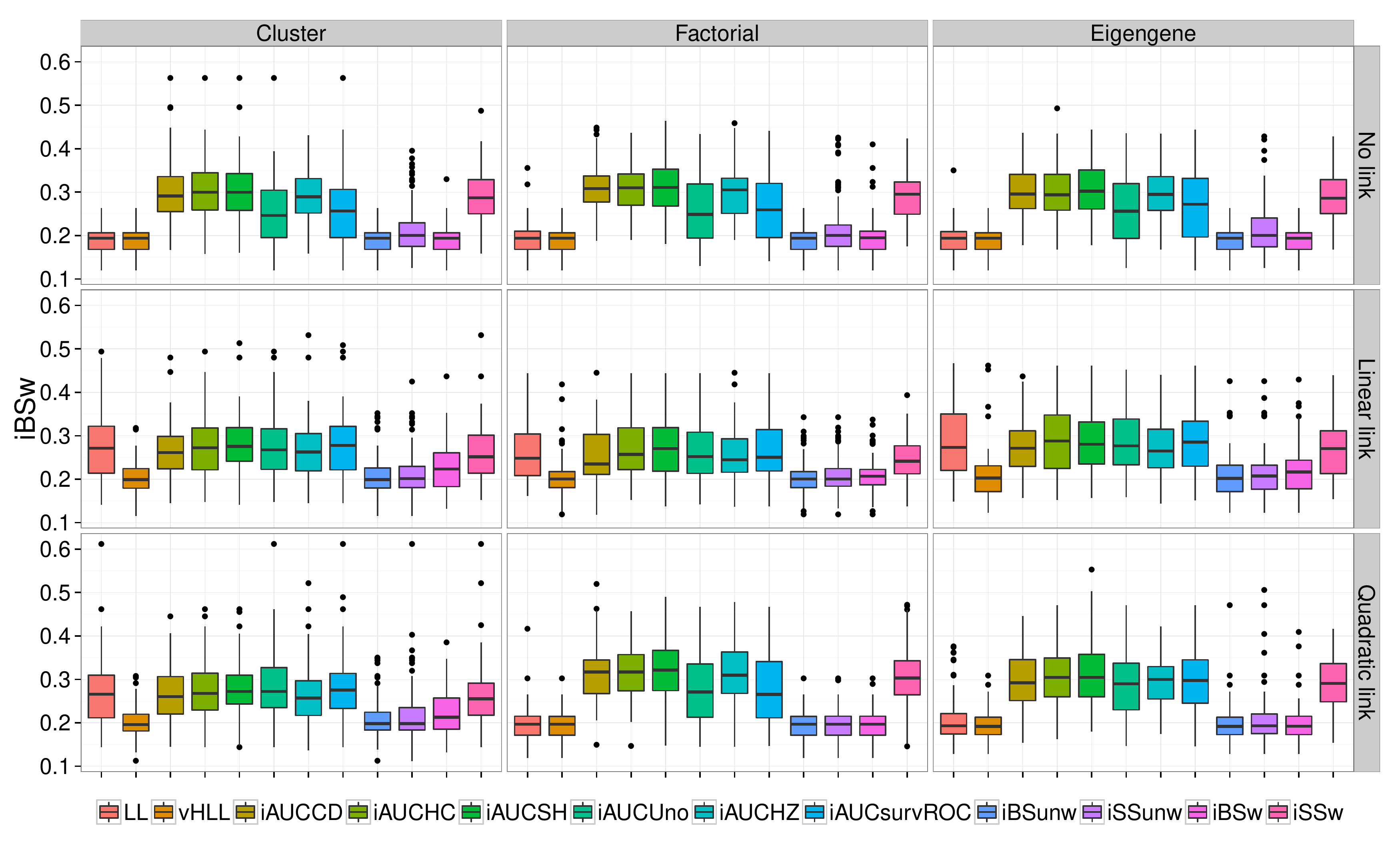}\phantomcaption\label{iBSWplsRcox}}\qquad{\includegraphics[width=.75\columnwidth]{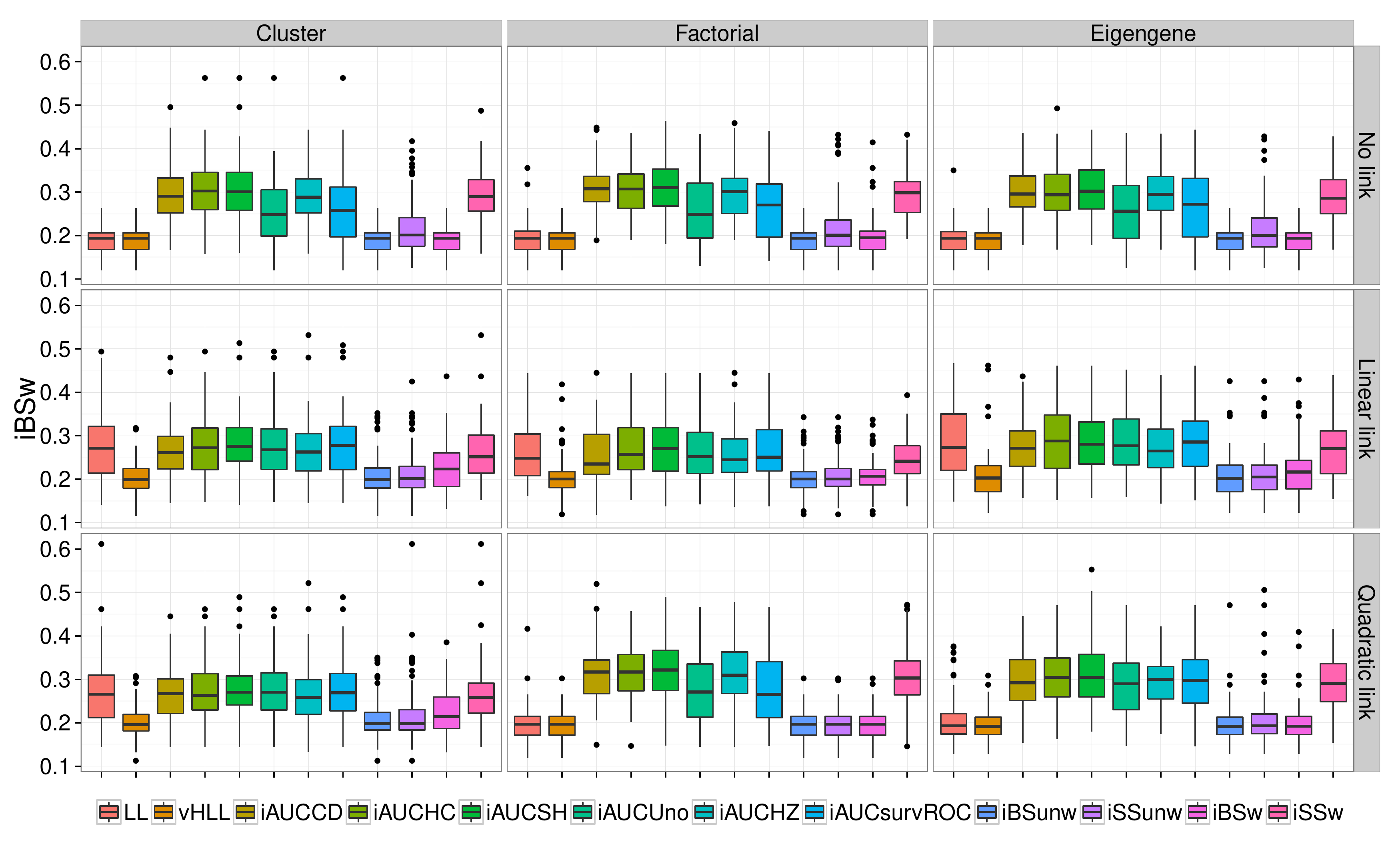}\phantomcaption\label{iBSWautoplsRcox}}}
\vspace{-.5cm}
\caption*{\hspace{-1.25cm}\mbox{Figure~\ref{iBSWplsRcox}:  iBSW vs CV criterion. plsRcox. \qquad\qquad Figure~\ref{iBSWautoplsRcox}:  iBSW vs CV criterion. autoplsRcox.}}
\end{figure}

\begin{figure}[!tpb]
\centerline{{\includegraphics[width=.75\columnwidth]{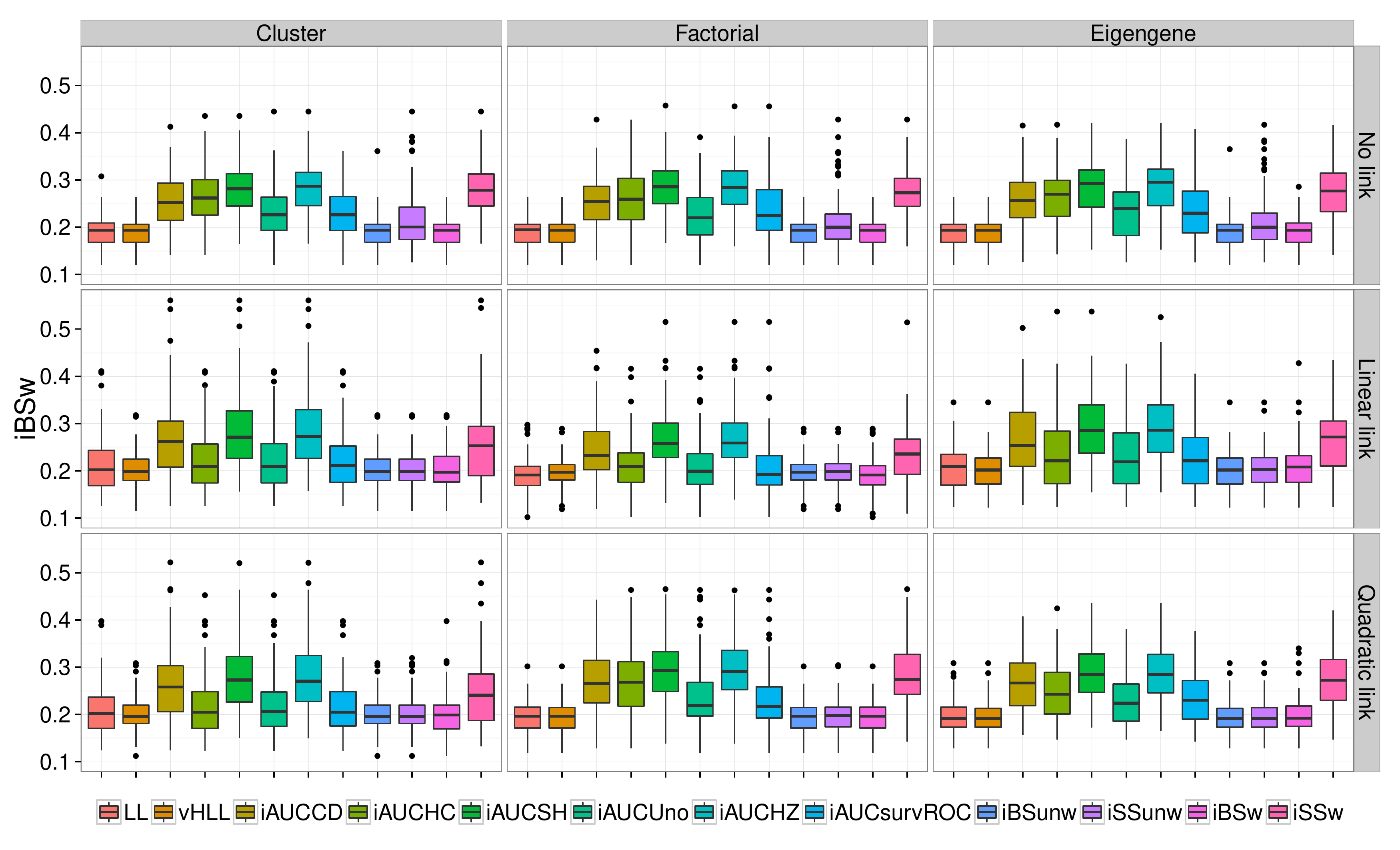}\phantomcaption\label{iBSWcoxpls}}\qquad{\includegraphics[width=.75\columnwidth]{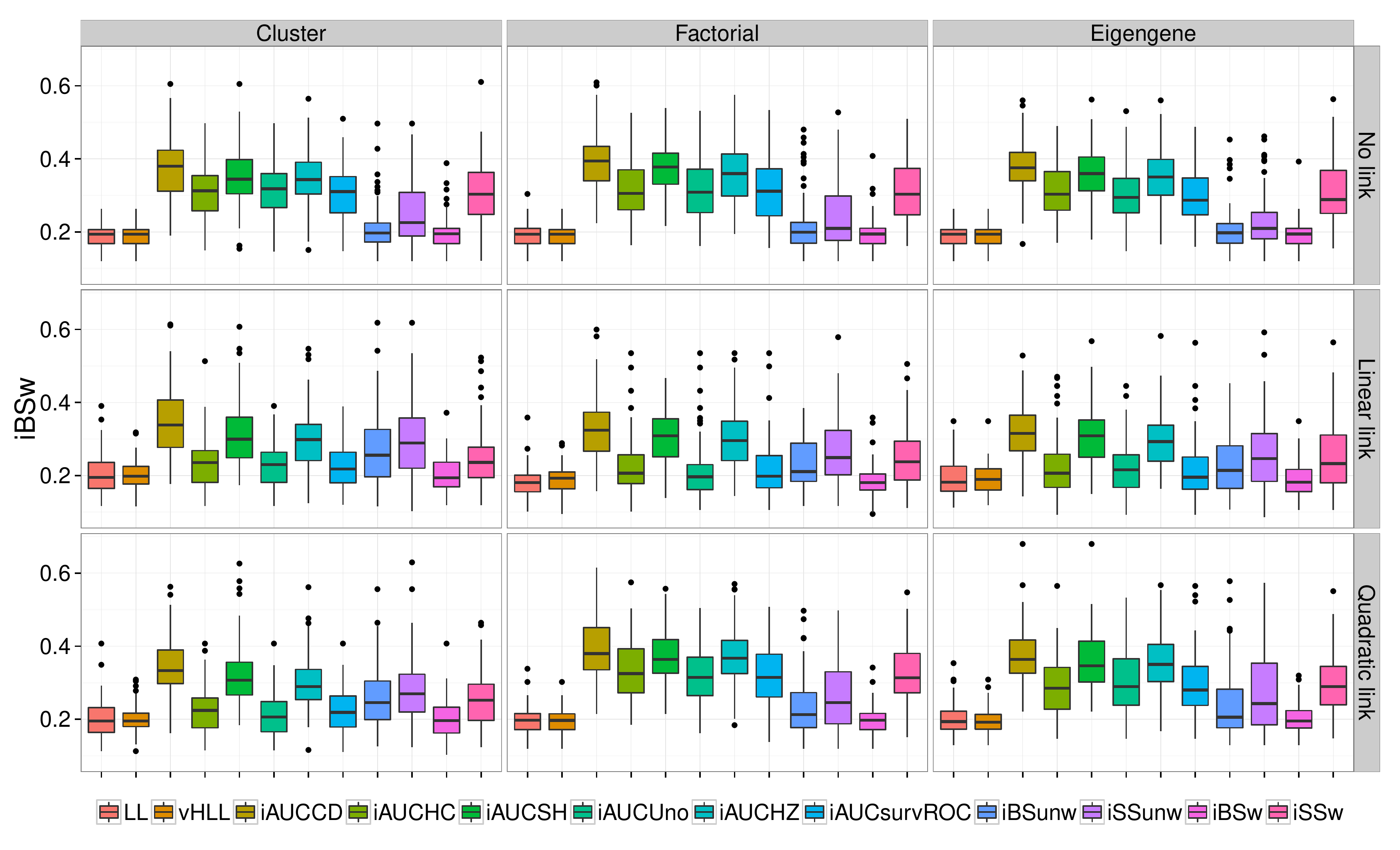}\phantomcaption\label{iBSWcoxsplsDR}}}
\vspace{-.5cm}
\caption*{\hspace{-1cm}\mbox{Figure~\ref{iBSWcoxpls}:  iBSW vs CV criterion. coxpls. \qquad\qquad Figure~\ref{iBSWcoxsplsDR}:  iBSW vs CV criterion. splsDR.}}
\end{figure}


\begin{figure}[!tpb]
\centerline{\includegraphics[width=.75\columnwidth]{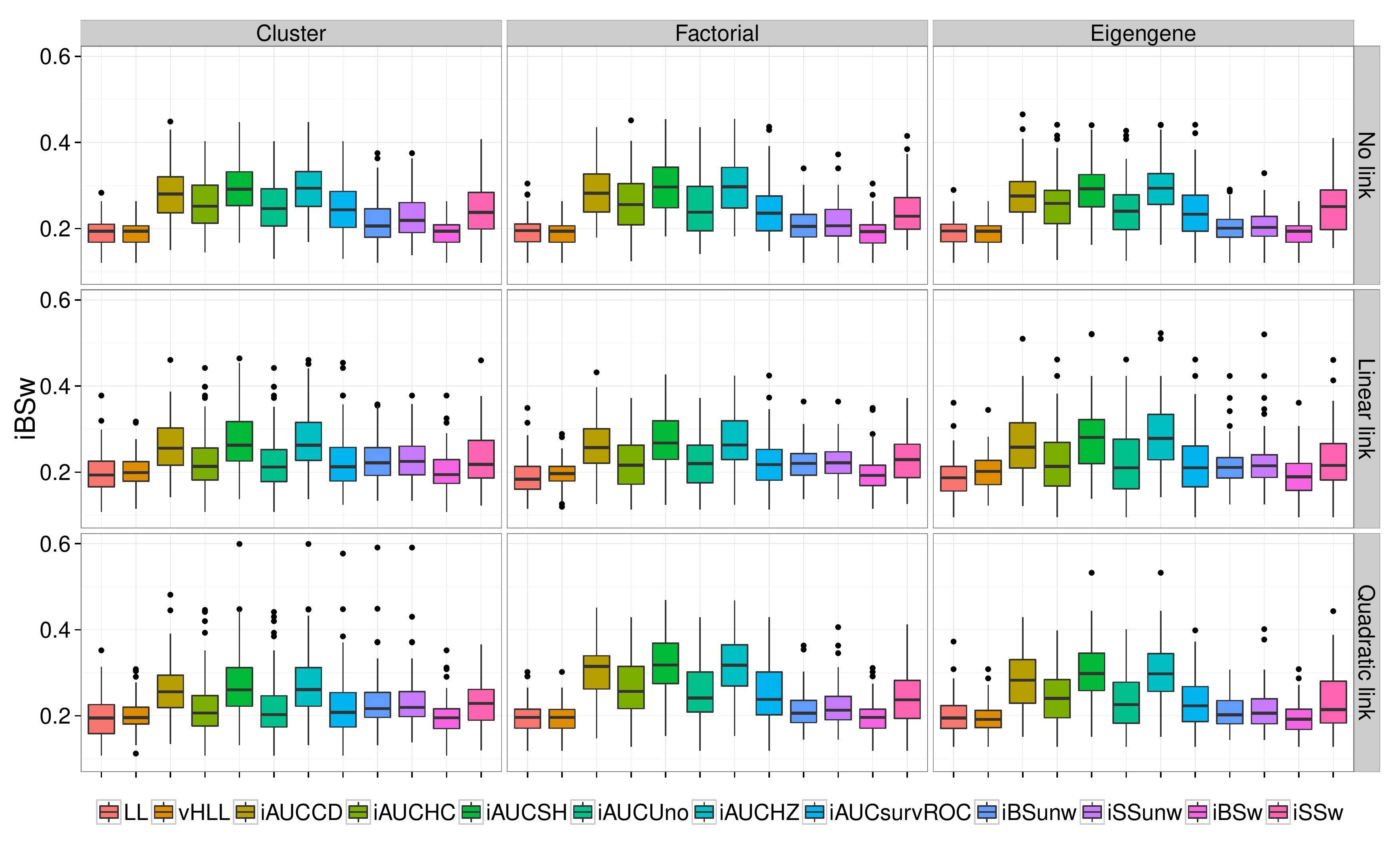}}
\vspace{-.5cm}
\caption{iBSW vs CV criterion. DKsplsDR.}\label{iBSWcoxDKplsDR}
\end{figure}



\clearpage

\begin{figure}[!tpb]
\centerline{{\includegraphics[width=.75\columnwidth]{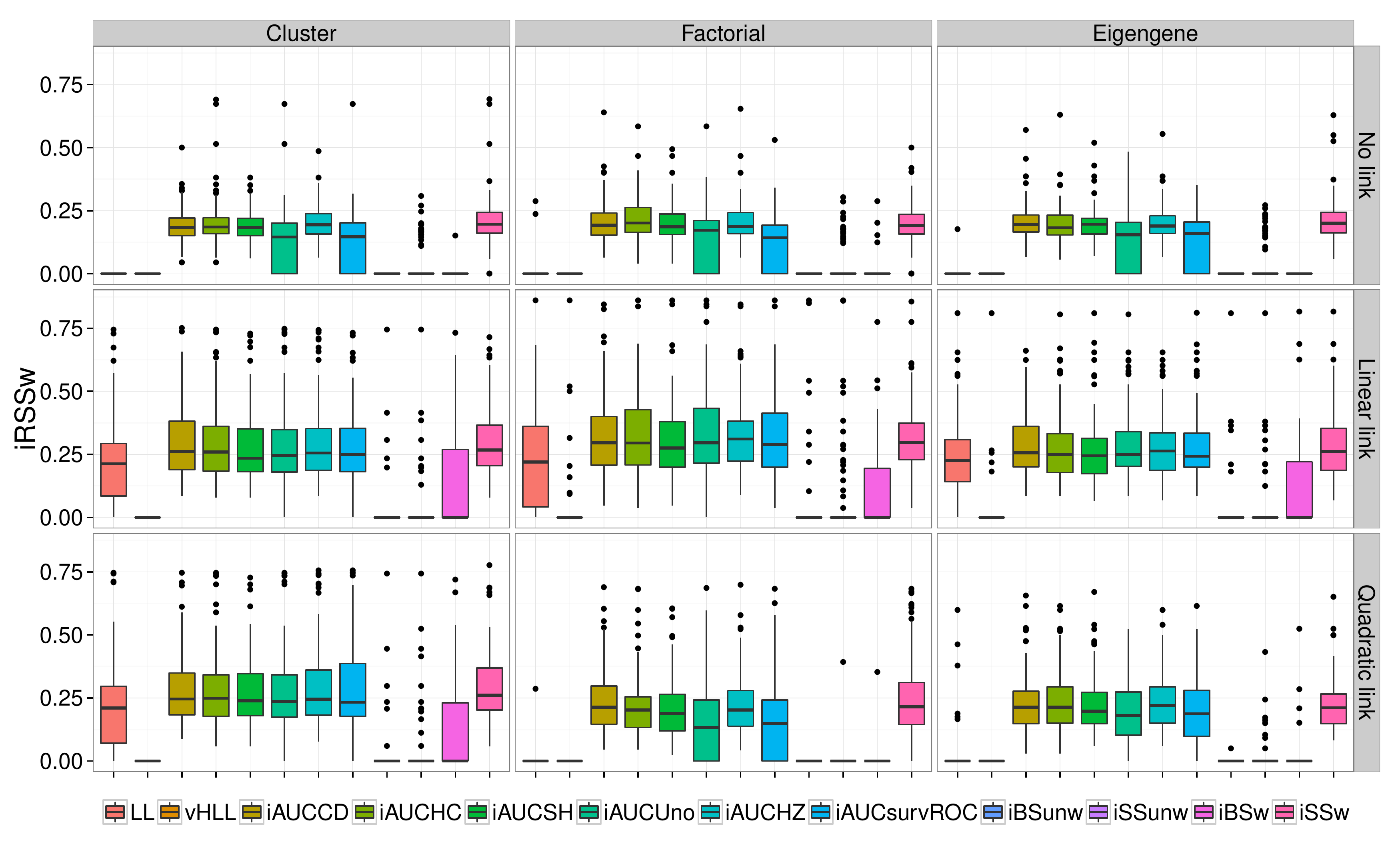}\phantomcaption\label{iRSSWplsRcox}}\qquad{\includegraphics[width=.75\columnwidth]{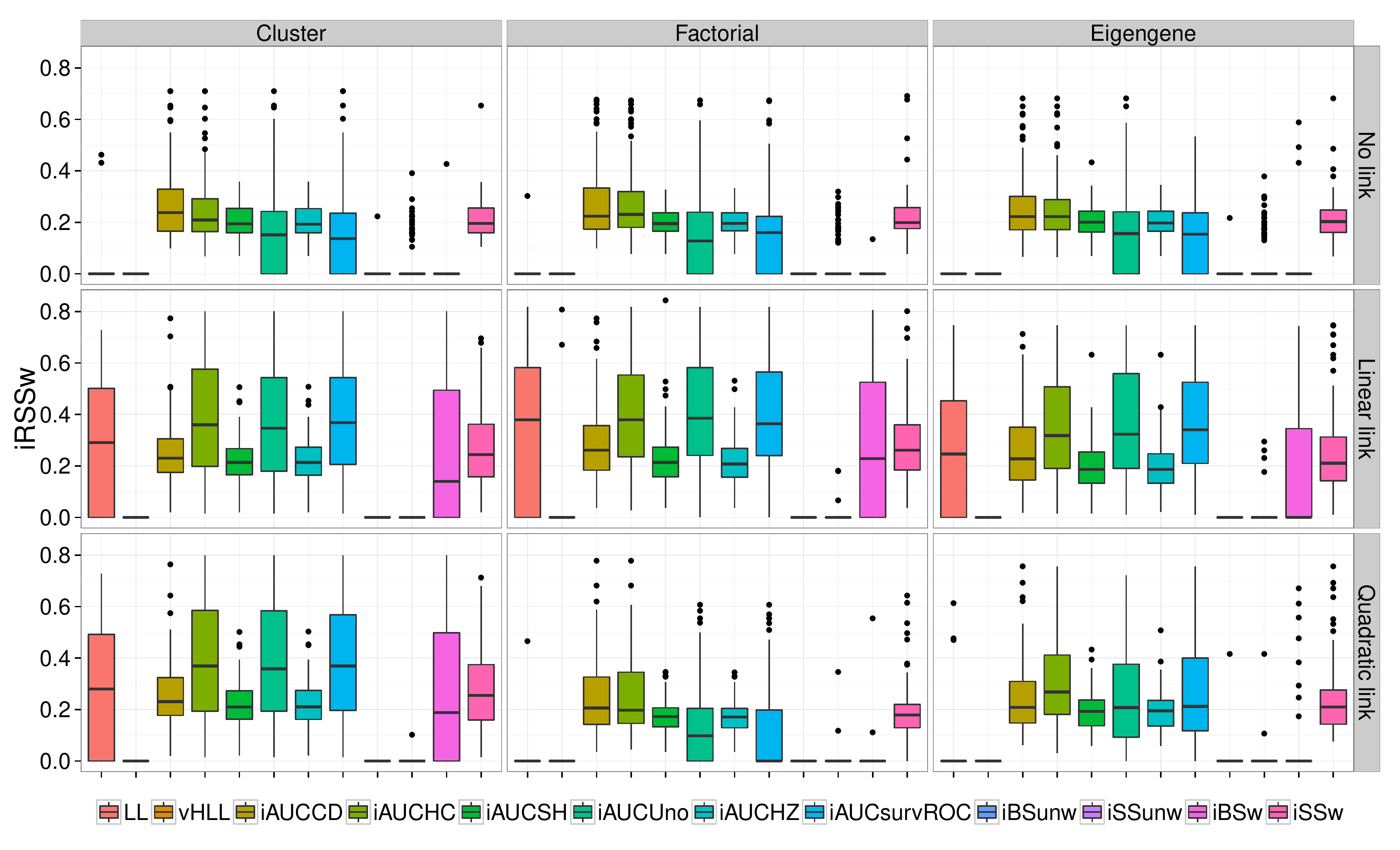}\phantomcaption\label{iRSSWcoxpls}}}
\vspace{-.5cm}
\caption*{\hspace{-1.5cm}\mbox{Figure~\ref{iRSSWplsRcox}:  iRSSW vs CV criterion. plsRcox. \qquad\qquad Figure~\ref{iRSSWcoxpls}:  iRSSW vs CV criterion. coxpls.}}
\end{figure}

\begin{figure}[!tpb]
\centerline{{\includegraphics[width=.75\columnwidth]{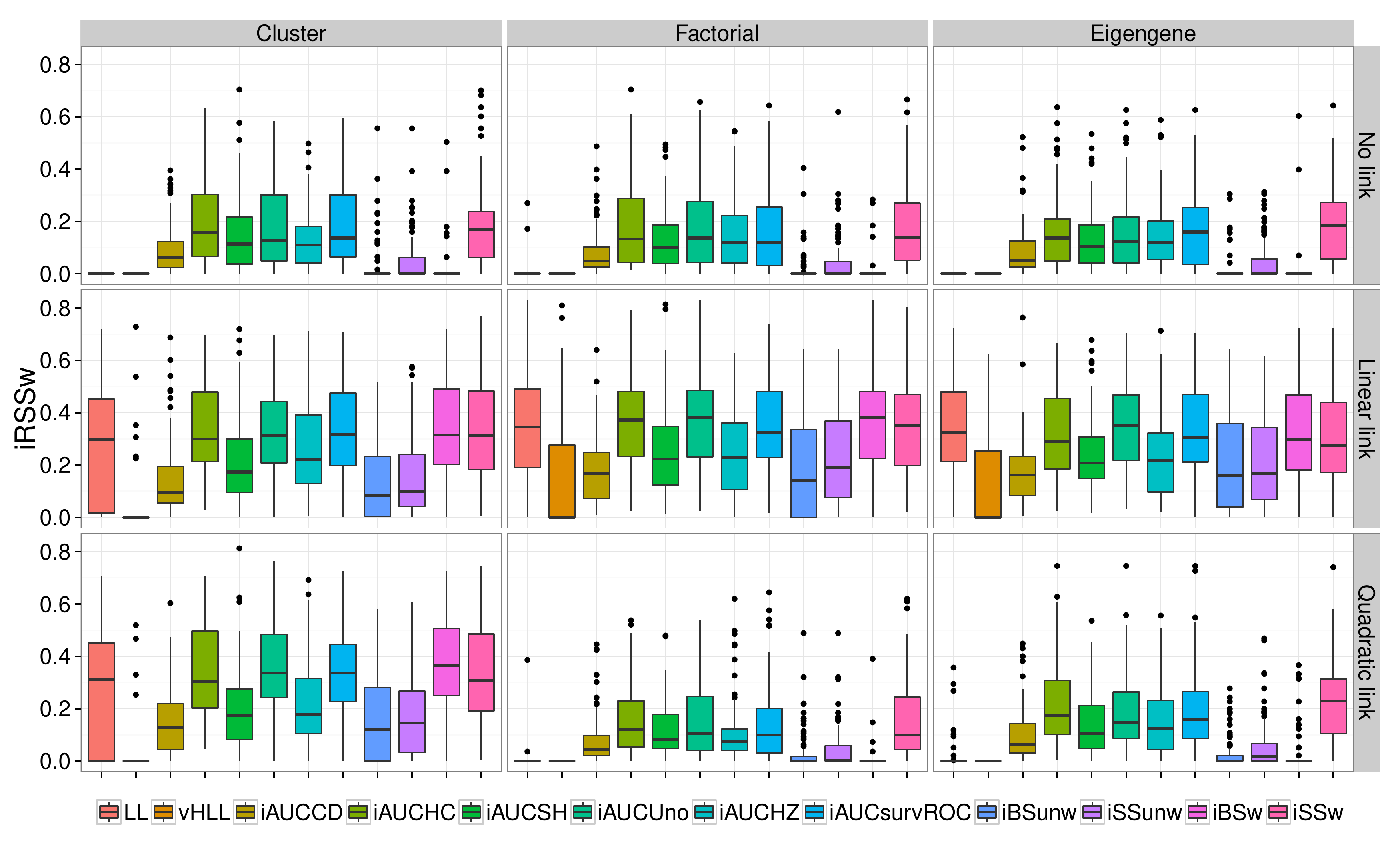}\phantomcaption\label{iRSSWcoxsplsDR}}\qquad{\includegraphics[width=.75\columnwidth]{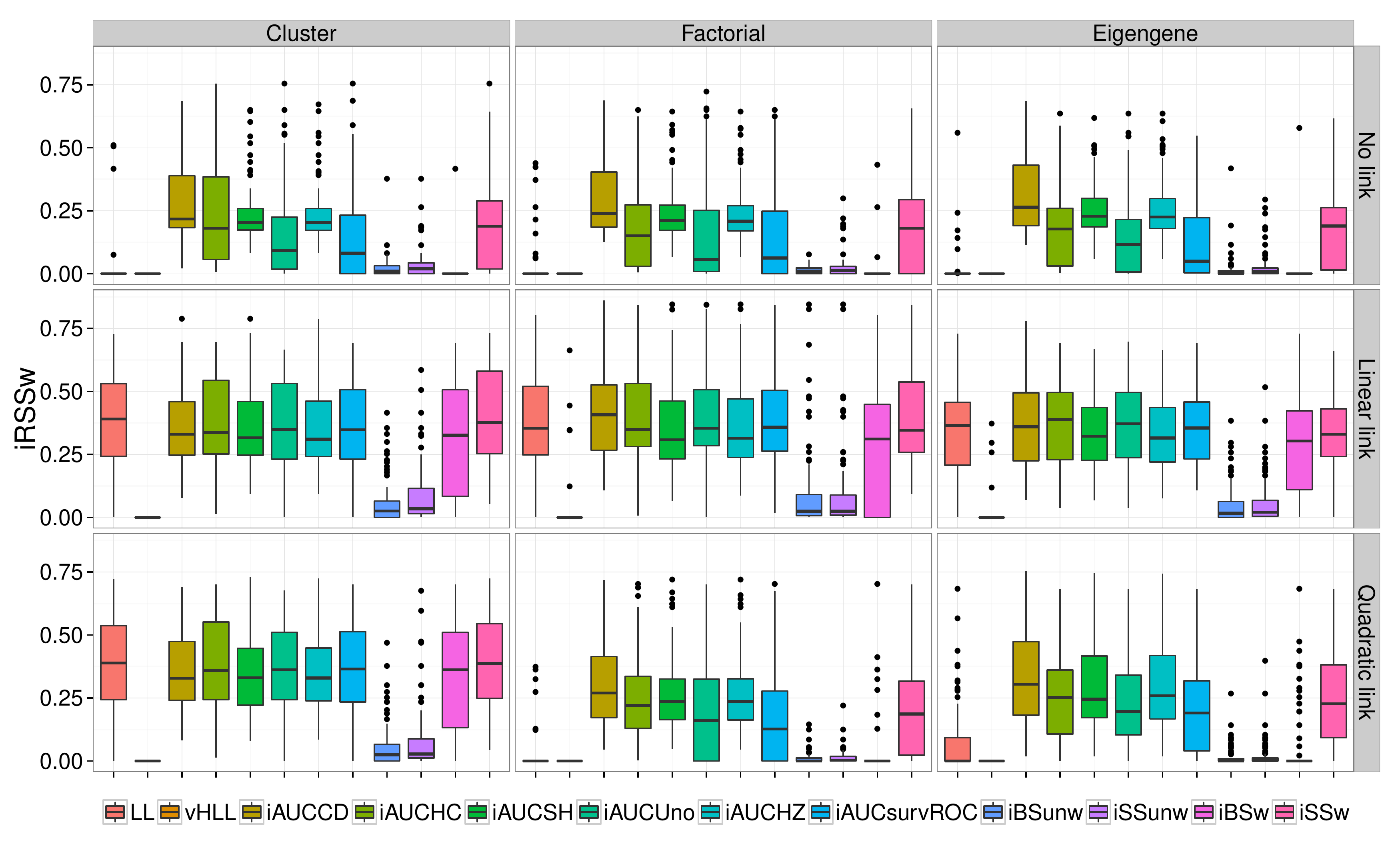}\phantomcaption\label{iRSSWcoxDKplsDR}}}
\vspace{-.5cm}
\caption*{\hspace{-1.5cm}\mbox{Figure~\ref{iRSSWcoxsplsDR}:  iRSSW vs CV criterion. coxsplsDR. \qquad\qquad Figure~\ref{iRSSWcoxDKplsDR}:  iRSSW vs CV criterion. DKplsDR.}}
\end{figure}



\begin{figure}[!tpb]
\centerline{\includegraphics[width=.75\columnwidth]{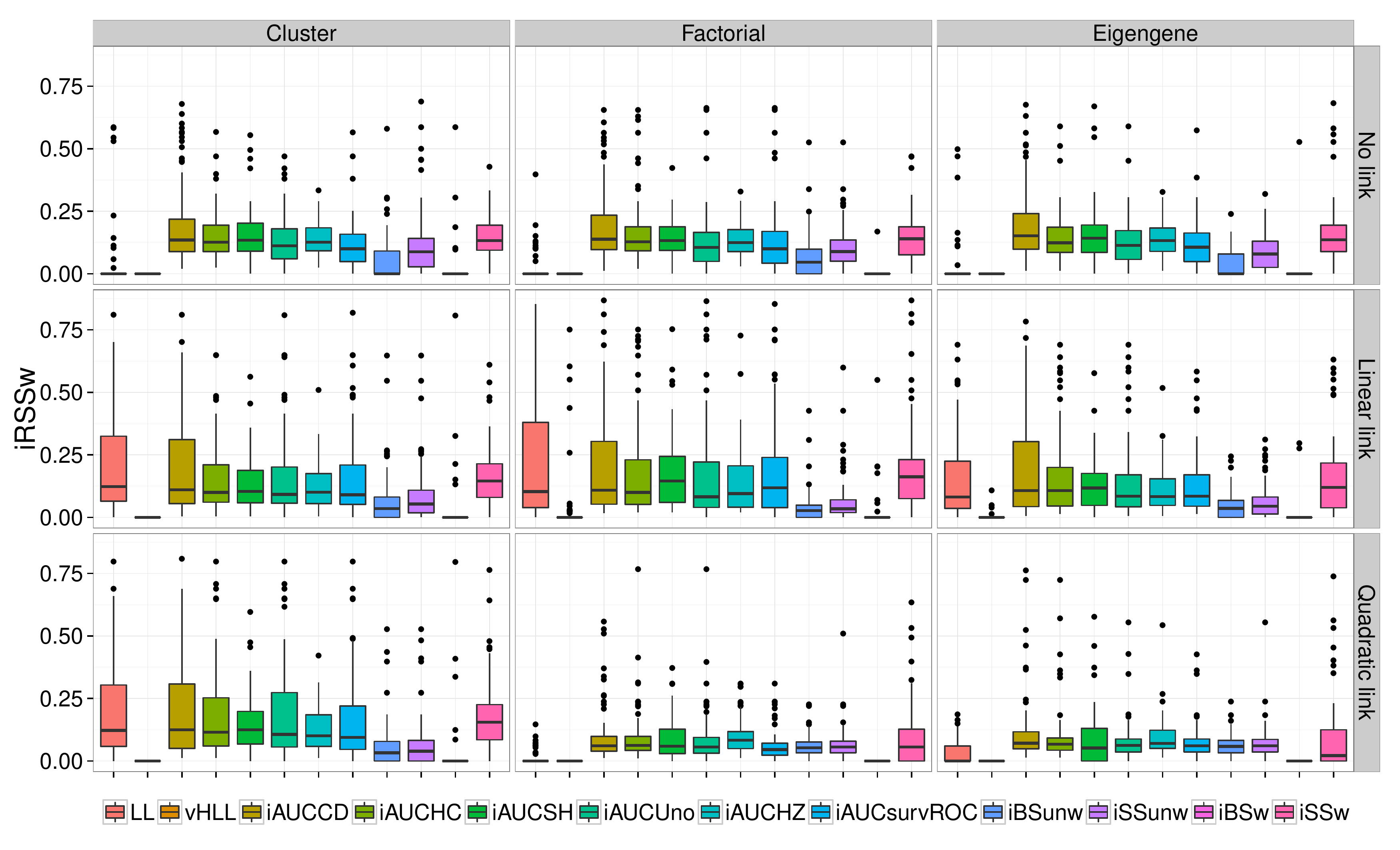}}
\vspace{-.5cm}
\caption{iRSSW vs CV criterion. DKsplsDR.}\label{iRSSWcoxDKsplsDR}
\end{figure}

\clearpage

\begin{figure}[!tpb]
\centerline{{\includegraphics[width=.75\columnwidth]{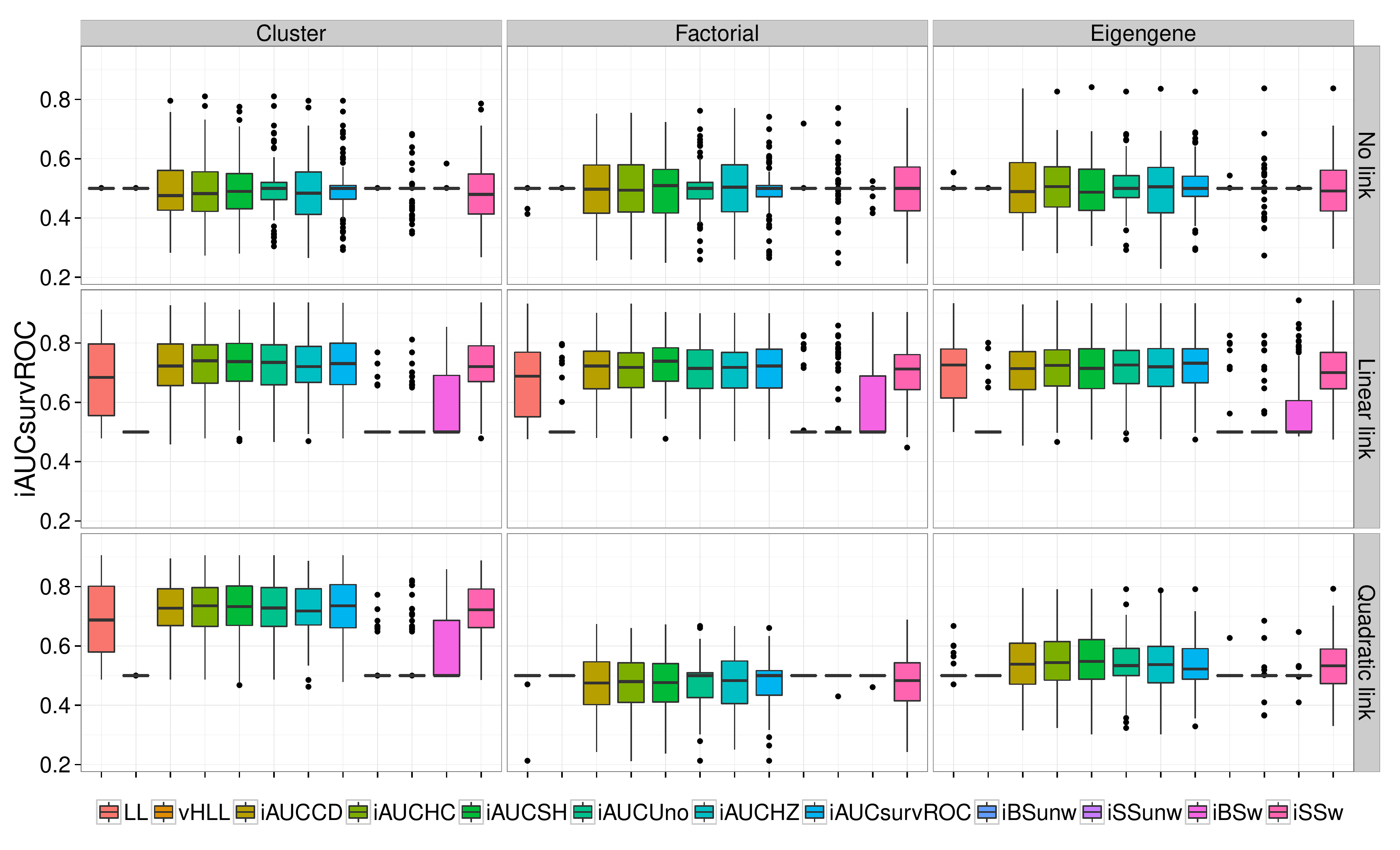}\phantomcaption\label{SurvROCplsRcox}}\qquad{\includegraphics[width=.75\columnwidth]{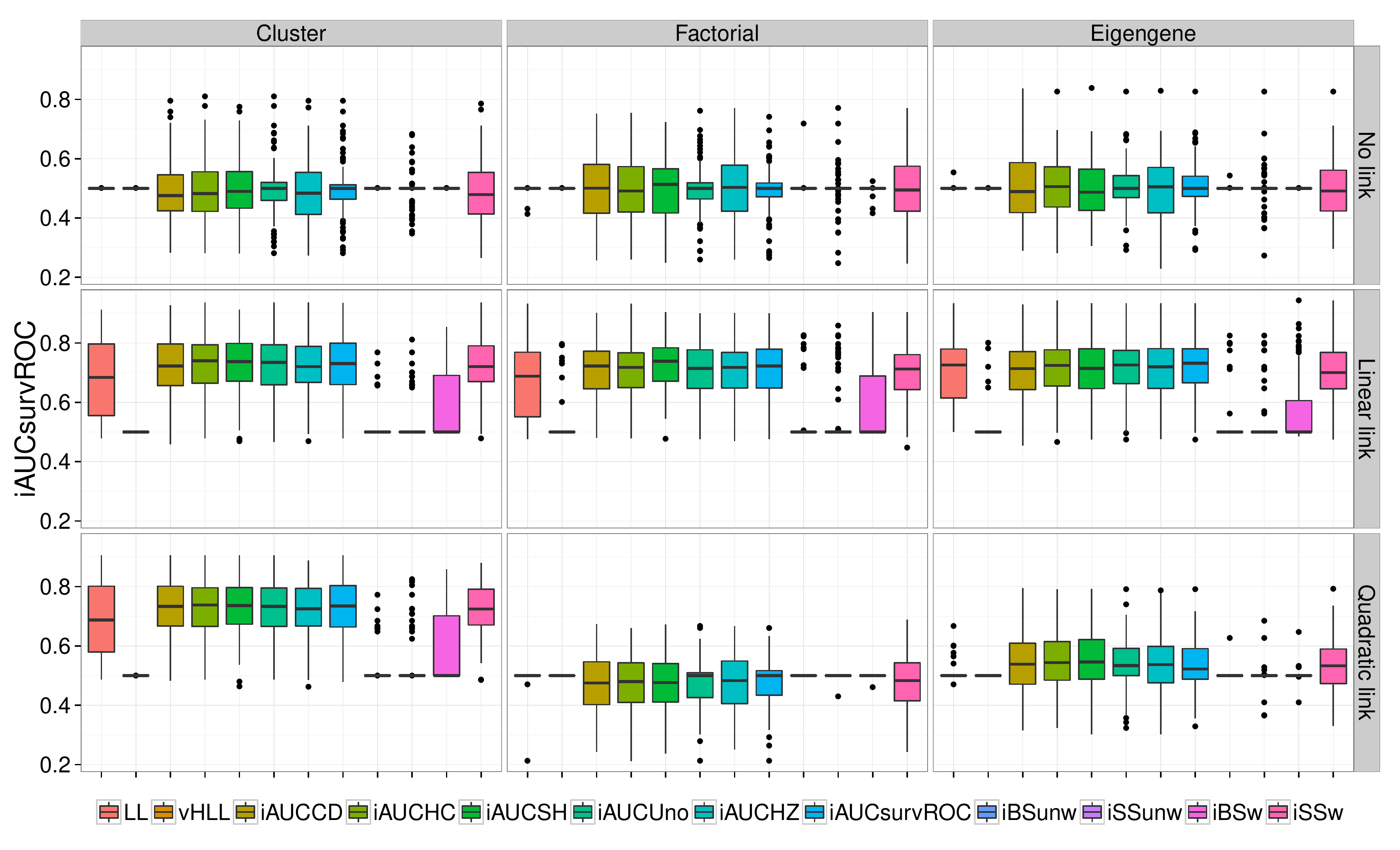}\phantomcaption\label{SurvROCautoplsRcox}}}
\vspace{-.5cm}
\caption*{\hspace{-1.5cm}\mbox{Figure~\ref{SurvROCplsRcox}:  SurvROC vs CV criterion. plsRcox. \qquad\qquad Figure~\ref{SurvROCautoplsRcox}:  SurvROC vs CV criterion. autoplsRcox.}}
\end{figure}

\begin{figure}[!tpb]
\centerline{{\includegraphics[width=.75\columnwidth]{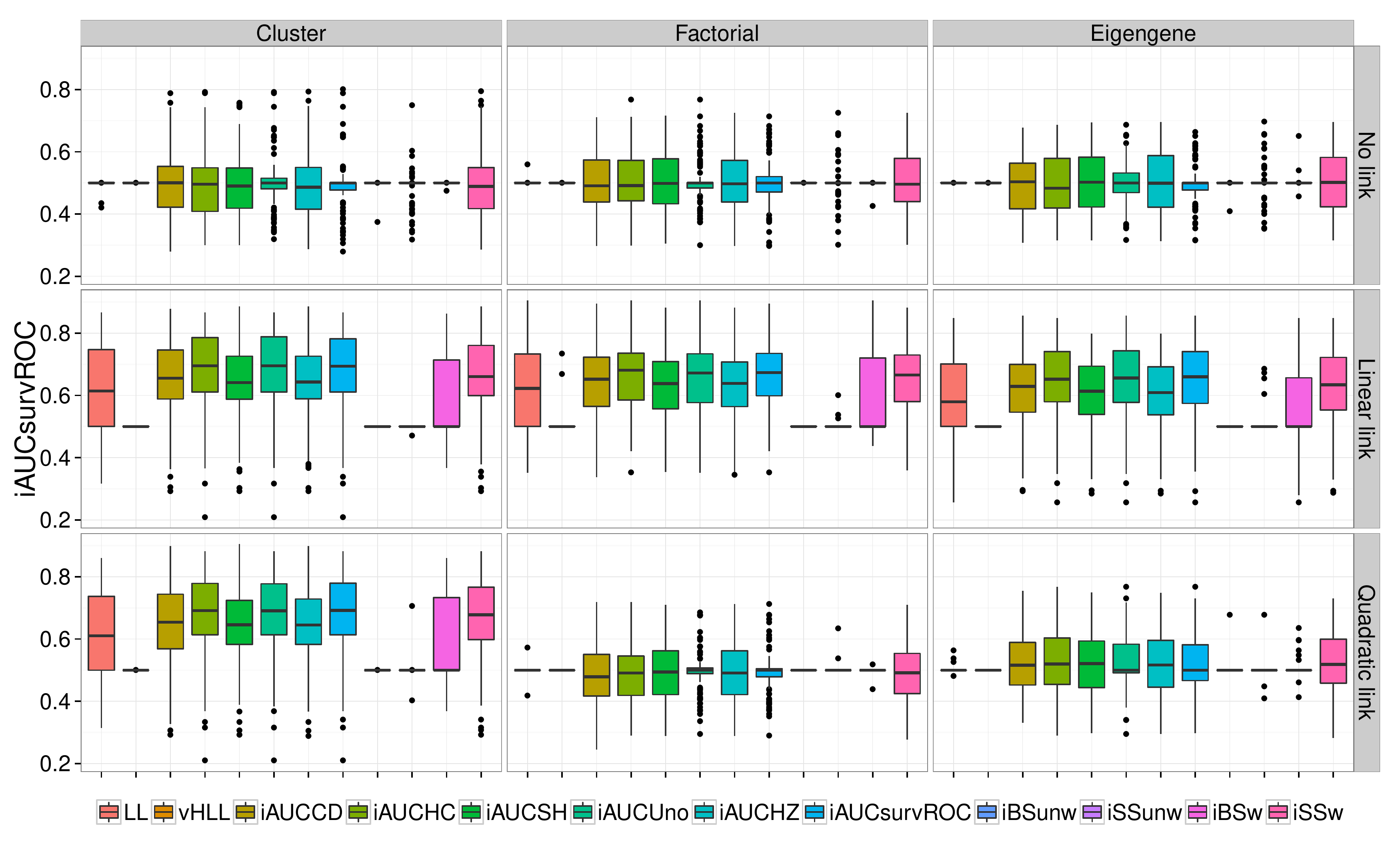}\phantomcaption\label{SurvROCcoxpls}}\qquad{\includegraphics[width=.75\columnwidth]{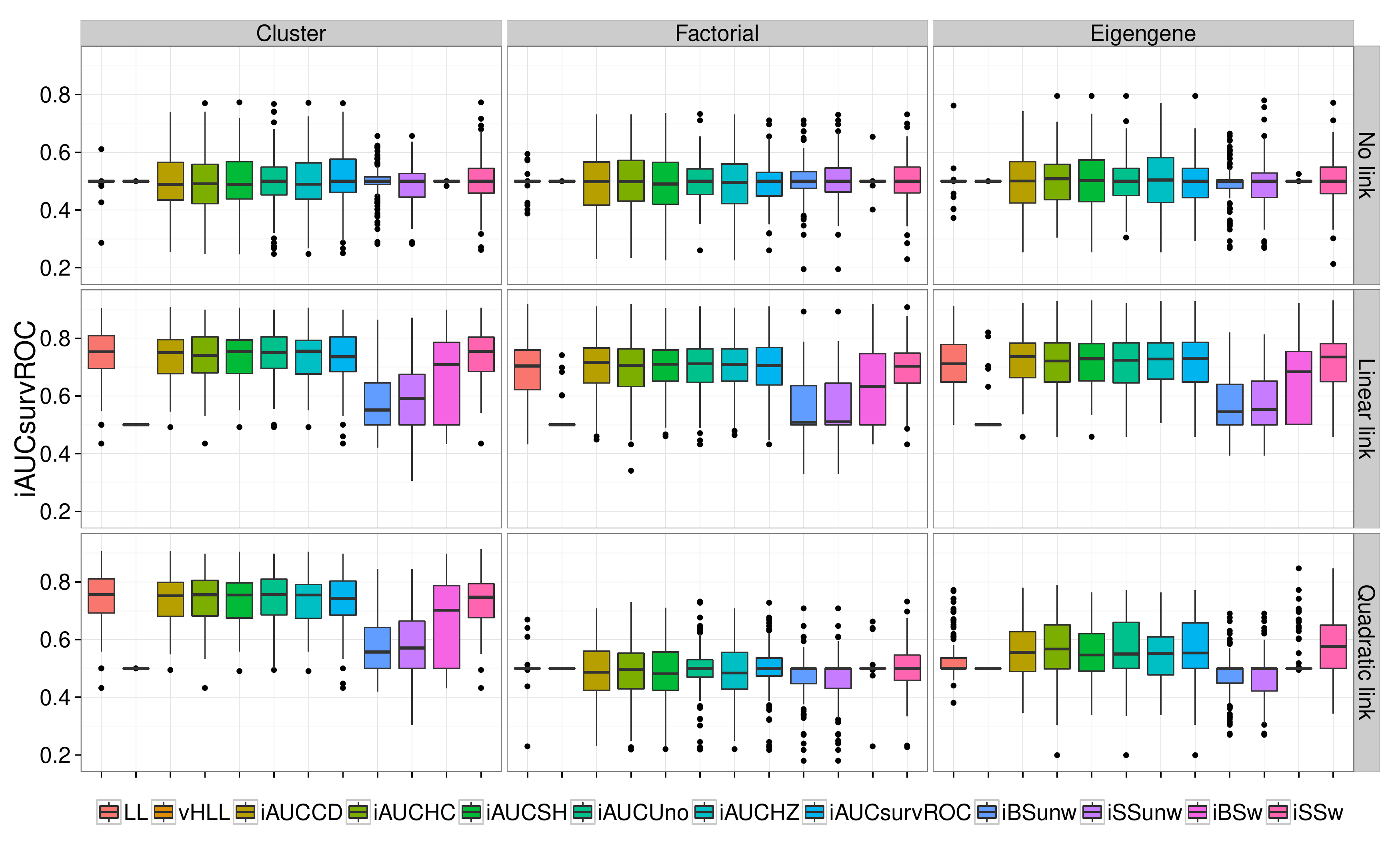}\phantomcaption\label{SurvROCDKplsDR}}}
\vspace{-.5cm}
\caption*{\hspace{-1.5cm}\mbox{Figure~\ref{SurvROCcoxpls}:  SurvROC vs CV criterion. coxpls. \qquad\qquad Figure~\ref{SurvROCDKplsDR}:  SurvROC vs CV criterion. DKplsDR.}}
\end{figure}



\begin{figure}[!tpb]
\centerline{\includegraphics[width=.75\columnwidth]{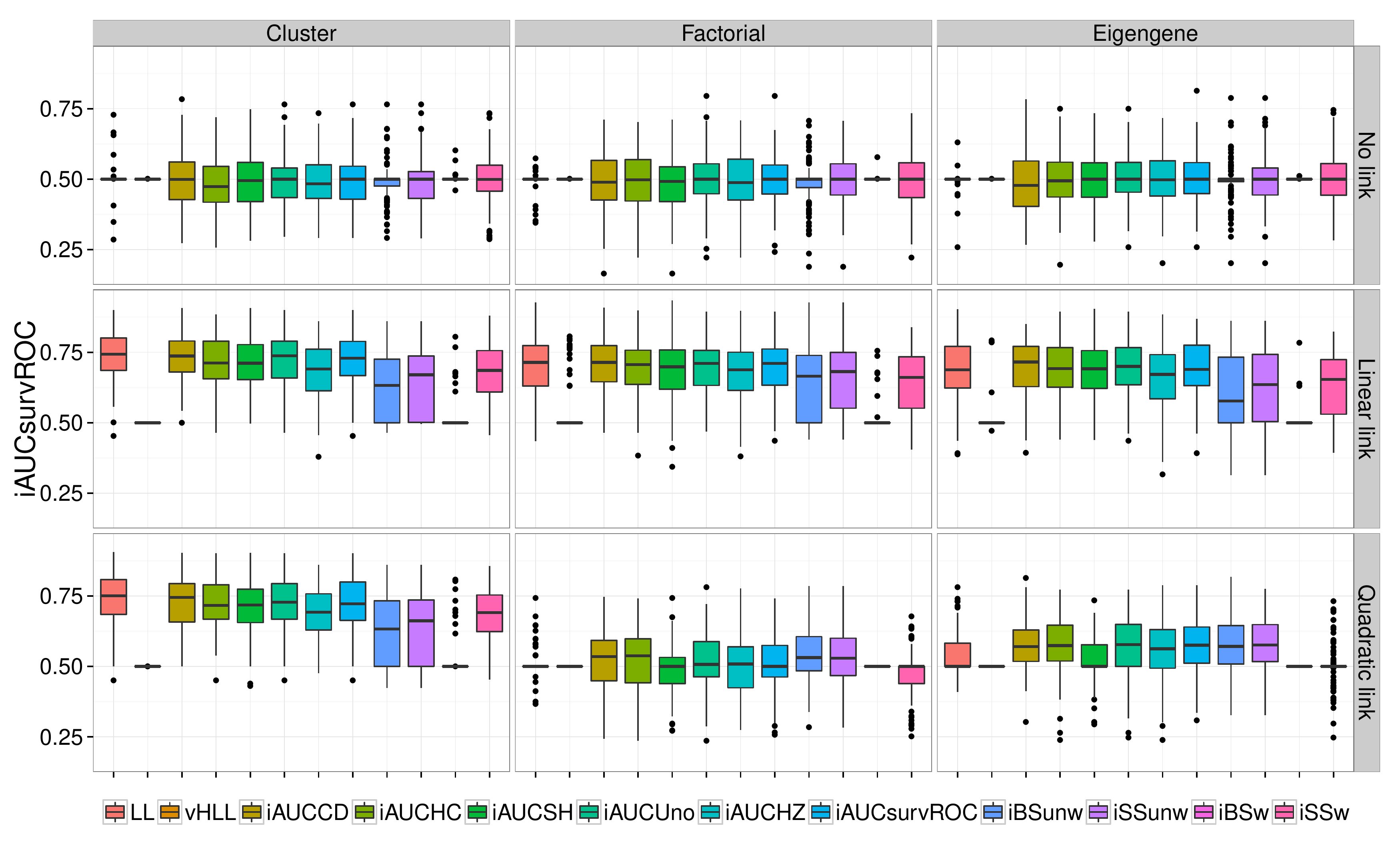}}
\vspace{-.5cm}
\caption{iAUCSurvROC vs CV criterion, DKsplsDR.}\label{SurvROCDKsplsDR}
\end{figure}

\clearpage


\begin{figure}[!tpb]
\centerline{{\includegraphics[width=.75\columnwidth]{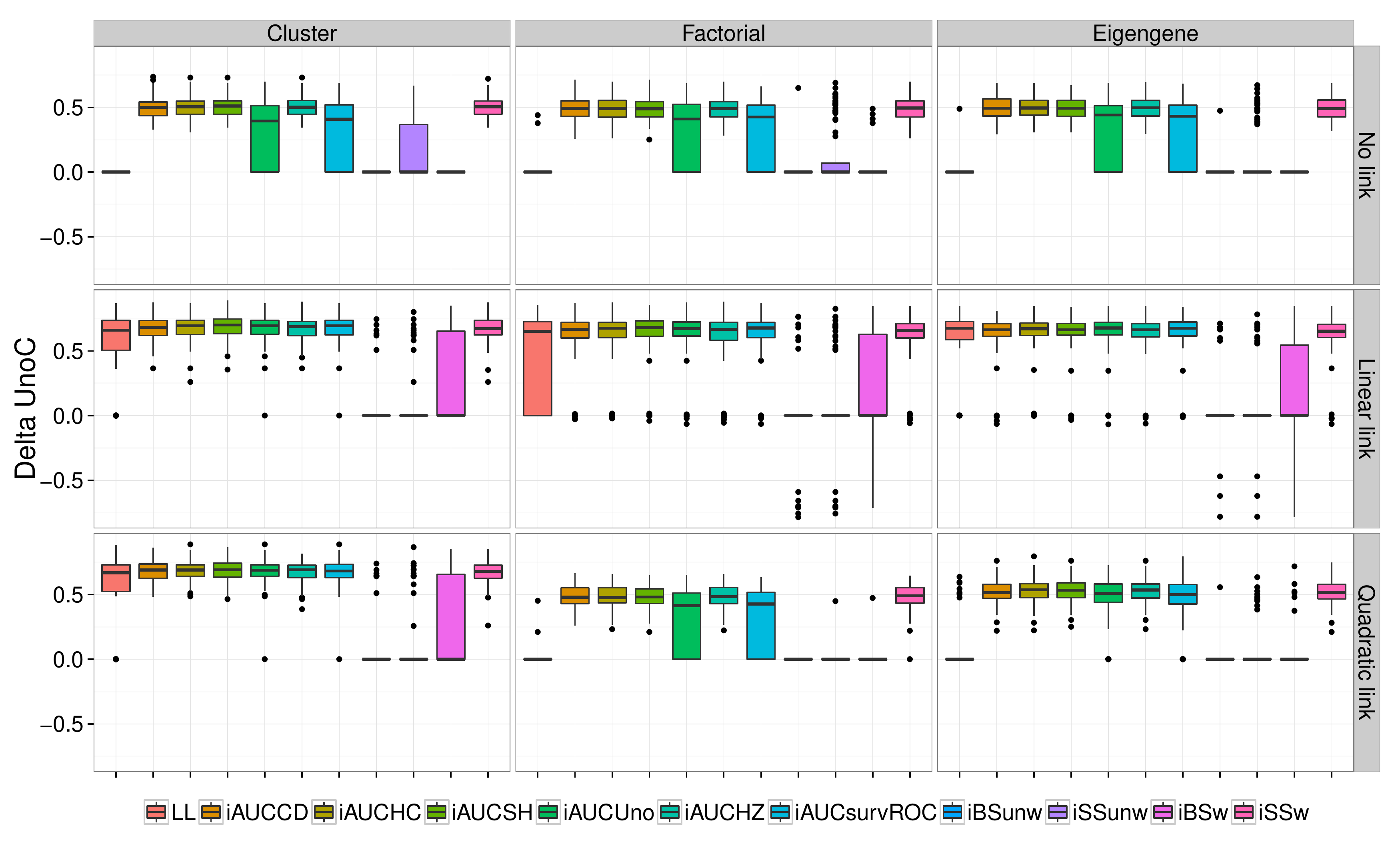}\phantomcaption\label{incautoplsRcoxUnoCstat}}\qquad{\includegraphics[width=.75\columnwidth]{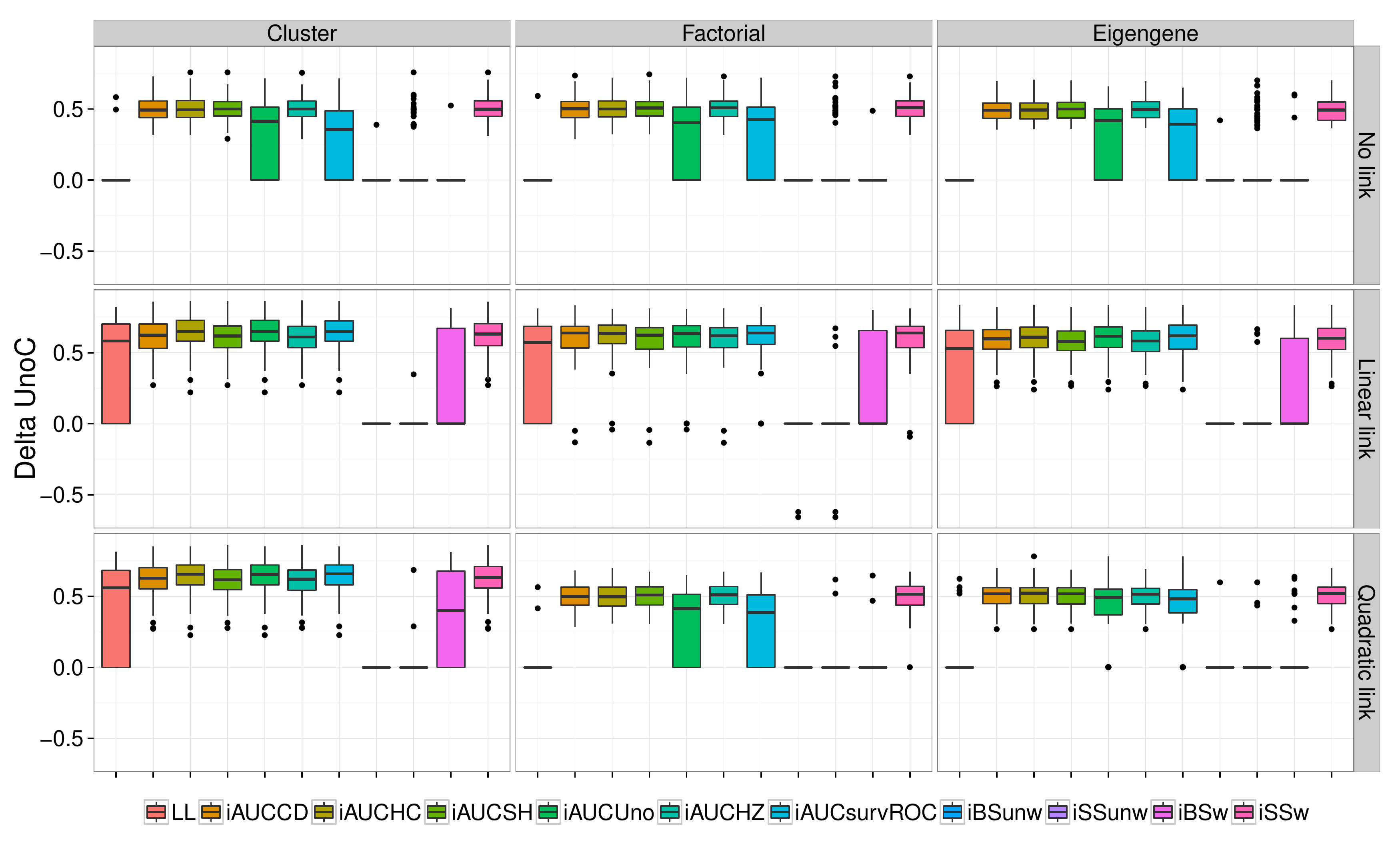}\phantomcaption\label{inccoxplsUnoCstat}}}
\vspace{-.5cm}
\caption*{\hspace{-2.5cm}\mbox{Delta of UnoC (CV criteria $-$ vHCVLL value).\quad Figure~\ref{incautoplsRcoxUnoCstat} (left): autoPLS$-$Cox. Figure~\ref{inccoxplsUnoCstat} (right): Cox$-$PLS.}}
\end{figure}

\begin{figure}[!tpb]
\centerline{{\includegraphics[width=.75\columnwidth]{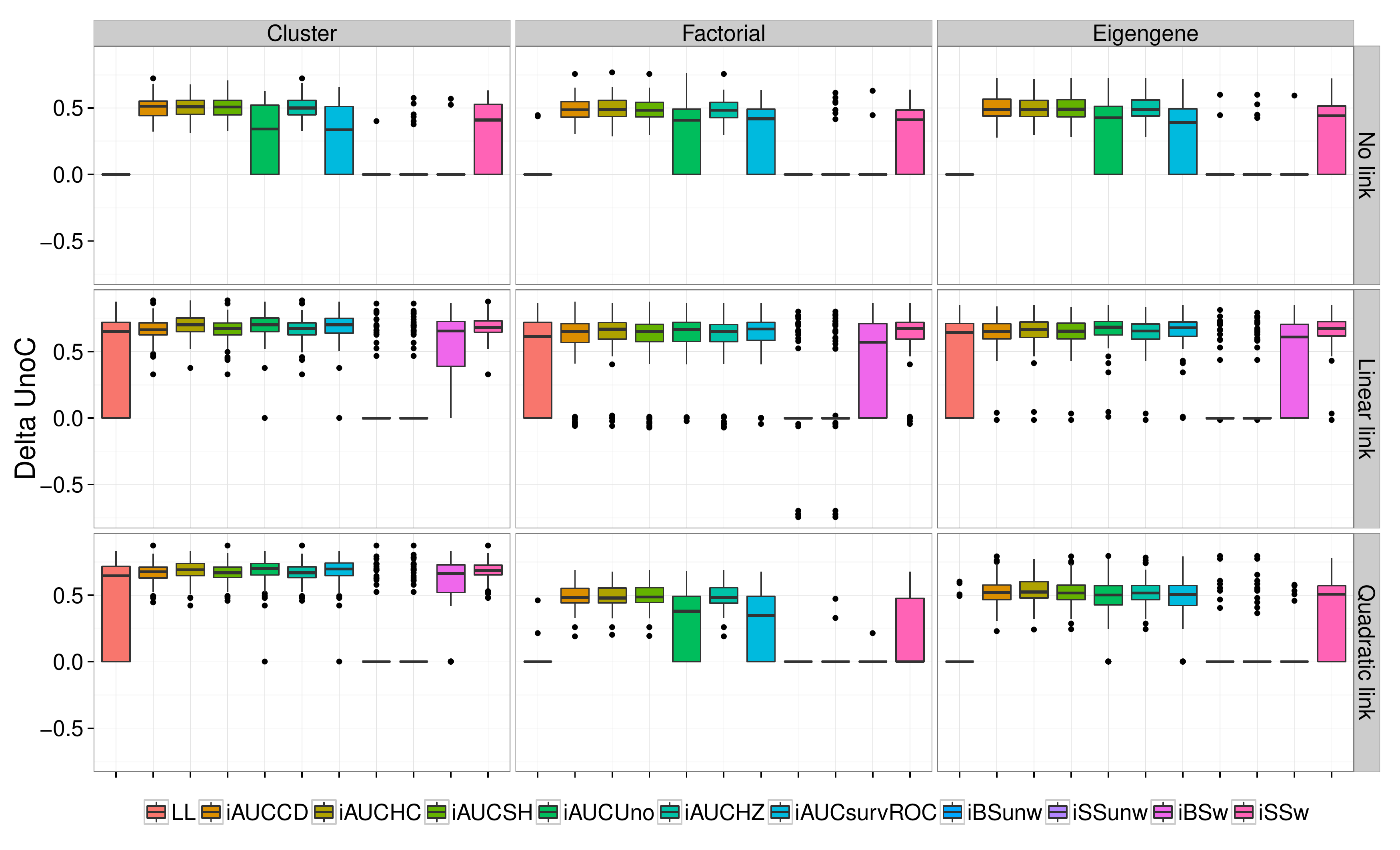}\phantomcaption\label{inccoxplsDRUnoCstat}}\qquad{\includegraphics[width=.75\columnwidth]{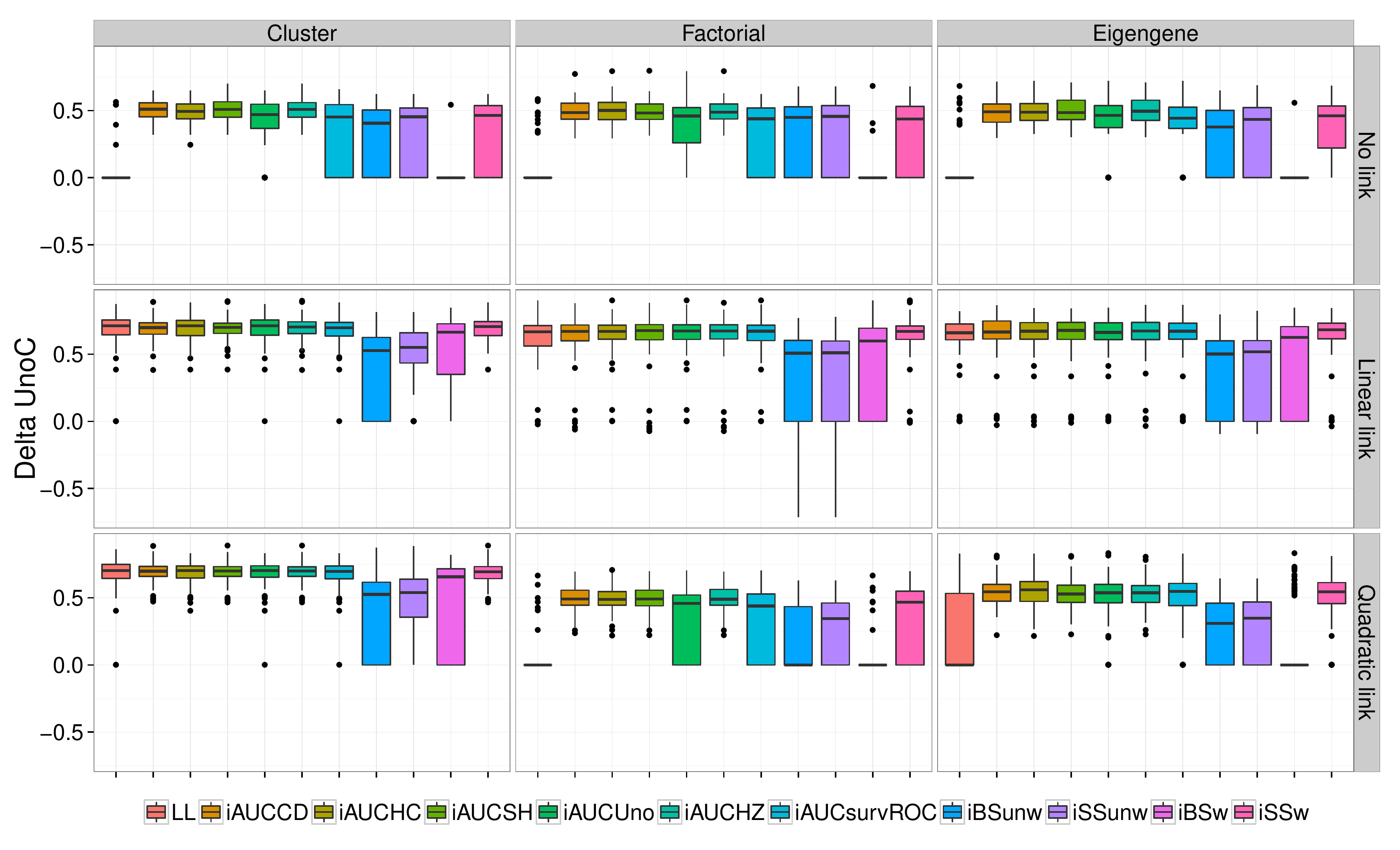}\phantomcaption\label{inccoxDKplsDRUnoCstat}}}
\vspace{-.5cm}
\caption*{\hspace{-2.5cm}\mbox{Delta of UnoC (CV criteria $-$ vHCVLL value).\quad Figure~\ref{inccoxplsDRUnoCstat} (left): PLSDR. Figure~\ref{inccoxDKplsDRUnoCstat} (right):  DKPLSDR.}}
\end{figure}



\begin{figure}[!tpb]
\centerline{\includegraphics[width=.75\columnwidth]{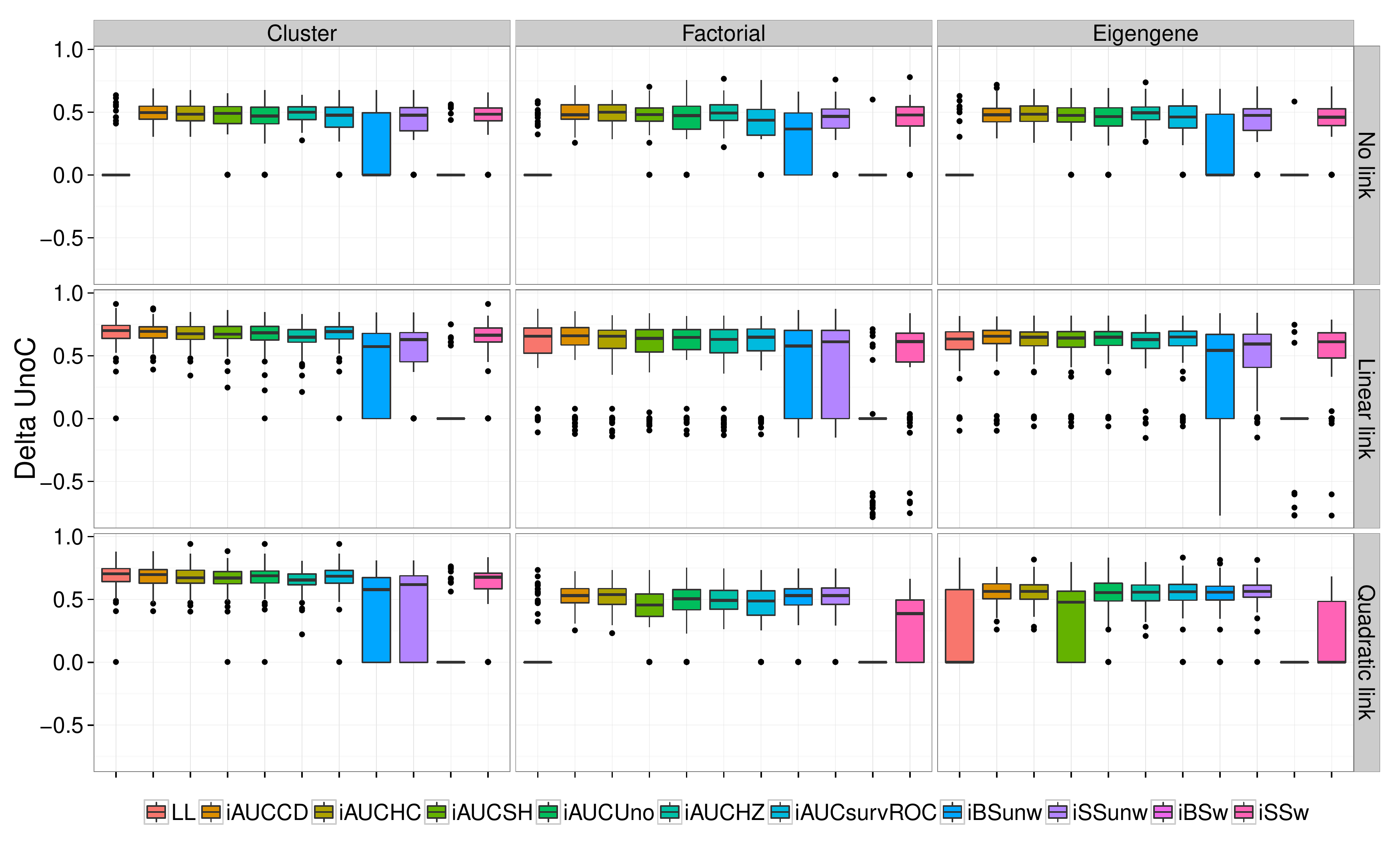}}
\vspace{-.5cm}
\caption{Delta of UnoC (CV criteria $-$ vHCVLL value). DKsPLSDR.}\label{inccoxDKsplsDRUnoCstat}
\end{figure}

\clearpage

\begin{figure}[!tpb]
\centerline{{\includegraphics[width=.75\columnwidth]{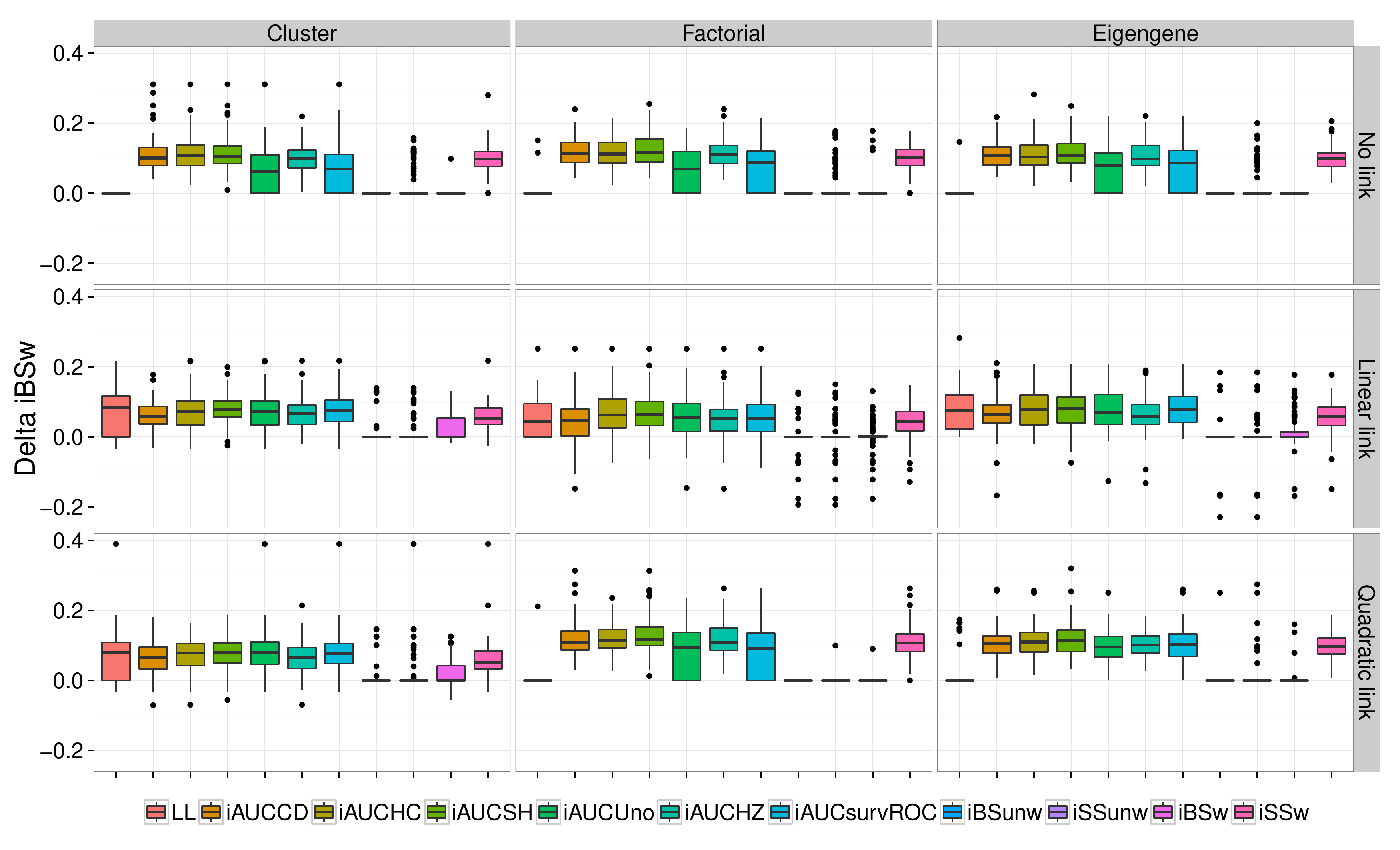}\phantomcaption\label{inciBSWplsRcox}}\qquad{\includegraphics[width=.75\columnwidth]{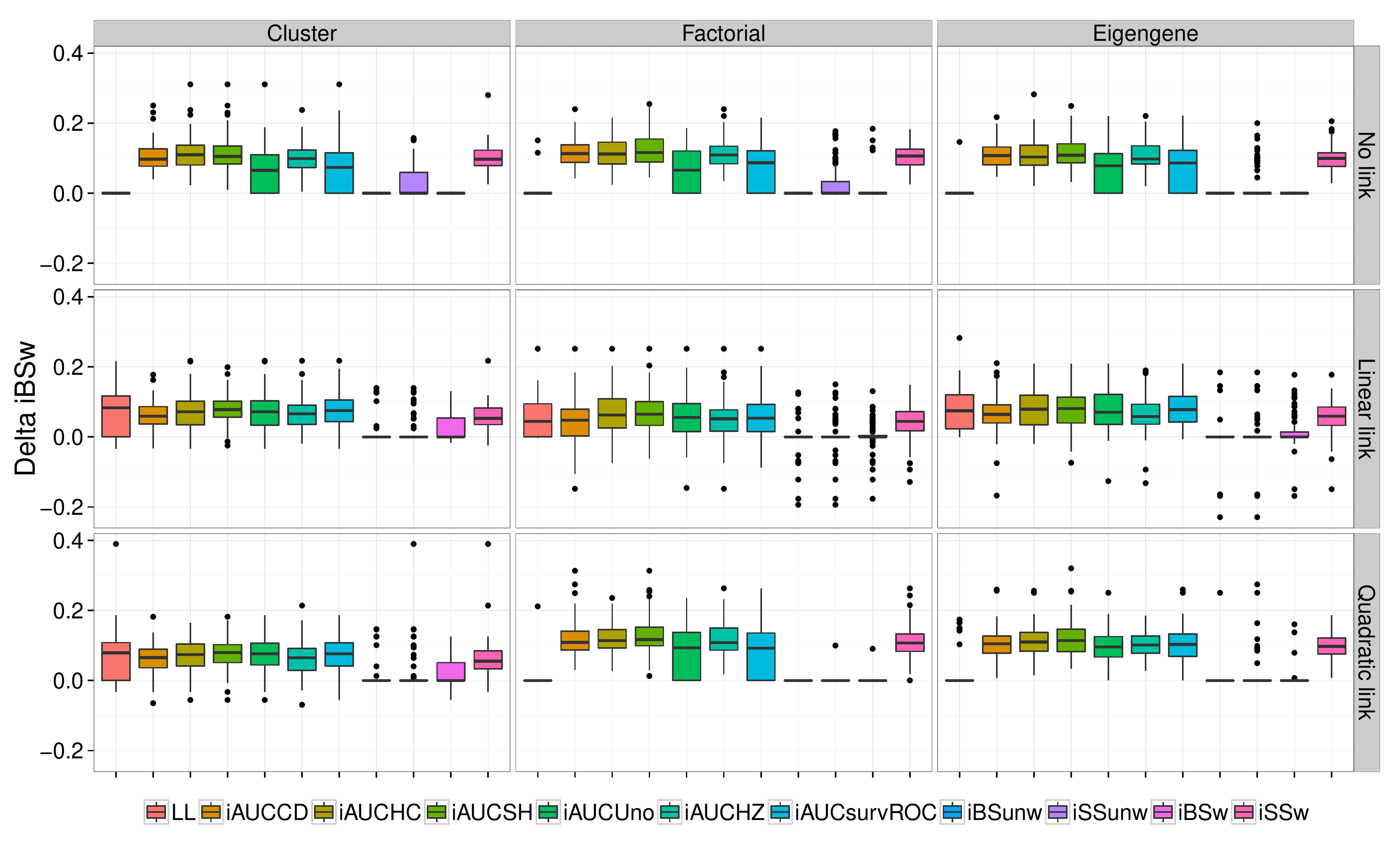}\phantomcaption\label{inciBSWautoplsRcox}}}
\vspace{-.5cm}
\caption*{\hspace{-2.5cm}\mbox{Delta of iBSW (CV criteria $-$ vHCVLL value).\quad Figure~\ref{inciBSWplsRcox} (left): PLS$-$Cox. Figure~\ref{inciBSWautoplsRcox} (right):  autoPLS$-$Cox.}}
\end{figure}

\begin{figure}[!tpb]
\centerline{{\includegraphics[width=.75\columnwidth]{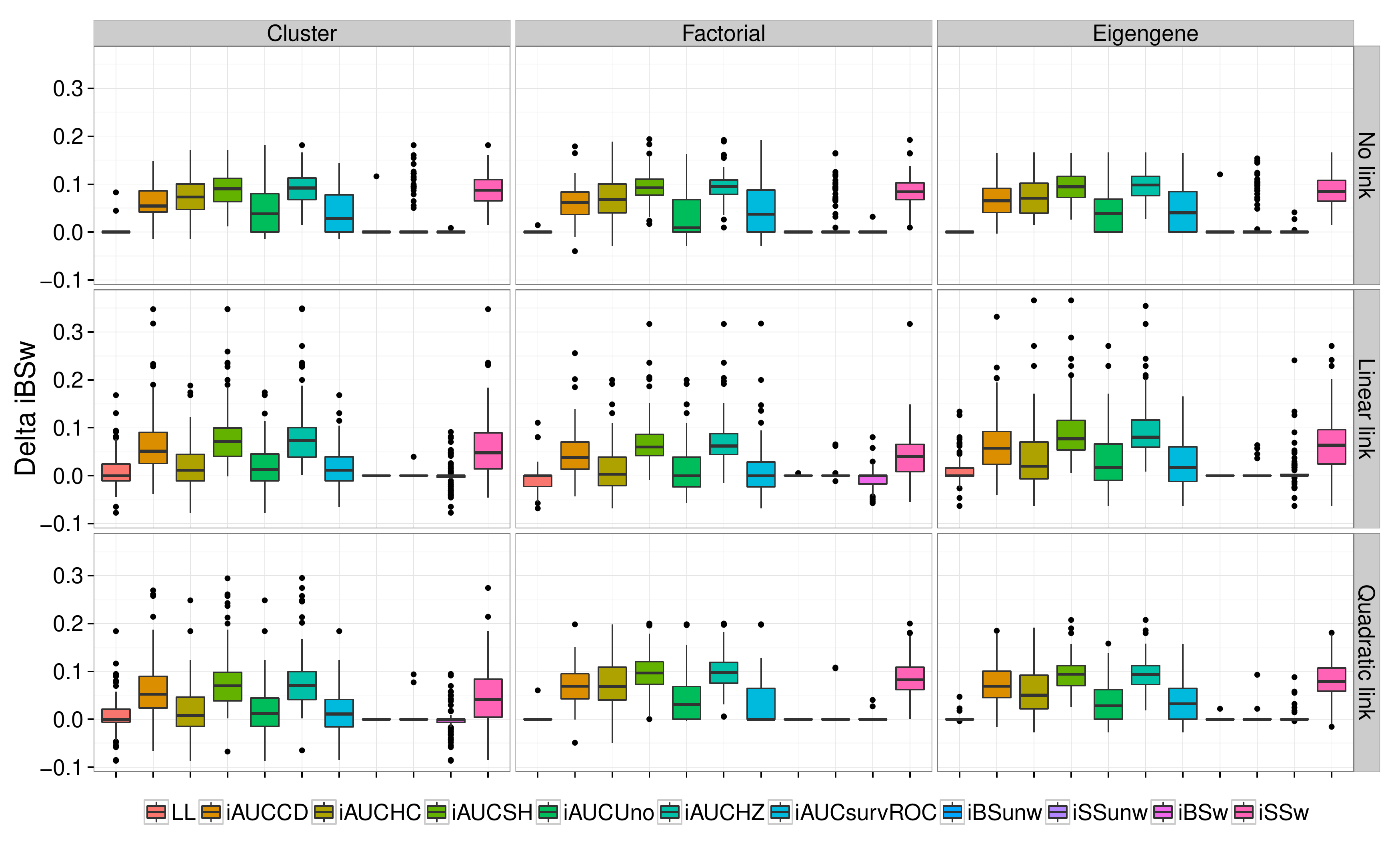}\phantomcaption\label{inciBSWcoxpls}}\qquad{\includegraphics[width=.75\columnwidth]{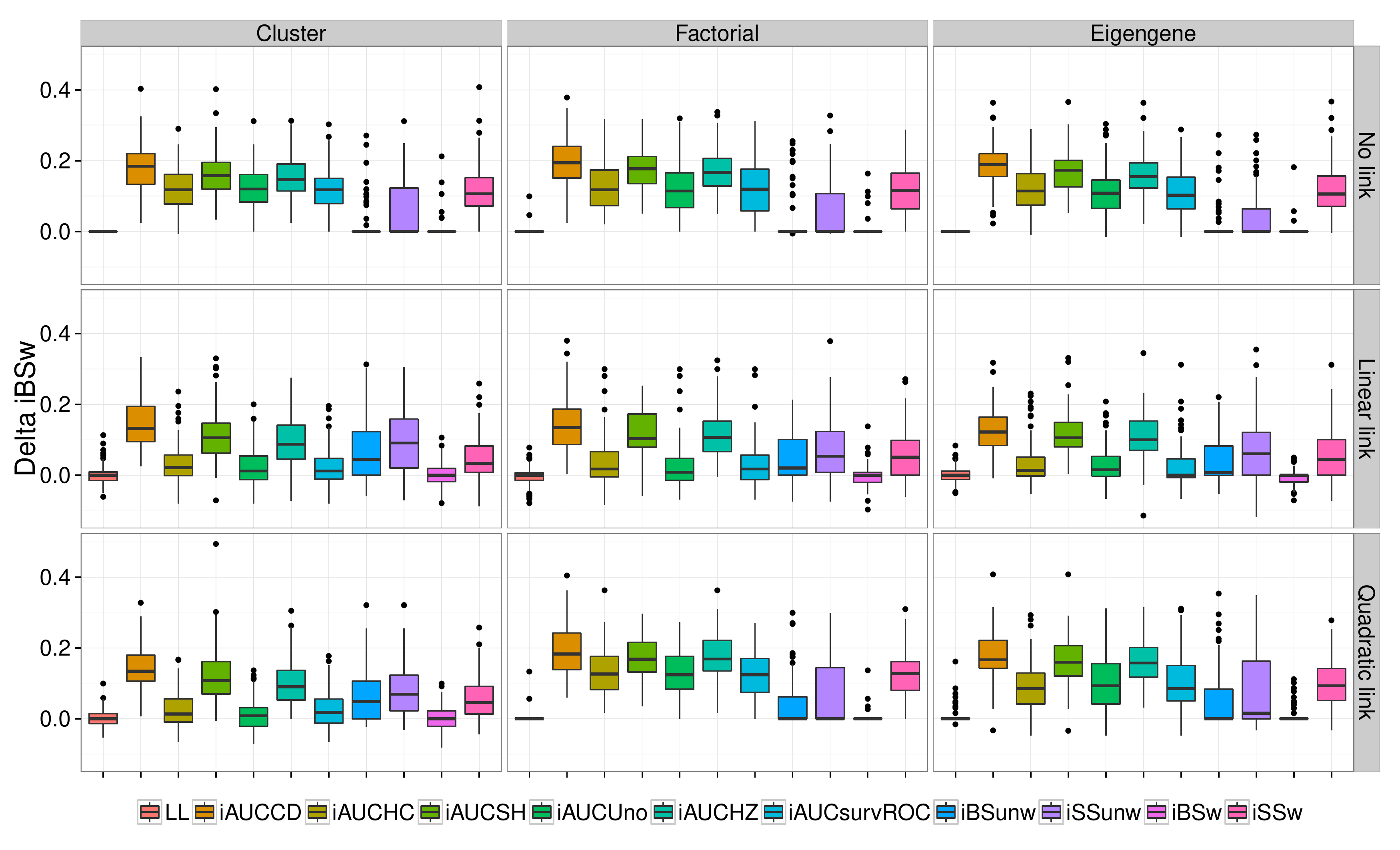}\phantomcaption\label{inciBSWcoxsplsDR}}}
\vspace{-.5cm}
\caption*{\hspace{-2.5cm}\mbox{Delta of iBSW (CV criteria $-$ vHCVLL value).\quad Figure~\ref{inciBSWcoxpls} (left): Cox$-$PLS. Figure~\ref{inciBSWcoxsplsDR} (right):  sPLSDR.}}
\end{figure}


\begin{figure}[!tpb]
\centerline{\includegraphics[width=.75\columnwidth]{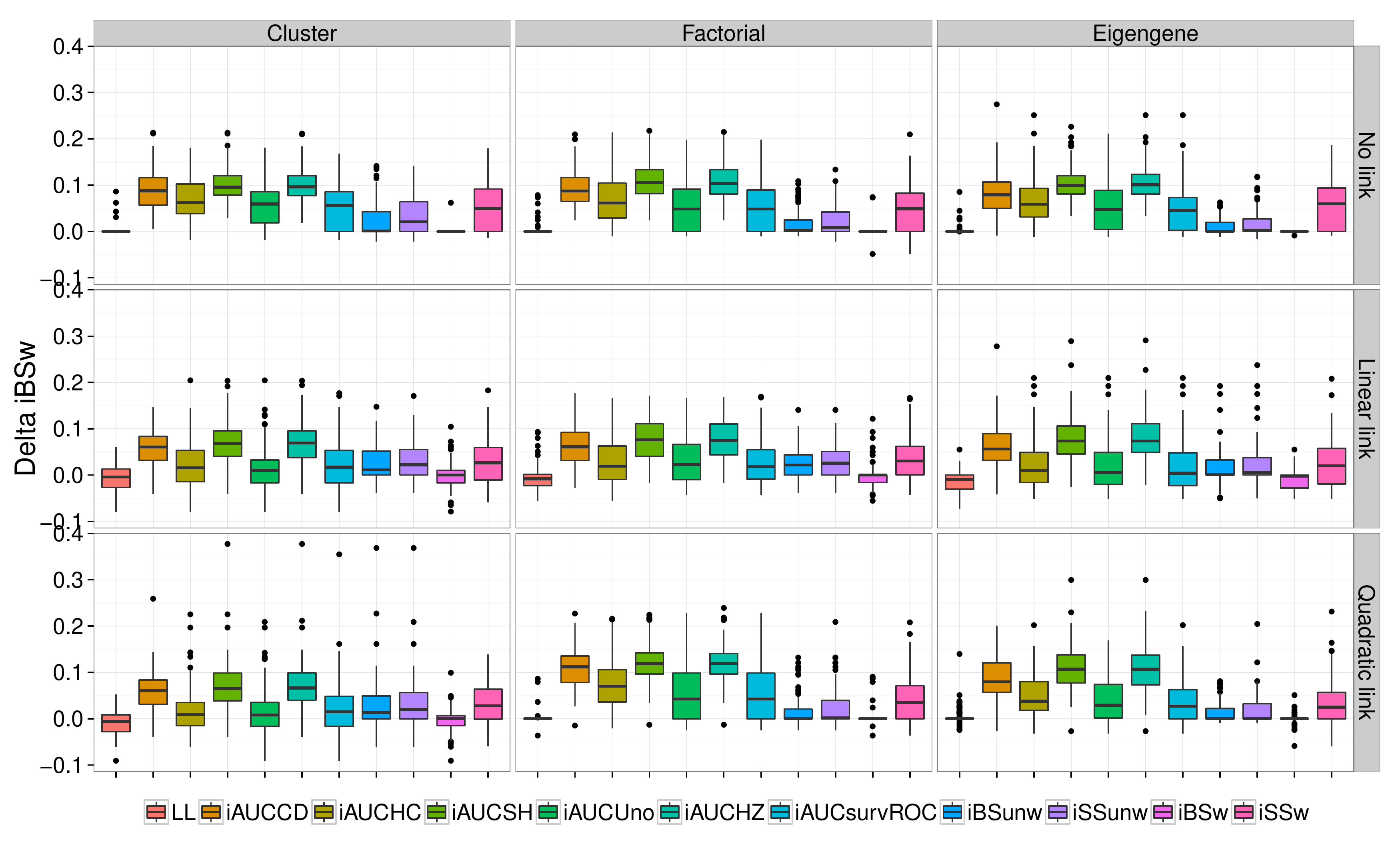}}
\vspace{-.5cm}
\caption{Delta of iBSW (CV criteria $-$ vHCVLL value). DKsPLSDR.}\label{inciBSWcoxDKplsDR}
\end{figure}



\clearpage

\begin{figure}[!tpb]
\centerline{{\includegraphics[width=.75\columnwidth]{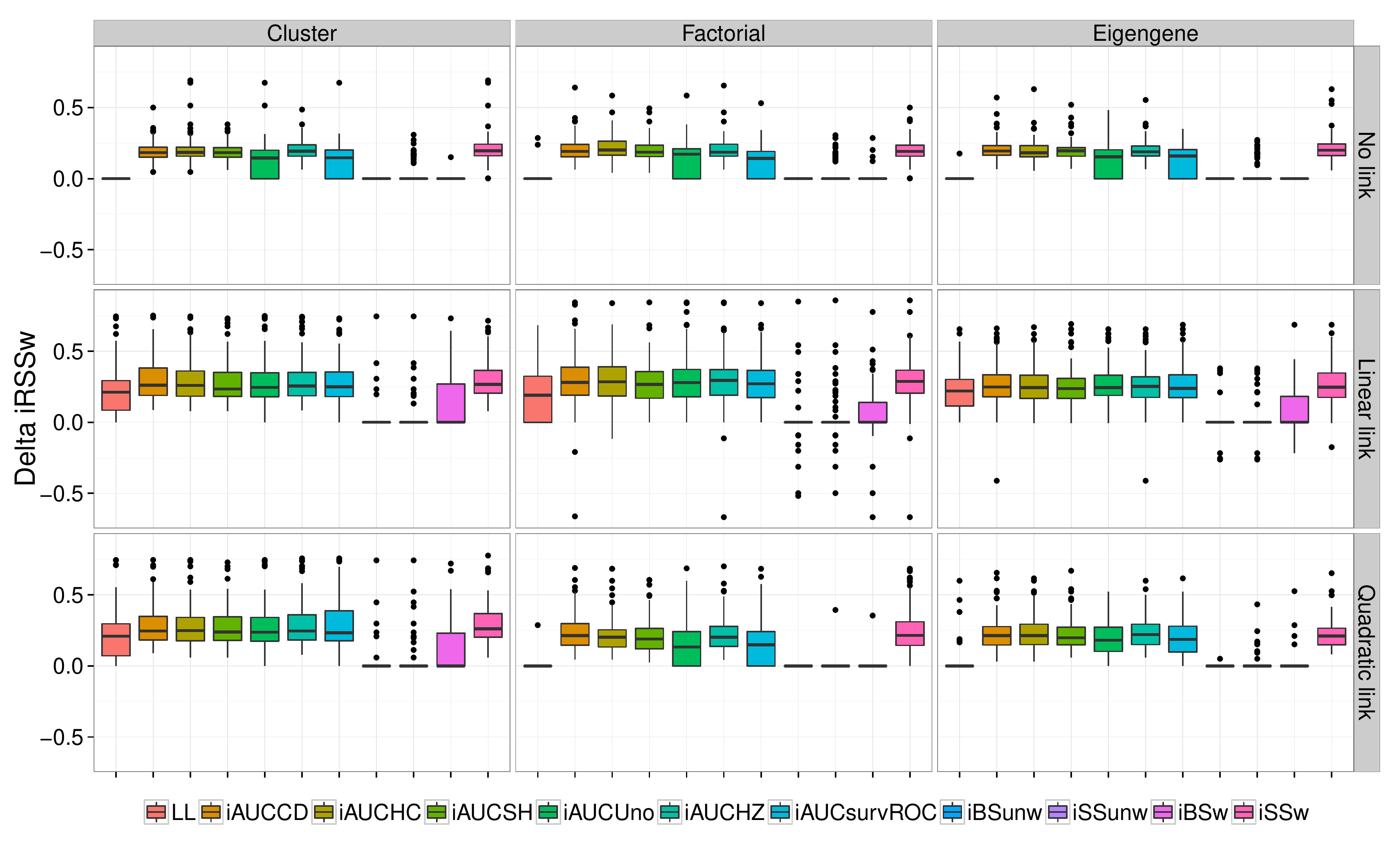}\phantomcaption\label{inciRSSWplsRcox}}\qquad{\includegraphics[width=.75\columnwidth]{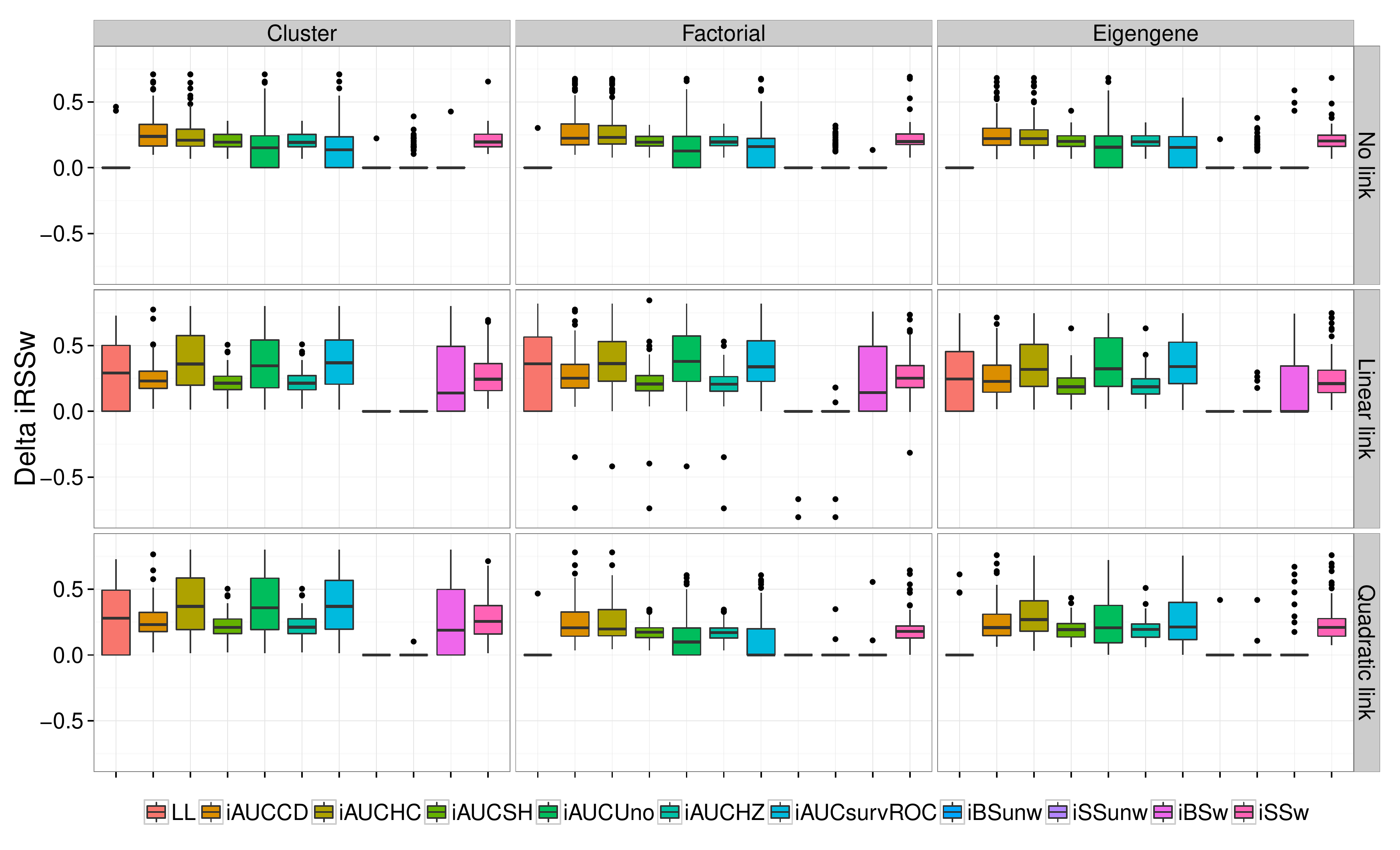}\phantomcaption\label{inciRSSWcoxpls}}}
\vspace{-.5cm}
\caption*{\hspace{-2.5cm}\mbox{Delta of iRSSW (CV criteria $-$ vHCVLL value).\quad Figure~\ref{inciRSSWplsRcox} (left): PLS$-$Cox. Figure~\ref{inciRSSWcoxpls} (right):  Cox$-$PLS.}}
\end{figure}

\begin{figure}[!tpb]
\centerline{{\includegraphics[width=.75\columnwidth]{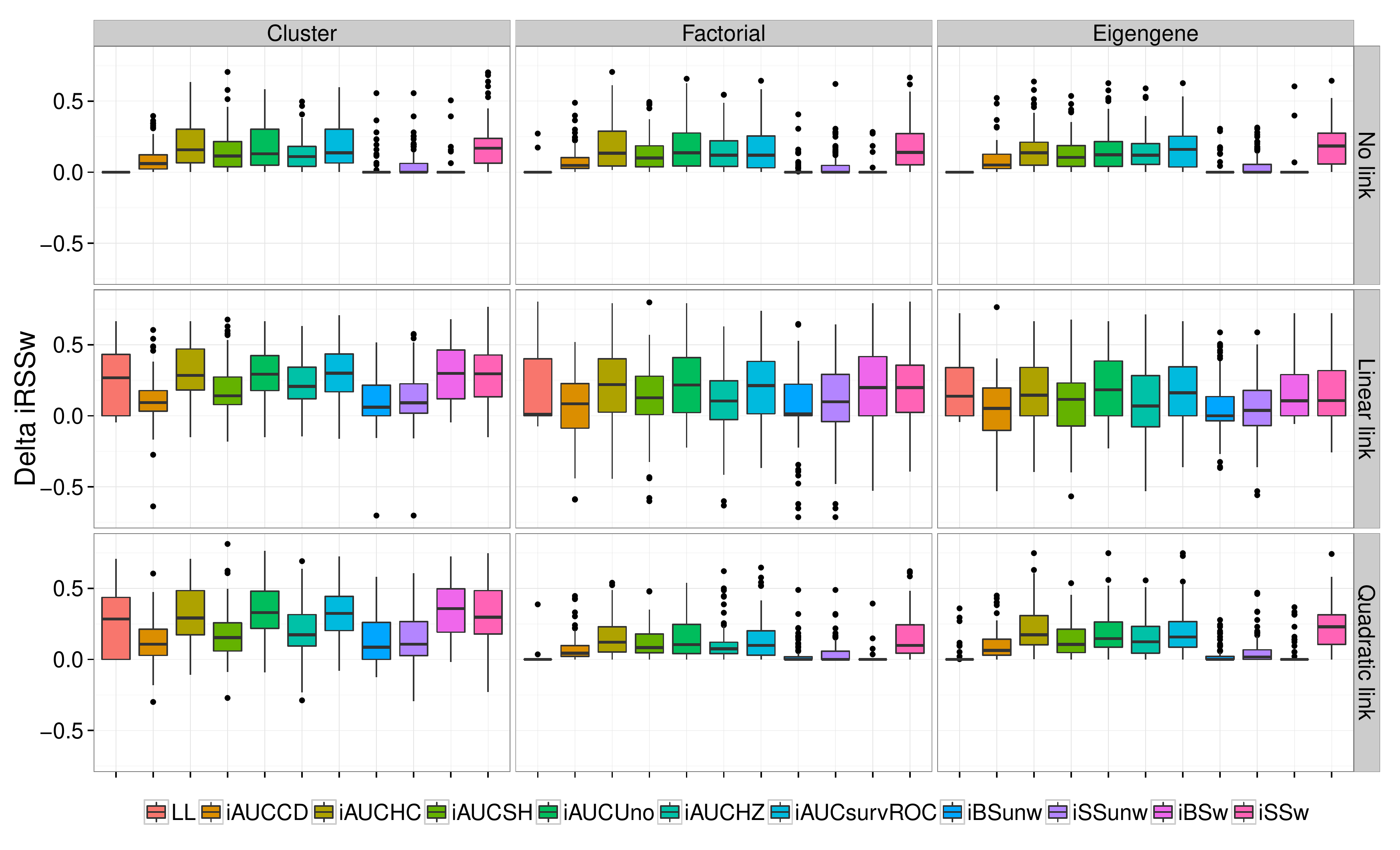}\phantomcaption\label{inciRSSWcoxsplsDR}}\qquad{\includegraphics[width=.75\columnwidth]{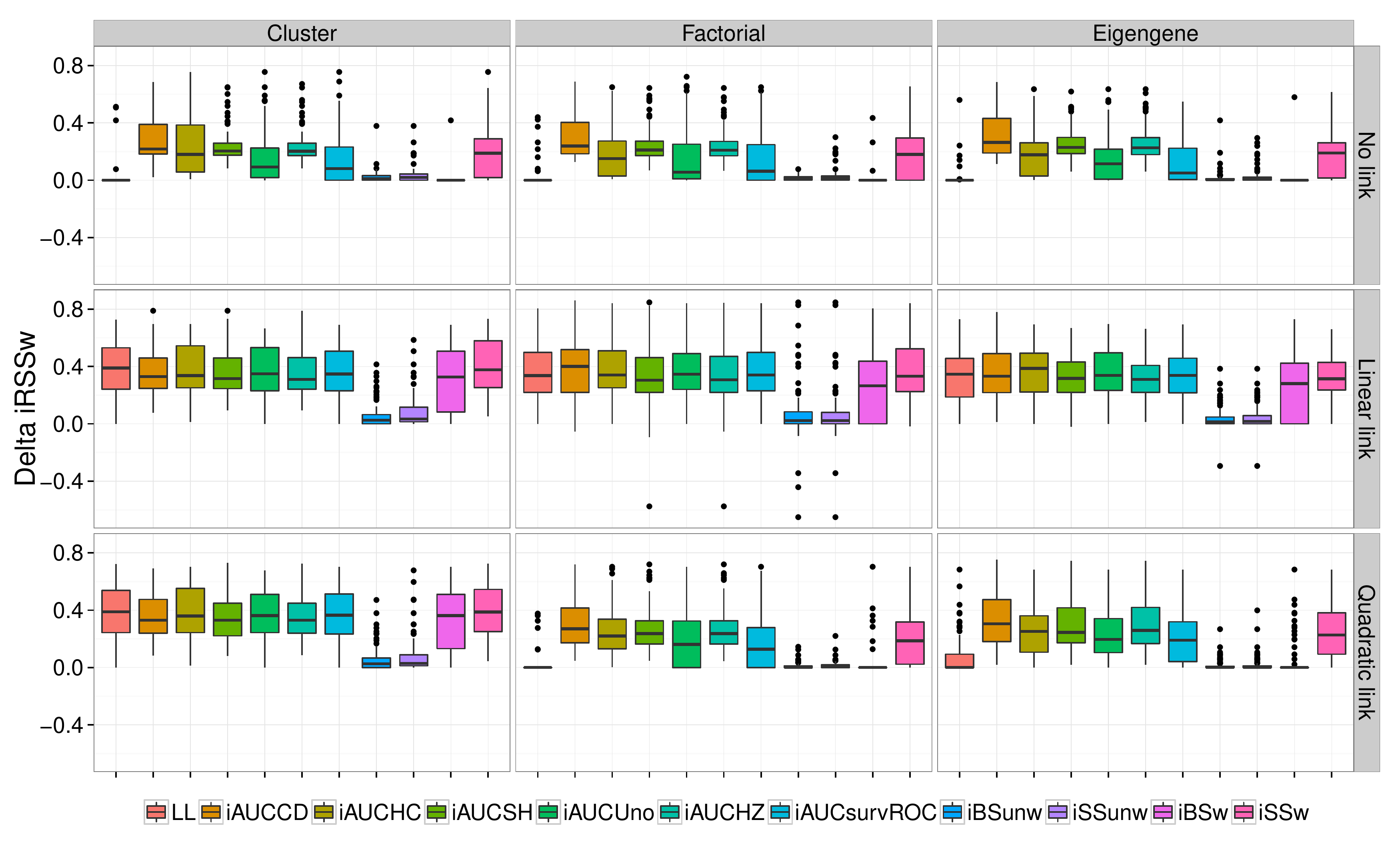}\phantomcaption\label{inciRSSWcoxDKplsDR}}}
\vspace{-.5cm}
\caption*{\hspace{-2.5cm}\mbox{Delta of iRSSW (CV criteria $-$ vHCVLL value).\quad Figure~\ref{inciRSSWcoxsplsDR} (left): sPLSDR. Figure~\ref{inciRSSWcoxDKplsDR} (right):  DKPLSDR.}}
\end{figure}



\begin{figure}[!tpb]
\centerline{\includegraphics[width=.75\columnwidth]{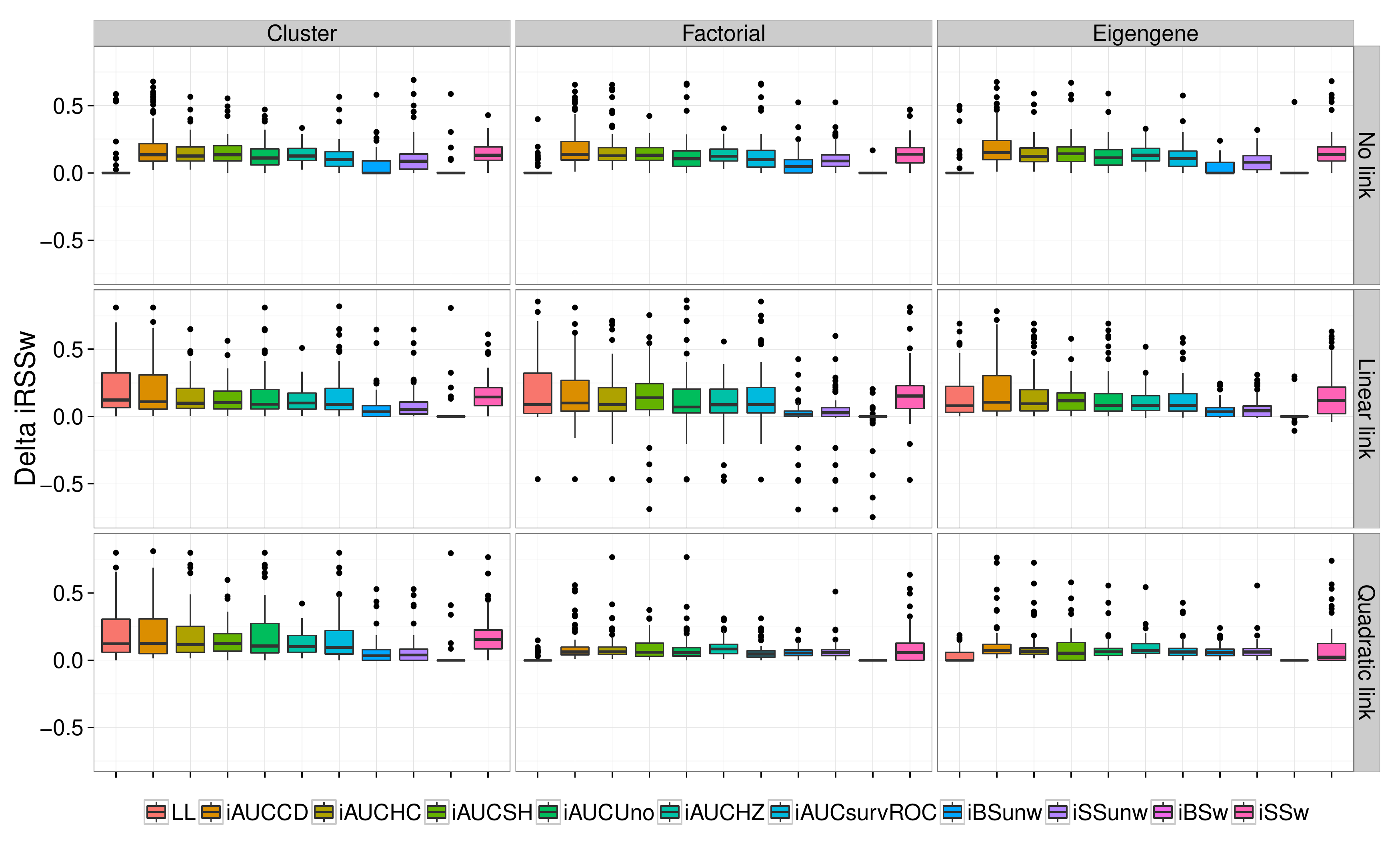}}
\vspace{-.5cm}
\caption{Delta of iRSSW (CV criteria $-$ vHCVLL value). DKsPLSDR.}\label{inciRSSWcoxDKsplsDR}
\end{figure}

\clearpage

\begin{figure}[!tpb]
\centerline{{\includegraphics[width=.75\columnwidth]{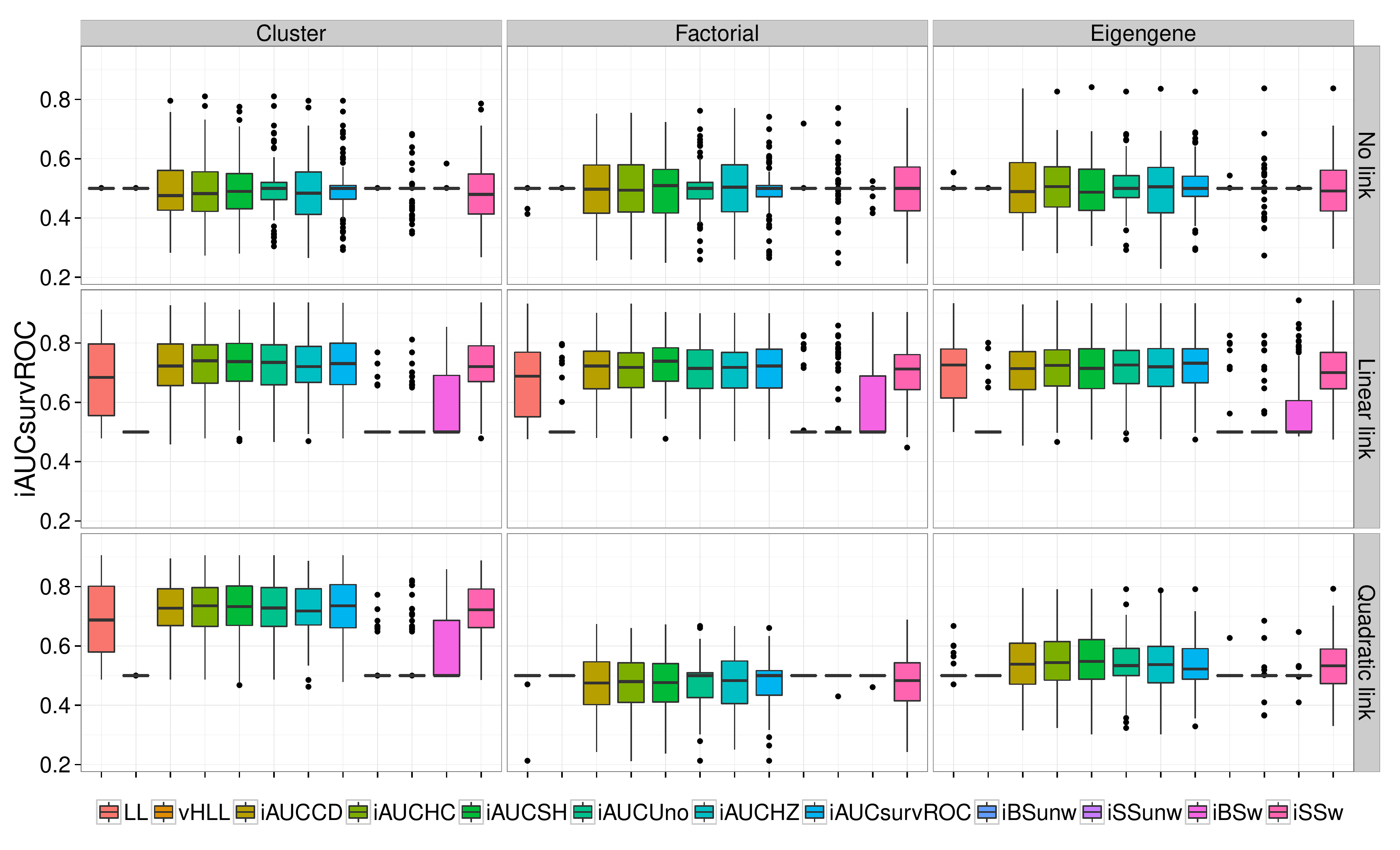}\phantomcaption\label{incSurvRocplsRcox}}\qquad{\includegraphics[width=.75\columnwidth]{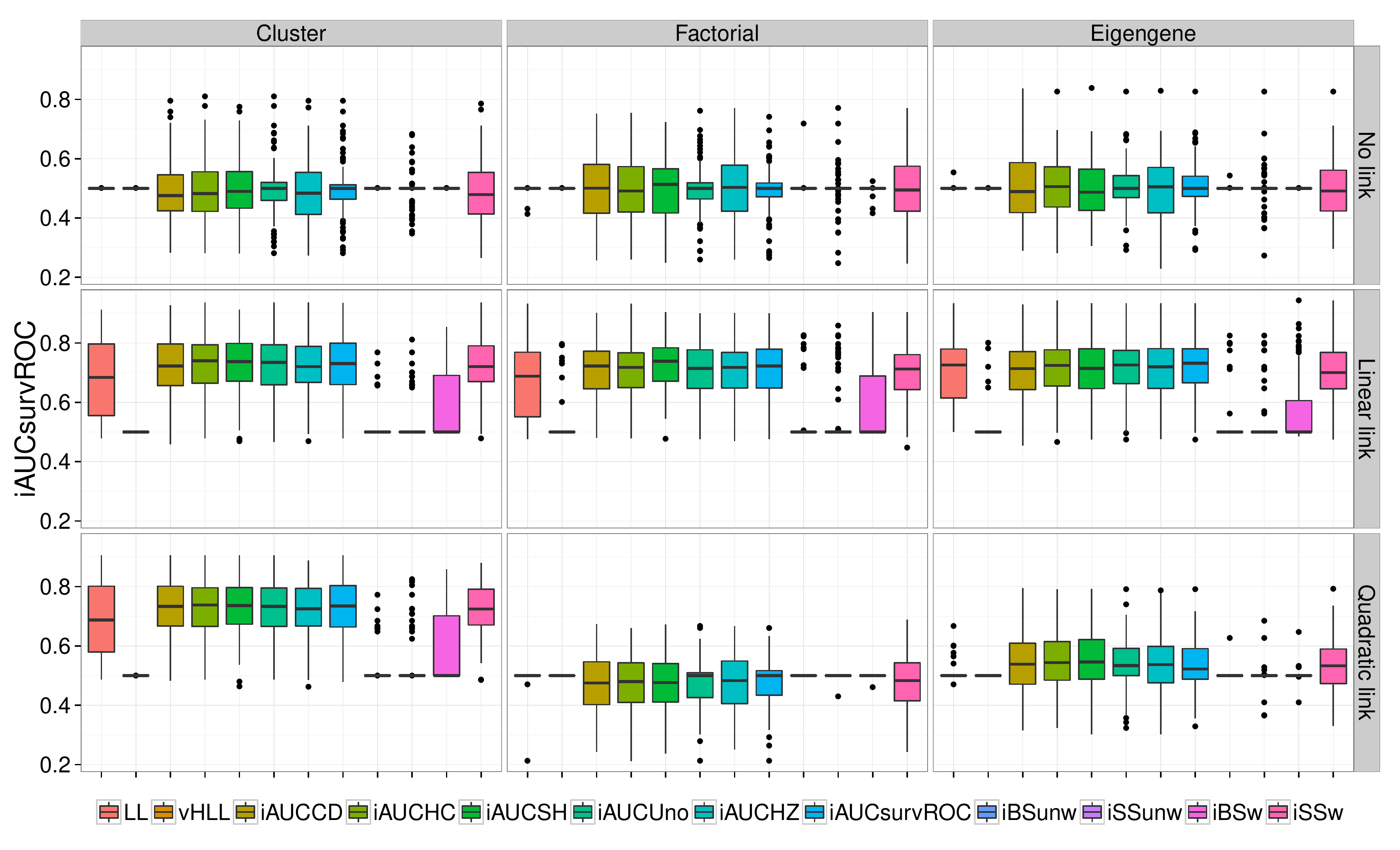}\phantomcaption\label{incSurvRocautoplsRcox}}}
\vspace{-.5cm}
\caption*{\hspace{-2.5cm}\mbox{Delta of SurvROC (CV criteria $-$ vHCVLL value).\quad Figure~\ref{incSurvRocplsRcox} (left):  PLS$-$Cox. Figure~\ref{incSurvRocautoplsRcox} (right):  autoPLS$-$Cox.}}
\end{figure}

\begin{figure}[!tpb]
\centerline{{\includegraphics[width=.75\columnwidth]{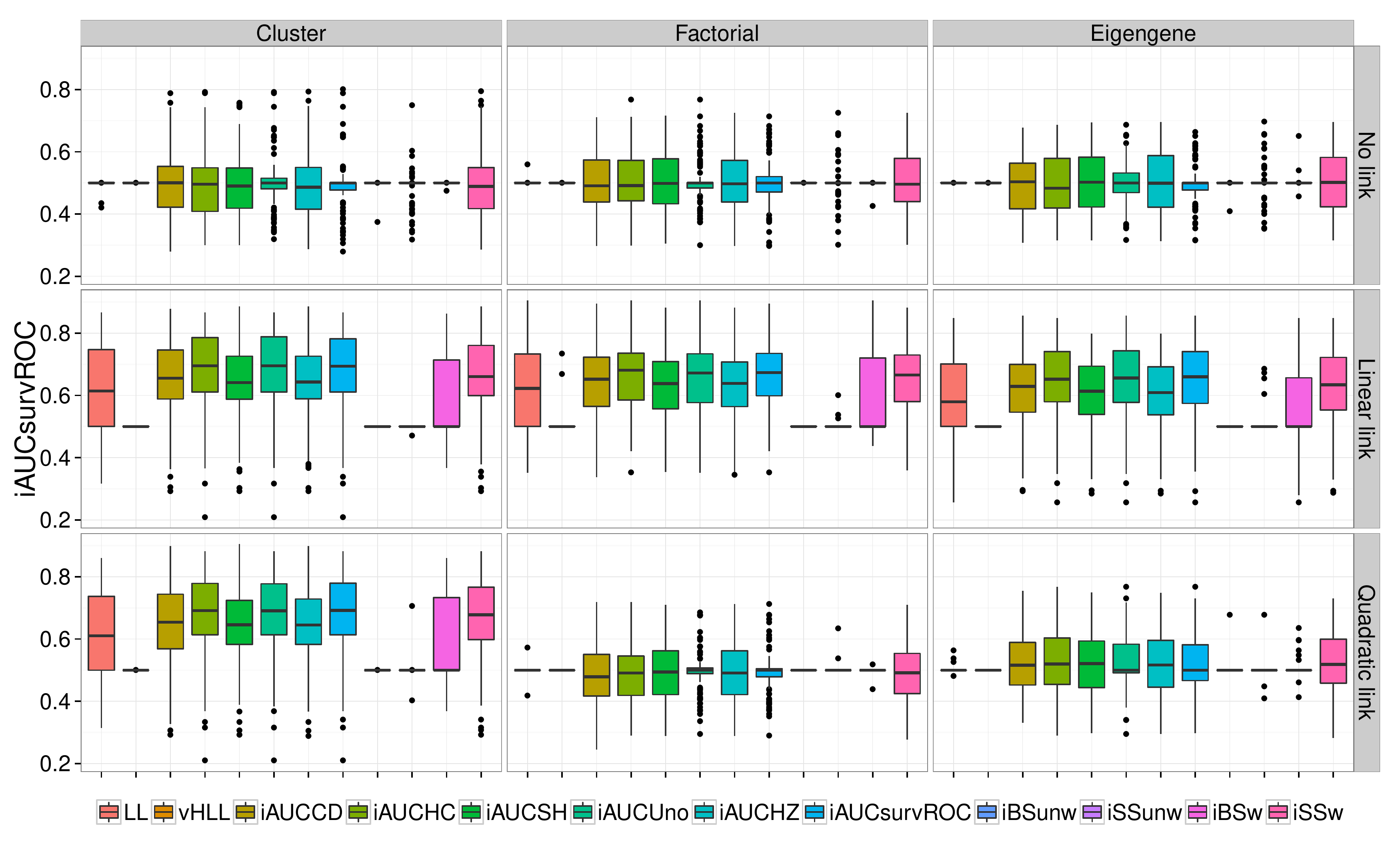}\phantomcaption\label{incSurvRoccoxpls}}\qquad{\includegraphics[width=.75\columnwidth]{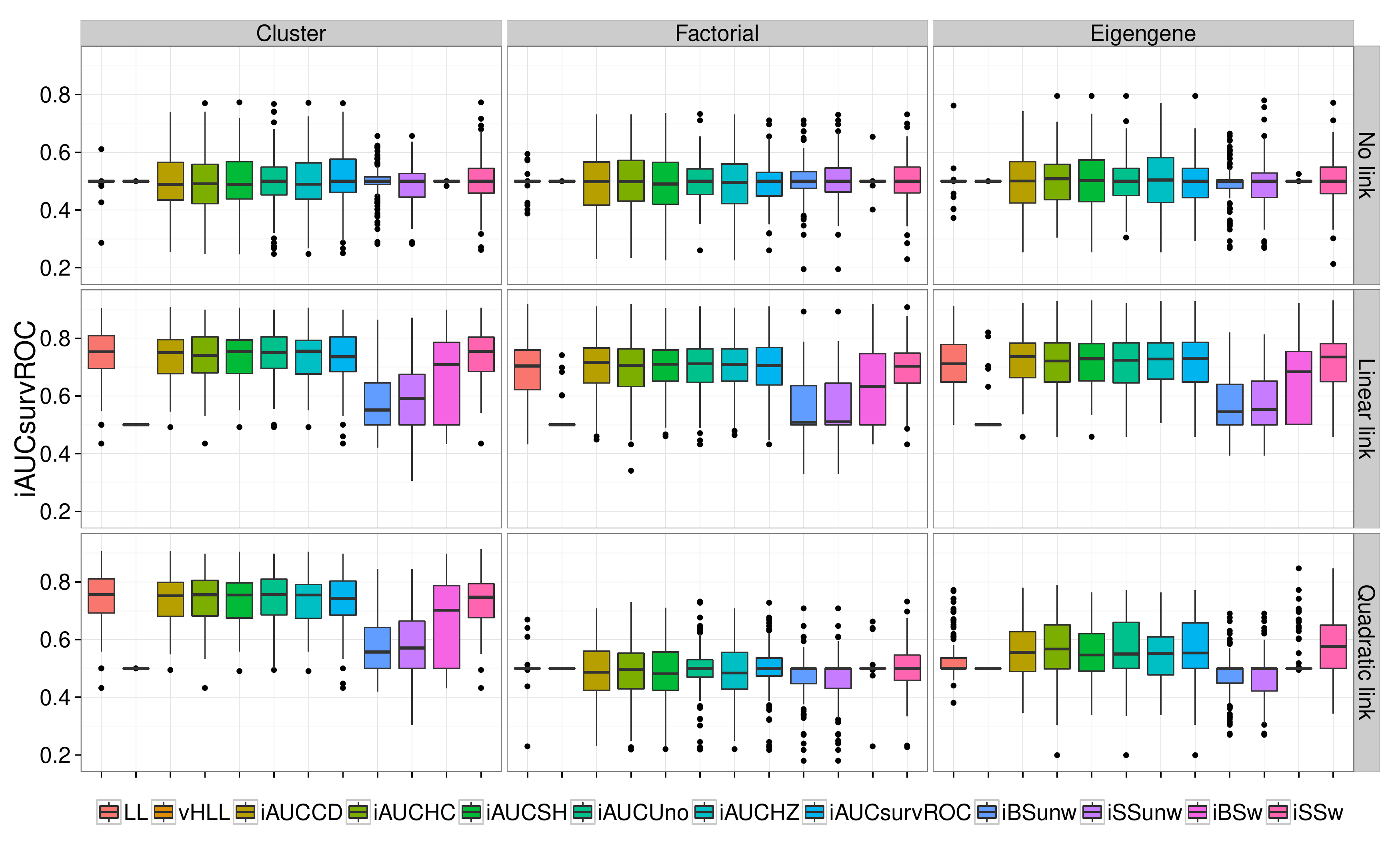}\phantomcaption\label{incSurvRocDKplsDR}}}
\vspace{-.5cm}
\caption*{\hspace{-2.5cm}\mbox{Delta of SurvROC (CV criteria $-$ vHCVLL value).\quad Figure~\ref{incSurvRoccoxpls} (left):  CoxPLS. Figure~\ref{incSurvRocDKplsDR} (right): DKPLSDR.}}
\end{figure}



\begin{figure}[!tpb]
\centerline{\includegraphics[width=.75\columnwidth]{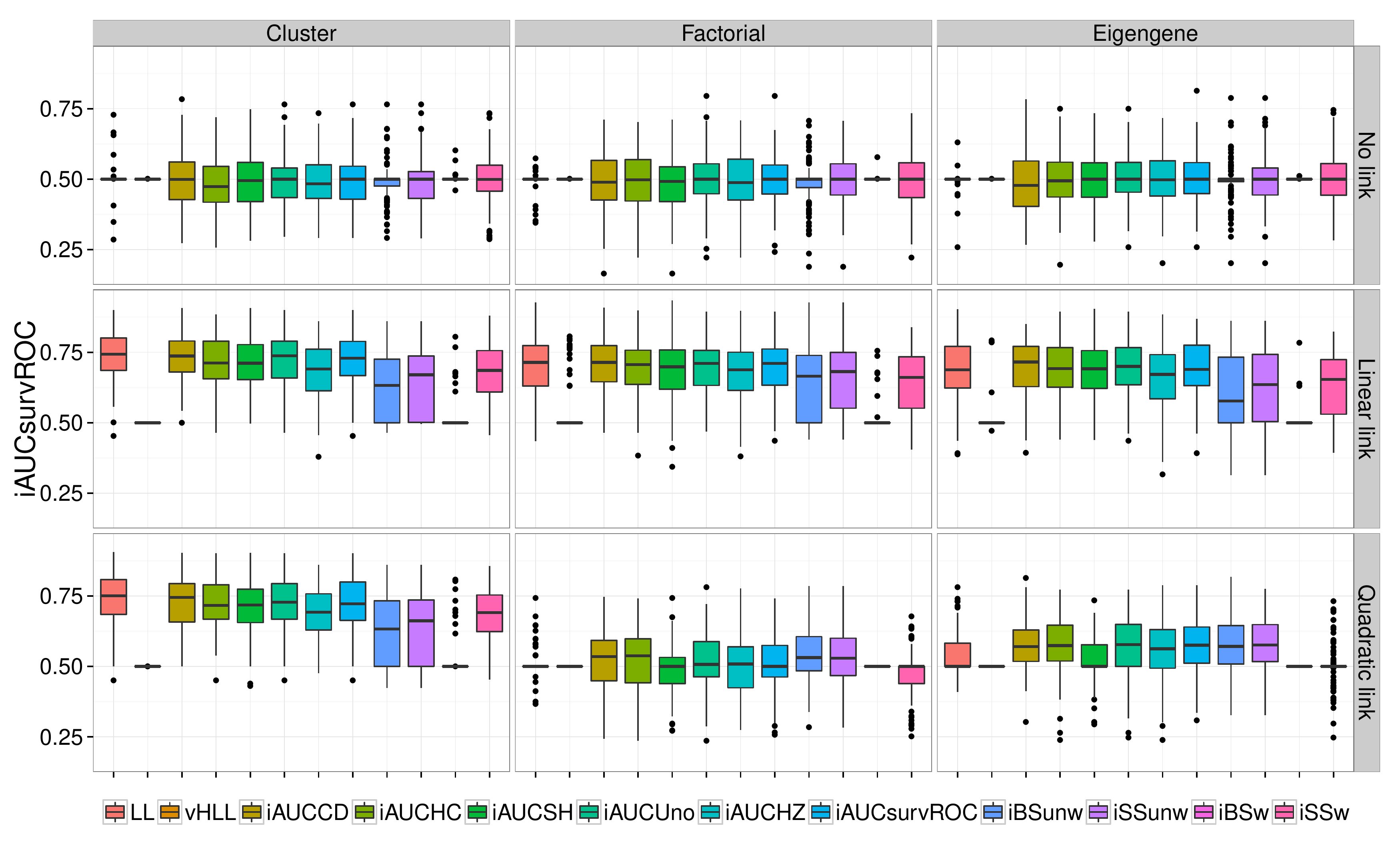}}
\vspace{-.5cm}
\caption{Delta of SurvROC (CV criteria $-$ vHCVLL value), DKsPLSDR.}\label{incSurvRocDKsplsDR}
\end{figure}

\end{document}